\documentclass[twoside,openright,11pt]{book}
\usepackage[utf8]{inputenc}
\usepackage[b5paper,layout=b5paper,left=3cm,right=2cm,bottom=2.5cm,top=3cm,twoside]{geometry}
\pdfoutput=1
\usepackage[spanish,USenglish]{babel}
\decimalpoint
\usepackage{moresize}
\usepackage{graphicx}
\usepackage{float}
\usepackage{booktabs} 
\usepackage{lscape}
\usepackage{dpfloat}
\usepackage{rotating}
\usepackage{amsmath}
\usepackage{amssymb}
\usepackage{upgreek}
\usepackage{mathabx} 
\usepackage{wasysym} 
\usepackage{multicol}
\usepackage{soul} 
\usepackage{enumerate} 
\usepackage{mdwlist} 
\usepackage[font=small]{caption}
\usepackage{url}
\usepackage{hyperref}
\usepackage{bookmark}
\usepackage[usenames,dvipsnames]{color}
\usepackage[cmyk]{xcolor}
\usepackage{anyfontsize} 
\usepackage{xspace} 
\usepackage[colorinlistoftodos, textwidth=2cm]{todonotes} 
\usepackage{tikz} 
\usepackage{eso-pic}
\usepackage[explicit,raggedright]{titlesec}
\usepackage{setspace}
\usepackage[acronym]{glossaries}
\usepackage{caption}
\usepackage{fancyhdr}
\usepackage{natbib}
\bibpunct{(}{)}{;}{a}{}{,}
\usepackage{pdfpages}
\usepackage{chemformula}

\pdfmapfile{+kurier.map}
\usepackage[T1]{fontenc}
\usepackage{calligra}
\usepackage[math]{kurier}
\usepackage[varg]{txfonts} 

\DeclareCaptionFormat{captionformat}{{\bf #1}#2#3}
\titleformat{\chapter}[block]
	{\setstretch{1.0}\usefont{T1}{kurier}{b}{n}\selectfont\huge\bfseries}
	{\parbox{2cm}{\includegraphics[width=2.2cm]{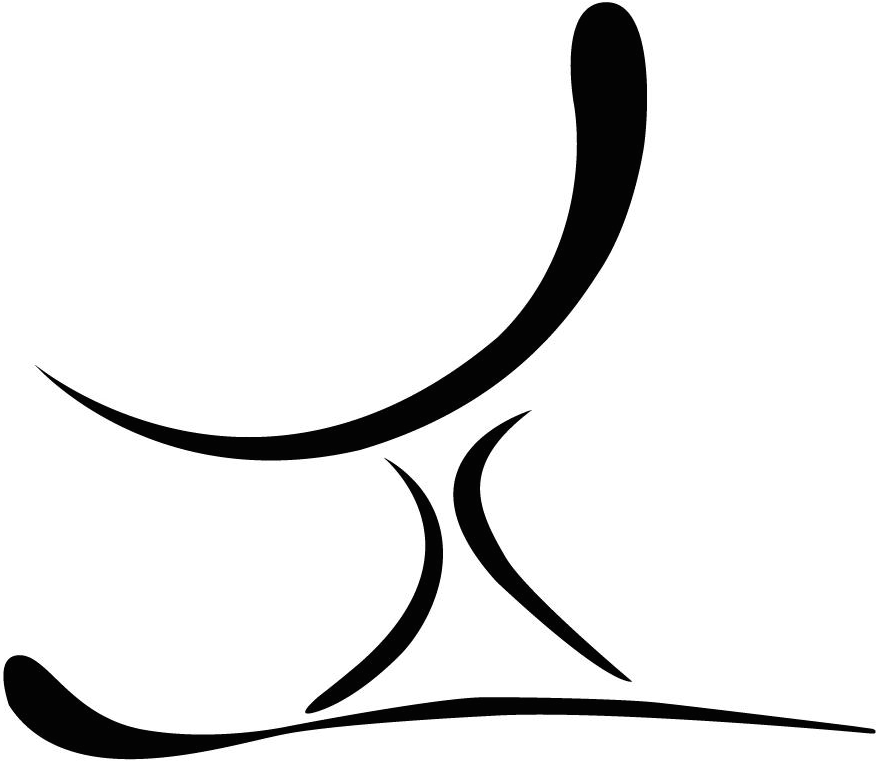}}}
	{0pt}
	{
		\begin{minipage}{0.8\textwidth}
			{\sc\Large\chaptername\ {\LARGE\thechapter}}\\[-5pt]\filright#1
	}[\end{minipage}]

\titleformat{\section}{\usefont{T1}{kurier}{b}{n}\selectfont\Large}{\thesection.}{10pt}{#1}[\hrule]

\titleformat{\subsection}{\usefont{T1}{kurier}{b}{n}\selectfont\large}{\thesubsection.}{8pt}{#1}
\titleformat{\subsubsection}{\usefont{T1}{kurier}{b}{n}\selectfont\large}{}{}{#1}

\renewcommand{\sectionmark}[1]{\markright{\color{gray}\thesection.\ #1}{}}
\renewcommand{\chaptermark}[1]{\markboth{\color{gray}\thechapter.\ #1}{}}

\newcommand{\fullpagecover}[6]{
    \clearpage
    \thispagestyle{empty}
    \AddToShipoutPicture*{\put(0,0){\includegraphics[#6]{#3}}}
    \begin{tikzpicture}[remember picture,overlay]
        \node[text width=0.5\textwidth,fill=white!30,fill opacity=#5,text opacity=1,inner sep=10pt,rounded corners=0pt, align=justify,#2] at (#1) {\small #4};
    \end{tikzpicture}
\newpage
}

\renewcommand{\labelitemi}{$\circ$}

\newcommand{\aap}{A\&A}
\newcommand{\araa}{ARAA}
\newcommand{\mnras}{MNRAS}
\newcommand{\apjl}{ApJL}
\newcommand{\apjs}{ApJS}
\newcommand{\apj}{ApJ}
\newcommand{\aj}{ApJ}

\newcommand{\nat}{Nature}
\newcommand{\aapr}{A\&AR}
\newcommand{\pasp}{PASP}
\newcommand{\aaps}{A\&AS}
\newcommand{\apss}{Ap\&SS}
\newcommand{\pasa}{Publ.\ Astron.\ Soc.\ Australia}
\newcommand{\na}{New Astron.}
\newcommand{\physrep}{Phys.\ Rep.}

\renewcommand{\vec}[1]{\mathbf{#1}}

\renewcommand{\bibname}{References}
\renewcommand{\refname}{References}
\def\g{$\upgamma$\xspace}
\def\D{\mathrm{d}}
\def\arcsec{\mathrm{arcsec}\xspace} 
\def\mjy{{\rm mJy}\xspace}
\def\ujy{{\rm \upmu Jy}\xspace}

\def\mjybeam{\mathrm{mJy\ beam^{-1}}\xspace}
\def\ujybeam{\mathrm{\upmu Jy\ beam^{-1}}\xspace}
\def\tsys{$T_{\rm sys}$\xspace}
\def\sun{\tiny\astrosun}

\newcommand{\new}[1]{{\color{red}#1}}

\title{Non-thermal emission from high-energy binaries through interferometric radio observations}
\author{Benito Marcote Mart\'in}
\date{June 2015}

\makeatletter
\hypersetup{
	pdfborder={0 0 0 0},
	pdftitle={\@title},
	pdfauthor={\@author},
	pdfcreator={\@author},
	pdfsubject={PhD thesis, astrophysics, radio, high-energy, binaries},
	colorlinks=true,
	linkcolor=blue,
	urlcolor=blue,
	citecolor=blue,
	linktoc=page
}

\def\colorexpandedcite{\color{blue}}

\begin{document}

\pagestyle{empty}  
\includepdf{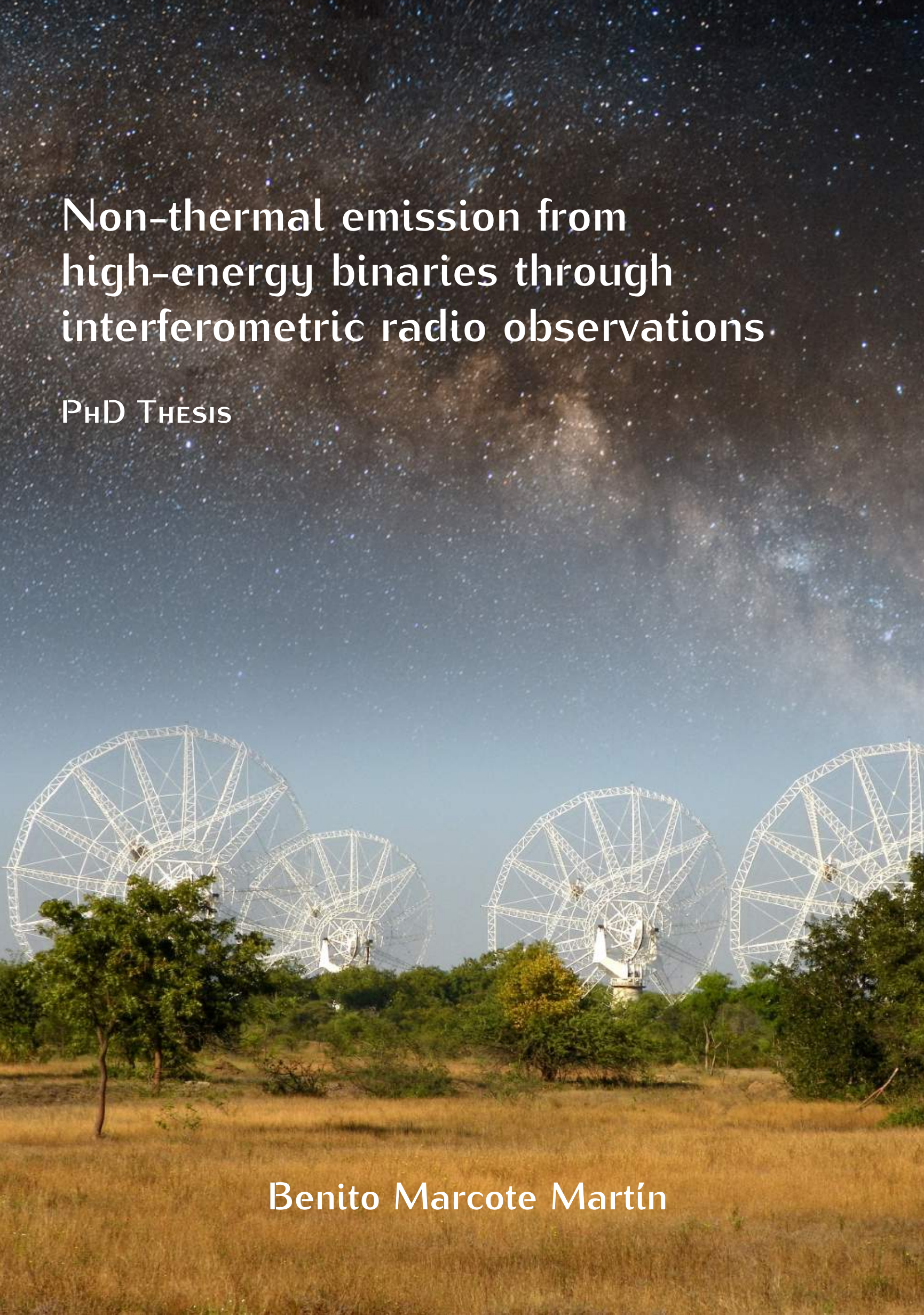}

\include{headers/portada-internal}
\begin{titlepage}
\begin{center}

\makeatletter

\Large

{\sc Programa de doctorat en F\'isica}\\[.3cm]
{\sc L\'inia de Recerca en Astronomia i Astrof\'isica}\\[.3cm]
{\sc 2011--2015}

\vfill

{Mem\`oria presentada per {\bf \@author}\\ per optar al grau de Doctor en F\'isica}

\vfill

{\sc Directors}\\[+.25cm]
Dr. Marc Rib\'o Gomis\\[+.1cm]
Dr. Josep Maria Paredes Poy\\[+1cm]


\end{center}
\end{titlepage}

\newpage

\ 
\vfill
\noindent\includegraphics[scale=0.5]{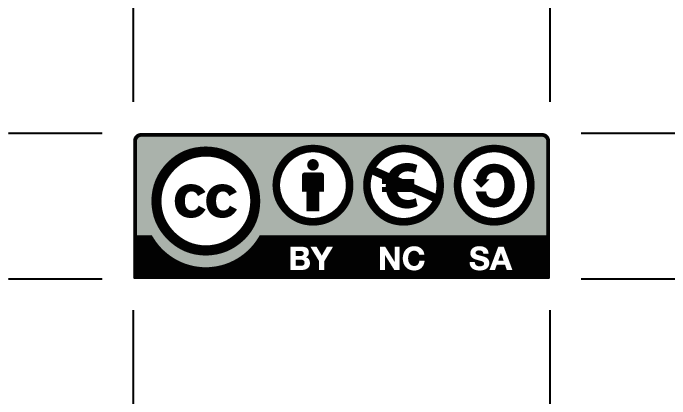}\\
2015. Benito Marcote Mart\'in.\\
{\em Cover illustration:} composition of an author's photograph of the Giant Metrewave Radio Telescope (GMRT) and a photograph of the sky taken by Rogelio Bernal Andreo under a Creative Commons {\em by-nc-nd} license.\\
{\em Bookmark illustration:} same as cover illustration plus an image of a bison of the cave of Altamira, Cantabria (Spain) under Public Domain (CC-0 1.0).\\
{\em Chapter header:} Sketch of an antenna from part of the logo of the Centro de Radioastronomía y Astrofísica (CRyA) from the Universidad Nacional Autónoma de México (UNAM).\\
All the figures and images shown in this thesis have been made by the author unless otherwise specified.

%
%
%
%


\cleardoublepage
\thispagestyle{empty}

\ \\[+0.3\textheight]
{\raggedleft \Huge\em\fontfamily{calligra}\selectfont\em A mis padres\\}

\cleardoublepage
\ \\[+0.3\textheight]
{\raggedleft {\em ``Stay hungry, stay foolish''}\\[+4pt]  {\bf Steve Jobs\\[+3cm]}
{\em ``La verdadera Ciencia ense\~na,\\ por encima de todo,\\ a dudar y a ser ignorante''}\\[+4pt]
{\bf Miguel de Unamuno}\\}
\vfill

\cleardoublepage



\frontmatter
\pagestyle{fancy}
\titleformat{\chapter}[block]{\setstretch{1.0}\usefont{T1}{kurier}{b}{n}\selectfont\huge\bfseries}{}{}{\noindent #1}
\renewcommand{\sectionmark}[1]{\markright{\color{gray}#1}{}}
\renewcommand{\chaptermark}[1]{\markboth{\color{gray}#1}{}}

\phantomsection
\hypertarget{tableofcontents}{}
\bookmark[level=chapter,dest=tableofcontents]{Contents}
\tableofcontents

%
%
%
%


\chapter{Resumen de la tesis}

\begin{otherlanguage*}{spanish}

\section*{Introducción}

En un cielo nocturno se pueden llegar a distinguir unas mil estrellas a simple vista, aunque el número total de estrellas presentes en nuestra galaxia realmente asciende a unos cien mil millones. Las características de estas estrellas pueden ser muy diferentes de unas a otras, en particular, presentando una masa que puede ir desde $\sim$0.1 hasta $\sim$150 veces la masa del Sol, lo que produce a su vez un gran rango de tamaños, luminosidades o temperaturas.

Un gran porcentaje de las estrellas presentes en el Universo forman sistemas binarios, donde hay dos estrellas que están orbitando entre sí. Las órbitas que describen estas estrellas pueden presentar enormes diferencias de un sistema a otro. Desde órbitas muy compactas, donde ambas estrellas están muy próximas entre sí, hasta órbitas muy abiertas, donde las estrellas se encuentran muy alejadas. A su vez, estas órbitas pueden ser desde casi circulares, donde la distancia entre las dos estrellas se mantiene prácticamente constante a lo largo del tiempo, hasta ser altamente elípticas, donde la distancia de mayor acercamiento entre estrellas, el {\em periastro}, es notablemente más pequeña que la distancia de mayor alejamiento, el {\em apastro}. Todo esto causa que estos sistemas binarios, o simplemente binarias, puedan mostrar un comportamiento muy dispar.

En esta tesis nos vamos a centrar exclusivamente en las binarias que producen emisión a muy altas energías, es decir, que producen emisión de rayos X o rayos gamma (\g). Para que esta emisión se pueda producir se requiere un entorno suficientemente violento y energético, que permita la aceleración de partículas (electrones y protones) hasta velocidades relativistas, próximas a la velocidad de la luz. Estas partículas son las responsables, a través de diversos mecanismos, de producir la radiación (luz) de alta energía, como son los rayos X y rayos \g. A su vez, también producen emisión en el otro extremo del espectro, en las ondas de radio.

A pesar que las ondas de radio son la radiación menos energética posible, en estos sistemas está conectada con la emisión de alta energía debido a que está originada por la misma población de partículas aceleradas.
Estudiar esta emisión radio nos permite estudiar los fenómenos de emisión y absorción que se producen en estos sistemas, o resolver la morfología de la región emisora si observamos con interferómetros de muy larga base, es decir, utilizando conjuntamente antenas que están muy distantes entre sí, lo que nos permite alcanzar una gran resolución. 
Por todo lo anterior, esta tesis está centrada en el estudio a radio frecuencias de estas binarias de alta energía.

\subsection*{Binarias con emisión en rayos \g}

Las binarias que emiten en rayos \g y que se han descubierto hasta la fecha se pueden clasificar en varios tipos, pero todas ellas constan, al menos, de una estrella muy masiva (más de 10 veces la masa del Sol). El otro componente del sistema binario puede ser otra estrella masiva o un objeto compacto. Este último puede ser tanto una estrella de neutrones como un agujero negro.

En el caso de tener dos estrellas muy masivas con una órbita relativamente compacta, los vientos que producen estas dos estrellas interaccionan entre sí fuertemente, chocando entre sí. Este choque entre ambos vientos es un potente y eficiente acelerador de partículas, pudiendo generar una emisión intensa en todo el espectro electromagnético, desde las ondas de radio hasta los rayos \g. A estos sistemas se los suele denominar {\em binarias con colisión de vientos} (o {\em colliding wind binaries}). La región de choque entre los dos vientos es fácilmente detectable en radio, donde se observa una emisión con forma de arco, curvada hacia la estrella con un viento más débil. De todas las binarias con colisión de vientos que se conocen, únicamente una se ha detectado en rayos \g: $\upeta$-Carinae. De hecho, ésta es la binaria con colisión de vientos más masiva que se conoce, compuesta por dos estrellas de 90--120 y unas 30 masas solares que orbitan entre sí con un periodo orbital de unos 5.5 años.

Sin embargo, las estrellas muy masivas tienen vidas muy cortas (del orden de unos pocos millones de años, a comparar con los 10\,000 millones de años que se estima para la vida del Sol). Cuando una de estas estrellas llega al final de su vida, estalla como {\em supernova} y se convierte en una estrella de neutrones o un agujero negro (en función de la masa que presente en el momento de producir la supernova). En cualquiera de ambos casos, estamos hablando de un objeto extremadamente compacto. 

Durante el tiempo en que la otra estrella del sistema todavía no ha llegado al final de su vida, tendremos un sistema binario formado por una estrella masiva y un objeto compacto. Este tipo de sistemas puede originar también una emisión de alta energía, radiando en rayos X y/o rayos \g. En función de cómo sea el sistema y qué tipo de interacción se produzca, se suele hablar de dos tipos diferentes de binarias: binarias de rayos X y binarias de rayos \g.

Las {\em binarias de rayos X} están compuestas por una estrella masiva y un objeto compacto, como acabamos de comentar, y exhiben una emisión dominada por los rayos X (es decir, en esta región es donde se radia mayor energía). En estos sistemas, la emisión observada puede ser explicada por la presencia de un disco de acreción en torno al objeto compacto, que radia intensamente en rayos X. También pueden producirse jets (chorros) colimados por el campo magnético, los cuales aceleran partículas y emiten fuertemente en radio. Únicamente se han detectado tres binarias de rayos X que emitan también en rayos \g: Cygnus~X-1, Cygnus~X-3 y SS~433.

Por otro lado, las {\em binarias de rayos $\gamma$} exhiben una emisión dominada por los rayos \g (de ahí su nombre). Aunque estos sistemas también están compuestos por una estrella masiva y un objeto compacto, presumiblemente una estrella de neutrones, el origen de la emisión de alta energía es completamente distinto. En estos casos se piensa que existe un choque entre el fuerte viento de la estrella masiva y el viento relativista producido por la estrella de neutrones. La interacción de estos dos vientos sería análoga a la mencionada en el caso de binarias con colisión de vientos, aunque en este caso bastante más energética.
Hasta la fecha se han descubierto cinco binarias de rayos \g: PSR~B1259--63, LS~5039, LS~I~+61~303, HESS~J0632+057 y 1FGL~J1018.6$-$5856. Únicamente en la primera, PSR~B1259--63, se ha confirmado que el objeto compacto es una estrella de neutrones. En el resto de sistemas la naturaleza del objeto compacto permanece desconocida (pudiendo ser tanto una estrella de neutrones como un agujero negro).

\section*{Resultados de la tesis}

Como se ha mencionado anteriormente, esta tesis se centra en el estudio de la emisión radio de binarias de alta energía, con el objetivo de aumentar el conocimiento que tenemos de estos objetos.
Para ello, se han explorado tres binarias de rayos \g con observaciones radio, tanto a bajas frecuencias, con el fin de obtener su espectro radio y la evolución de sus curvas de luz a lo largo de la órbita, como a muy alta resolución (con interferómetros de muy larga base), con el objetivo de analizar cómo cambia la morfología de la región emisora a lo largo de la órbita. También se presenta el descubrimiento de una nueva binaria con colisión de vientos a partir de observaciones radio de muy alta resolución y observaciones ópticas. Por último, se presentan los resultados obtenidos para dos fuentes que se postularon inicialmente como posibles binarias de rayos \g. A continuación se realiza un breve resumen de los resultados obtenidos para cada una de estas fuentes:
\begin{itemize}
    \item {\em LS~5039.} Se ha determinado que la emisión radio de esta binaria de rayos \g no está modulada orbitalmente, siendo el único caso de todas las binarias de rayos \g conocidas en el que esto ocurre. También se ha obtenido el espectro promedio de la fuente en el rango 0.15--15~GHz, observando que la emisión es persistente a lo largo del tiempo (en escalas de tiempo desde días hasta años) y que presenta un cambio de pendiente a $\sim$0.5~GHz. La presencia del efecto Razin, detectado por primera vez en una binaria de rayos \g, da un soporte adicional al escenario de un pulsar y una región con choque de vientos. También se han obtenido dos espectros con datos casi simultáneos en el rango 0.15--5~GHz, similares al promedio pero que muestran ligeras diferencias entre sí, evidenciando cambios en los procesos de absorción. Estos espectros nos han permitido estimar varias propiedades de la región emisora utilizando un modelo sencillo, como su tamaño, campo magnético, densidad electrónica o la tasa de pérdida de masa de la estrella.
    \item {\em LS~I~+61~303.} Analizando datos de radio a bajas frecuencias se han podido obtener las curvas de luz de esta binaria de rayos \g a 150, 235 y 610~MHz, observando que la emisión está modulada casi sinusoidalmente a lo largo de la órbita. El máximo de esta emisión depende tanto de la frecuencia como de la modulación superorbital que ya se conocía a frecuencias más altas. Una interpretación de los retrasos vistos en los máximos de emisión a diferentes frecuencias asumiendo una región emisora en expansión nos ha permitido realizar una estimación de su velocidad de expansión, observando que ésta es compatible con la velocidad del viento estelar, apoyando por tanto un escenario de choque de vientos, como en el caso anterior.
    \item {\em HESS~J0632+057.} Se llevó a cabo una observación radio de muy alta resolución para determinar la morfología de la región emisora durante el máximo secundario que se observa en rayos X. Sin embargo, la fuente no se detectó en estos datos, evidenciando una reducción de al menos un orden de magnitud con respecto a anteriores observaciones. La falta de datos simultáneos en rayos X no permite concluir si esta reducción se debe a que la fuente presentaba una emisión mucho menor en ese periodo orbital o a que el aumento de emisión todavía no había tenido lugar.
    \item {\em HD~93129A.} La reducción de observaciones radio con muy alta resolución, junto con observaciones ópticas detalladas, ha confirmado que esta fuente es una binaria con colisión de vientos. Se ha observado una emisión radio curvada, típica de una región donde los vientos de las dos estrellas chocan, y coincidente con la posición y dirección que se esperaría en el caso de ser producida por las dos estrellas del sistema. También se ha observado un aumento de su emisión radio durante los últimos años, compatible con las evidencias de que las dos estrellas se están aproximando al periastro, durante el cual se prevé un aumento de la emisión como consecuencia de un choque más intenso entre ambos vientos.
    \item {\em TYC~4051-1277-1.} Cruzando catálogos de estrellas masivas y fuentes radio, se obtuvo una posible coincidencia para esta estrella con una fuente radio que exhibe una emisión de origen no térmico. Sin embargo, observaciones radio más precias han determinado que la emisión radio no está realmente asociada a la estrella.
    \item {\em MWC~656.} Este sistema es la primera binaria que se conoce que está compuesta por una estrella de tipo espectral Be y un agujero negro. La búsqueda de su contrapartida radio no ha dado resultados positivos, aunque se espera una emisión ligeramente por debajo de los actuales límites superiores.
\end{itemize}

\section*{Conclusiones}

Los resultados presentados en esta tesis han permitido un mejor conocimiento de las binarias de alta energía y su emisión radio. Es la primera vez que se ha estudiado la emisión radio a bajas frecuencias de binarias de rayos \g, en donde varios procesos de absorción comienzan a ser relevantes. Los resultados obtenidos parecen apoyar la idea de que la emisión de estos objetos se produce por colisión de vientos, en lugar del escenario de acreción/eyección observado en las binarias de rayos X.
El descubrimiento de una binaria con colisión de vientos evidencia también la importancia de observaciones de radio de muy larga base, donde resolver la morfología de la fuente radio es crucial para poder entender qué fenómenos están ocurriendo en el sistema.
Por todo lo anterior, y aunque pueda resultar extraño, se puede concluir que las observaciones radio son una herramienta que puede contribuir notablemente a la comprensión de las binarias de alta energía.

\end{otherlanguage*}

\chapter{Summary}

High-mass binary systems involve extreme environments that produce non-thermal emission from radio to gamma rays. Only three types of these systems are known to emit persistent gamma-ray emission: colliding-wind binaries, high-mass X-ray binaries and gamma-ray binaries. This thesis is focused on the radio emission of high-mass binary systems through interferometric observations, and we have explored several of these sources with low- and high-frequency radio observations, and very high-resolution VLBI ones.

We have studied the gamma-ray binary LS~5039 at low and high frequencies, and we have determined its spectra and light-curves in the frequency range of 0.15--15~GHz by analyzing radio observations from the VLA, GMRT and WSRT. We have observed that its spectrum is persistent along the time on day, month and year timescales, exhibiting a turnover at $\sim$0.5~GHz. The obtained quasi-simultaneous spectra reveal subtle differences between them. Synchrotron self-absorption can explain the observed spectra, but the Razin effect is necessary at some epochs. This is the first time that this effect is reported in a gamma-ray binary. With all these data and a simple model, we have estimated the physical properties of the radio emitting region, providing an estimation of its size, the magnetic field, the electron density, and the mass-loss rate of the companion star.

We have also explored the low-frequency emission of the gamma-ray binary LS~I~+61~303 through GMRT and LOFAR observations. We have detected for the first time a gamma-ray binary at a frequency as low as 150~MHz. We have also determined the light-curves of the source at 150, 235 and 610~MHz. These light-curves are modulated with the orbital and superorbital period, with a quasi-sinusoidal modulation along the orbital phase. The shifts observed between the orbital phases at which the maximum emission takes place at different frequencies have been modeled with a simple model, suggesting an expanding emitting region, with an expansion velocity close to the stellar wind one.

The gamma-ray binary HESS~J0632+057 has been explored with a very high-resolution EVN observation to unveil the evolution of its radio emission along the orbit. However, the source was not detected, setting an upper-limit which is one order of magnitude below the radio emission detected in previous observations.

We have discovered a new colliding wind binary (HD~93129A) through a multiwavelength campaign with optical and LBA radio data. We have resolved the radio emission from the wind collision region, observing the expected bow-shaped structure. This source is one of the earliest, hottest, and most massive binary systems discovered up to now. We provide a rough estimation of the wind momentum rates ratio based on the observed structure. We have also observed an increase of the radio emission during the last years, as the system approaches to the periastron passage, which is estimated to take place in $\sim$2024.

Finally, we performed radio observations on two new sources that were hypothesized to be gamma-ray binaries. On one hand, the star TYC~4051-1277-1 was initially proposed to be associated with a non-thermal radio source, but he have concluded that the radio emission is originated by a quasar. On the other hand, MWC~656 has been discovered to be the first Be/BH binary system. However, its radio emission remains undetected.

Based on these results, we have improved the knowledge of several high-energy binary systems through their radio emission, conducting for the first time detailed low-frequency studies on these types of sources.

\titleformat{\chapter}[block]
	{\setstretch{1.0}\usefont{T1}{kurier}{b}{n}\selectfont\huge\bfseries}
	{\parbox{2cm}{\includegraphics[width=2.2cm]{figures/section-item.png}}}
	{0pt}
	{
		\begin{minipage}{0.8\textwidth}
			{\sc\Large\chaptername\ {\LARGE\thechapter}}\\[-5pt]\filright#1
	}[\end{minipage}]

\mainmatter
\renewcommand{\sectionmark}[1]{\markright{\color{gray}\thesection.\ #1}{}}
\renewcommand{\chaptermark}[1]{\markboth{\color{gray}\thechapter.\ #1}{}}

%
%
%


\chapter{Introduction}

High-energy binary systems are powerful astronomical sources displaying emission from radio to X-rays or \g-rays. In this thesis we will study the radio emission of some of these sources at low and high radio frequencies with different facilities and with different angular resolutions. In this Chapter we make an introduction to the thesis. In Sect.~\ref{sec:hea} we introduce high-energy astrophysics and the most common types of sources displaying high-energy emission. In Sect.~\ref{sec:binaries} we focus on the high-energy binary systems. Finally, in Sect.~\ref{sec:motivation} we state the motivation of this thesis, which is the study of the physical properties that can be determined from the radio emission of these high-energy binary systems.

\section{High-energy astrophysics} \label{sec:hea}

High-energy astrophysics is dedicated to the study of astronomical sources which exhibit emission in X-rays or \g-rays. Although some of these sources show thermal X-ray emission, the majority of them exhibit non-thermal emission, involving extreme environments.
This non-thermal emission requires efficient mechanisms to accelerate particles up to relativistic energies, being these particles the fuel for the processes that originate the observed X-ray or \g-ray emission.
Since all these high-energy photons are absorbed in the Earth atmosphere, the high-energy astrophysics has been closely linked to the development of rocket or satellite-based science, or indirect ground-based observatories, being one of the youngest fields in astrophysics. The use of sensitive instruments observing at X-rays and \g-rays has allowed us the exploration of new kinds of sources during the last decades.

The \g-ray emission is usually separated in three domains: the 0.5--100~MeV, the high-energy one (HE; $\sim$$0.1$--$100~\mathrm{GeV}$) and the very high-energy one (VHE; $>100~\mathrm{GeV}$). The HE band is explored by satellites such as {\em AGILE} and {\em Fermi}, both observing the range of 0.3--300~GeV, whereas to observe the VHE band indirect methods are required. Given the low number of VHE photons that arrive to the Earth, the use of satellites is not feasible due to their reduced collecting areas. Atmospheric Cherenkov telescopes, that use the Earth atmosphere as a calorimeter, are widely used for these purposes. The High Energy Stereoscopic System (H.E.S.S.), the Major Atmospheric Gamma-Ray Imaging Cherenkov (MAGIC) Telescopes or the Very Energetic Radiation Imaging Telescope Array System (VE\-RI\-TAS) are the current observatories exploring the VHE sky with this method, with the addition of the Cherenkov Telescope Array (CTA) in the coming years.
\begin{figure}
    \centering
		\includegraphics[width=\textwidth]{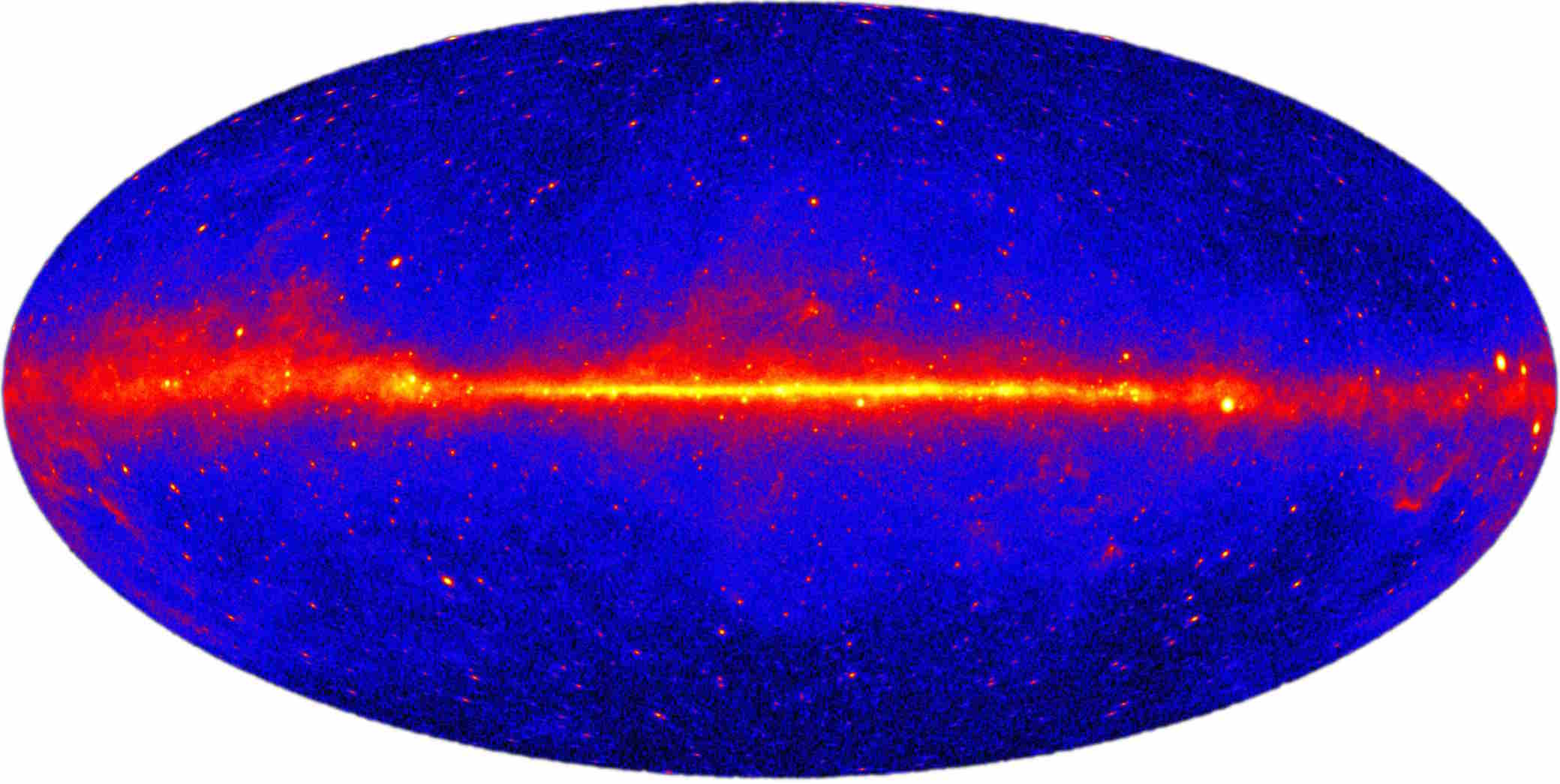}\\[+20pt]
		\includegraphics[width=\textwidth]{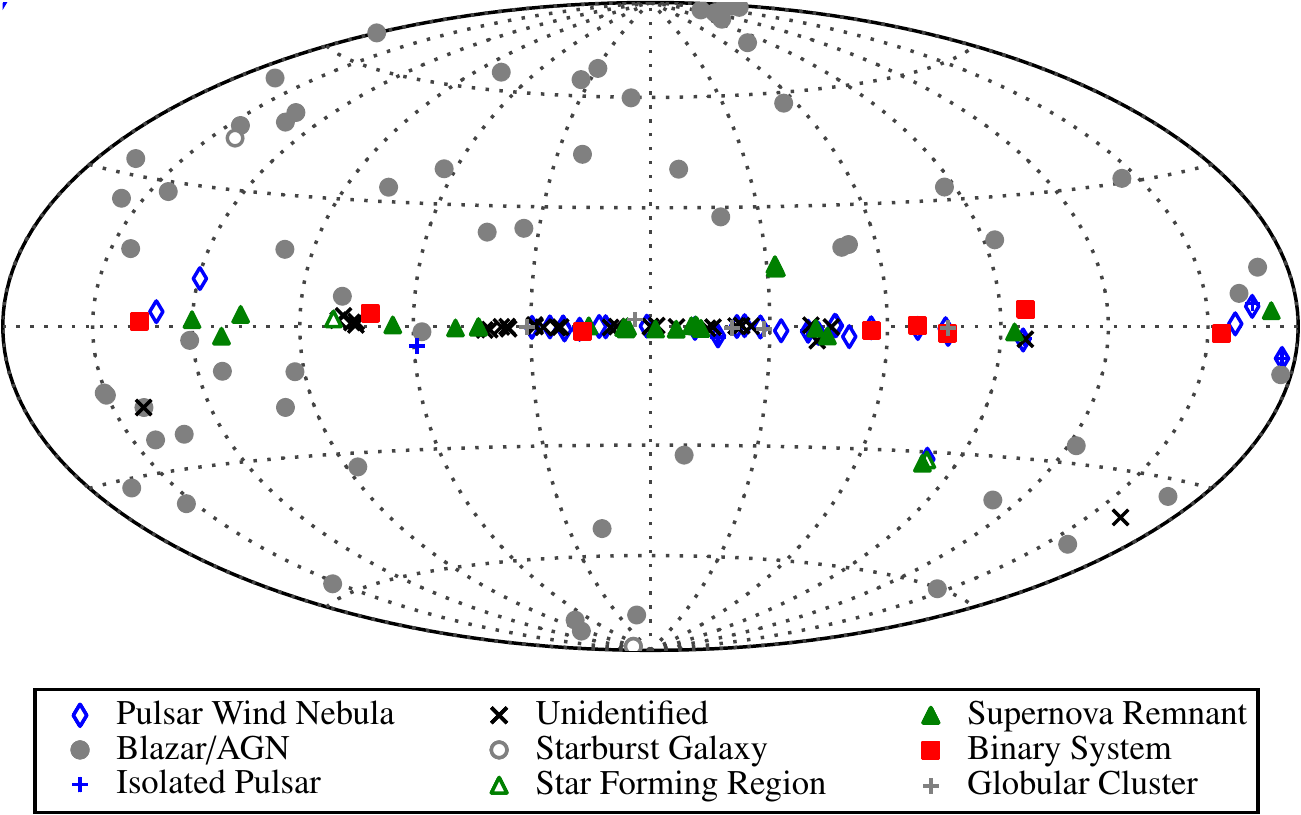}\\[+15pt]
		\caption[High-energy and very high-energy \g-ray all-sky map.]{{\em Top:} high-energy \g-ray all-sky map obtained with the Large Area Telescope (LAT) detector on board the {\em Fermi} satellite above 1~GeV after five years of integration time.
		{\em Bottom:} very-high energy \g-ray all-sky map showing the 161 sources discovered up to now emitting at TeV energies. Binary systems are represented by red squares. Image Credit: NASA/DOE/{\em Fermi}/LAT Collaboration (\url{http://fermi.gsfc.nasa.gov/ssc}). TeV source list obtained from the TevCat catalog (\url{http://tevcat.uchicago.edu}).}
		\label{fig:high-energy-all-sky}
\end{figure}
A large number of sources ($\sim$3\,000; \citealt{acero2015}) emitting at HE are known, while the number of known VHE emitting sources remains relatively small ($\sim$150; \citealt{lorenz2012}). Figure~\ref{fig:high-energy-all-sky} shows the HE and VHE \g-ray sky. The VHE sources can be clustered in few kinds of sources \citep[see][for a review]{rieger2013}. In the following we briefly introduce the most common ones. However, binary systems are thoroughly described in \S\,\ref{sec:binaries}.

\paragraph{Supernova Remnants (SNRs).}

When a massive star ($M \gtrsim 10\ \mathrm{M_{\sun}}$) reaches the end of its life, produces an iron core that can no longer support the star itself. Therefore, a gravitational collapse of the whole star cannot be avoided. During this collapse, the material can reach velocities up to a fraction of the speed of light, compressing the core and originating an explosion that liberates a huge amount of energy in a small fraction of time. The material is, in this process, blown off from the star into the space. A supernova appears then in the sky.
After the explosion of a star as supernova, the accelerated material quickly expands, producing a shock wave that interacts with the interstellar medium (ISM), and a shell which is known as supernova remnant. The shocks produced between the expanding material and the ISM are one of the most efficient particle accelerators ever known. \g-ray emission has been widely observed and studied in this kind of systems \citep{aharonian2013}. The accelerated particles can typically produce photons of up to $\sim$100~TeV \citep{rieger2013}.

\paragraph{Isolated pulsars.}

A rapidly rotating and highly magnetized neutron star produces a collimated beam of radiation along the magnetic axis, which can emit at different wavelengths. If the magnetic axis is not aligned with the spin axis, the beam of radiation swaps different regions of the sky.
If due to this rotation the beam points towards the direction of the Earth, a pulsed emission is observed, hence the name of pulsar. The intense magnetosphere of these systems is thought to accelerate particles up to relativistic energies, producing \g-ray emission at least via synchrotron-curvature radiation \citep[see][for a review]{deona2013}.

\paragraph{Pulsar Wind Nebulae (PWNe).}

A young neutron star with a rapid spin originates a relativistic wind, which strongly interacts with the original SNR. Even in old pulsars, where their SNRs have already disappeared, the nebula originated by the pulsar wind remains, interacting with the surrounding medium. This interaction, either with the SNR or the medium, accelerates particles up to relativistic velocities. These particles interact with the magnetic field and/or the low-energy radiation, producing emission up to $\sim$100~TeV.
The Crab Nebula is the best-known and brightest member of this kind of sources, considered as the ``standard candle'' in HE/VHE astrophysics \citep[see][for a review]{deona2013}.

\paragraph{Galactic Center.}

The center of the Milky Way hosts a supermassive black hole (SMBH) of $\sim$$4 \times 10^6~\mathrm{M_{\sun}}$, Sagittarius~A$^*$ \citep{gillessen2009}. It is known to be a strong point-like radio emitter. A point-like \g-ray emission, coincident with its position, has been observed up to 30~TeV \citep{aharonian2009}. An extended \g-ray emission of about few parsec (pc) in the direction of the Galactic plane is also detected. However, it remains unconfirmed the direct association between this emission and Sagittarius~A$^*$ \citep{aharonian2006gc}.

\paragraph{Active Galactic Nuclei (AGNs).}

A fraction of the SMBHs located at the center of galaxies are fueled by the surrounding material, leading to the formation of an accretion disk and launching relativistic jets, which can originate emission from radio to VHE. AGNs are classified in a wide number of subclasses, being {\em Blazars} the most relevant in VHE astrophysics. These objects present a jet directly pointing to the Earth, and thus we can observe their emission amplified by Doppler boosting. Blazars and quasars are known to exhibit a fast VHE variability on timescales down to minutes \citep{aleksic2014}.

\paragraph{Gamma-Ray Bursts (GRBs).}

The brightest and fastest explosions detected in the Universe are the GRBs. These events exhibit a strong flash, dominated by the \g-ray emission, which in a short interval of time (from milliseconds to minutes) radiates the same energy than the Sun in its entire lifetime. These events are usually located at far extragalactic distances.
Although the GRBs have been observed at GeV energies, TeV emission is also expected for sources located at low redshifts. However, the instruments must be repositioned very quickly to detect the flash during the first minutes, when most part of the emission is radiated.
The most common framework to explain these events establishes two different types of GRBs. One is originated by the explosion of a supermassive star as a supernova or hypernova, and the other one by the merger of two neutron stars (or a neutron star and a black hole), which were previously forming a binary system. See \citet{kumar2015} for a review.

\paragraph{Unidentified sources.}

Some VHE sources have no clear counterpart, and thus their natures remain unknown. Most of them are located in the Galactic plane, strongly suggesting their Galactic origin. However, the absence of a known counterpart at other wavelengths does not allow us to clearly identify the origin of the VHE emission.

\section[Binary systems with \g-ray emission]{Binary systems with \boldmath{\g}-ray emission} \label{sec:binaries}

A small fraction of the discovered HE and/or VHE sources are binary systems, usually displaying a broadband non-thermal spectral energy distribution (SED) from radio to \g-rays. Many of the properties of these systems are still poorly understood and we see large differences in their behaviors from one system to another.
For that reason, we try to classify them as a function of their observed behavior or their physical properties.

Binaries showing X-ray and/or \g-ray emission are found in binary systems composed of two massive stars (with masses $\gtrsim$$8~\mathrm{M_{\sun}}$) or a compact object and a massive star (in both cases we refer to them as high-mass binary systems) or finally a compact object and a low-mass star (typically $\lesssim$$1~\mathrm{M_{\sun}}$, hence low-mass binary systems).
None of the known low-mass binary systems exhibit persistent VHE emission.
However, there are tens of millisecond pulsars which are located in binary systems with a low-mass companion that are known to show pulsed \g-ray emission (like fast rotating isolated pulsars), and continuum VHE emission is also expected to be detectable with more sensitive observatories such as CTA \citep{bednarek2014}. These systems exhibit compact orbits (with orbital periods smaller than 24~h), and most of them are classified as {\em black widows} or {\em redbacks}, as a function of the mass of the companion star \citep[see][for a review]{roberts2013}.

In this thesis we only consider high-mass binary systems, which are the only ones that are reported to be VHE emitters. First, we briefly mention the main properties of the high-mass companion stars found in these systems. Secondly, we describe the three main kinds of high-mass binary sources which are known to display \g-ray emission: colliding wind binaries, high-mass X-ray binaries, and gamma-ray binaries.

\subsection{The high-mass companion stars}

Most parameters of a star, such as its luminosity, $L$, its radius, $R$, or the effective temperature on its surface, $T_{\rm eff}$, are related to its mass, $M$.
The wide range of masses shown by stars (from $\sim$$0.1~\mathrm{M_{\sun}}$ to $\sim$$150~\mathrm{M_{\sun}}$) has lead to several methods to classify them. One of the most useful methods is the Morgan-Keenan system \citep[MK;][]{morgan1973}, which is based on the spectral properties of the stars. This system introduces seven main classes, labeled with the letters O, B, A, F, G, K, and M, from the hottest to the coolest star, respectively. Each class is divided in 10 subclasses, labeled with digits from 0 to 9, being 0 the hottest and 9 the coolest star.
In addition to its spectral type, the luminosity information is also considered in the MK system, adding a Roman numeral from I to V to distinguish between supergiant (I), giant (III), and main-sequence stars (V), or intermediate classes in between. The labels sd or D are assigned in the case of subdwarf or dwarf stars, respectively. In this way, a star such as the Sun is classified as a G2~V star (main-sequence star with an effective temperature of $\sim$5\,800~K).

All binary systems discussed in this thesis present high-mass early-type stars, with spectral types O or B. These stars have masses larger than $\gtrsim$$10~\mathrm{M_{\sun}}$, surface effective temperatures $\gtrsim$$15\,000~\mathrm{K}$, and luminosities $\sim$$10^4$--$10^6~\mathrm{L_{\sun}}$ (where $\mathrm{L_{\sun}}$ is the solar luminosity). These high luminosities also involve large mass-loss rates, typically $\dot M \sim 10^{-5}$--$10^{-8}~\mathrm{M_{\sun}\, yr^{-1}}$ (for comparison, the Sun exhibits a mass-loss rate of $\dot M_{\sun} \sim 10^{-14}~\mathrm{M_{\sun}\, yr^{-1}}$).

O stars have masses in the range of $\sim$10 and $\sim$$150~\mathrm{M_{\sun}}$, with surface temperatures between $\sim$30\,000 and $\sim$50\,000~K. Therefore, their emission peaks in the ultraviolet. This kind of stars presents a really short lifetime: only 10~Myr in the main-sequence branch, although after $\sim$5~Myr the stars would loss enough mass to become an early B star. The population of O stars in our Galaxy is quite small (only 1 in $3 \times 10^{6}$ main-sequence stars are O-type stars), implying that typical studies about isolated O stars are made considering only $\sim$50 stars. Many uncertainties are thus still present in the knowledge of these systems. This type of stars exhibits strong winds, with velocities $\sim$$2 \times 10^3~\mathrm{km\ s^{-1}}$ \citep{krticka2001}. The interaction of these winds with the surrounding medium produces X-ray emission even in isolated O stars, being this emission variable in an important fraction of the known X-ray emitting O stars \citep{chlebowski1989}. A fraction of O-type stars, those with $M \gtrsim 20~\mathrm{M_{\sun}}$, can loss all their outer Hydrogen shell along their lives, evolving into Wolf-Rayet stars \citep[WR;][]{crowther2007}.

Be stars are a large subset of non-supergiant B-type stars that show, or have shown at some time, emission lines in the Balmer series \citep{slettebak1988}. The Be-star population represents up to $\sim$30\% of the total population of B stars \citep{grudzinska2015}. This type of stars exhibits a near-infrared excess, which is associated with a circumstellar decretion disk. This disk consists of a cool plasma ($T \gtrsim 10^4~\mathrm{K}$), with typical electron densities of $\sim$$4 \times 10^{11}~\mathrm{cm^{-3}}$ in quasi-Keplerian rotation \citep{porter2003}, whereas the stellar wind exhibits velocities of $\sim$$10^3~\mathrm{km\ s^{-1}}$ \citep{krticka2001}.
Although the origin of this circumstellar disk is not well understood yet, the high rotational velocity of Be stars is one of the mechanisms that could contribute to its formation.

\subsection{Colliding wind binaries}

A significant fraction of massive stars are found in binary, or higher multiplicity, systems \citep[e.g.][]{sana2014}. In such configurations, their powerful stellar winds are likely to collide, producing strong shocks. 
The current standard model for particle acceleration in massive binaries considers the diffuse shock acceleration in the hydrodynamic shocks produced by the colliding winds \citep{pittard2006,reimer2006}. These colliding wind binaries (CWBs) are excellent laboratories to study particle acceleration up to relativistic energies, exhibiting the same physical processes as observed in SNRs, but at higher mass, photon and magnetic energy densities.

About 40 systems of this category are known to be non-thermal radio emitters \citep{debecker2013catalogue}. This radio emission is thought to be synchrotron emission from relativistic electrons accelerated in shocks \citep{white1985}.
For some of these systems, the radio emission has been resolved into a bow shape, characteristic of the presence of a wind-collision region (WCR).
Since one of the stellar winds would typically be weaker than the other one, these systems exhibit this bow-shaped structure curved towards the star with the weakest wind. The WCR can be spatially resolved with high resolution radio observations, which are the most common mechanism to discover this kind of sources \citep{debecker2013catalogue}.

Apart from the synchrotron emission detected in the radio domain, these objects are also expected to be non-thermal emitters in the high-energy domain. The WCRs heat the shocked material up to temperatures of $10^7~\mathrm{K}$, naturally emitting in X-rays, but also producing non-thermal X-ray emission through IC scattering. However, X-ray emission from CWBs is difficult to be identified due to the fact that isolated O or WR stars are also X-ray emitters.

$\upeta$-Carinae, which is the most massive CWB system known, comprising a $\sim$$90$--$120~\mathrm{M_{\sun}}$ Luminous Blue Variable (LBV) star and a $\sim$$30~\mathrm{M_{\sun}}$ O star, is the only case in which HE \g-ray emission has been detected \citep{tavani2009etacar,abdo2009etacar,reitberger2015}. This source is located at 2.3~kpc from Earth, exhibiting an orbital period of $\approx$5.5~yr \citep{damineli2008} and a high eccentricity of 0.9 \citep{teodoro2012}. The separation between stars at periastron is only 1.7~AU.
$\upeta$-Carinae shows an X-ray and GeV light-curve modulated with the orbital phase.
At X-rays, we observe an outburst before the periastron passage, followed by a fast drop of the emission that lasts several months. This is thought to be originated by the occultation of the shock from the observer's line of sight, due to the higher opacity of the region dominated by the wind of the main star, and/or by a disruption or collapse of the shock during the maximum approach of the two stars. This collapse could also imply a shell ejection of material, which is also supported by optical spectral line observations \citep[see e.g.][]{damineli2008}. At HE, $\upeta$-Carinae exhibits a smoother light-curve than at X-rays, with a low-energy component and a high-energy one. The low-energy light-curve (0.2--10~GeV) exhibits a quasi-sinusoidal modulation, with its maximum emission at periastron. However, the high-energy one (10--300~GeV) shows a modulation similar to the X-ray one: a maximum emission during the periastron passage, followed by a fast drop in the emission. After several months, the emission increases again \citep{reitberger2015}. At gigahertz (GHz) radio frequencies, a quasi-sinusoidal modulation is observed, with minimum emission during the periastron passage \citep{duncan2003}.

Massive stars have very short lifetimes. When they reach the end of their lives, they evolve into a neutron star (NS) or a black hole (BH). In the case of binary systems, the supernova explosion will affect the system, sometimes disturbing the orbit or completely disrupting the system. However, when the systems remain bounded, the system will be comprised of a massive star and a compact object. The powerful wind from the massive star, together with the strong gravitational field originated by the compact object, produce an environment that is able to accelerate particles up to relativistic energies and produce \g-ray emission.
Depending on the kind of interaction produced within the system we typically talk about high-mass X-ray binary systems (HMXBs) or gamma-ray binaries.

\subsection{High mass X-ray binaries}

HMXBs are binary systems composed of a compact object, either a neutron star or a black hole, and a massive early-type star, with either an O or B spectral type. These sources exhibit a non-thermal SED dominated by the X-ray emission, which is explained by the presence of an accretion disk around the compact object.
Only three HMXBs have been reported to display \g-ray emission up to now: Cygnus X-1, Cygnus X-3, and SS~433. In any case, these systems display a non-thermal SED dominated by the X-ray emission, and the \g-ray emission appears occasionally.

\paragraph{Cygnus~X-1}\hspace{-10pt} is located 1.9~kpc away from the Sun and composed of an O9.7~Iab star with a mass of $\approx$$19~\mathrm{M_{\sun}}$ and a BH of $\approx$$15~\mathrm{M_{\sun}}$ orbiting it every 5.6~d \citep{grinberg2015}. MAGIC reported evidences of TeV emission during a flare that was also detected by {\em INTEGRAL}, {\em Swift}/BAT and {\em RXTE}/ASM \citep{albert2007}. At HE, {\em AGILE} also has reported significant transient emission above 100~MeV \citep{sabatini2010}.
\citet{bodaghee2013} analyzed {\em Fermi}/LAT data and found three low-significance excesses (at $\sim$3--4$\sigma$) on daily timescales that are contemporaneous with some low-significance \g-ray flares reported by {\em AGILE}.

\paragraph{Cygnus~X-3}\hspace{-10pt} is located at $\sim$10~kpc and comprised of a WR star with $\approx$$ 10~\mathrm{M_{\sun}}$ and a compact object of $\approx$$2.4~\mathrm{M_{\sun}}$ orbiting it every 4.8~h \citep{zdziarski2013}. The nature of the compact object remains unknown. Cygnus~X-3 exhibits flaring HE emission detected by {\em AGILE} and {\em Fermi}/LAT \citep{tavani2009,abdo2009}, although it remains undetected at VHE \citep{aleksic2010}.

\paragraph{SS~433}\hspace{-10pt} is located at $\sim$5~kpc \citep{fabrika2004} and composed of an A3--7 star with a mass of $\approx$$10~\mathrm{M_{\sun}}$ \citep{fabrika1990} and a black hole of $10$--$20~\mathrm{M_{\sun}}$ orbiting it every 13~d \citep{gies2002}.
Persistent \g-ray emission, spatially compatible at 3-$\sigma$ level with the position of SS~433, has been recently found with {\em Fermi}/LAT data \citep{bordas2014,bordas2015talk}.

\subsection{Gamma-ray binaries}

A new population of binary systems exhibiting $\upgamma$-ray emission has been discovered in the last decades. This kind of systems also consists of a young massive star and a compact object. However, in contrast to HMXBs, gamma-ray binaries are extreme high-energy systems exhibiting a non-thermal SED clearly dominated by the MeV--GeV photons.
Only five gamma-ray binaries are known at present \citep[see][for a detailed review]{dubus2013}: 

\paragraph{PSR~B1259$-$63.}

The first gamma-ray binary discovered at VHE, after the detection reported by \citet{aharonian2005psr}. Previously, the source was found through a radio pulsar search \citep{johnston1992} and postulated to belong to a new kind of sources, the gamma-ray binaries, by \citet{tavani1997} and \citet{kirk1999}. PSR~B1259$-$63 is the only gamma-ray binary which is confirmed to host a pulsar as a compact object.
With an orbital period of $\approx$$3.4~\mathrm{yr}$ and a high eccentricity of 0.87, most of its multiwavelength emission is originated around the periastron passage, where the pulsar interacts with the circumstellar disk of the O9.5~Ve companion star.
The X-ray emission clearly exhibits a two-peaked emission around periastron passage, coincident with the two epochs at which the compact object is thought to cross the circumstellar disk \citep{chernyakova2014}. The second peak, slightly brighter, is coincident with the position of the pulsar in front of the massive star (from the point of view of the observer). The VHE and radio emission is similar to the X-ray one, with a peak before periastron passage, and a subsequent second one after periastron passage, which is significantly stronger than the first one. The GeV emission only shows one outburst, which is delayed with respect to the second peak detected at the other wavelengths \citep{abdo2011}. Radio pulsations from the pulsar are detected in most part of the orbit, although they are absorbed during the periastron passage \citep{wang2004}.

\paragraph{LS~5039.}

Initially identified as a high-mass X-ray binary from a cross-correlation of unidentified {\em ROSAT} X-ray sources with OB star catalogs \citep{motch1997}, subsequent observations after the discovery of its \g-ray emission \citep{paredes2000} leaded to the confirmation of a new gamma-ray binary \citep{aharonian2005ls5039}.
LS~5039 is the most compact gamma-ray binary, with an O6.5~V star and a compact object orbiting it every 3.9~d in a relatively low eccentric orbit of 0.35 \citep{casares2005ls5039}. Whereas the TeV and X-ray emission is modulated with the orbital period, exhibiting a quasi-sinusoidal light-curve \citep{takahashi2009}, the GHz radio emission is not orbitally modulated and it only shows a small variability \citep{ribo1999,clark2001}. The GeV emission also shows a quasi-sinusoidal modulation, but anticorrelated with the TeV or X-ray ones \citep{abdo2009}. This source and its radio behavior is discussed in Chapter~\ref{chap:ls}. 

\paragraph{LS~I~+61~303.}

This source was originally identified to be the counterpart of a  HE \g-ray source detected by {\em COS~B} \citep{hermsen1977}. However, its VHE emission was not observed until several decades later by \citet{albert2006}. The system consists of a B0~Ve star and a compact object in an orbit with an orbital period of 26.5~d and an eccentricity of 0.72, obtaining a system similar to PSR~B1259$-$63 but much more compact \citep{casares2005lsi61303}.
The emission of LS~I~+61~303 is orbitally modulated at all wavelengths, exhibiting an outburst per orbital cycle, which takes place between orbital phases 0.5 and 0.9 at TeV, X-rays, or GHz radio frequencies \citep[see e.g.][]{paredesfortuny2015}. The GeV light-curve shows a more sinusoidal shape, with the minimum located close to the apastron (orbital phases $\approx$0.7--0.8). The light-curves of LS~I~+61~303 exhibit a superorbital period of $\approx$$4.6~\mathrm{yr}$, modulating not only the amplitude of the emission, but also the orbital phases at which the maximum emission takes place. This source and its behavior at low radio frequencies is discussed in Chapter~\ref{chap:lsi}.

\paragraph{HESS~J0632+057.}

Serendipitously detected as a VHE emitter by the H.E.S.S. Collaboration \citep{aharonian2007}, the source was afterwards associated to the B0~pe star MWC~148 \citep{hinton2009}. The orbit shows a period of 315~d, with a high eccentricity of 0.83 \citep{bongiorno2011,casares2012}.
The behavior of HESS~J0632+057 is thought to be similar to the LS~I~+61~303 one, but with larger timescales due to the wider orbit. The source exhibits a main outburst at TeV and X-rays at orbital phases $\approx$0.3, followed by a dip at 0.45, and a secondary broad and less clear outburst at orbital phases $\sim$0.6--0.9 \citep{aliu2014}. This source and the results obtained from a new high-resolution radio observation are discussed in Chapter~\ref{chap:hess}.

\paragraph{1FGL~J1018.6$-$5856.}

Detected in a search of periodic variable \g-ray sources from the {\em Fermi}/LAT catalog \citep{corbet2011}, the source was confirmed to be a \g-ray binary by the \citet{fermi2012}. A reduced set of physical parameters are known for this source. The compact object orbits an O6~V star every 16.6~d, although the eccentricity and the semimajor axis remain unknown.
The emission of 1FGL~J1018.6$-$5856 is orbitally modulated, exhibiting a maximum emission at orbital phases $\sim$0.0 at TeV and GeV energies. At X-rays a double-peaked light-curve is detected: a sharp peak is observed at orbital phases $\approx$0.0, superimposed to a quasi-sinusoidal emission with a maximum located at $\sim$0.4. The radio emission is also orbitally modulated, with the maximum emission taking place at $\approx$0.3, and a spectral index of zero between 5 and 9~GHz \citep[see][and references therein]{hess20151fgl}.\\

\begin{sidewaystable}
\small
\caption[Properties of the known gamma-ray binaries.]{Properties of the known gamma-ray binaries and their emission. For each binary system we list the spectral type of the companion, the distance to the system, $d$, the orbital period $P_{\rm orb}$, the eccentricity, $e$, the periastron separation, $a(1-e)$, and the apastron separation, $a(1+e)$. We also quote the general behavior of each binary system: the flux density range at 5~GHz, the spectral index $\alpha$ around this frequency, and the shape of the light-curve at radio and other wavelengths. P means that the light-curve is orbitally modulated, S that the emission is persistent, and U that the source remains undetected. P$^\ast$ means that the light-curve is orbitally modulated but variations from cycle to cycle are also observed.}
\label{tab:ls5039-other-binaries}
\def\ts{~~~~}
\begin{tabular}{l@{\ts}l@{~\ts}c@{~\ts}c@{~\ts}c@{~\ts}c@{\ts}c@{\ts}c@{\ts}c@{~~~}c@{~~~~}c@{~~~~}c@{~~}c}
	\hline\\[-10pt]
	Source &  Spectral & $d$ & $P_{\rm orb}$ & $e$ & $a(1-e)$ & $a(1+e)$ & $S_{5\,\mathrm{GHz}}$ & $\alpha$ & \multicolumn{4}{l}{\hspace{-6pt}Multiwavelength behavior}\\[+2pt]
	\cline{10-13}\\[-8pt]
	& type & (kpc) & (d) && (AU) & (AU) & (mJy) && radio & X-ray & GeV & TeV\\[+2pt]
	\hline\\[-8pt]
	LS~5039$^{(1)}$			&O6.5\,V	& 2.9	 	&3.9		&0.35	&0.09&0.19	&15--30 	&$\sim-0.5$	& S\hphantom{1} &P\hphantom{1}  & P\hphantom{1} & P\hphantom{1}\\
1FGL~J1018.6--5856$^{(2)}$&O6\,V	& 1.9	 	&16.6 	&--		&--   	&--		&1--7 	&$\sim 0.0$		& P\hphantom{1} &P\hphantom{1}  & P\hphantom{1} & P*\\
LS~I~+61~303$^{(3)}$	&B0\,Ve	& 2.0 	&26.5 	&0.72 	&0.1 &0.7		&20--200	&$-0.5$ to $+0.5$		& P*	        &P*             & P*            & P*\\
HESS~J0632+057$^{(4)}$	&B0\,Vpe&$\sim1.4$	&315   	&0.83	&0.40&4.35 	&0.1--0.4 	&$\sim-0.6$	        & P*   	        &P*             & U\hphantom{1} & P*\\
PSR~B1259--63$^{(5)}$ 	&O9.5\,Ve	& 2.3	 	&1237  	&0.87	&0.93&13.4	&1--100 	&$-1.0$ to $0.0$	& P*	        &P\hphantom{1}  & P\hphantom{1} & P\hphantom{1}\\

	\hline\\
\end{tabular}
{{(1)}~\citet{aharonian2005ls5039}, \citet{casares2005ls5039}, \citet{kishishita2009}, \citet{abdo2009}, \citet{casares2012}, \citet{zabalza2013};
	{(2)}~\citet{napoli2011}, \citet{fermi2012}, \citet{an2013} \citet{bordas2013};
	{(3)}~\citet{frail1991},  \citet{paredes1997}, \citet{strickman1998}, \citet{harrison2000}, \citet{gregory2002}, \citet{casares2005lsi61303}, \citet{aragona2009}, \citet{acciari2011}, \citet{hadasch2012}, \citet{ackermann2013};
	{(4)}~\citet{skilton2009}, \citet*{aragona2010}, \citet{bongiorno2011}, \citet{casares2012}, \citet{bordas2012}, \citet{aliu2014};
	{(5)}~\cite{johnston1994}, \citet*{wang2004}, \citet{negueruela2011}, \citet*{dembska2012}, \citet{chernyakova2014}.
}
\end{sidewaystable}
Table~\ref{tab:ls5039-other-binaries} summarizes the main properties of the known gamma-ray binaries. We observe that all these systems host a late O or B0-type star. However, their orbital parameters exhibit a huge range of values, from orbital periods of 3.9~d to 3.4~yr, from apastron distances of 0.19 to 13~AU, and from relatively low eccentric orbits (with $e \approx 0.35$) to high eccentric ones ($\approx$$0.87$). Figure~\hyperlink{fig:representation-orbits}{1.2} shows a representation of the orbits that are known for the population of gamma-ray binaries (all the mentioned systems with the exception of 1FGL~J1018.6$-$5856).

The behavior of this kind of systems still represents a challenge for the high-energy astrophysical community. Although the particle acceleration mechanisms remain unclear, two different scenarios have been proposed. On the one hand, the microquasar scenario sets the existence of an accretion disk, corona and relativistic jets in the system, with the particle acceleration driven by accretion/ejection processes \citep{mirabel2006}. This scenario predicts the existence of bipolar relativistic jets, whose electrons would upscatter stellar UV photons (according to leptonic models, \citealt{bosch-ramon2009a}); or whose protons would interact with wind ions (according to hadronic models, \citealt{romero2003}). Any of both models would explain the observed HE/VHE emission.
On the other hand, the young non-accreting pulsar scenario establishes the existence of a shock between the relativistic wind of the pulsar and the non-relativistic wind of the companion star, which would accelerate particles up to relativistic energies. The \g-ray emission would be produced by IC scattering of stellar photons by the most energetic electrons. The same population of electrons would produce the synchrotron emission observed at radio frequencies \citep{dubus2006,durant2011,bosch-ramon2013}. The most recent results seem to favor the young non-accreting pulsar scenario for the four gamma-ray binaries deeply explored up to now \citep[see][and references therein]{dubus2013}.

\section{Motivation and overview of the thesis} \label{sec:motivation}

The high-mass binary systems displaying high-energy emission are a really interesting field to explore, as we have seen in the previous section. They involve extreme environments where particles are accelerated up to relativistic energies and subsequent interactions produce the observed emission through the whole electromagnetic spectrum, from radio to VHE \g-rays. In these kinds of systems, particle\fullpagecover{current page.north west}{xshift=12cm,yshift=-9.5cm}{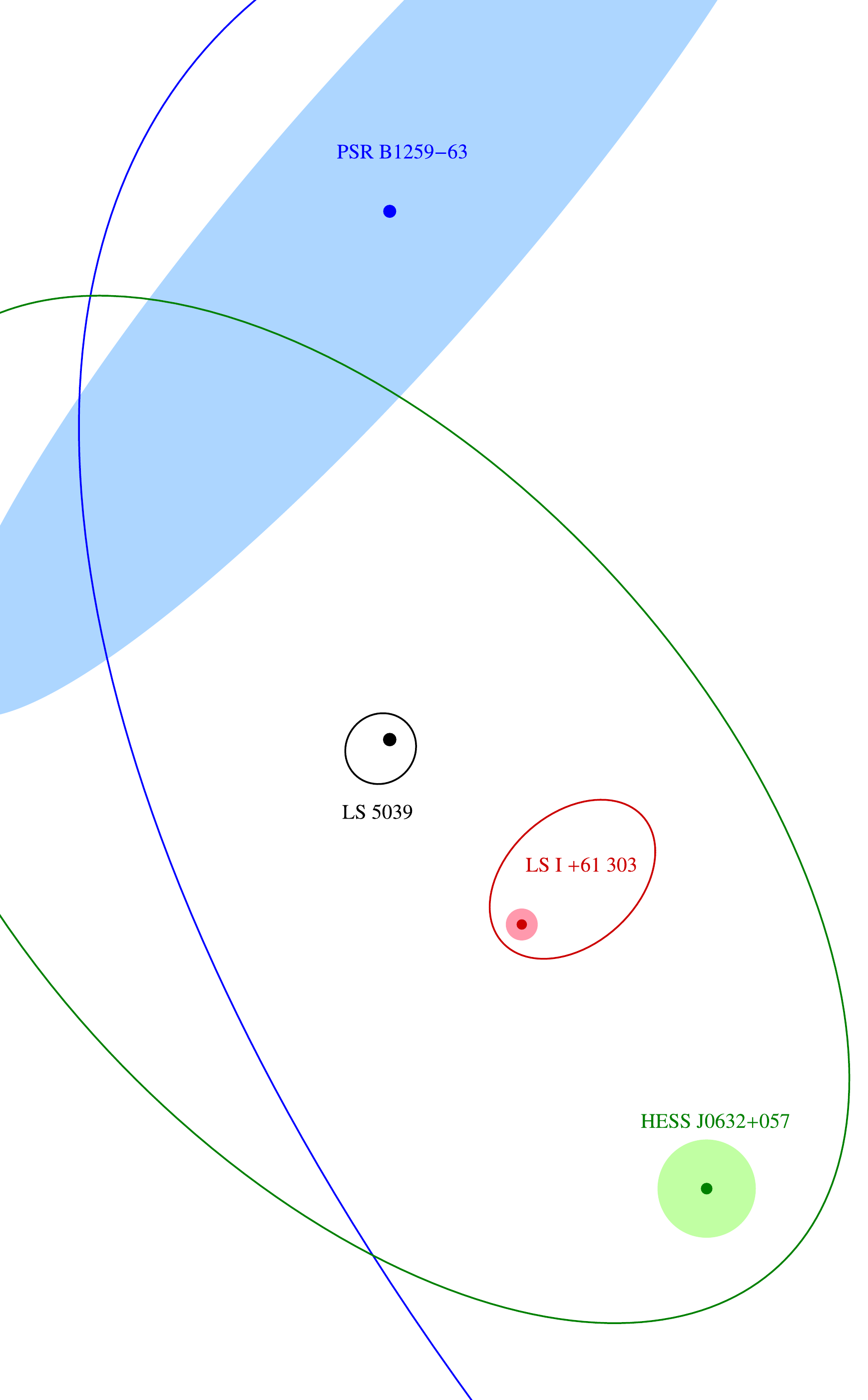}{\hypertarget{fig:representation-orbits}{}\label{fig:representation-orbits}\addtocounter{figure}{1}\addcontentsline{lof}{figure}{\number\value{chapter}.\number\value{figure}\quad Representation at scale of the four orbits that are known for the population of gamma-ray binaries.}{\bf Figure~\number\value{chapter}.\number\value{figure}.} Representation at scale of the four orbits that are known for the population of gamma-ray binaries. Dark circles represent the star and the lighter ellipses around them represent a rough estimation of the circumstellar disks, both at scale. We do not show the orbit of 1FGL~J1018.6$-$5856 as it remains unconstrained, but its semimajor axis would be between the ones of LS~5039 and LS~I~+61~303. We note that the orientation and inclination for most of the orbits are poorly constrained.}{0.97}{height=\paperheight}
\noindent physics is clearly connected to astrophysics.

The reduced number of sources known up to now, together with the fact that all of them have been discovered at VHE in the last decade, constitutes a breeding ground for big breakthroughs in our understanding of the physical processes and particle acceleration mechanisms involved in these systems. The discovery of additional sources is necessary for statistical purposes. More sensitive and with higher resolution multiwavelength observations are thus required to improve the knowledge of these systems, discard predictions or establish the roadmap of future models.
The non-thermal processes observed at the two extremes of the electromagnetic spectrum, at radio frequencies and at X- or \g-rays, are usually connected, as they involve the same population of particles. Therefore, radio observations emerge, interestingly, as a powerful tool to study in detail these high-energy systems. 
At low radio frequencies we can unveil the presence of some absorption mechanisms in the spectrum.
Radio observations with very high resolution allow us to resolve on the emission, and thus distinguish the different regions that produce the observed multiwavelength emission. A more easy comparison between different multiwavelength observations is also possible.

Based on this, this thesis is focused on the radio emission of high-mass binary systems through interferometric observations. In general, the radio emission of these sources shows a maximum located at GHz frequencies, which is the band that has traditionally been used. Thanks to the addition of radio observatories sensitive enough at much lower frequencies, in this thesis we show the first detailed observations of gamma-ray binaries at frequencies between $150~\mathrm{MHz}$ and $\sim$$1~\mathrm{GHz}$. To obtain a full overview of the emission, observations at GHz frequencies have also been analyzed, embracing thus frequencies from 150~MHz to 15~GHz.
For some systems, high resolution radio observations are critical to understand the emission, as happens in CWBs. For that reason, we also include this kind of radio observations.

The aim of this thesis, therefore, is to derive physical properties of the high energy binaries from their light-curves and spectra in a frequency range that has been poorly studied, as well as to obtain the radio structure through high-resolution radio observations. In the following we briefly give an overview of the topics covered in this thesis.

In Chapter~\ref{chap:radio} we introduce the technique of radio interferometry, the observatories that we have been used to obtain the data presented in this thesis, and we provide an explanation about how these data are reduced and analyzed.

In Chapter~\ref{chap:emission} we show the most common emission and absorption processes observed at radio frequencies for high-energy binary systems.

In the subsequent three chapters we present the results for three gamma-ray binaries:

In Chapter~\ref{chap:ls} we show the spectra and light-curves obtained for LS~5039 in the range of 0.15--15~GHz, and a simple model that can explain these spectra.

In Chapter~\ref{chap:lsi} we discuss the light-curves at low frequencies (0.15--0.61~GHz) obtained for LS~I~+61~303. We also discuss a reasonable model to explain the shifts in orbital phase observed between the maximum emission at different frequencies.

In Chapter~\ref{chap:hess} we show the results obtained from a high-resolution radio observation of HESS~J0632+057.

In Chapter~\ref{chap:hd} we focus on a new colliding wind binary: HD~93129A. We present the results that have confirmed its nature through optical and high-resolution radio observations.

In Chapter~\ref{chap:new} we present the results for two sources that were candidates to be gamma-ray binaries at the beginning of this thesis: TYC~4051-1277-1 (\S\,\ref{sec:tyc}) and MWC~656 (\S\,\ref{sec:mwc}). The first one was observed at radio frequencies and we discard here its possible association with a Be star. The second one, MWC~656, is now known to be the first high-mass binary hosting a Be star and a black hole. Here we show the radio observations conducted to unveil its possible radio emission.

Finally, we present the concluding remarks of this thesis in Chapter~\ref{chap:conclusions}.


%
%
%
%

\chapter{Radio interferometry} \label{chap:radio}

This thesis is focused on the study of the radio emission of high-energy binary systems. In this Chapter we introduce the topic of radio astrophysics. In Sect.~\ref{sec:intro-radio} we summarize the development of the radio astronomy along the history, and the main properties of radio interferometry. In Sect.~\ref{sec:radio-facilities} we detail the radio facilities that have been used along this thesis. Finally, in Sect.~\ref{sec:datareduction} we introduce the common tasks that are required to reduce and analyze the radio data.

\section{An introduction to radio astronomy} \label{sec:intro-radio}

\subsection{Everything has a beginning} 

Until 1800, mankind ignored that the light was composed of more ``colors'' than the ones that our eye can detect.
That year, Friedrich W.~Herschel (Germany, 1738--1822) made a discovery that would represent the tip of the iceberg. He was measuring the light of the stars with a telescope and a prism to determine how much radiation was being emitted at the different colors. At that point, he realized that the thermometers\footnote{At that epoch, was usual to use thermometers to quantify how much light was arriving from a star. The incident light produces an increase in the temperature of the thermometers that can be measured.} located beyond the red color reported even higher temperatures, when they should not. He had just discovered that there was more light arriving beyond the red color, the now called {\em infrared light}.

In a completely different way, The Dynamical Theory of Electromagnetism Field, published by James C.~Maxwell (Britain, 1831--1879) in 1865, demonstrated the existence of coupled electric and magnetic (electromagnetic) waves moving at the speed of light. He also predicted that light itself was an electromagnetic wave. This meant the starting point in the race for the discovery of radio waves, being originated by electrical instruments.
Heinrich R.~Hertz (Germany, 1857--1894) was the first one to transmit and detect electromagnetic waves in 1887. Later on, he also determined the velocity of these waves, confirming that it was the same than the speed of light. The common nature of the light and the {\em radio waves} was demonstrated.

The completion of the electromagnetic spectrum was done with two discoveries on the other side of the spectrum: at the high energies. First, the discovery of the X-rays in 1895 by W.~C.~R\"{o}ntgen (Germany, 1845--1923), making the first radiography in the History. Secondly, Paul Villard (France, 1860--1934) discovered the gamma rays in 1900 by studying the radioactive decay of Ra atoms. The electromagnetic spectrum was established.

Going back to the radio waves, after their discovery people immediately tho\-ught about the possibility of using their transmission and reception as a new long-distance wireless communication system. Sir Oliver J.~Lodge (Britain, 1851--1940) showed the transmission and reception of Morse code through radio waves in 1894. The same year, Jagadish C.~Bose (India, 1858--1937) also demonstrated publicly the use of radio waves. He transmitted a signal (showing that this can pass through walls and people) to a receiver located in another room, which activated a bell ringing, discharged a pistol and exploded a mine. Although the emitter was not really strong, Bose designed an instrument to focalize the signal, which could be denominated as a precursor of the future antennas.
In 1895, Alexander S.~Popov (Russia, 1859--1906) made another public demonstration of transmission and reception of radio waves to use them as a communication system.
In 1898, Nikola Tesla (Austrian Empire, 1856--1943) demonstrated the first remote control at the Madison Square Garden, in New York City, controlling a miniaturized boat by radio waves. After that, he focused on the transmission of electrical energy, being granted with some patents for his successfulness in 1900. Roberto L.~de~Moura (Brazil, 1861--1928) worked in wireless phonetic transmissions, conducting the first public demonstration in 1900 in Sao Paulo. He could transmit an audio signal up to 8~km away.
Almost simultaneously, Guglielmo Marconi (Italy, 1874--1937) made the first transatlantic radio transmission in 1901. He discovered the ionosphere layer and its effects in the radio waves, which affect specially at the low radio frequencies, and he could model its reflective and refractive properties. This study helped him to reach the goal of transmitting transatlantic signals, which requires multiple reflections along the ionosphere.
The battle for the invention of the {\em radio} (as instrument) took place focused between Marconi and Tesla, mainly because of their patents, but only the first one won the Nobel Prize in Physics in 1909 for his contributions.

During the following decades, the radio instruments exhibited a dramatic development, faster than almost any other new technology that has appeared. This development was intimately connected to the Army and the two World Wars.
In 1904, Christian H\"ulsmeyer (Germany, 1881--1957) demonstrated how, through radio waves, a ship could be detected even under a dense fog. It was the origin of the RAdio Detection And Ranging (RADAR or radar). This technology was strongly developed during the previous years to the World War II. The ability to detect an enemy (ship or plane) that was still kilometers away was promising enough to invest on it a huge fraction of Army funds. After the war, a significant understanding of these systems was reached, with a large number of highly developed antennas and sensitive receivers.
Many of these instruments where quickly adapted to study the cosmic radiation, helping the astronomers in their research during the following years. Hence, they were used for a completely different purpose that they were conceived.

\subsection{The old days}

Although light outside the optical range was already discovered many decades before, as we have already seen, till the first years of XX century almost nobody thought that celestial sources could in fact emit at radio frequencies. Only some attempts to detect the Sun at radio wavelengths were performed during the last years of the XIX century and the first ones of the XX century, without success.
The starting of the radio astronomy became, surprisingly, in the decade of 1930 from the transatlantic service of the Bell Telephone Company. Some technical problems they were obtaining with the functioning of their systems ended up in an exhaustive search of possible interferences, conducted by Karl G.~Jansky (USA, 1905-1950).
He designed a rotating antenna that observed at 20.5~MHz. After several months, he realized that one of the sources that produced the observed interferences appeared with the same period than the orbital period of the Earth (i.e.\ every 23~h and 56~min). Therefore, the signal came from the sky. Karl Jansky concluded in 1932 (although published in 1935) that this source was placed at the direction of the Sagittarius constellation, around the center of the Milky Way. He had just detected for the first time the Galactic Center.

During the following years, the Sun was also detected at radio wavelengths, coinciding with its maximum activity. In 1944 Grote Reber (USA, 1911--2002), considered as the father of the radio astronomy, obtained the first radio map of the sky at a frequency of 160~MHz with a 9-m antenna. He reported the radio emission of the Galactic Plane, and several sources such as Cygnus~A or Cassiopeia~A. He also determined the non-thermal nature of the radio emission with multifrequency observations. Instead of the expected emission proportional to the square of the frequency (Raleigh-Jeans law, see \S\,\ref{sec:bb-emission}), he found negative spectral indexes, which undoubtedly were originated by non-thermal processes.

One year later, in 1945, Hendrik C.~van~de~Hulst (Netherlands, 1918--2000) suggested for the first time a 21~cm (1.4~GHz) radio emission originated by the hyperfine transition of neutral interstellar Hydrogen ({\sc Hi}). He predicted that, although this transition is highly improbable, the large amount of neutral Hydrogen in the Universe could produce an emission that should be detected by the contemporaneous radio antennas. This prediction, and its later confirmation by \citet{ewen1951}, ended in one of the revolutions in the radio astronomy.
The discovery of the {\sc Hi} line allowed astronomers to map the whole structure of the Milky Way and its rotation curve. The deepest study of our Galaxy ever conducted could be carried on thanks to its radio emission, which is not absorbed by the dust, as it happens for optical light.

Along the XX century the radio antennas improved, and parabolic dishes with increasing diameters were designed. More sensitive and with higher resolution instruments were built. Antenna dishes with diameters up to 100~m (as Effelsberg in Germany or Green Bank in USA) became close to the physical limits of having orientable structures with small deformations.
The largest antenna ever built was finished in 1963 in Arecibo, Puerto Rico. This dish presents a diameter of 305~m, and rests over a natural depression in the terrain. It is thus not steerable and it can only observe around the zenith.

\subsection{Aperture synthesis technique, a clever solution}

The biggest problem that radio astronomy has suffered since its beginning is the resolution of its instruments.
The angular resolution that any telescope can reach is physically limited by the diameter of its collecting area, $D$. The fact that this area is finite produces that a point-like source is seen, due to the diffraction of the light, as an extended shape in the final image described by the Airy disk \citep{airy1835}. The greatest angular resolution that any telescope can reach (without considering additional effects which could decrease this value such as the atmosphere) is proportional to the observed frequency, $\lambda$, and inversely proportional to the diameter of the collecting area, $D$:
\begin{equation}
	\theta \approx 1.22 \frac{\lambda}{D},
	\label{eq:airy}
\end{equation}
where $\theta$ is the angular radius at which the first minimum of the Airy disk takes place. This relationship implies that for the same diameter, an optical telescope (which observes at $\lambda \sim 10^{-7}~{\rm m}$) produces an image with a resolution of around five orders of magnitude better than a radio telescope (considering that it observes at $\lambda \sim 10^{-2}~{\rm m}$)\footnote{We note that at radio frequencies the effect of the atmosphere is much less relevant than at the optical range. That produces that commonly the radio telescopes reach the diffraction limit. That is, the resolution that we usually obtain with a radio telescope is close to the theoretical value given by equation (\ref{eq:airy}). In contrast, the optical telescopes obtain a much poorer resolution, compared with the theoretical one, due to the distortion produced by the atmosphere.}. As an example, the largest existing radio antenna (Arecibo, 305~m) can reach a resolution of $\sim$2~arcmin observing at $2~\mathrm{GHz}$.
That is, the resolution of Arecibo is comparable to the human eye, and thus much worse than almost any optical telescope, which usually have resolutions of arcseconds.

Even with the increased sensitivity of the radio telescopes, their poor resolution remained as a big problem. In addition to the difficulty of resolving the radio emission produced by sources located close in the sky, it was difficult to determine the optical counterpart of the detected radio sources. Typically, inside of a radio detection (with a resolution of arcminutes) there were tens or hundreds of optical sources.

Sir Martin Ryle (Britain, 1918--1984) applied the concepts of interferometry to the radio antennas to significantly increase their resolution \citep{ryle1946}. Instead of observing with only one antenna, he proposed to observe simultaneously the same position of the sky with two or more antennas. Properly combining their signals, the output results in a single image with a resolution that is now limited by the distance between antennas, $b$, and not only by their diameter, $D$. This distance $b$ is widely known as the {\em baseline} between the two antennas. Given that the antennas can be separated large distances, one could easily reach $b\gg D$. Therefore, the resolution of the instrument would dramatically increase. Whereas a dish of, for example, 10~km, can not be built, it is possible to combine the signals from small antennas spread up to 10~km away. Figure \ref{fig:interferometry-explanation} represents a sketch of this idea.
\begin{figure}[t]
	\begin{center}
		\includegraphics[width=10cm,trim=2cm 1.5cm 2cm 0.7cm,clip]{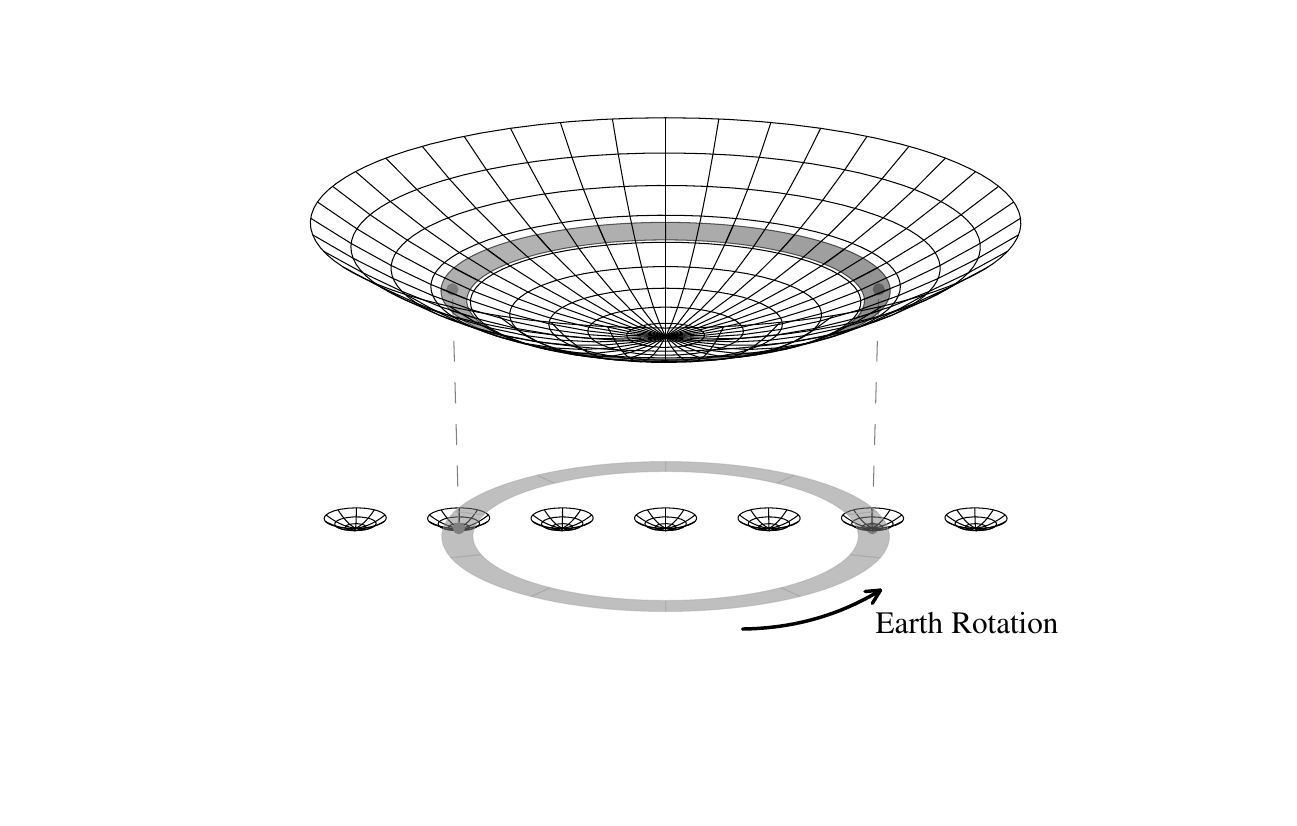}
		\caption[Sketch of the synthesis imaging technique.]{Sketch of the synthesis imaging technique. The combination of signals from many small telescopes (bottom), with the help of the rotation of the Earth, produces a result equivalent to the one obtained by a huge telescope (top). In this example, the synthesis of the 7 small dishes produces, after 12~h of observation, an image with a resolution equivalent to the one which would be obtained by the huge dish. We note that the sensitivity would in any case increase proportional to the total collecting area and would not be comparable.}
		\label{fig:interferometry-explanation}
	\end{center}
\end{figure}
The image obtained with the array of antennas is not, obviously, completely equivalent to the one obtained with a huge individual antenna. The sensitivity of the final result scales as the collecting area, and thus proportional to the number of antennas and the squares of their diameters. We note that in these cases the response of the array to a point-like source will not be anymore an Airy disk, and it can even show a very complex pattern, which is known as the {\em dirty beam}. In radio images, an approximation of the central part of this dirty beam as a two-dimensional gaussian is usually given as an equivalence to the resolution of the image, which is known as the {\em synthesized beam}. In these cases, the widths of the gaussian are provided by the widths at which the dirty beam exhibits half of its power, or Half-Power Beam Width (HPBW).

\subsection{Very Long Baseline Interferometry}

Once antennas spread over a long distance are connected, the only limitation is the Earth size, and the precision of the distances between antennas and the arrival times of the radio signal at each antenna.
Within the Earth, we can only separate antennas up to projected distances of $\sim$10\,000~km, which are close to the diameter of our planet and the longest distance at which two antennas can observe the same astronomical source. For distances larger than that an antenna into the space must be positioned. However, the atmosphere plays a role here. Although we have previously mentioned that the atmosphere presents much smaller effects to the radio waves compared to the optical range, these effects are not negligible, specially at long baselines. Antennas located at different places (typically in different countries) will usually have different weather conditions. These different conditions produce a different optical path for each antenna, which need to be compensated to recover the constructive interferometry between antennas. Whereas at frequencies $\gtrsim$10~GHz) the troposphere is the most problematic atmospheric layer, at low frequencies ($\lesssim$1~GHz) the ionosphere is the dominant one (because we are observing at a frequency close to the characteristic frequency of the plasma located in the ionosphere).

Therefore, if we want to combine antennas located at large distances we need to compensate the mentioned effects in each antenna. The Very Long Baseline Interferometry (VLBI) is a technique to perform radio observations using antennas which are separated very long distances, typically of the order or more than 1\,000~km.
Historically, the atmosphere has been the main limitation to VLBI observations. As the behavior of the troposphere was better known than the one of the ionosphere, this kind of observations was firstly developed at high frequencies. Only in the last decades observations at low frequencies with VLBI techniques have been developed.

As a summary, radio astronomy has become as an important tool in modern astrophysics. The fact that it has reached incredibly high resolutions (currently milliarcseconds, mas, and even microarcseconds, $\upmu$as), which are not accessible at any other wavelength; that the radio emission is not absorbed by the dust in the Galaxy; and that we observe completely different processes (mainly non-thermal ones) than at other wavelengths such as the optical range, has established radio astronomy as a discipline that is used in edge research astrophysics.

\subsection[Correlation, the $uv$-plane and the $uvw$-space]{Correlation, the \boldmath{$uv$}-plane and the \boldmath{$uvw$}-space}

Considering the most general case of a radio emitting source located at a position $\vec R$, we can disentangle the electric and magnetic field of the incoming radiation in $\vec E (\vec R, t)$ and $\vec B  (\vec R, t)$, respectively. The receivers from our instruments, located at $\vec r$, will be only sensitive to the incoming electric field, and assuming a propagation through empty space we can deduce the magnetic field arriving to the receiver:
\begin{equation}
	\vec E_{\nu} (\vec r) = \int \varepsilon_\nu (\vec R) \frac{e^{2\pi i \nu |\vec R - \vec r|/c}}{|\vec R - \vec r|} \D S,
\end{equation}
where $\varepsilon_\nu (\vec R)$ is the distribution of the electric field on the surface $\D S$.
The signal arriving at each pair of antennas, located at positions $\vec r_1$ and $\vec r_2$, is combined in a process called correlation, which produces
\begin{equation}
    V_{\nu} (\vec r_1, \vec r_2) = \langle \vec E_{\nu} (\vec r_1) \vec{E}_{\nu}^* (\vec r_1)\rangle,
\end{equation}
where the asterisk represents the complex conjugate.
It can be shown that this leads into \citep{clark1999}
\begin{equation}
    V_{\nu} (\vec r_1, \vec r_2) \approx \int I_{\nu} (\vec s) e^{-2\pi i\,\nu\,\vec s \cdot \vec b\, c^{-1}} \D\Omega, \label{eq:vis1}
\end{equation}
where $I_{\nu} (\vec s)$ is the observed intensity at a frequency $\nu$ in the solid angle $\D\Omega$, located at a direction of the unit vector $\vec s$, and $\vec b = \vec r_1 - \vec r_2$ is the baseline defined by the two antennas. We note that $I_{\nu} (\vec s)$ is not directly the intensity emitted by the astrophysical source. This magnitude is affected by the response of our receivers at the direction $\vec s$ (known as the attenuation beam pattern), the general response of each antenna pair, and a large number of effects related with the calibration of each antenna.

During an observation, we usually consider a position of the sky as reference (e.g.\ pointing the antennas to the position of a target source), which is known as {\em phase tracking center}, $\vec s_0$. In this case, there is a privileged reference system, defined by $\vec s_0$ and a plane normal to this direction. Furthermore, we can measure the distances in terms of the wavelength $\lambda = c / \nu$ at which the observations have been performed (or the central frequency in the case of observing with a bandwidth $\Delta \lambda$). In this case, we translate the coordinates $\vec b = \vec r_1 - \vec r_2 = \lambda (u, v, w)$, defining the {\em $uvw$-space}. With these changes we can redefine the unit vector $\vec s$ as $(l, m, n\equiv\sqrt{1 - l^2 - m^2})$, being $(l, m, n)$ the direction cosines, and rewrite equation (\ref{eq:vis1}) in to
\begin{equation}
    V_{\nu} (u, v, w) = \iint I_{\nu} (l, m) e^{-2\pi i\,(lu + mv + nw)} \D l\,\D m. \label{eq:vis2}
\end{equation}
The $w$ component only becomes relevant in VLBI observations, where the antennas follow the curvature of the Earth, or in the case of observing wide field of views (usual in low frequency observations). In other cases, we can consider that all antennas are coplanar, and thus $w \equiv 0$. The plane defined by this condition, which is the one previously mentioned and normal to $\vec s_0$, is known as the {\em $uv$-plane}. In these cases, the equation (\ref{eq:vis2}) can be reduced to
\begin{equation}
    V_{\nu} (u,v) \equiv |V_\nu| e^{i\phi} = \iint I_{\nu} (l, m) e^{-2\pi i\,(lu + mv)} \D l\,\D m. \label{eq:visibilities}
\end{equation}
This quantity, $V_{\nu}$, is known as the {\em complex visibility}, defined by an amplitude $|V_\nu|$ and a phase $\phi$. Given that $I_\nu \in \mathbb{R}$, then $V_\nu(-u, -v) = V_\nu^*(u, v)$.
The visibilities are the raw data recorded by any radio interferometer, and they are calibrated by the astronomers to recover an image of the sky.
We note that equation (\ref{eq:visibilities}) is a Fourier transform between $V_\nu$ and $I_\nu$, and thus the image of the sky can be recovered by an inverse Fourier transform of the recorded visibilities $V_\nu$.
However, these recorded visibilities are actually (following the notation of \citealt{fomalont1999}):
\begin{equation}
	\tilde{V}_{ij} (u, v) = \mathcal{G}_{ij} V_{ij} + \epsilon_{ij} + \eta_{ij}
\end{equation}
where $\mathcal{G}_{ij}$ is the baseline-based complex gain, $\epsilon_{ij}$ is the baseline-based complex offset, and $\eta_{ij}$ is the stochastic complex noise. That is, the obtained visibilities are affected by different phenomena or problems that must be corrected for to recover the real signal from the sky. We will see in \S\,\ref{sec:datareduction} how to do this in our data.

We note that in any interferometric observation, we use a finite number of antennas (and thus of baselines). Therefore, we do not obtain a full coverage of the $uv$-plane, implying that we do not recover all the spatial frequencies from the field of view. The better $uv$-coverage we get, the more accurate image we obtain. This can be easily understood looking at Figure~\ref{fig:uv-coverage}. Most of the short-baseline interferometers are composed by a large number of antennas, in order to provide a good $uv$-coverage. However, the long-baseline interferometers present antennas that are spread long distances, and thus we obtain a poorer $uv$-coverage. In these cases, the resulting data must be carefully analyzed. A problem, or bad calibration, in one or some antennas can produce fake morphologies in the final image, which are only the result of the calibration process.
\begin{figure}[t]
	\begin{center}
	\includegraphics[width=\textwidth]{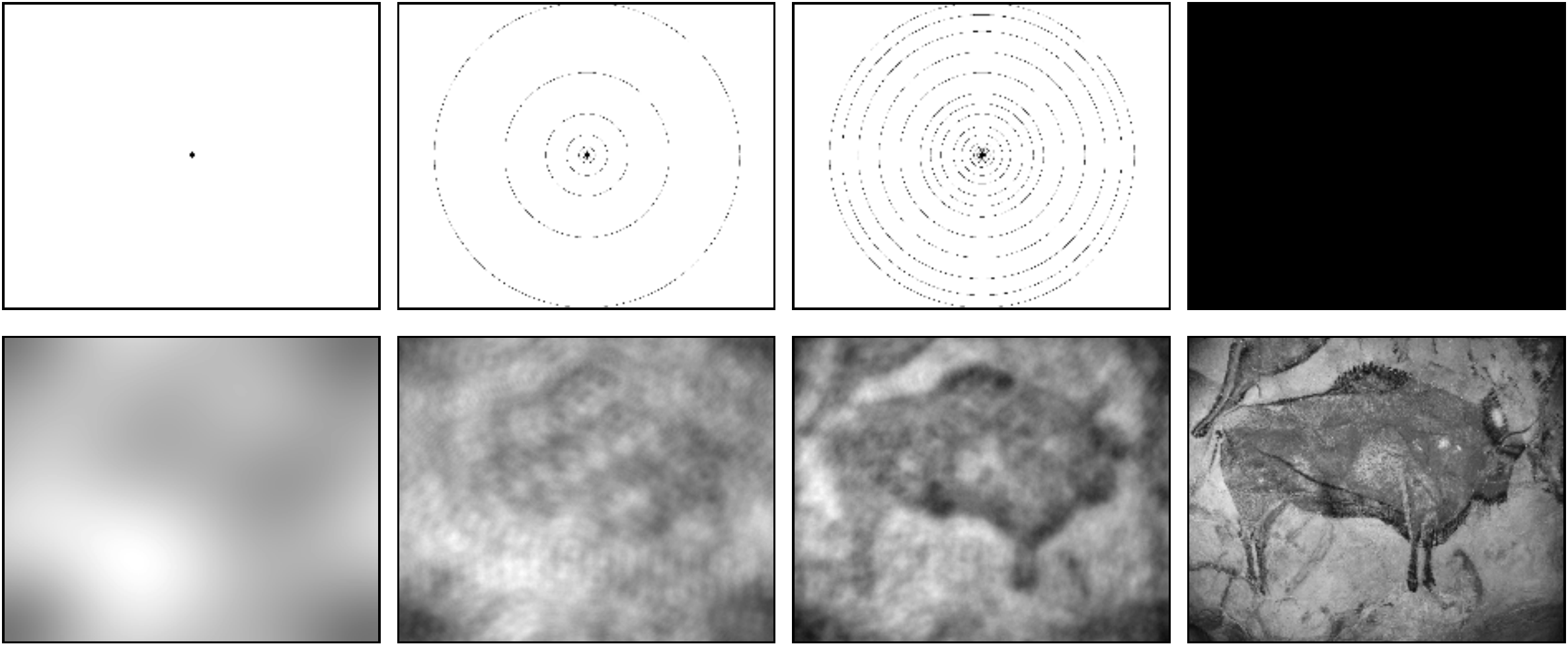}
	\caption[Representation of the effect of the $uv$-coverage in the recovered image.]{Representation of how an image is reconstructed as a function of the $uv$-coverage. On top we show different coverages, from the equivalent to a single dish observation (left), to a total $uv$-coverage that recovers the full image (right). On bottom we show the image that we would recover for each $uv$-coverage. The \href{https://en.wikipedia.org/wiki/File:AltamiraBison.jpg}{original image} is a reproduction of a bison of the cave of Altamira, Cantabria (Spain), under \href{https://creativecommons.org/publicdomain/zero/1.0/}{Public Domain (CC-0 1.0)}}
	\label{fig:uv-coverage}
	\end{center}
\end{figure}

\section{Radio facilities} \label{sec:radio-facilities}

Lots of antennas have been built around the world, covering either single-dish or interferometric observations, a huge range of frequencies (10~MHz--900~GHz) and different spatial resolutions (with baselines of few meters to thousands of kilometers). They allow the astronomers to choose the better instrument to perform their research.
Along this section we describe the facilities used for the work presented in this thesis. These facilities cover high frequency interferometer arrays (VLA and WSRT), low frequency arrays (GMRT and LOFAR), and also very long baseline interferometers (EVN and LBA). Therefore, we have covered almost all the baseline range available for interferometric observations at low and high frequencies.

\subsection{Very Large Array}

The Karl G. Jansky Very Large Array (VLA; \citealt{perley2011})\footnote{\url{https://science.nrao.edu/facilities/vla}}, see Figure~\ref{fig:vla-photo}, is a connected radio interferometer located in the plain of San Agustin, at an elevation of 2\,100~m and near to the city of Socorro, New Mexico (USA).
Built and operated by the National Radio Astronomy Observatory (NRAO)\footnote{\url{htts://nrao.edu}}, the VLA became one of the main references in radio astronomy because of its sensitivity and resolution since its inauguration in 1980.
The array consists of 27 antennas with a diameter of 25~m each spread in a Y-shaped configuration. The antennas are arranged over railway lines which allow them to be moved and change their positions within the full array. There are four standard configurations labeled from A to D. A configuration is the widest one, displaying a largest baseline of 36.4~km. D configuration, the most compact one, shows a largest baseline of only 1~km. However, as each antenna can be moved independently from the others, there are also intermediate, or hybrid, configurations. These hybrid configurations take place during the change from one configuration to another one, and they are labelled with the combined name of the previous and next configurations (AnB, BnC, CnD, or DnA).

\begin{figure}[t]
\begin{center}
	\includegraphics[width=\textwidth]{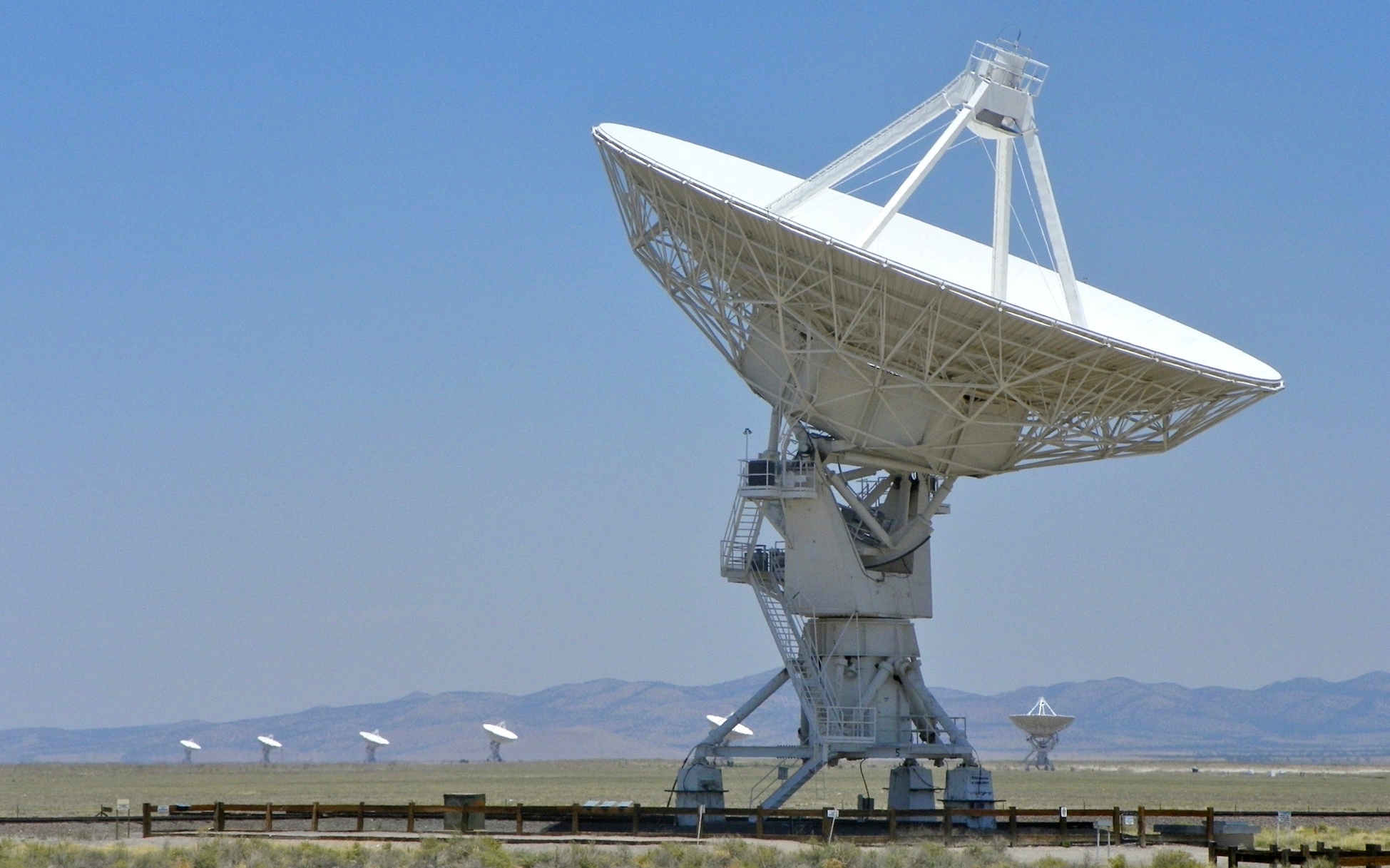}
	\caption[The Karl G. Jansky Very Large Array (VLA).]{The Karl G. Jansky Very Large Array (VLA) is composed of 27 antennas with a diameter of 25~m each. The largest separation between them is 36.4~km.}
	\label{fig:vla-photo}
\end{center}
\end{figure}
After an important upgrade during 2011--2013, the VLA can observe the full band between 1 and 50~GHz (30~cm--6~mm) with up to 8-GHz bandwidth per polarization. The point-like source sensitivity is between $2$ and $6~\mathrm{\upmu Jy}$ after one hour of observation. The VLA can reach a resolution of about a hundredth of arcsec for its widest configuration (A) and highest frequency.
From 1998 to 2008 the VLA also had detectors for low-frequency observations: at 74~MHz and 330~MHz (4~m and 90~cm). These detectors were removed, but in 2014 the VLA has started again observations at P-band (230--470~MHz) with new and more sensitive detectors.

\subsection{Westerbork Synthesis Radio Telescope}

\begin{figure}[t]
\begin{center}
	\includegraphics[width=\textwidth]{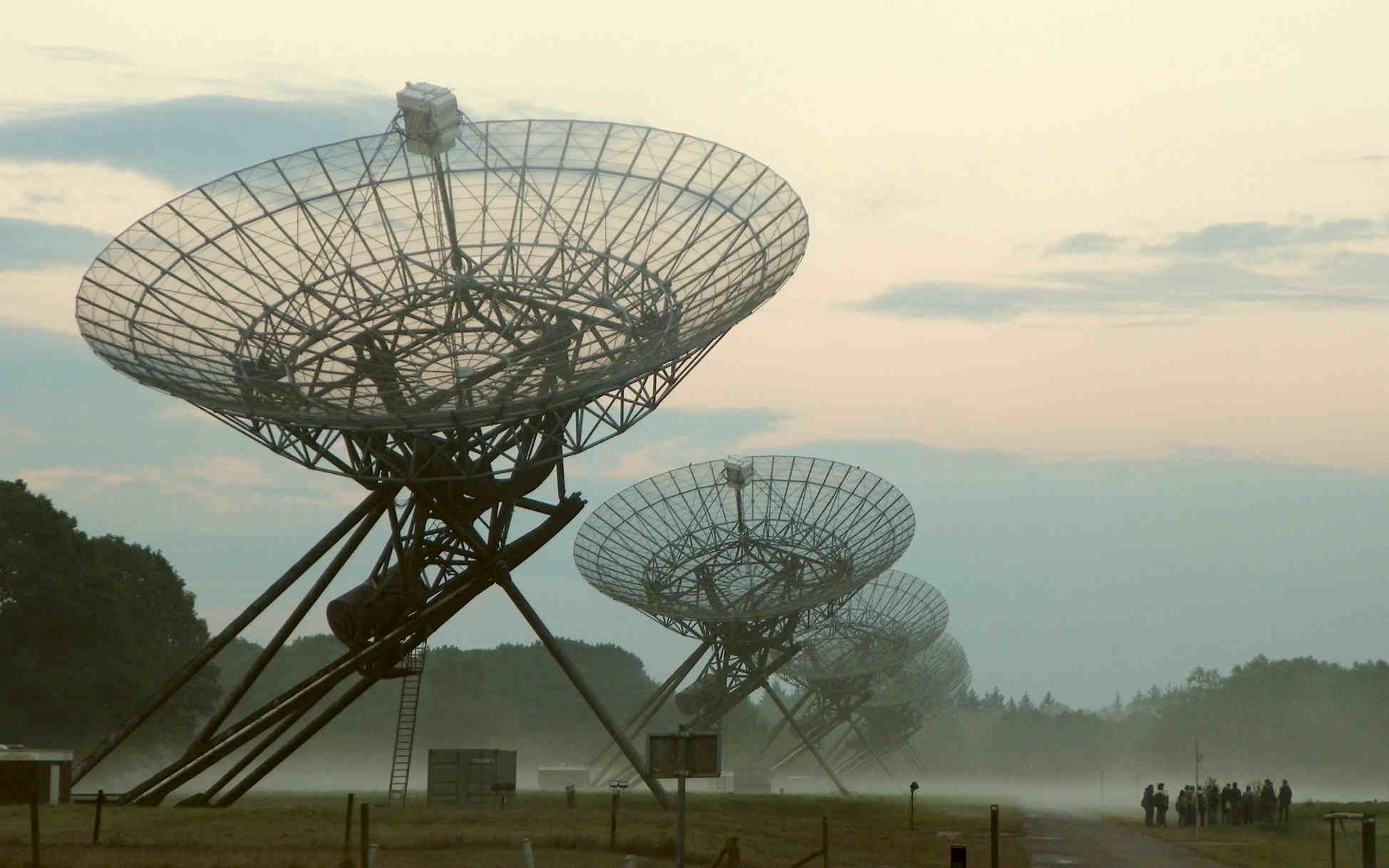}
	\caption[The Westerbork Synthesis Radio Telescope (WSRT).]{The Westerbork Synthesis Radio Telescope (WSRT) is composed of 14 antennas with a diameter of 25~m each. The largest separation between them is 2.7~km.}
	\label{fig:wsrt-photo}
\end{center}
\end{figure}
The Westerbork Synthesis Radio Telescope (WSRT; \citealt{baars1974})\footnote{\url{http://www.astron.nl/radio-observatory/astronomers/wsrt-astronomers}}, see Figure~\ref{fig:wsrt-photo}, is a linear array built near of Westerbork (The Netherlands) operated by the Netherlands Institute for Radio Astronomy (ASTRON), with support from the Netherlands Foundation for Scientific Research (NWO)\footnote{\url{http://www.nwo.nl}}. WSRT consists of 14 antennas of 25~m diameter each one spread in a linear configuration, east-to-west oriented, with a maximum baseline of 2.7~km. 10 antennas have a fixed location and are equally-spaced. The remaining 4 antennas can be moved to set different array configurations.

WSRT operates at frequencies from 330~MHz to 8.3~GHz, with dual polarization (recording linear polarization at all frequencies except at 2.3~GHz, where it records circular polarization). Given the linear shape of the array, only long observations (of about 12~h) can properly cover the $uv$-plane. The root-mean-square (rms) of a 12-hr observation lies in the range of $0.25$--$0.085$ $\mathrm{mJy\ beam^{-1}}$, at frequencies 0.33--8.3~GHz, respectively.
Given the redundancy of baselines, the recorded phases are stable enough to only require calibrator runs at beginning and end of the observation. Interleaved runs every few minutes, as typically happens in other arrays, are not necessary in WSRT observations.

Nowadays, WSRT is living its second youth, being the base for testing new detector prototypes for the Square Kilometre Array (SKA; \citealt{schilizzi2012})\footnote{\url{https://www.skatelescope.org}}. A new system which uses focal-plane arrays, called Apertif \citep{verheijen2008}, is being implemented in all dishes. Apertif will increase the field of view of WSRT and will allow to survey large areas of the sky much faster than with the previous system.

\subsection{Giant Metrewave Radio Telescope}

\begin{figure}[t]
\begin{center}
	\includegraphics[width=\textwidth]{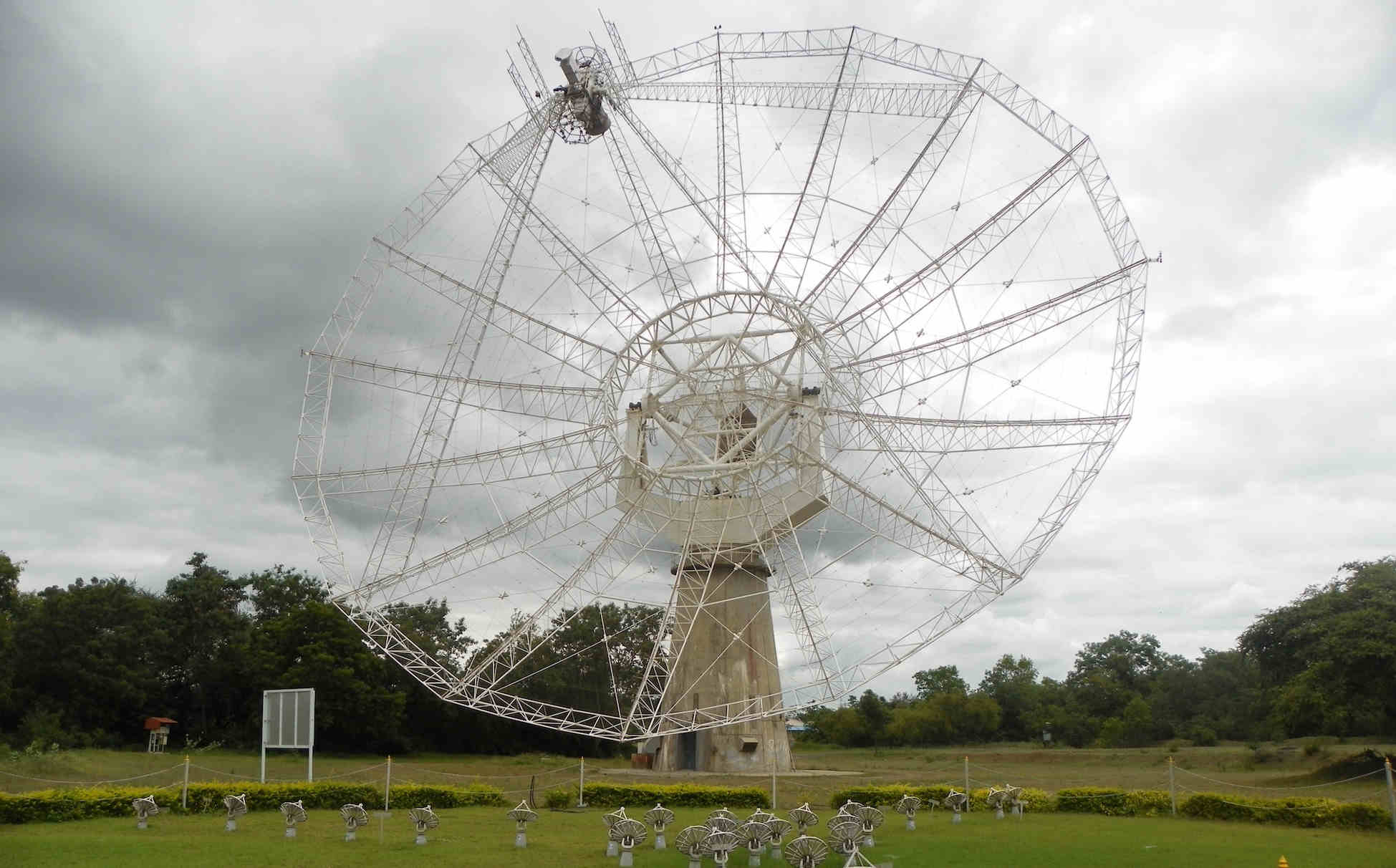}
	\caption[The Giant Metrewave Radio Telescope (GMRT).]{The Giant Metrewave Radio Telescope (GMRT) is composed of 30 antennas with a diameter of 45~m each. The largest separation between them is 25~km.}
	\label{fig:gmrt-photo}
\end{center}
\end{figure}
The Giant Metrewave Radio Telescope (GMRT; \citealt{ananthakrishnan1995})\footnote{\url{http://gmrt.ncra.tifr.res.in}}, see Figure~\ref{fig:gmrt-photo}, is a connected radio interferometer placed in a region about 80~km to the North of Pune, India, operated by the National Centre for Radio Astrophysics (NCRA)\footnote{\url{http://ncra.tifr.res.in/ncra}}, which is part of the Tata Institute of Fundamental Research (TIFR)\footnote{\url{http://www.tifr.res.in}}. GMRT consists of 30 dishes of 45-m diameter each. Fourteen of these dishes are located in a compact region of $\sim$$1~\mathrm{km^2}$. The other sixteen ones are spread over distances of up to 25~km along three arms describing roughly a Y-shaped configuration.

GMRT observes at six frequency bands centered at 50, 153, 233, 325, 610 and 1420~MHz, with a bandwidth of 16 or 32~MHz and dual polarization. Observations at two simultaneous frequency bands (235 and 610~MHz) are also available, using only one polarization at each frequency. GMRT can reach a resolution between 60 and 2~arcsec, considering the lowest and the highest frequency bands, respectively.

\subsection{LOw Frequency ARray}

\begin{figure}[t]
\begin{center}
	\includegraphics[width=\textwidth]{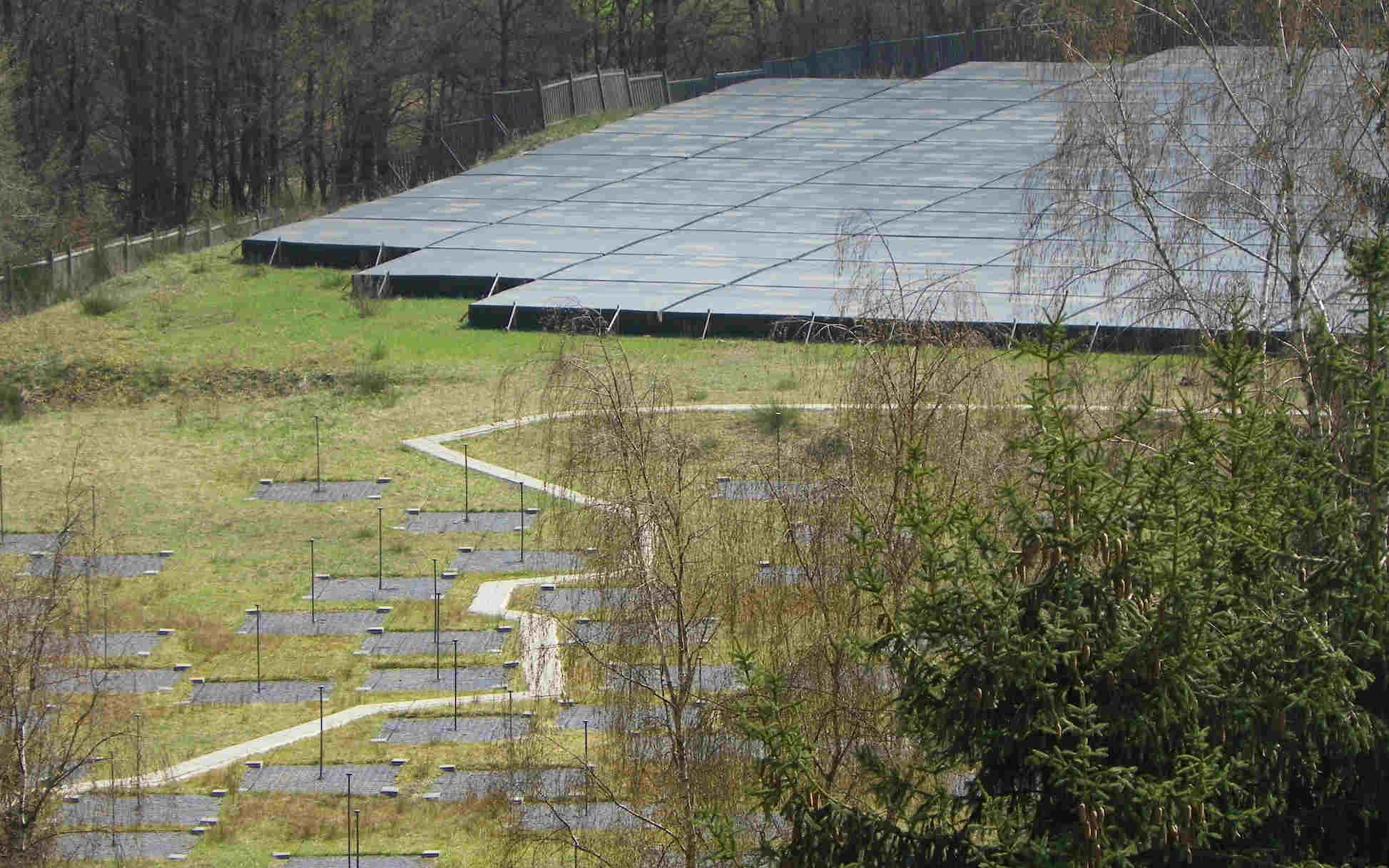}
	\caption[The Low Frequency Array (LOFAR).]{The Low Frequency Array (LOFAR) is a digital radio interferometer with stations in The Netherlands, United Kingdom, France, Germany, Sweden, Poland and Ireland. The photo shows the LOFAR station in Effelsberg, Germany. The Low Band Antennas are located in the front and the High Band Antennas in the back.}
	\label{fig:lofar-photo}
\end{center}
\end{figure}
The Low Frequency Array (LOFAR; \citealt{vanhaarlem2013})\footnote{\url{http://www.lofar.org}}, see Figure~\ref{fig:lofar-photo}, is a new generation digital radio interferometer that started its regular observations in 2012. LOFAR was designed and constructed by ASTRON, and is developed and operated by the International LOFAR Telescope (ILT) foundation under a joint scientific policy, which is a consortium of several institutions, universities and industrial parties from different European countries: The Netherlands, United Kingdom, France, Germany, Sweden, Poland and Ireland.

There are two distinct antenna types: the Low Band Antenna (LBA), which operates between 10 and 90~MHz and the High Band Antenna (HBA), which operates between 110 and 250~MHz. Both are low-cost omni-directional dipole antennas with no moving parts and each individual antenna observes all the sky at the same time. 
LOFAR consists of several stations spread over all the countries mentioned above. These stations are distributed in 24 core and 14 remote ones, all of them in The Netherlands, and 8 international stations (additional ones could be added in the future), displaying baselines from 100~m to 1\,500~km. Each core or remote station is composed of 48 HBAs and 96 LBAs, whereas the international stations consist of 96 HBAs and 96 LBAs.
LOFAR can theoretically reach a resolution of 0.65~arcsec and a sensitivity of $10~\mathrm{mJy\, beam^{-1}}$ in one hour of observation with the LBA and a resolution of 0.2~arcsec and a sensitivity of $0.3~\mathrm{mJy\, beam^{-1}}$ with the HBA, using the whole array.

LOFAR presents some science drivers to pursue fundamental science goals during its operation. These drivers, known as Key Science Projects, are the following ones: Epoch of Reionisation, deep extragalactic surveys, transient sources \citep{fender2008}, ultra high energy cosmic rays, solar science and space weather, and cosmic magnetism.
The Transients Key Science Project (TKP) conducts one ambitious project to investigate different sources for possible transients: the LOFAR Radio Sky Monitor (RSM; \citealt{fender2008}). This project will become as one of the deepest, largest-volume, and multi-epoch surveys dedicated to finding low-frequency transient objects ever performed.
The RSM regularly maps a large field of the sky to search for radio transient sources using the LBA and HBA antennas.

LOFAR is the strongest attempt of having a long baseline, high sensitivity array for the low frequencies, and is the most developed pathfinder of SKA.

\subsection{European VLBI Network}

\begin{figure}[t]
\begin{center}
	\includegraphics[width=0.6\textwidth]{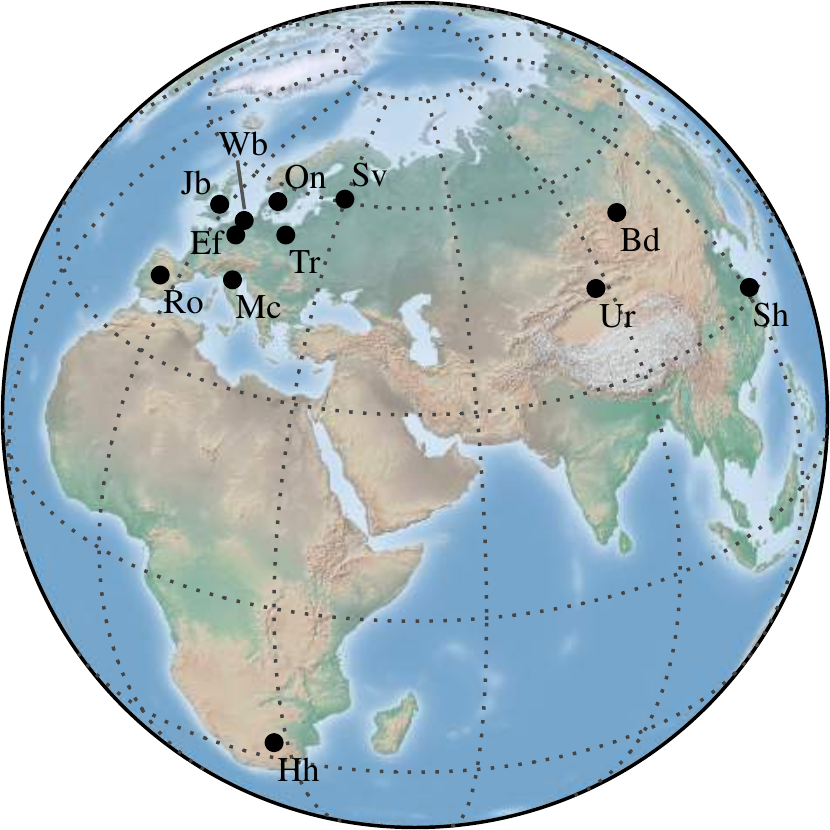}
	\caption[The European VLBI Network (EVN).]{Location of the antennas from the European VLBI Network (EVN) used in this work. Most of them are located in Europe, but some are located in Asia or Africa. The shown antennas are, ordered from West to East: Robledo (Ro), Jodrell Bank (Jb), Westerbork (Wb), Effelsberg (Ef), Medicina (Mc), Onsala (On), Torun (Tr), Hart (Hh), Svetloe (Sv), Urumqi (Ur), Badary (Bd) and Shanghai (Sh). The dotted lines denote the Earth longitude and latitude in intervals of 30$^{\degree}$.}
	\label{fig:evn-photo}
\end{center}
\end{figure}
The European VLBI Network (EVN; \citealt{zensus2014})\footnote{\url{http://www.evlbi.org}} is an European facility and the most sensitive VLBI array in the world. The network is composed of 21 telescopes, 17 of them located in Europe, three in Asia and one in South Africa. Figure~\ref{fig:evn-photo} shows the location of the antennas used in this thesis. The network is operated by several national agencies and institutes in Europe, China and South Africa. The Joint Institute for VLBI in Europe (JIVE)\footnote{\url{http://www.jive.nl}} was created to provide scientific and technical support for EVN observations.
In most EVN observations the data are recorded on each telescope, and then transported and correlated at JIVE. A mode in which the data are transferred from the telescopes to the correlator in real time (e-VLBI technique) is also available for EVN, known as e-EVN.

Effelsberg (Germany) is the largest dish, with 100-m diameter, all the antennas from WSRT are used as a single tied array, and most of the other antennas have diameters between 20--40~m.
The EVN can observe in the range of 1.6--22~GHz with a bandwidth of 64 MHz with all the mentioned antennas. Observations at 330~MHz, 610~MHz, 1.4~GHz, 15~GHz and 43~GHz can be performed with only part of the array.

\subsection{Australian Long Baseline Array}

\begin{figure}[t]
\begin{center}
	\includegraphics[width=\textwidth]{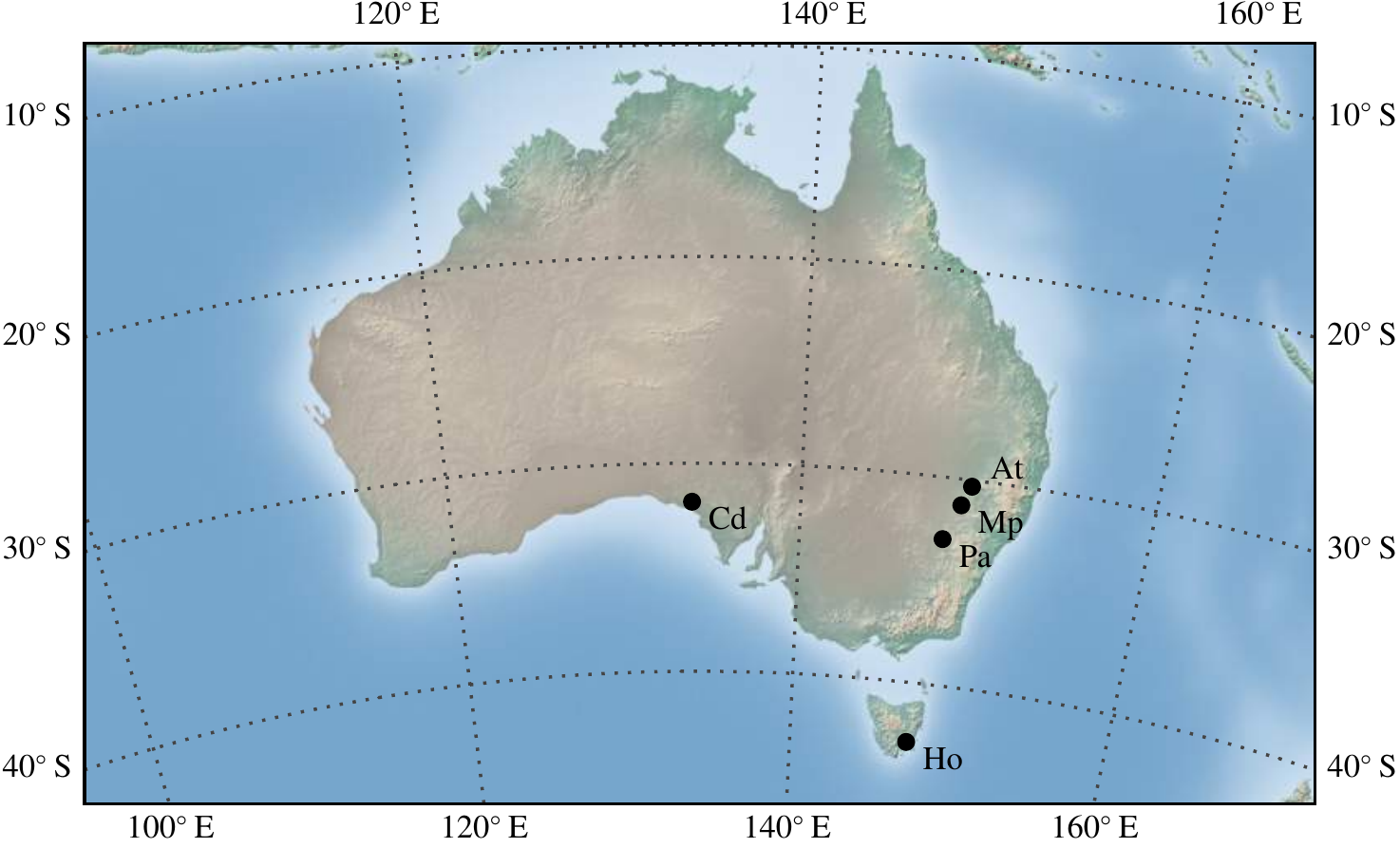}
	\caption[The Long Baseline Array (LBA).]{Location of the antennas from the Long Baseline Array (LBA) network that have been used in this work. The antennas shown are, ordered again from West to East:Ceduna (Cd), Hobart (Ho), Parkes(Pa), Mopra (Mp), and ATCA (At).}
	\label{fig:lba-photo}
\end{center}
\end{figure}
The Australian Long Baseline Array (LBA; \citealt{tzioumis1997})\footnote{\url{http://www.atnf.csiro.au/vlbi}} is the only VLBI network fully located in the southern hemisphere. The array combines antennas from different institutions which operate together as a National Facility managed by the Commonwealth Scientific and Industrial Research Organisation (CSIRO)\footnote{\url{http://atnf.csiro.au}}.
The core is located in the South-East of Australia and the full array is composed of the following antennas:
Parkes (64~m), the Australia Telescope Compact Array (ATCA; composed of 6 antennas of 22~m each), Mopra (22~m), all of them from the Australia Telescope National Facility (ATNF), and Hobart (26~m) and Ceduna (30~m), from the University of Tasmania. Additionally, more antennas can be requested into the array for special observations:  the Tidbinbilla antenna (70~m, from the Deep Space Network, NASA), the Hartebeesthoek antenna (26~m, from the National Research Foundation in South Africa), the Warkworth dish (12~m, from the Auckland University of Technology, New Zealand) and a single ASKAP antenna (from the Australian Square Kilometre Array Pathfinder). Figure~\ref{fig:lba-photo} shows the location of the antennas used in this thesis. The LBA, as in the case of EVN, presents two modes for data taking: record the data on each antenna and then transport them to the correlator, or correlate the data in real time. The LBA can observe at 2.3 and 8.4~GHz with all the antennas, and in the range of 1.4--22~GHz with most of them.
The raw data is transported to a correlator facility and correlated in real time. Since 2008, the main VLBI correlation for the LBA is done at the Curtin University of Technology using the DiFX software correlator, under a service agreement with ATNF.


\section{Data reduction and analysis}\label{sec:datareduction}

\subsection{A brief overview on radio data reduction}

The reduction of radio interferometric data includes the use of several common tasks, independently of the type of interferometer used, or the setup of the observation, and are briefly summarized in the following\footnote{All the data used along this thesis have been taken in continuum mode, and thus all the procedures mentioned here are referred to this type of observational mode.}. Almost all the data presented in this thesis have been calibrated and analyzed using standard procedures mainly within AIPS\footnote{The Astronomical Image Processing System, AIPS, is a software produced and maintained by the National Radio Astronomy Observatory (NRAO), a facility of the National Science Foundation operated under cooperative agreement by Associated Universities, Inc.
\url{http://aips.nrao.edu}}. Other tools such as Obit\footnote{\url{http://www.cv.nrao.edu/~bcotton/Obit.html}} \citep{cotton2008},
ParselTongue\footnote{\url{http://www.jive.nl/jivewiki/doku.php?id=parseltongue:parseltongue}} \citep{kettenis2006}
and SPAM\footnote{\url{https://safe.nrao.edu/wiki/bin/view/Main/HuibIntemaSpam}} \citep{intema2009} have also been used mainly to run scripts that call AIPS tasks to reduce the data. In the case of WSRT or recent VLA data (after its upgrade in 2011--2013), we have used CASA\footnote{The Common Astronomy Software Applications, CASA, is also a software produced and maintained by the NRAO. \url{http://casa.nrao.edu}} to perform the data reduction and analysis.

\subsubsection{Editing and flagging the data}

The raw visibilities obtained from the correlator usually exhibit data that are corrupted or useless. Actually, keeping these data would end up in a high noise in the final image, which would be thus of poor quality, or even in completely wrong results (hence the popular sentence: ``{\em no data are better than bad data}'').
These bad data, which must be flagged, can be a result of different problems encountered during the data taking. In particular, we could have data recorded while one or more antennas were not pointing to the expected source yet (known as telescope off-source data); instrumental problems, such as having one or some antennas that were not properly recording the data; or radio frequency interference (RFI): radio signals that are not coming from astrophysical sources but from different radio emitters, such as radio or tv channels, mobile phones, planes passing through the sky, or additional ground-based phenomena. Any of these phenomena can produce a signal that is orders of magnitude brighter than the signals coming from astrophysical sources. RFI is strongly dependent on the frequency: whereas at frequencies above $\sim$1~GHz we do not observe a relevant amount of RFI, at frequencies as low as 150~MHz we can loose an important fraction of the recorded data due to this effect, even up to $\sim$$30\%$ in some of the observations reduced in this thesis. Frequencies between 80 and 110~MHz are, for example, extremely contaminated of RFI because of the emission of radio channels.

In new-generation radio interferometer arrays, such as LOFAR, or in pipelines developed for arrays such as the VLA, automatic tasks are performed to remove the RFI. However, a manual examination and flagging is still required in all observations to completely remove the bad signals.

\subsubsection{Calibrating the data}

After removing the bad data, we need to calibrate the remaining ones. The raw amplitudes and phases of the visibilities exhibit in general slightly wrong values for each baseline, and we must recover the right ones by applying gain factors.
In this process, an {\em amplitude calibrator} is usually observed during a short run to properly scale the flux densities of the sources. The sources used as amplitude calibrators must be strong radio sources exhibiting a constant emission, well known at any epoch and frequency. This allow us to correctly scale the obtained raw amplitudes for all the observed sources. Typically, a {\em phase calibrator} is also observed. The sources observed as phase calibrators must be compact (ideally, point-like sources) and bright enough to be clearly detected. These sources must also be located close to the target source (a few degrees or less), being observed in short runs interleaved with the target source ones. During the calibration process, the phases are corrected for the phase calibrator, setting them to zero, and then the solutions are transferred to the target source.
Therefore, all phase solutions and astrometric positions are referred to the phase calibrator position. We note that the phase calibrator position is not the real position in the sky of the source used as phase calibrator, but the value set during the correlation as such position. This implies that in case of using a position slightly displaced from the real position of the phase calibrator (e.g.\ because at observing time the position of the phase calibrator was not well known), we will need to compensate this displacement after the calibration to obtain accurate astrometric measurements.
With the use of both, amplitude and phase calibrators, we can correct for amplitudes and phases in the target source data. For a more detailed description of the calibration process, see \citet{fomalont1999}.

In the case of VLBI observations, we usually reach resolutions of milliarcseconds. The only sources that are compact and bright enough are highly variable, therefore they cannot be used as amplitude calibrators. In these observations, the amplitudes are usually calibrated by fitting the system temperatures, $T_{\rm sys}$, of the receivers. In this fit we look for the solution that produces consistent visibility amplitudes between all antennas.
Two additional steps are also performed in VLBI data: ionospheric corrections and {\em fringe fitting}. Given that the antennas are separated long distances, we find completely different atmospheric conditions from one antenna to another, resulting as we mentioned in the previous sections, in different optical paths for the incoming signal. In the case of the ionosphere, nowadays we know the state of this layer at any location in the Earth thanks to GPS based measurements and we can compensate its effects in the data. The long separations between antennas also imply that any small error in their positions produces significant errors in the phases of the visibilities. To properly find phase and amplitude solutions we conduct a least-square fit, using the data from all baselines, which is known as {\em fringe fitting} \citep{cotton1995}.

\subsubsection{Cleaning and imaging}

Once we have flagged and calibrated the visibilities, in amplitude and phase, we are ready to obtain an image of the field of view. From equation (\ref{eq:visibilities}) we can see that the brightness distribution of the field of view can be obtained by applying an inverse Fourier transform.
However, the process is not straightforward given that the $uv$-plane is not fully covered. The most common algorithm to perform this process is CLEAN \citep{hogbom1974}. It assumes that the sky is mainly dark, and it is composed of a certain number of point-like sources. Then it subtracts a few per cent (typically $1$--$10\%$) of the brightest source in the field of view, assuming that it is a point-like source. Therefore, it subtracts the dirty beam centered on that position and normalized to the mentioned flux density. In next iterations, it subtracts recursively a few per cent of the brightest source. When no more sources are detected above the noise level, the algorithm stops. From the initial image, known as {\em dirty image}, we end up with an image, which basically shows the residual noise in the field of view, and a list of clean components, which consist of the positions of the point-like components that have been removed and the flux density removed in each one. To obtain the final image, and recover the flux density distribution of the sky, we add again all these components to the residual noise image. But this time we add the point-like sources, convolving them with the synthesized beam. It can be proven that the CLEAN algorithm obtains an accurate distribution of the sky, even with the presence of extended sources (which are interpreted as a combination of point-like sources).

Most of the interferometer arrays provide a non-uniform $uv$-coverage, with a larger number of antennas at short baselines and fewer at long baselines. During the cleaning process we can play with the weight assigned to different baselines. Two typical choices are used: {\em natural} or {\em uniform weighting}. The first one assigns constant weight to all visibilities, providing the optimal point-like source sensitivity in the image. The second one assigns weights inversely proportional to the density of visibilities in the $uv$-plane (i.e.\ to the local coverage of the $uv$-plane), producing a higher noise level but increasing the resolution of the final image. Intermediate weightings between these two choices are considered by using a Briggs robustness parameter, or robust, in the weighting \citep{briggs1995}. Different packages use different definitions of robust, but in AIPS a robust of $+5$ ($+2$ in CASA) is equivalent to a natural weighting, whereas a robust of $-5$ ($-2$ in CASA) is equivalent to a uniform weighting. A robust of zero is usually a good trade-off between sensitivity and resolution.

In general, the cleaning process is conducted manually. That is, at each step the user decides what regions of the sky must be cleaned, or when the cleaning process should stop. The largest difference with respect to an automatic cleaning appears when there are strong sources together with a poor $uv$-coverage. In these cases, several artifacts will be present in the final image, and automatic tasks are still not good enough to distinguish that they are not real sources and that they should not be cleaned as such.
However, in arrays such as LOFAR, where the amount of data is remarkable, a manual cleaning is usually unfeasible.

\subsubsection{Self-calibration}

Once we have obtained the cleaned image, the calibration process ends. However, this image is not the most optimal one. We can still produce a more accurate one with a lower noise. The initial calibration has transferred the solutions from the amplitude and phase calibrators to the target source, which is located in a different direction in the sky. Therefore, there can be effects that have not been properly corrected. The next step is to use the model of the sky produced during the cleaning process (i.e. the list of clean components) to calibrate again the data, in a process known as {\em self-calibration}.

New solutions assuming that the cleaned model is the true flux density distribution of the sky are computed. At this step, it is very important to consider a correct model. An incorrect model will produce errors in the solutions and thus in the new images. For that reason, it is recommended that the model created during the first cleaning considers only unambiguous components (i.e.\ no sources in the detection limit, or components that could be artifacts). The first model is an {\em approximate} model of the sky. We calibrate the data based on that model, usually correcting only the phases, and clean them again. Additional cycles of self-calibration/cleaning can be carried out, each time with a more accurate model of the sky. A self-calibration correcting the amplitudes can also be performed. However, this step is a bit more risky, as it can introduce wrong flux density values in the image. Therefore, it is only performed when we are sure that the clean model is accurate enough, and we check after this calibration that the flux densities of the sources have not significantly changed. After some cycles we will not improve anymore the model or the final image, and we finish the data reduction.

For data at high frequencies, one or two self-calibration cycles are enough to obtain the optimal results. However, at low frequencies, because of the large field of view of the images, we usually conduct about five or six self-calibration cycles.

\subsubsection{Measuring flux densities} \label{sec:reduction-measuringfluxes}

The final data should show, if the calibration has been done without problems, the most accurate image of the observed field of view, with correct flux density values. Now we need to estimate the noise level of the image, or of the part of the image in which we are interested, and the flux densities from the detected sources. The two used packages (AIPS and CASA) provide two main different methods to measure flux densities. {\tt tvstat} in AIPS, which is called {\tt imstat} in CASA, allows us to define an arbitrary region in the image and obtain the statistics from this region. Among other parameters, we obtain the maximum and minimum flux density from the region, the mean or median value, the integrated flux density, the standard deviation, or the rms for the whole region. These tasks are useful to estimate the rms of the whole image or around an interesting source. They can also be used to determine the flux density from any source that can not be fit by a gaussian component. The other method is performed by the task {\tt jmfit} in AIPS, or its equivalent {\tt imfit} in CASA. These tasks fit a gaussian component to the predefined region. In the case of the existence of a point-like source on it, we obtain the parameters from the gaussian fit to the source: the peak and the integrated flux density, and the position of the peak. Therefore, with this fit we obtain the position of the source and its flux density.

\subsection[$T_{\rm sys}$ corrections in the GMRT data]{\boldmath{$T_{\rm sys}$} corrections in the GMRT data}\label{app:tsys}

The power measured by a radio antenna contains the radio emission from a wide number of sources. It is common to separate this power in two components: $P\,\propto\,T_{\rm a} + T_{\rm sys}$, where $T_{\rm a}$ is the {\em antenna temperature}, which includes the contribution from the target sources, and \tsys is the {\em system temperature}. \tsys includes the contributions from the Galactic diffuse emission ($T_{\rm sky}$ or sky brightness temperature), the receiver noise, feed losses, spillover and atmospheric emission, all of them considered as the effective receiver temperature, $T_{\rm erc}$ \citep{crane1989,sirothia2009}:
\begin{equation}
	T_{\rm sys}=T_{\rm sky} + T_{\rm erc} \label{eq:tsys}
\end{equation}
While at high frequencies the contribution of the Galactic diffuse emission, $T_{\rm sky}$, is negligible compared to $T_{\rm erc}$, it becomes dominant at low frequencies, thus producing very high system temperatures. This is particularly relevant when observing close to the Galactic plane. 

To recover the emission from the target sources, \tsys must be determined and removed from the measured power $P$. For single-dish antennas and most interferometers, \tsys can be accurately determined through injected calibration signals and is automatically subtracted for each antenna. 
In the case of GMRT, the observations are, in general, conducted with the Automatic Gain Controller (AGC) enabled to avoid possible limitations such as saturation produced by RFI or excess power. However, this rescales the gain of the system in the presence of strong emission (such as the one of the Galactic plane).
In these kind of observations, the determination of \tsys is not accurate enough and it requires to be accounted for during the calibration of the radio data. The amplitudes of the complex visibilities are proportional to the ratio of the antenna temperature to the system temperature ($T_{\rm a}$/\tsys). In a typical radio observation, the amplitude calibrator allows us to recover the real flux densities of the sky and the target source. In the case of GMRT we will end up with the following visibilities for the calibrator (C) and the target source (S):
\begin{equation}
	V_{\rm C} \,\propto\, \frac{T_{\rm a} ({\rm C})}{T_{\rm sys} ({\rm C})},      \qquad V_{\rm S} \,\propto\, \frac{T_{\rm a} ({\rm S})}{T_{\rm sys} ({\rm S})}
\end{equation}
During the calibration process, the amplitudes of the complex visibilities are correctly scaled (or calibrated) to flux densities by applying a scaling factor $G\,\propto\,T_{\rm sys}({\rm C})$. The scaling factor is determined with the amplitude calibrator, which presents a known constant flux density along the time. Therefore, in this process the contribution of $T_{\rm sys}({\rm C})$ is evaluated and properly removed.
Schematically, after the calibration process we recover the calibrated visibilities, $\tilde{V}$, of the amplitude calibrator:
\begin{equation}
	\tilde{V}_{\rm C} = G\, V_{\rm C}
\end{equation}
However, for the target source we apply the same scaling factor:
\begin{equation}
	\tilde{V}_{\rm S} = G\, V_{\rm S}
\end{equation}
which will only provide properly calibrated visibilities for the target source, $\tilde{V}_{\rm S}$, if $T_{\rm sys}({\rm S}) = T_{\rm sys}({\rm C})$.
For target sources at high Galactic latitudes and/or at high frequencies the Galactic diffuse emission is negligible, and thus \tsys is similar for C and S\footnote{Provided that other effects dependent on the antenna position, such as elevation, are considered during the calibration process.}.
For target sources close to the Galactic plane and/or at low frequencies
$T_{\rm sys}({\rm S}) \gg T_{\rm sys}({\rm C})$ (because the amplitude calibrator is never located close to the Galactic plane). In this case, we will need to quantify the ratio $T_{\rm sys}({\rm S}) / T_{\rm sys}({\rm C})$, or \tsys correction, to obtain the correct flux density values in the target source field.
Since the gamma-ray binaries LS~5039 and LS~I~+61~303 lie very close to the Galactic plane, an accurate correction of \tsys is mandatory to properly estimate the flux densities of these sources at each frequency (see Chapters~\ref{chap:ls} and \ref{chap:lsi}).

\subsubsection{The Haslam approximation}

A common method to implement this correction is based on estimating the sky temperature values, $T_{\rm sky}$, from the measurements made by the all-sky 408~MHz survey, which was conducted with several single-dish radio telescopes \citep{haslam1982}.

Assuming that the Galactic diffuse emission follows a power-law spectrum, we can determine the sky temperature at a frequency $\nu$ for any position in the sky ($\alpha,\delta$) from the emission reported at 408~MHz by
\begin{equation}
    T_{\mathrm{sky}}^{\,(\nu)} (\alpha,\delta) = T_{\mathrm{sky}}^{\mathrm{(408)}} (\alpha,\delta)\,\left(\frac{\nu}{\mathrm{408~MHz}}\right)^{\gamma}\label{eq:tsky}
\end{equation}
The spectral index for the Galactic diffuse emission is usually assumed to be $\gamma = -2.55$ \citep{roger1999}.
The sky brightness at the positions of the amplitude calibrator and the target source, convolved with the synthesized beam of our interferometer, allows us to estimate the ratio of the sky temperatures between both positions. From equation~(\ref{eq:tsys}) we have seen that \tsys can be divided in two terms, with $T_{\rm sky}$ being dominant at low frequencies and/or close to the Galactic Plane. The $T_{\rm erc}$ values can be obtained from the tabulated data available in the specification documents of GMRT and the $T_{\rm sky}$ values from equation~(\ref{eq:tsky}). With these data we determine the $T_{\rm sys}({\rm S}) / T_{\rm sys}({\rm C})$ ratio, which is the \tsys correction.

However, this method has some problems. First, it assumes a constant spectral index for the Galactic diffuse emission across all the sky. Secondly, it assumes that the response of our instrument to the Galactic diffuse emission is the same as those of the radio telescopes used for the Haslam survey. Thirdly, it does not take into account the \tsys dependence on the elevation of the Galactic diffuse emission.

\subsubsection{Direct measurement of the self-power for each antenna}

A more accurate method to implement the \tsys correction should only involve measurements conducted with the same radio telescope.
In the case of an interferometer, one can obtain the power measured at any given frequency and elevation for each antenna in the array by self-correlating the corresponding data from that antenna.
Although the sky brightness is the same for all the antennas, the power that comes from $T_{\rm erc}$ is different for each antenna, given its internal origin. Thus, considering the full array as a collection of isolated single-dish antennas, the problem is reduced to determine the power of each antenna separately, the so-called self-power, and then average all the self-powers to derive the power of the interferometer.

The self-correlated data of each antenna is proportional to the system temperature. Therefore, the ratio between the self-powers for each antenna for the target source and the amplitude calibrator is equal to the ratio between the system temperatures at these two positions. This ratio is thus a more accurate measurement of $T_{\rm sys}({\rm S}) / T_{\rm sys}({\rm C})$ than the Haslam method to estimate the contribution from the Galactic diffuse emission.

As GMRT can not produce self-correlated data at the same time than standard correlated data, additional observations have to be conducted. The self-power for each antenna is measured at the position of the flux calibrator and the position of the source with the same configuration than in the standard correlated observations. 
To avoid possible dependencies with the elevation of the sources, both types of observations should be conducted when the sources are at similar elevations.

The self-correlated data show stronger RFI than the standard correlated data, and thus a large number of channels must be removed through a dedicated flagging process.
The obtained amplitudes in different channels show a large dispersion. For that reason, it is recommended to take first the average of the ratios for each antenna and each channel, to minimize the dispersion, and then average all the ratios. The use of the median, instead of the mean, is also recommended and it has been used along our data reduction process. The derived value of the ratio is the \tsys correction that must be applied to the image of the target source field to obtain the correct flux density values.

The main source of error in this method is the large dispersion of all the self-powers measured. This can be reduced by accumulating more data obtained in different epochs, because the Galactic diffuse emission is persistent.

\citet{sirothia2009} determined the \tsys corrections for GMRT with direct measurements of the full sky at 240~MHz, and compared the obtained results with the Haslam approximation at that frequency. Although the two methods report similar values for the whole map, these authors found a high scatter in the comparison of the corrections obtained using both methods, with an rms in these differences of $\sim$56\%. Dependencies on the elevation and diurnal/nocturnal time have also been found.

\begin{table}[!t]
\small
\caption[\tsys corrections applied for the GMRT data.]{\tsys corrections ($C$) determined in this work for the position of LS~5039 and LS~I~+61~303 using the amplitude calibrators 3C~286 and 3C~48 with the Haslam approximation ($C_{\rm H}$) and with the direct measurement of the self-powers ($C_{\rm SP}$). Uncertainties are reported at 1-$\sigma$ level. In this work we have used $C_{\rm SP}$ to correct our data.}
\label{tab:tsys-corrections}
\centering
\begin{tabular}{c@{\hspace{+20pt}}c@{\hspace{+20pt}}c@{~~}c@{\hspace{+20pt}}c@{\hspace{+20pt}}c}
	\hline\\[-10pt]
	$\nu$ & \multicolumn{2}{c}{LS~5039} & & \multicolumn{2}{c}{LS~I~+61~303}\\[+2pt]
	\cline{2-3}\cline{5-6}\\[-9pt]
	$(\mathrm{GHz})$ & $C_{\rm H}$ & $C_{\rm SP}$ && $C_{\rm H}$ & $C_{\rm SP}$\\[+2pt]
	\hline\\[-10pt]
	154 & $\sim$4.0 & $1.9\pm0.2$ & & $\sim$1.4 & $1.5 \pm 0.2$\\
	235 & $\sim$3.0 & $1.9\pm0.3$ & & $\sim$1.4 & $1.522 \pm 0.016$\\
	610 & $\sim$1.6 & $1.39\pm0.06$&& $\sim$1.1 & $1.328 \pm 0.010$\\
	\hline
\end{tabular}
\end{table}
Table~\ref{tab:tsys-corrections} shows the \tsys corrections ($C$) found with the two discussed methods for the two gamma-ray binaries observed with GMRT in this thesis: LS~5039 and LS~I~+61~303.
In the case of LS~5030, we observe similar corrections at 610~MHz, but different ones at 235~MHz (of about 60\%), although the $C_{\rm H}$ and $C_{\rm SP}$ values are roughly compatible at 3-$\sigma$ level in both cases. At 154~MHz the corrections are very different and clearly incompatible. We have seen that the Haslam approximation overestimates the \tsys correction for the field of LS~5039 at all frequencies. At 235~MHz we have obtained an overestimation by an amount in agreement with the rms observed by \citet{sirothia2009} at 240~MHz. As the direct measurements of the self-powers provide more reliable values of the \tsys corrections, and allow us to compute the corresponding uncertainties, we have used $C_{\rm SP}$ to correct our GMRT data.

In the case of the gamma-ray binary LS~I~+61~303, we observe a smaller contribution of the Galactic emission, and both methods provide similar values. However, with the direct measurements we obtain more precise corrections, which have been the ones considered in this thesis.

%
%
%

\chapter{Radio emission and absorption processes} \label{chap:emission}

In this Chapter we summarize the equations considered in the thesis for the emission and absorption processes. First, we introduce some fundamental magnitudes such as the luminosity, intensity or flux in Sect.~\ref{sec:fundamental-magnitudes}, and we derive the flux density of an emitting region as a function of its emissivity and absorption coefficients in Sect.~\ref{sec:radiative-transport}.
Second, we present the most common emission processes observed at radio frequencies relevant to this work in Sect.~\ref{sec:emission-processes}. Finally, we present the absorption mechanisms considered in this thesis in Sect.~\ref{sec:absorption-processes}. We note that the notation and derivations of the equations presented here are based on \citet{longair2011} and \citet{rybicki1979}.

\section{Fundamental magnitudes} \label{sec:fundamental-magnitudes}

The {\em luminosity}, $L$, is defined as the energy irradiated per time unit by any celestial source. It is thus the power emitted by the source, measured in $\mathrm{erg\ s^{-1}}$ in the centimeter-gram-second, cgs, system (or in W in the International System of Units, SI). The {\em monochromatic luminosity}, $L_\nu$, is the luminosity per unit of frequency $\nu$, and both can be obviously related through:
\begin{equation}
    L  = \int_0^\infty L_\nu\, \D \nu. \label{eq:luminosity}
\end{equation}
For a source located at a distance $D$, we define the {\em flux} as the power received per area unit:
\begin{equation}
	F_\nu = \frac{L_\nu}{4\pi\, D^2}.
\end{equation}

The {\em surface brightness}, or {\em intensity}, $I_\nu$, is defined as the power received (or emitted) per area unit and per solid angle unit. Therefore, it can be related to the flux by
\begin{equation}
    F_\nu = \int_{\Omega} I_\nu \cos \theta\ \D \Omega,
\end{equation}
where $\theta$ denotes the incident angle between the direction of the particles and the normal to the incident area, and we integrate over the solid angle $\Omega$. We note that we can approximate this relation to $F_\nu \approx \Omega\, I_\nu$ for a source covering a small solid angle.

%
%
%
%

\subsection{Radiative transport} \label{sec:radiative-transport}

The radiation usually crosses a certain amount of material before reaching the Earth. In this path, part of the radiation can be absorbed. We can define the {\em absorption coefficient}, $\kappa_\nu$, as the decrease of the intensity when the radiation passes through a path of length $\D s$:
\begin{equation}
    \D I_\nu = - \kappa_\nu\, I_\nu\, \D s.
\end{equation}
At the same time, the source is radiating with an {\em emissivity}\footnote{The emissivity is defined as the energy radiated per volume unit, per time unit, per frequency unit, and per solid angle unit.} $J_\nu$, and thus contributes to increase the intensity by
\begin{equation}
    \D I_\nu = J_\nu\, \D s.
\end{equation}
Therefore, the equation of the radiative transport becomes:
\begin{equation}
    \frac{\D I_\nu}{\D s} = J_\nu - \kappa_\nu\, I_\nu,
\end{equation}
which can be integrated to obtain the incident intensity as:
\begin{equation}
    I_\nu = \frac{J_\nu}{\kappa_\nu}\left( 1 - e^{-\kappa_\nu\, \D s} \right).
\end{equation}
It is convenient here to introduce the {\em opacity}, $\tau_\nu$, defined as
\begin{equation}
	\tau_\nu = \int_0^{\ell} \kappa_\nu \D s
\end{equation}
Assuming an isotropic emitting source subtending a small solid angle $\Omega$, the {\em flux density} results:
\begin{equation}
    S_\nu = \int_{\Omega} I_\nu\, \D \Omega \approx \Omega \frac{J_\nu}{\kappa_\nu}\left( 1 - e^{-\kappa_\nu \ell} \right), \label{eq:flux-density}
\end{equation}
where $\ell$ is the linear size of the source.
At radio frequencies, due to the extremely low values obtained for common sources, the flux density is usually measured in Jansky (Jy) units, which are defined as
$$1~\mathrm{Jy} \equiv 10^{-23}~\mathrm{erg\ cm^{-2}\ s^{-1}\ Hz^{-1}} = 10^{-26}~\mathrm{W\ m^{-2}\ Hz^{-1}}.$$

\section{Radio emission} \label{sec:emission-processes}

The radio emission observed from astrophysical sources can be produced by a wide number of mechanisms. In this section we detail the ones that are relevant for the work presented in this thesis. Given that we are exclusively focused in the radio continuum emission, we will not mention emission or absorption processes corresponding to transition lines in atoms or molecules.

\subsection{Thermal black body emission} \label{sec:bb-emission}

We must start with one of the oldest known radiative processes, the black body (BB) emission. A BB is defined as an ideal body that absorbs all the incoming electromagnetic energy. If this body is in thermal equilibrium with its medium, it must also radiate energy (otherwise it will not be in thermal equilibrium as it is absorbing energy).
Although the BB is an ideal case, most of the thermal bodies in the Universe are very close to a BB and this is an accurate approximation.

Every object with a temperature $T$ thus radiates electromagnetic energy according to the {\em Planck's law of black-body radiation}:
\begin{equation}
    I_\nu(T) = \frac{2 h \nu^3}{c^2} \frac{1}{e^{\,h\nu / k_{\rm B}T} - 1},
\end{equation}
where $I_\nu(T)$ is the power radiated per area unit, per solid angle unit, per frequency unit by a body at a temperature $T$, $h$ is the Planck constant, $c$ is the speed of light in the vacuum, $k_{\rm B}$ is the Boltzmann constant, and $\nu$ the frequency.
We note that the frequency at which the maximum emission takes place is given by the {\em Wien's displacement law}:
\begin{equation}
    \nu_{\rm max} \approx \frac{1}{4.965} \frac{h}{k_{\rm B} T}.
\end{equation}
For typical main-sequence stars (with temperatures between 2\,000--50\,000~K), the maximum of the emitted radiation peaks in the optical range. To obtain the maximum emission at radio frequencies, the source must be at a temperature $\lesssim 10~\mathrm{K}$. Therefore, only some really cold molecular clouds or the Cosmic Microwave Background (CMB) are objects where their BB emission peaks at radio frequencies.

In general, we observe sources with $T \gg 100~\mathrm{K}$. Given that at radio frequencies $\nu \sim 10^9~\mathrm{GHz}$, we obtain $\frac{h \nu}{k_{\rm B} T} \sim 4.8 \times 10^{-4} \ll 1$. This allow us to approximate the Planck's law by the Rayleigh-Jeans law:
\begin{equation}
    I_\nu (T) \approx \frac{2 k_{\rm B} T \nu^2}{c^2}\ \propto\ \nu^2.
\end{equation}
Even for the CMB, which displays a temperature $\approx$$2.7~\mathrm{K}$, and is the coolest observable object in the Universe in thermal equilibrium, we obtain values of $\frac{h \nu}{k_{\rm B} T} \approx 0.018$ at 1~GHz (or 0.18 at 10~GHz). The Rayleigh-Jeans approximation is thus valid for almost any radio observation in where we detect BB thermal emission. Therefore, we observe this emission at radio frequencies as a quadratic function of the frequency.


\subsection{Synchrotron emission} \label{sec:synchrotron-emission}

When the first spectra at radio frequencies were taken, an inverted spectrum was observed, i.e. $\propto\, \nu^{\alpha}$ with $\alpha < 0$, contrary to the $\propto\, \nu^2$ expected from a thermal BB emission. This was the first and clear signature that non-thermal emission processes were taking place.

The most common non-thermal process observed at radio frequencies is the synchrotron emission, produced by relativistic electrons moving in presence of a magnetic field $\vec B$. These electrons are accelerated by $\vec B$, producing a helicoidal movement and thus a continuous angular acceleration. The force produced by the magnetic field over the particle is known as the {\em Lorentz force}:
\begin{equation}
	\dot{\vec p} = \frac{e}{c} \vec v \times \vec B,
\end{equation}
where $\vec p = \gamma m_{\rm e} \vec v$, $e$ is the electron charge, $m_{\rm e}$ is the electron mass, $\vec v$ is the velocity of the electron, and $\gamma$ the {\em Lorentz factor}, which is defined as
\begin{equation}
	\gamma \equiv \frac{1}{\sqrt{1 - (v/c)^2}} = \frac{1}{\sqrt{1 - \beta^2}}.
\end{equation}
While this particle is being accelerated, it emits electromagnetic radiation with a power \citep{longair2011}
\begin{equation}
	P_{\rm syn} = \frac{2 e^2}{3c^3} \gamma^2 \frac{v_\perp^2 e^2 B^2}{m_{\rm e}^2 c^2} = 2 \beta^2 \gamma^2 c \sigma_{\rm T} U_{\rm B} \sin^2\alpha,
\end{equation}
where we have introduced the {\em Thomson-cross section} $\sigma_{\rm T} = \frac{8\pi e^4}{3 m_{\rm e}^2 c^4}$, and the energy density of the magnetic field $U_{\rm B} = B^2 / 8\pi$. Assuming an isotropic velocity distribution, we can integrate over all directions, obtaining $\langle \sin^2\alpha \rangle = 2/3$, and thus
\begin{equation}
	\langle P_{\rm syn} \rangle = \frac{4}{3}\beta^2 \gamma^2 c\, \sigma_{\rm T} U_{\rm B} \approx 2.56 \times 10^{-14}\left( \frac{B^2}{8\pi} \right) \left( \frac{E}{m_{\rm e} c^2} \right)^2.
\end{equation}

In many astrophysical scenarios we observe a particle population that exhibits a power-law energy distribution. In the general case, we assume that the energy distribution is given by:
\begin{equation}
    N(E) \D E = K E^{-p} \D E,
\end{equation}
where $N(E)$ is the number of particles with an energy between $E$ and $E + \D E$, and $K$ is the normalization of the spectrum.
The emission from this distribution of particles results \citep{longair2011}:
\begin{equation}
    J_\nu = a(p)\frac{\sqrt{3} e^3 B K}{(4\pi)^2 \epsilon_0\, c\, m_{\rm e}} \left( \frac{3 e B}{2\pi\, m_{\rm e}^3 c^4} \right)^{(p-1)/2} \nu^{(1-p)/2},
\end{equation}
where
\begin{equation}
    a(p) = \frac{\sqrt{\pi}}{2} \frac{\Gamma\left( \frac{p}{4}+\frac{19}{12}\right)\Gamma\left( \frac{p}{4}-\frac{1}{12}\right)\Gamma\left( \frac{p}{4}+\frac{5}{4}\right)}{(p+1)\Gamma\left( \frac{p}{4}+\frac{7}{4}\right)}.
\end{equation}
For typical values of $p \sim 2$, we obtain the typical synchrotron spectrum, where the emission is $\propto\ \nu^{-1/2}$. It is usual to define the spectral index as $\alpha \equiv (1 - p)/2$, and then $J_\nu\ \propto\ \nu^{\alpha}$.

\section{Absorption processes at low frequencies} \label{sec:absorption-processes}

\subsection{Synchrotron self-absorption} \label{sec:ssa}

The synchrotron emission is also balanced with an absorption process at low frequencies, the so-called {\em synchrotron self-absorption} (SSA). As the synchrotron emission increases at lower frequencies, at a certain point the absorption mechanisms become dominant over emission, producing a cutoff in the spectrum.
The absorption coefficient, $\kappa_{\nu}^{\rm SSA}$, for a random magnetic field results in \citep{longair2011}:
\begin{equation}
    \kappa_{\nu}^{\rm SSA} = \frac{\sqrt{3} e^3 c}{8\pi^2 \epsilon_0 m_{\rm e}}K B^{(p+2)/2}\left( \frac{3e}{2\pi m_{\rm e}^3 c^4} \right)^{p/2} b(p)\, \nu^{-(p+4)/2},
\end{equation}
where
\begin{equation}
    b(p) = \frac{\sqrt{\pi}}{8} \Gamma\left( \frac{3p+22}{12}\right)\Gamma\left( \frac{3p+2}{12}\right)\Gamma\left( \frac{p+6}{4}\right)\Gamma^{-1}\left( \frac{p+8}{4}\right).
\end{equation}
From these equations we can we can determine the SSA opacity as
\begin{equation}
    \tau_{\nu}^{\rm SSA} = 3.354 \times 10^{-11} (3.54 \times 10^{18})^{p} K B^{(p+2)/2} b(p) \nu_{\rm GHz}^{-(p+4)/2}\, \ell_{\rm AU}.
    \label{eq:taussa}
\end{equation}
We note that in the optically thick part of the spectrum, we obtain a flux density
\begin{equation}
	S_\nu\ \propto\ \nu^{5/2},
\end{equation}
a positive spectral index, independent of the value of $p$.

\subsection{Free-free absorption} \label{sec:ffa}

When radiation passes through a plasma, part of the radiation is absorbed by the charged particles in the plasma, the so-called {\em free-free absorption} or {\em Bremsstrahlung} (FFA). In this case, the absorption coefficient is related to the properties of the plasma as \citep{rybicki1979}:
\begin{equation}
    \kappa_{\nu}^{\rm FFA} \approx 30 \dot M_{-7}^2 \nu_{\rm GHz, max}^{-2} \ell_{\rm AU}^{-4} v_{\rm W, 8.3}^{-2} T_{\rm W, 4}^{-3/2} \quad {\rm AU^{-1}}
\label{eq:kappaff}
\end{equation}
where $\dot M_{-7}$ is the mass-loss rate of the companion star in a binary system in units of $10^{-7}~\mathrm{M_{\sun}\ yr^{-1}}$, $\nu_{\rm GHz, max}$ is the frequency at which the maximum emission takes place (or turnover frequency), in GHz units, $\ell_{\rm AU}$ is the radius in the spherically symmetric case, measured in AU, $v_{\rm W, 8.3}$ is the velocity of the stellar wind, in units of $2 \times 10^{8}~\mathrm{cm\ s^{-1}}$ and $T_{\rm W, 4}$ is the wind temperature, in units of $10^4~\mathrm{K}$.
From this relation we can obtain directly the free-free opacity:
\begin{equation}
    \tau_{\nu}^{\rm FFA} \approx 30 \dot M_{-7}^2 \nu_{\rm GHz, max}^{-2} \ell_{\rm AU}^{-3} v_{\rm W, 8.3}^{-2} T_{\rm W, 4}^{-3/2}.
\label{eq:tauff}
\end{equation}

\subsection{Razin effect} \label{sec:razin-effect}

The Razin-Tsytovitch effect (also known as Tsytovitch-Eidman-Razin or simply Razin effect, as we will refer to it in the following; \citealt{hornby1966}), is an absorption effect produced when the relativistic particles pass through a non-relativistic, or thermal, plasma. The presence of this plasma implies a refractive index $n < 1$, that can be related with the plasma frequency, $\nu_{\rm p}$, by:
\begin{equation}
    n^2 = 1 - \left( \frac{\nu_{\rm p}}{\nu} \right)^2,\qquad \nu_{\rm p} \equiv \frac{n_{\rm e} e^2}{\pi m_{\rm e}},
\end{equation}
where $n_{\rm e}$ is the electron density.

Meanwhile, the synchrotron emission from relativistic particles is strongly beamed along the direction of motion of these particles, producing a boosting of the emitted power $\propto\ \gamma^2$. This emission is then concentrated only around a small angle along the direction of motion given by $\theta_{\rm b}\ \propto\ \gamma^{-1}$. However, this $\gamma$ factor depends on the refractive index, and therefore, this beaming effect can be written as
\begin{equation}
    \theta_{\rm b} \sim \gamma^{-1} = \sqrt{1 - n^2 \beta^2}.
\end{equation}
At low radio frequencies we approach the plasma frequency, and the refractive index drops quickly to zero. In that case, the beaming effect is suppressed ($\theta_{\rm b} \approx 1$). We observe then a large decrease in the synchrotron emission. This decrease produces a quasi-exponential cutoff, which can be approximated by an exponential cutoff in the emissivity \citep{dougherty2003}:
\begin{equation}
    J_\nu \leadsto J_\nu\, e^{-\nu_{\rm R}/\nu}, \qquad  \nu_{\rm R} \equiv 20 n_{\rm e} B^{-1}. \label{eq:razin-effect}
\end{equation}
We note that at high frequencies this effect is irrelevant given that $n \sim 1$.

The Razin effect is widely observed in solar bursts studies or in systems such as colliding wind binaries, where the presence of a thermal plasma (the stellar wind) is significant and originates a detectable absorption at the low frequencies.

%
%
%
\def\path{figures/ls5039_figures}



\chapter[Physical properties of the gamma-ray binary LS~5039 through low- and high-frequency observations]{\huge Physical properties of the gamma-ray binary LS 5039 through low- and high-frequency observations\vspace{-25pt}} \label{chap:ls}

We have studied in detail the 0.15--15~GHz radio spectrum of the gamma-ray binary LS~5039 to look for a possible turnover and absorption mechanisms at low frequencies, and to constrain the physical properties of its emission. We have analyzed two archival VLA monitorings, all the available archival GMRT data and a coordinated quasi-simultaneous observational campaign conducted in 2013 with GMRT and WSRT. The data show that the radio emission of LS~5039 is persistent on day, week and year timescales, with a variability $\lesssim$$25\%$ at all frequencies, and no signature of orbital modulation. The obtained spectra reveal a power-law shape with a curvature below 5~GHz and a turnover at $\sim$$0.5~\mathrm{GHz}$, which can be reproduced by a one-zone model with synchrotron self-absorption plus Razin effect. We obtain a coherent picture for the size of the emitting region of $\sim$$0.85~\mathrm{mas}$, setting a magnetic field of $B \sim 20~\mathrm{mG}$, an electron density of $n_\mathrm{e} \sim 4 \times 10^5~\mathrm{cm^{-3}}$ and a mass-loss rate of $\dot M \sim 5 \times 10^{-8}~\mathrm{M_{\sun}\ yr^{-1}}$. These values imply a significant mixing of the stellar wind with the relativistic plasma outflow from the compact companion. At particular epochs the Razin effect is negligible, implying changes in the injection and the electron density or magnetic field. The Razin effect is reported for the first time in a gamma-ray binary, giving further support to the young non-accreting pulsar scenario. This work has been published in \citet{marcote2015ls5039}.

\section{Introducing LS~5039}

LS~5039 is a binary system composed of a young O6.5\,V star of $\approx$$23~\mathrm{M_{\sun}}$ and a compact object of $1$--$5~\mathrm{M_{\sun}}$ \citep{casares2005ls5039} with coordinates
$$ \alpha = 18^{\rm h}\ 26^{\rm m}\ 15.0593^{\rm s}\ (\pm 1.4~\mathrm{mas}),\qquad \delta = -14^{\circ}\ 50'\ 54.301''\ (\pm 1.9~\mathrm{mas})$$
at the epoch of MJD~$53797.064$, being located at $2.9 \pm 0.8~{\rm kpc}$ with respect to the Sun \citep{moldon2012lspsr}. This system presents a compact orbit with a period of 3.9~d and an eccentricity of 0.35 \citep{casares2005ls5039}.
The source exhibits a photometric visual magnitude of 11.3, which is known to be optically stable within 0.1~mag over the years and 0.002~mag on orbital timescales \citep{sarty2011}.

\citet{paredes2000} proposed for the first time that LS~5039 was associated with an EGRET \g-ray source. H.E.S.S. detected the binary system in the 0.1--4~TeV energy range \citep{aharonian2005ls5039}, and later reported variability at those energies along the orbital period \citep{aharonian2006}. {\em Fermi}/LAT confirmed the association with the EGRET source by clearly detecting the GeV counterpart, which showed also an orbital modulation \citep{abdo2009}. The system has been observed and detected with different X-ray satellites, showing moderate X-ray variability and a power-law spectrum with low X-ray photoelectric absorption consistent with interstellar reddening \citep{motch1997,ribo1999,reig2003,bosch-ramon2005,martocchia2005}. A clear orbital variability of the X-ray flux, correlated with the TeV emission, has been reported by \citet{takahashi2009}. However, the GeV emission is anticorrelated with these ones. Figure~\ref{fig:ls5039-lihgtcurves-multiwavelength} shows the light-curve of LS~5039 at TeV, GeV, and X-rays. Extended X-ray emission at scales up to 1--2~arcmin has also been detected \citep{durant2011}. X-ray pulsations have also been searched for, with null results \citep{rea2011}.
\begin{figure}
        \centering
        \includegraphics[width=9cm]{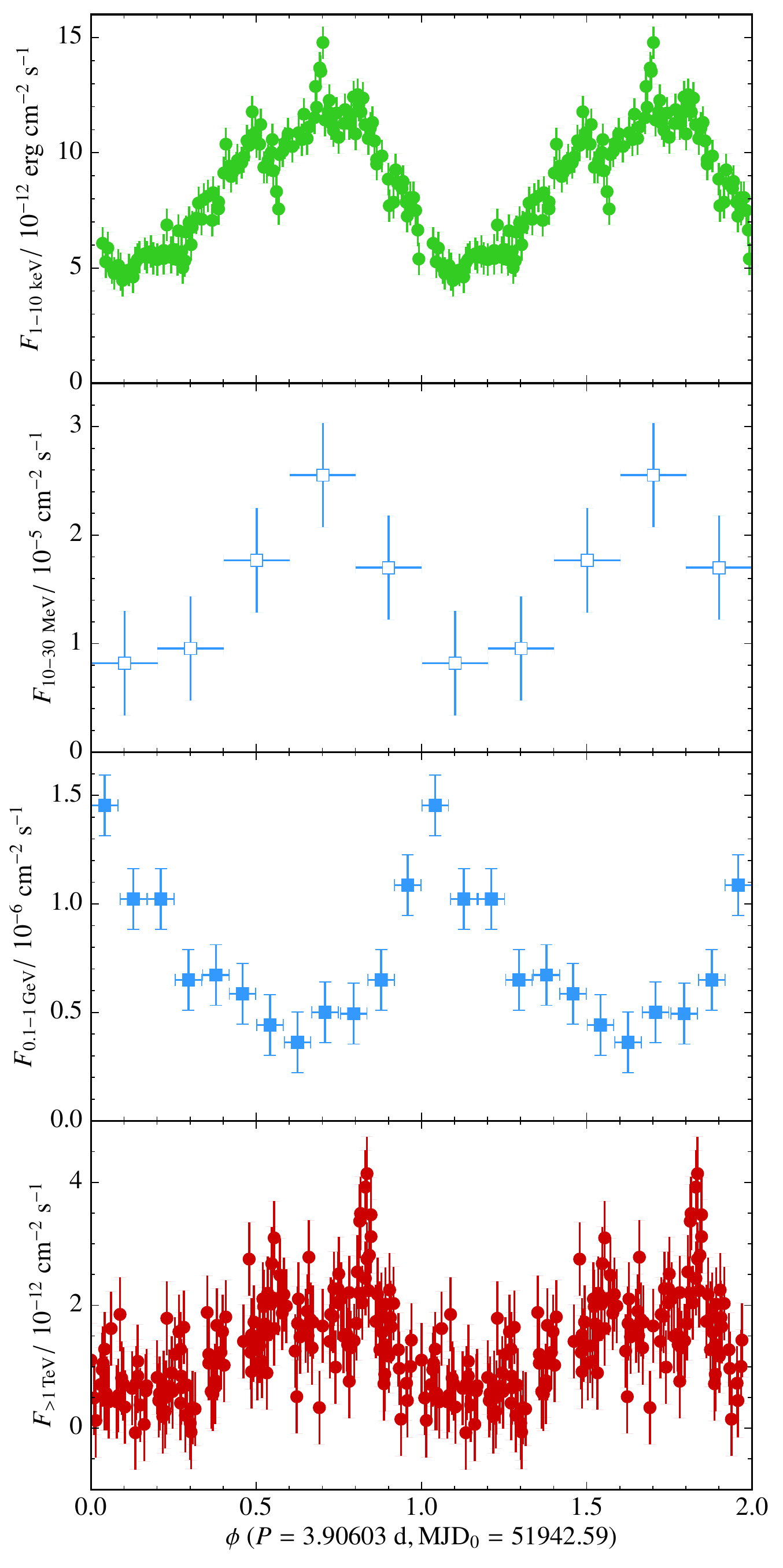}
        \caption[Multiwavelength light-curves of LS~5039 at X-rays, MeV, GeV and TeV.]{Light-curves of LS~5039 folded with the orbital period at X-rays, MeV, GeV, and TeV, from top to bottom, respectively. Data adapted from \citet{kishishita2009}, \citet{collmar2014}, \citet{abdo2009}, and \citet{aharonian2005ls5039}, respectively.}
        \label{fig:ls5039-lihgtcurves-multiwavelength}
\end{figure}

At radio frequencies, \citet{marti1998} conducted a multifrequency study of LS~5039 with the VLA from 1.4 to 15~GHz with four observations spanning 3 months (one run per month). The radio emission of the source has a non-thermal origin, and can be described by a power-law with a spectral index of $\alpha = -0.46\pm 0.01$. These authors reported a flux density variability below 30\% with respect to the mean value.
\citet{ribo1999} and \citet{clark2001} presented the results of a radio monitoring campaign conducted on a daily basis during approximately one year with the Green Bank Interferometer (GBI) at 2.2 and 8.3~GHz frequencies. The emission was similar to the one reported in \citet{marti1998}, exhibiting a mean spectral index of $\alpha = -0.5_{-0.3}^{+0.2}$ \citep{ribo1999}. \citet{clark2001} constrained any orbital modulation to be $<4$\% at 2.2~GHz. Searches of pulsed radio emission have been also conducted with null results, probably because of strong FFA by the dense wind of the companion at the position of the compact object \citep{mcswain2011}.

The radio emission of the source is resolved at milliarcsecond scales using VLBI observations. \citet{paredes2000,paredes2002} found an asymmetrical bipolar extended emission on both sides of a bright dominant core, suggesting that LS~5039 was a microquasar. Later on, \citet{ribo2008} reported persistent emission from the dominant core and morphological changes of the extended emission on day timescales, which were difficult to reconcile with a microquasar model. Finally, \citet{moldon2012ls5039} discovered morphological variability modulated with the orbital phase, and pointed out that a scenario with a young non-accreting pulsar is more plausible to explain the observed behavior. In this case the relativistic wind from the pulsar shocks with the stellar wind of the massive companion and a cometary tail of accelerated particles forms behind the pulsar (see \citealt{moldon2012ls5039} and references therein for further details).

At low radio frequencies ($\lesssim$$1~\mathrm{GHz}$) only a few observations have been published up to now. \citet{pandey2007} performed two observations with GMRT simultaneously at 235 and 610~MHz, and reported a power-law spectrum with a spectral index of $\alpha \approx -0.8$ at these frequencies. On the other hand, \citet{godambe2008} conducted additional GMRT observations and reported a turnover at $\sim$$1~\mathrm{GHz}$, with a positive spectral index of $\alpha \approx +0.75$ at the same frequencies (further discussed in \citealt{bhattacharyya2012}). These results would indicate, surprisingly, that the flux variability of LS~5039 increases significantly below 1~GHz (see a possible explanation in the framework of the microquasar model in \citealt{bosch-ramon2009ls5039}). However, as discussed in this work, the change in the spectrum is the result of using different and uncomplete calibration procedures.

The radio spectrum of gamma-ray binaries is dominated by the synchrotron emission, which is expected to be self-absorbed at low frequencies. The spectrum could also reveal the presence of other absorption mechanisms such as FFA or the Razin effect. 
Given that the Razin effect is widely reported in CWBs, we would also expect to observe this effect in gamma-ray binaries. Both types of systems probably share the origin of their emission: the collision between the winds of the two components of a binary system and the presence of shocks \citep[see][for the case of CWBs]{dougherty2003}. Therefore, similar absorption processes must take place.

Due to the surprising behavior of the radio emission of LS~5039 at low frequencies obtained with a limited number of observations, we decided to make an in-depth study of the source. We have reduced and analyzed two monitorings of LS~5039 at high frequencies that we conducted several years ago, most (if not all) of the publicly available low-frequency radio data, and conducted new coordinated observations to get a complete picture of the source behavior at low and high radio frequencies. This has allowed us to monitor the variability of LS~5039 on different timescales (orbital and long-term), and determine its spectrum using simultaneous and non-simultaneous data.
This Chapter is organized as follows. In the next section we present the radio observations analyzed in this work. In Sect.~\ref{sec:ls5039-data-reduction} we detail the data reduction and analysis of these data, and the results are presented in Sect.~\ref{sec:ls5039-results}. We detail the emission models used to describe the observed spectra in Sect.~\ref{sec:ls5039-models}, and discuss the observed behavior in Sect.~\ref{sec:ls5039-discussion}. We summarize the obtained results and state the conclusions of this work in Sect.~\ref{sec:ls5039-conclusions}.







\section{Radio observations}\label{sec:ls5039-radio-observations}

The data presented along this Chapter include two high-frequency VLA monitorings, all the available archival low-frequency radio observations performed with the GMRT, and a new coordinated campaign that we conducted in 2013 using GMRT and WSRT.\footnote{The VLA and ATCA data archives contain several isolated observations of LS~5039 above 1~GHz which have not been analyzed in this work, because sporadic flux density measurements are not useful for the study of short term and orbital flux density modulations. These heterogeneous data sets could be useful in future spectral studies.}$^{,}$\footnote{We have also analyzed two unpublished VLA observations conducted on 2006 December 16 and 18 at 330~MHz (project code AM877). However, we only obtained upper limits, because the noise is too high to clearly detect the source according to the GMRT data presented below. For this reason we will not explicitly include these observations and the obtained results in the rest of this work.}$^{,}$\footnote{We have also analyzed a LOFAR observation performed in 2011 at 150~MHz during its commissioning. However, the noise and the quality of the image at this stage were not accurate enough for our purposes.}
The details of all these observations, including the used facilities and configurations, the observation dates and frequencies, can be found in the first six columns of Table~\ref{tab:ls5039-results}. Figure~\ref{fig:ls5039-summary-obs} summarizes these observations in a frequency versus time diagram.

\begin{figure}[t]
        \centering
        \includegraphics[width=10cm]{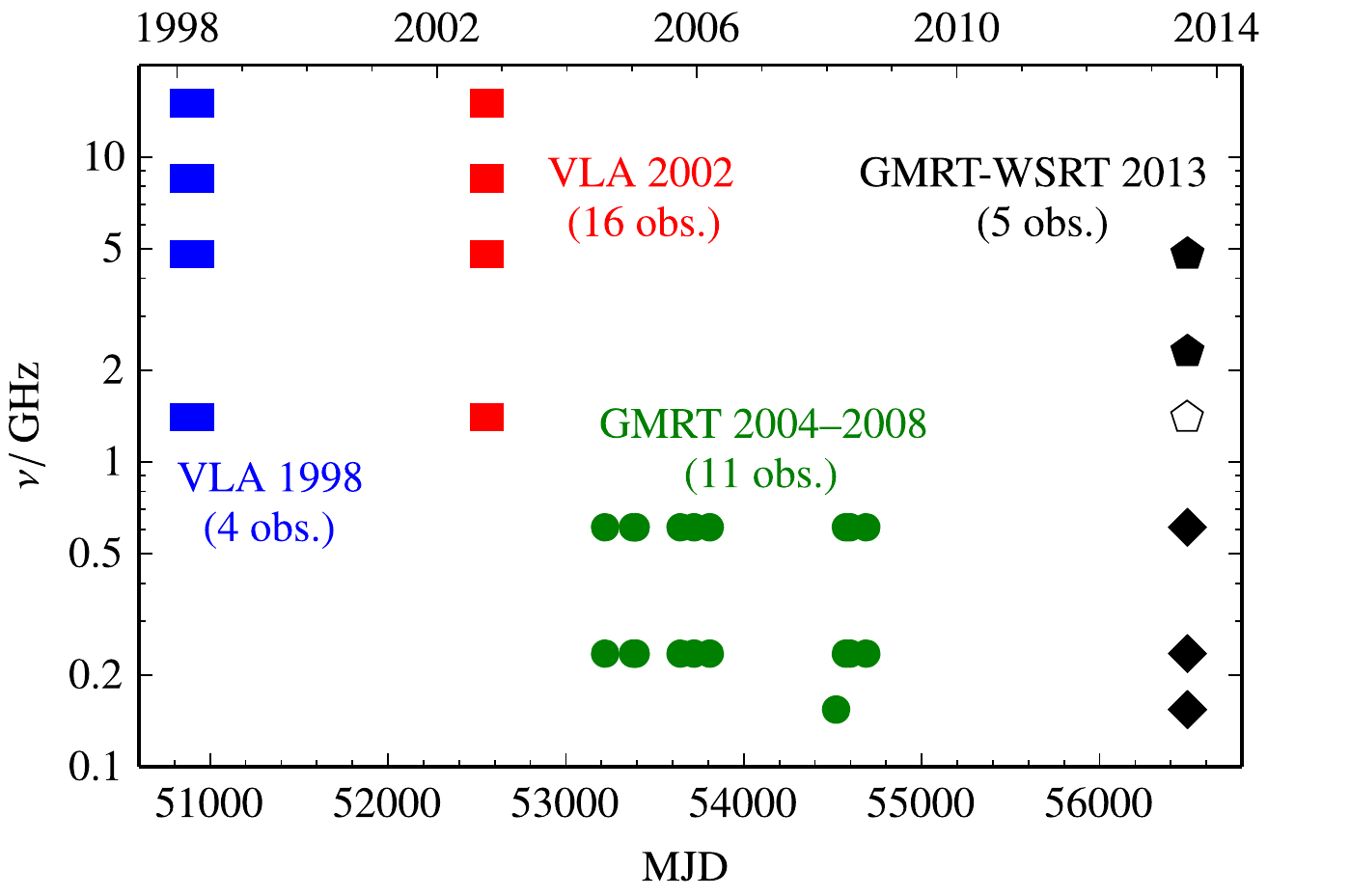}
        \caption[Summary of all the data of LS~5039 presented in this work.]{Summary of all the data presented in this work in a frequency versus time diagram, including the Modified Julian Date (MJD, bottom axis) and the calendar year (top axis). Squares represent the VLA monitorings in 1998 (blue) and 2002 (red), green circles correspond to the archival GMRT observations, black markers represent our coordinated GMRT (diamonds) and WSRT (pentagons) quasi-simultaneously campaign (the open pentagon represents a data set that could not be properly calibrated, see text). The radio data presented here span 15~yr.}
        \label{fig:ls5039-summary-obs}
\end{figure}
\def\qq{~~~~~}
\begin{sidewaystable}
    \ssmall
    \centering    
        \caption[Summary of all the data of LS~5039 presented in this work.]{Summary of all the data of LS~5039 presented in this work. For each observation we show the facility used together with its array configuration (if any), the project code, the reference if previously published, the calendar date, the Modified Julian Date (MJD) of the central observation time, the corresponding orbital phase (using $P=3.90603~\mathrm{d}$ and $\mathrm{MJD_0} = 51942.59$), and the flux density values at each frequency with the 1-$\sigma$ uncertainty (3-$\sigma$ upper-limits in case of non-detection).}
        \label{tab:ls5039-results}
\begin{tabular}{l@{~}c@{~}c@{~~~}c@{\hspace{7pt}}c@{\hspace{10pt}}cr@{}c@{}l@{\qq}r@{}c@{}l@{\qq}r@{}c@{}l@{\qq}r@{}c@{}l@{\qq}r@{}c@{}l@{\qq}r@{}c@{}l@{\qq}r@{}c@{}l@{\qq}r@{}c@{}l}
        \hline\\[-7pt]
        Facility & Project code & Ref. & Calendar date & MJD & $\phi$ &
        \multicolumn{3}{c}{$S_{154\,\mathrm{MHz}}$} &
        \multicolumn{3}{c}{$S_{235\,\mathrm{MHz}}$} &
        \multicolumn{3}{c}{$S_{610\,\mathrm{MHz}}$} &
        \multicolumn{3}{c}{$S_{1.4\,\mathrm{GHz}}$} &
        \multicolumn{3}{c}{$S_{2.3\,\mathrm{GHz}}$} &
        \multicolumn{3}{c}{$S_{4.8\,\mathrm{GHz}}$} &
        \multicolumn{3}{c}{$S_{8.5\,\mathrm{GHz}}$} &
        \multicolumn{3}{c}{$S_{15\,\mathrm{GHz}}$}\\
        &&&(dd/mm/yyyy)&&&
        \multicolumn{3}{c}{$(\mathrm{mJy})$} &
        \multicolumn{3}{c}{$(\mathrm{mJy})$} &
        \multicolumn{3}{c}{$(\mathrm{mJy})$} &
        \multicolumn{3}{c}{$(\mathrm{mJy})$} &
        \multicolumn{3}{c}{$(\mathrm{mJy})$} &
        \multicolumn{3}{c}{$(\mathrm{mJy})$} &
        \multicolumn{3}{c}{$(\mathrm{mJy})$} &
        \multicolumn{3}{c}{$(\mathrm{mJy})$}\\[+2pt]
        \hline\\[-6pt]
        VLA-D & AP357 & (1) & 11/02/1998 & 50856.20 & 0.87 & &--& & &--& & &--& & 27 &$\pm$& 4 & &--& & 21.8 &$\pm$& 0.5 & 17.40 &$\pm$& 0.10 & 12.4 &$\pm$& 0.3\\
VLA-A & AP357 & (1) & 10/03/1998 & 50883.10 & 0.76 & &--& & &--& & &--& & 38.6 &$\pm$& 1.0 & &--& & 23.70 &$\pm$& 0.10 & 17.77 &$\pm$& 0.08 & 13.3 &$\pm$& 0.2\\
VLA-A & AP357 & (1) & 09/04/1998 & 50913.00 & 0.41 & &--& & &--& & &--& & 39.8 &$\pm$& 0.7 & &--& & 22.70 &$\pm$& 0.10 & 15.85 &$\pm$& 0.07 & 11.80 &$\pm$& 0.18\\
VLA-A & AP357 & (1) & 12/05/1998 & 50946.00 & 0.86 & &--& & &--& & &--& & 45.5 &$\pm$& 1.0 & &--& & 26.0 &$\pm$& 0.2 & 18.23 &$\pm$& 0.12 & 13.9 &$\pm$& 0.2\\
VLA-BC& AP444 & -- & 23/09/2002 & 52540.96 & 0.19 & &--& & &--& & &--& & 36.8 &$\pm$& 1.4 & &--& & 20.5 &$\pm$& 0.5 & 14.3 &$\pm$& 0.5 & 9.4 &$\pm$& 0.7\\
VLA-BC& AP444 & -- & 28/09/2002 & 52545.07 & 0.24 & &--& & &--& & &--& & &--& & &--& & 24.4 &$\pm$& 0.2 & 16.5 &$\pm$& 0.3 & &--&\\
VLA-BC& AP444 & -- & 29/09/2002 & 52546.96 & 0.73 & &--& & &--& & &--& & 29.2 &$\pm$& 1.7 & &--& & 22.3 &$\pm$& 0.4 & 14.7 &$\pm$& 0.3 & 8.4 &$\pm$& 0.7\\
VLA-BC& AP444 & -- & 02/10/2002 & 52549.04 & 0.26 & &--& & &--& & &--& & 27.5 &$\pm$& 1.3 & &--& & 15.6 &$\pm$& 0.5 & 9.7 &$\pm$& 0.5 & 6.5 &$\pm$& 0.6\\
VLA-BC& AP444 & -- & 05/10/2002 & 52552.03 & 0.03 & &--& & &--& & &--& & 33.6 &$\pm$& 1.2 & &--& & 18.8 &$\pm$& 0.3 & 11.3 &$\pm$& 0.4 & 8.4 &$\pm$& 0.4\\
VLA-BC& AP444 & -- & 07/10/2002 & 52554.98 & 0.78 & &--& & &--& & &--& & 32 &$\pm$& 2 & &--& & 15.6 &$\pm$& 0.6 & 6.3 &$\pm$& 0.5 & &--&\\
VLA-C & AP444 & -- & 12/10/2002 & 52559.01 & 0.81 & &--& & &--& & &--& & 34.5 &$\pm$& 1.9 & &--& & 20.0 &$\pm$& 0.4 & 16.7 &$\pm$& 0.3 & 13.3 &$\pm$& 0.4\\
VLA-C & AP444 & -- & 12/10/2002 & 52559.93 & 0.05 & &--& & &--& & &--& & 38.5 &$\pm$& 1.1 & &--& & 25.6 &$\pm$& 0.2 & 18.0 &$\pm$& 0.2 & 12.0 &$\pm$& 0.5\\
VLA-C & AP444 & -- & 14/10/2002 & 52561.03 & 0.33 & &--& & &--& & &--& & &--& & &--& & 20.25 &$\pm$& 0.19 & 14.5 &$\pm$& 0.3 & 10.8 &$\pm$& 0.5\\
VLA-C & AP444 & -- & 15/10/2002 & 52562.02 & 0.58 & &--& & &--& & &--& & &--& & &--& & 25.1 &$\pm$& 0.2 & 16.5 &$\pm$& 1.0 & 10 &$\pm$& 2\\
VLA-C & AP444 & -- & 15/10/2002 & 52562.98 & 0.83 & &--& & &--& & &--& & 42 &$\pm$& 2 & &--& & 29.6 &$\pm$& 0.2 & 20.8 &$\pm$& 0.3 & 15.3 &$\pm$& 0.6\\
VLA-C & AP444 & -- & 17/10/2002 & 52564.96 & 0.34 & &--& & &--& & &--& & 31.6 &$\pm$& 1.7 & &--& & 21.6 &$\pm$& 0.2 & 14.9 &$\pm$& 0.3 & 9.3 &$\pm$& 0.7\\
VLA-C & AP444 & -- & 19/10/2002 & 52566.01 & 0.60 & &--& & &--& & &--& & 35.0 &$\pm$& 1.8 & &--& & 23.6 &$\pm$& 0.5 & 13.0 &$\pm$& 0.6 & 7.9 &$\pm$& 0.7\\
VLA-C & AP444 & -- & 19/10/2002 & 52566.91 & 0.83 & &--& & &--& & &--& & 36.4 &$\pm$& 1.3 & &--& & 23.9 &$\pm$& 0.2 & 16.6 &$\pm$& 0.2 & 3.7 &$\pm$& 0.3\\
VLA-C & AP444 & -- & 20/10/2002 & 52567.95 & 0.10 & &--& & &--& & &--& & 26.2 &$\pm$& 1.5 & &--& & 18.9 &$\pm$& 0.2 & 14.8 &$\pm$& 0.2 & 11.2 &$\pm$& 0.6\\
VLA-C & AP444 & -- & 21/10/2002 & 52568.94 & 0.35 & &--& & &--& & &--& & 29.1 &$\pm$& 1.6 & &--& & 24.1 &$\pm$& 0.3 & 18.6 &$\pm$& 0.3 & 12.3 &$\pm$& 0.4\\
GMRT & 06PDA01 & (2) & 03/08/2004 & 53220.57 & 0.18 & &--& & \multicolumn{3}{l}{$<200$} & 51 &$\pm$& 5 & &--& & &--& & &--& & &--& & &--&\\
GMRT & 07PDA01 & (2) & 07/01/2005 & 53377.21 & 0.28 & &--& & 27 &$\pm$& 9 & 28 &$\pm$& 4 & &--& & &--& & &--& & &--& & &--&\\
GMRT & 07PDA01 & -- & 21/01/2005 & 53391.20 & 0.86 & &--& & \multicolumn{3}{l}{$<11$}& 38 &$\pm$& 3 & &--& & &--& & &--& & &--& & &--&\\
GMRT & 07PDA01 & -- & 22/01/2005 & 53392.20 & 0.12 & &--& & 18 &$\pm$& 6 & 49 &$\pm$& 3 & &--& & &--& & &--& & &--& & &--&\\
GMRT & 08SKC01 & -- & 01/10/2005 & 53644.60 & 0.75 & &--& & 28 &$\pm$& 9 & 46 &$\pm$& 3 & &--& & &--& & &--& & &--& & &--&\\
GMRT & 09PDA01 & -- & 16/12/2005 & 53720.17 & 0.09 & &--& & 26 &$\pm$& 8 & 41 &$\pm$& 2 & &--& & &--& & &--& & &--& & &--&\\
GMRT & 09SBB01 & (3, 4) & 15/03/2006 & 53809.04 & 0.84 & &--& & 31 &$\pm$& 9 & 49 &$\pm$& 3 & &--& & &--& & &--& & &--& & &--&\\
GMRT & 13MPA01 & -- & 23/02/2008 & 54519.00 & 0.63 & &$<6$& & &--& & &--& &&--& & &--& & &--& & &--& & &--&\\
GMRT & 14SVG02 & -- & 17/04/2008 & 54573.81 & 0.66 & &--& & 19 &$\pm$& 6 & 35 &$\pm$& 2 & &--& & &--& & &--& & &--& & &--&\\
GMRT & 14SVG02 & -- & 10/05/2008 & 54596.78 & 0.55 & &--& & 25 &$\pm$& 8 & 31.3 &$\pm$& 1.8 & &--& & &--& & &--& & &--& & &--&\\
GMRT & 14SVG02 & -- & 10/08/2008 & 54688.63 & 0.02 & &--& & \multicolumn{3}{l}{$<7$} & 40 &$\pm$& 2 & &--& & &--& & &--& & &--& & &--&\\
GMRT & 24\_001 & -- & 18/07/2013 & 56491.56 & 0.60 & &$<6$& & &--& & &--& & &--& & &--& & &--& & &--& & &--&\\
GMRT & 24\_001 & -- & 19/07/2013 & 56492.60 & 0.87 & &--& & 34 &$\pm$& 11 & 51 &$\pm$& 3 & &--& & &--& & &--& & &--& & &--&\\
WSRT & R13B012 & -- & 19/07/2013 & 56492.75 & 0.90 & &--& & &--& && --&& &-- & & 30.6 &$\pm$& 1.0 & 22.42 &$\pm$& 0.15 & &--& & &--&\\
GMRT & 24\_001 & -- & 20/07/2013 & 56493.56 & 0.11 & &$<12$& & &--& & &--& & &--& & &--& & &--& & &--& & &--&\\
GMRT & 24\_001 & -- & 21/07/2013 & 56494.57 & 0.37 & &--& & 24 &$\pm$& 7 & 37 &$\pm$& 2 & &--& & &--& & &--& & &--& & &--&\\
WSRT & R13B012 & -- & 21/07/2013 & 56494.75 & 0.42 & &--& & &--& & &--& & &--& & 27.2 &$\pm$& 0.7 & 18.68 &$\pm$& 0.11 & &--& & &--&\\
GMRT & 24\_001 & -- & 22/07/2013 & 56495.73 & 0.67 & &$<8$& & &--& & &--& & &--& & &--& & &--& & &--& & &--&\\

        \hline
        \end{tabular}
        
        \medskip{(1)~\citet{marti1998}; (2)~\citet{pandey2007}; (3)~\citet{godambe2008}; (4)~\citet{bhattacharyya2012}}
\end{sidewaystable}

\subsection{VLA monitorings}   \label{sec:ls5039-obs:vla}

Two multi-frequency monitorings have been conducted with the VLA: one in 1998 \citep{marti1998} and another one in 2002 (unpublished). In addition to the 2002 data set, we have reduced again the 1998 data set to guarantee that all the VLA data have been reduced using the same procedures.

The 1998 VLA monitoring consists on four observations conducted between February 11 and May 12 at 1.4, 4.8, 8.5 and 15~GHz using two Intermediate Frequency (IF) band pairs (RR and LL circular polarizations) of 50~MHz each (project code AP357). The amplitude calibrator used was 3C~286. The phase calibrators used were 1834$-$126 at 1.4~GHz, 1820$-$254 at 4.8 and 8.5~GHz, and 1911$-$201 at 15~GHz.

The 2002 VLA monitoring consists of 16 observations conducted between September 23 and October 21 (covering around seven orbital cycles) at the same frequency bands and using the same calibrators than the previous monitoring (pro\-ject code AP444).

\subsection{Archival GMRT observations}   \label{sec:ls5039-obs:gmrt}

We have reduced all the existing archival GMRT data of LS~5039. These data set includes 10 simultaneous observations at 235 and 610~MHz taken between 2004 and 2008: two published in \citet{pandey2007}, one reported in \citet{godambe2008}, and seven additional unpublished observations. There is an additional observation at 154~MHz performed in 2008. The GMRT data were obtained with a single IF, with dual circular polarization at 154~MHz (RR and LL), and single circular polarization at 235~MHz (LL) and 610~MHz (RR). The 154 and 235~MHz data were taken with a 8~MHz bandwidth and the 610~MHz data with 16~MHz (with 64 channels at 235~MHz and 128 channels at the other bands). In all these observations, 3C~286 and/or 3C~48 were used as amplitude calibrators and 1822$-$096 or 1830$-$360 as phase calibrators.  We have reduced again the published data to guarantee that all the GMRT data have been reduced using the same procedures.

\subsection{Coordinated GMRT-WSRT campaign in 2013}   \label{sec:ls5039-obs:coord}

To study in detail the radio emission and absorption processes in LS~5039, we obtained for the first time a quasi-simultaneous spectrum of the source at low and high radio frequencies with the GMRT (project code 24\_001) and WSRT (project code R13B012) on 2013 July 18--22.

We observed LS~5039 with the GMRT at 154~MHz on July 18, 20, and 22 in three 8-hr runs, and simultaneously at 235 and 610~MHz on July 19 and 21 in two 4-hr runs. We conducted WSRT observations at 1.4, 2.3 and 4.8~GHz on July 19 and 21 in two 10-hr runs, which took place just after the GMRT ones at 235 and 610~MHz. The GMRT IFs, channels and polarizations are the same as the ones described in Sect.~\ref{sec:ls5039-obs:gmrt}.
The total bandwidths are 16~MHz at 154 and 235~MHz, and 32~MHz at 610~MHz (double than in the earlier observations reported in Sect.~\ref{sec:ls5039-obs:gmrt}). The WSRT data were obtained with eight IFs of 46 channels each, with dual circular polarizations at 2.3~GHz (RR and LL) and dual linear polarizations at 1.4 and 4.8~GHz (XX and YY). The total bandwidth is 115~MHz at all frequencies.
We used 3C~286 and 3C~48 as amplitude calibrators for both observatories and 1822$-$096 as phase calibrator for GMRT (no phase calibrators are needed for WSRT due to the phase stability).

\section{Data reduction and analysis}\label{sec:ls5039-data-reduction}

The raw visibilities were loaded into AIPS and flagged removing telescope off-source times, instrumental problems or RFI. 
The VLA data have been reduced using amplitude and phase calibration steps. Self-calibration on the target source was successful at 1.4, 4.8 and 8.5~GHz, but failed for most of the 15~GHz observations due to the faintness of the source. For this reason, we preferred not to self-calibrate any VLA data of LS~5039 to compare the results at all frequencies with exactly the same data reduction process.

The GMRT data have been reduced as follows. First, we performed an a priori amplitude and phase calibration using an RFI-free central channel of the band. Afterwards, we performed the bandpass calibration. Finally, we calibrated all the channels of the band in amplitude and phase considering the bandpass calibration.
We imaged the target source and conducted several cycles of self-calibration/imaging to correct for phase and amplitude errors.
Given the huge field of view of the GMRT images (few degrees), we have not considered only the $uv$-plane but the full $uvw$-space during the imaging process. In the final images a correction for the primary beam attenuation has also been performed.

In addition, for the GMRT data we have also performed a correction of the system temperature, $T_{\mathrm{sys}}$, for each antenna to subtract the contribution of the Galactic diffuse emission, which is relevant at low frequencies (see \S\ref{app:tsys}).
This $T_{\mathrm{sys}}$ correction was performed not only using archival self-correlated observations of the target source and the calibrators, but also using recent data taken at our request after the campaign in 2013. The obtained \tsys corrections are directly applied to the flux densities of the final target images, noting that these corrections imply an additional source of uncertainty.

The WSRT data have been reduced using amplitude and phase calibration steps. The solutions for the amplitude calibrators were directly extrapolated to the target source. The data were imaged and self-calibrated a few times. We had calibration problems at 1.4~GHz, and we could not extract any result from the corresponding data.

During the imaging process we have used a Briggs robustness parameter of zero in all cases. For the VLA and GMRT data the synthesized beam had a major axis around two times the minor axis. For the VLA in A configuration we have obtained geometrical mean values for the synthesized beam between $2.3~\arcsec$ (1.4~GHz) and $0.18~\arcsec$ (15~GHz). In D configuration these values increase up to $58~\arcsec$ at 1.4~GHz. With GMRT we have obtained synthesized beams between $18$ and $5~\arcsec$ (at 154 and 610~MHz, respectively). The WSRT observations exhibit an elongated synthesized beam with a major axis around 10 times the minor axis as a consequence of the linear configuration of the array and the declination of the source. For WSRT the synthesized beam was $110 \times 10~{\arcsec}^2$ with a position angle (PA) of $5^{\degree}$ at 2.3~GHz, and $48 \times 5.7~\arcsec^2$ with PA of $-1^{\degree}$ at 4.8~GHz.

The measurement of the flux density value of LS~5039 in each image has been conducted using the {\tt tvstat} task of AIPS by considering a small region centered around the target.  We also measured the root-mean-square (rms) flux density of a region centered on the source, but clearly excluding it, and considered this to be the flux density uncertainty. However, in some cases the noise around the source is larger than the obtained rms (due to poor $uv$-plane coverage, calibration errors, etc.). In such cases, we repeated the flux density measurement a few times using different region sizes, and considered the dispersion of the obtained values as a measure of the flux density uncertainty. In any case, these uncertainties were never above 50\% of the corresponding rms values. The {\tt jmfit} task of AIPS has also been used to double check the flux densities of LS~5039.

To guarantee the reliability of the measured flux densities, we have monitored the flux densities of the amplitude calibrator for the VLA observations and the flux densities of several background sources detected in the field of view of LS~5039 for the GMRT observations. The constancy of the obtained VLA flux densities (below 3\% at 1.4 and 15~GHz and below 0.4\% at 4.8 and 8.5~GHz) and the lack of trends in the GMRT flux densities, allow us to ensure that the reported variability of LS~5039 (see below) is caused by the source itself, and not by additional effects during the observations or the calibration process.

\section{Results}\label{sec:ls5039-results}

LS~5039 appears as a point-like source in all the VLA, GMRT and WSRT images, and we summarize the results for all these observations in Table~\ref{tab:ls5039-results}. In this section we focus first on the VLA monitoring performed in 2002 to study the variability of LS~5039 at high frequencies along consecutive orbital cycles. Later on, we move to the GMRT observations to study the variability at low frequencies. We continue by reporting the high frequency spectra from the VLA monitoring of 2002. We then combine all the data to show the full radio spectrum of LS~5039 from 0.15 to 15~GHz. Finally, we report on the quasi-simultaneous spectra from 0.15 to 5~GHz from the coordinated GMRT-WSRT campaign in 2013.

\subsection{Light-curve at high frequencies ($>$1~GHz)}

\begin{figure}[t]
        \centering
        \includegraphics[width=10cm]{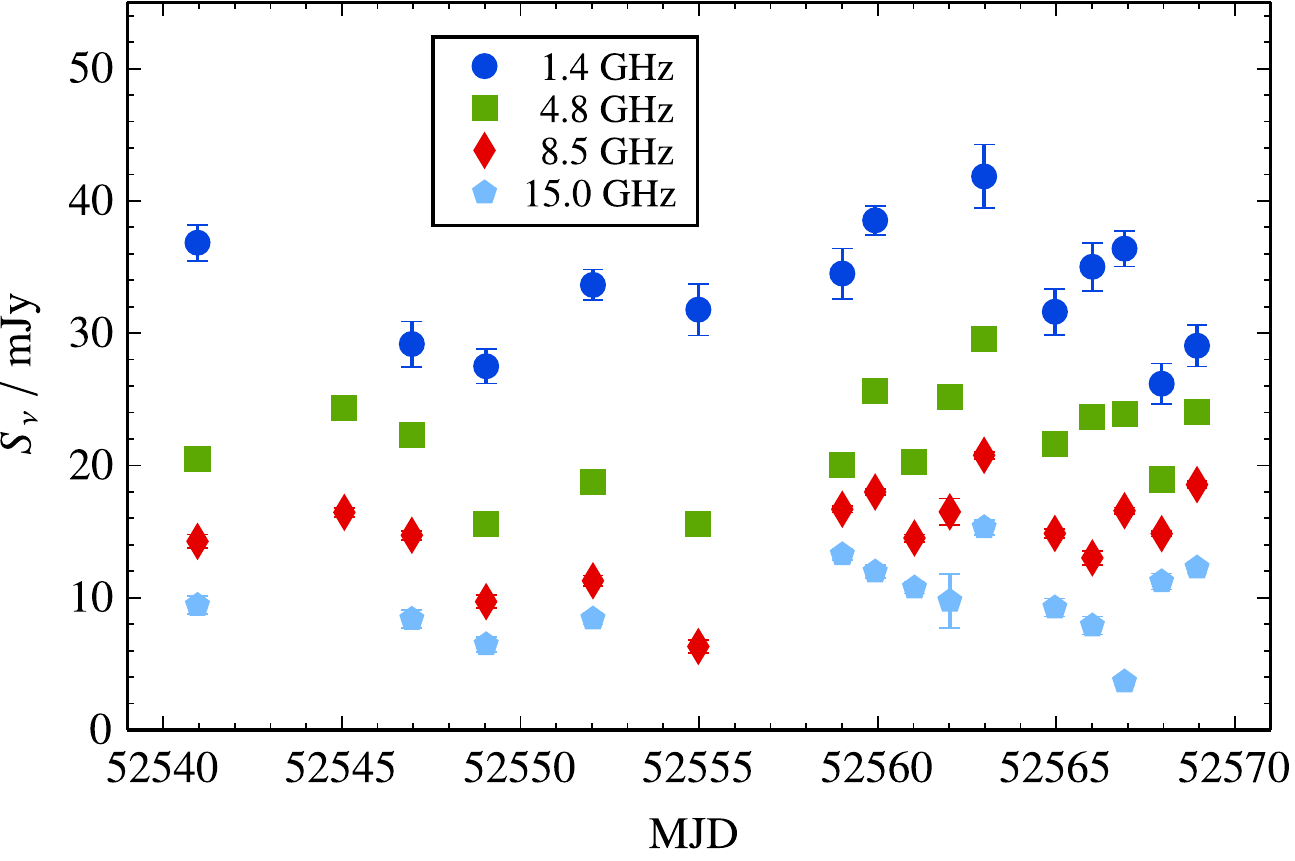}
        \caption[Multi-frequency light-curves of LS~5039 as a function of the MJD for the VLA monitoring in 2002.]{Multi-frequency light-curves of LS~5039 as a function of the Modified Julian Date (MJD) for the VLA monitoring in 2002. Error bars represent 1-$\sigma$ uncertainties.}
        \label{fig:ls5039-vla-mjd-flux}
\end{figure}
Figure~\ref{fig:ls5039-vla-mjd-flux} shows the multi-frequency light-curves obtained from the VLA monitoring in 2002 as a function of time. The flux of LS~5039 is persistent, and shows variability along all the observing period ($\approx 28~\mathrm{d}$) on timescales as short as one day. The trend of the variability is roughly similar at all frequencies. Apart from the daily variability, there is an evolution of the flux density on timescales of the order of 10~d. A $\chi^2$ test provides a probability of variability of 8--40$\sigma$, depending on the frequency. Since the flux density values encompass those of the 1998 monitoring (see \citealt{marti1998} and our values in Table~\ref{tab:ls5039-results}) and show similar average values at all frequencies, we conclude that the emission of LS~5039 is similar in both epochs (1998 and 2002).

In Figure~\ref{fig:ls5039-vla-phase-flux} we plot the same data as a function of the orbital phase. The reduced number of observations and the presence of some gaps ($\sim$0.35--0.55, $\sim 0.60$--$0.70$, $\sim 0.85$--$1.00$) do not allow us to clearly report on any significant orbital modulation on top of the reported daily and weekly variability. We note that at orbital phase $\sim 0.8$ there is a large dispersion of flux densities at all frequencies.
We have also binned the multi-frequency light-curves with different bin sizes (0.15, 0.20, 0.25) and different central bin positions to increase the statistics, but we have not seen any significant deviation with respect to the mean values or larger variabilities at specific orbital phases (not even at phase 0.8).
To further search for deviations at phase $\sim 0.8$, we have re-analyzed the full GBI data at 2.2 and 8.3~GHz, consisting on 284 observations spanning 340~d \citep{ribo1999,clark2001}, obtaining null results.

\begin{figure}[t]
        \centering
        \includegraphics[width=10cm]{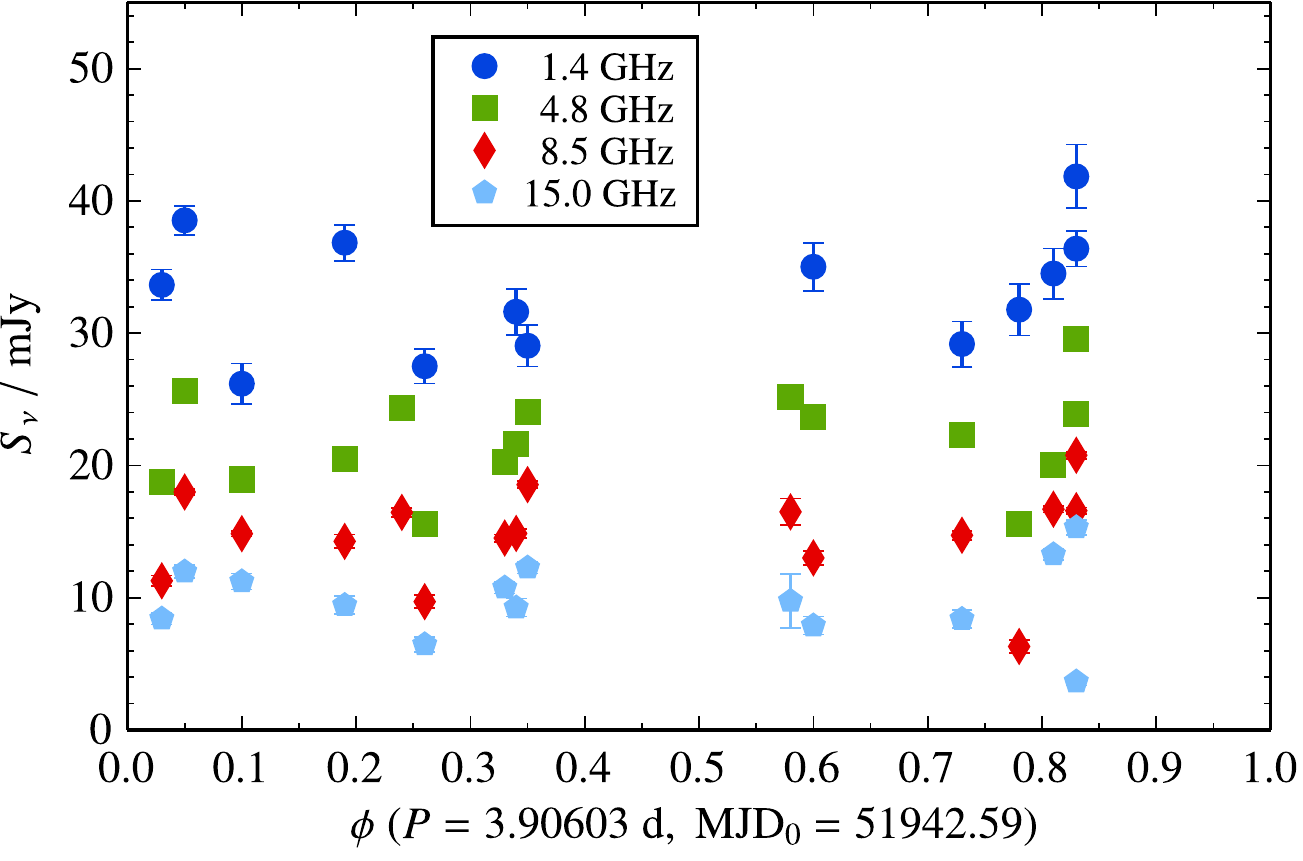}
        \caption[Folded multi-frequency light-curves of LS~5039 with the orbital period for the VLA monitoring in 2002.]{Folded multi-frequency light-curves of LS~5039 with the orbital period for the VLA monitoring in 2002. The campaign covers around 7 consecutive orbital cycles. Error bars represent 1-$\sigma$ uncertainties.}
        \label{fig:ls5039-vla-phase-flux}
\end{figure}
\begin{figure}[t]
        \centering
        \includegraphics[width=10cm]{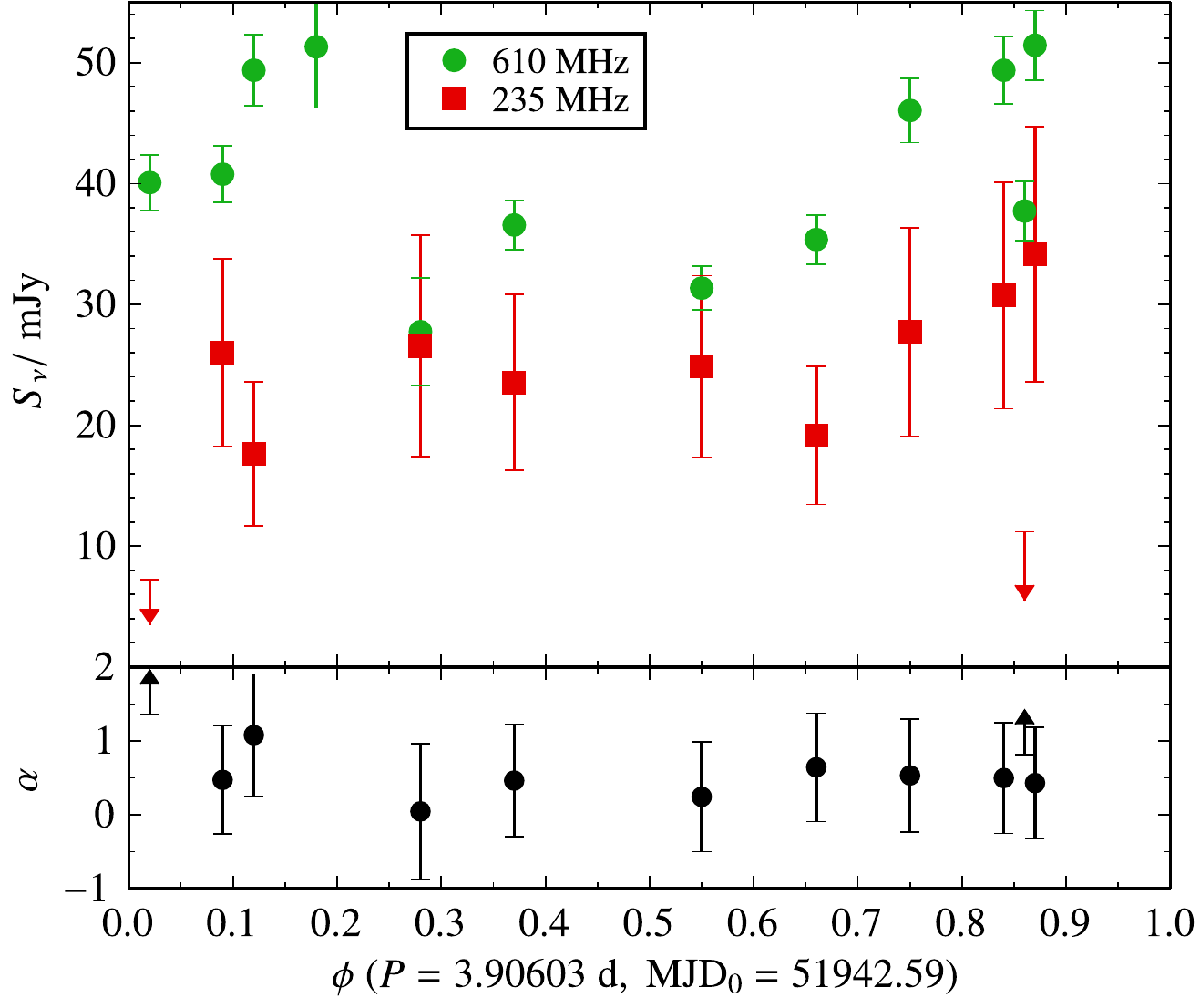}
        \caption[Folded multi-frequency light-curves of LS~5039 with the orbital period at 235 and 610~MHz of the archival GMRT observations and the derived spectral indexes.]{{\em Top:} folded light-curves of LS~5039 with the orbital period at 235 and 610~MHz of the archival GMRT observations from 2004 to 2008 and the two GMRT observations from the coordinated GMRT-WSRT campaign in 2013. Error bars represent 1-$\sigma$ uncertainties (resulting from the combination of the image noise and the \tsys correction uncertainties). Upper limits at 235~MHz are plotted at the 3-$\sigma$ level. {\em Bottom:} spectral indexes derived from the data above. We observe a mean spectral index of $\alpha \approx 0.5$ between 235 and 610~MHz.}
        \label{fig:ls5039-gmrt-phase-flux}
\end{figure}

\subsection{Light-curve at low frequencies ($<$1~GHz)}

Figure~\ref{fig:ls5039-gmrt-phase-flux} (top) shows the light curves of LS~5039 at 235 and 610~MHz as a function of the orbital phase from all the analyzed GMRT observations: the archival GMRT observations (2004--2008) and the two GMRT observations from the coordinated GMRT-WSRT campaign in 2013. The source exhibits a persistent radio emission at both frequencies during this 9-yr period. At 610~MHz LS~5039 shows clear variability ($6\sigma$). At 235~MHz the variability is not significant ($1\sigma$) for the most conservative decision of considering the upper-limits as detections with flux densities corresponding to three times the uncertainties.

These smaller significances in the variability with respect to the high frequency ones are produced mainly by the larger uncertainties of the flux densities, which are a combination of the image noise and the dominant \tsys correction uncertainties.
When we apply these corrections, we increase the typical 1--3\% uncertainties to $\approx 8$\% uncertainties in the flux densities at 610~MHz, and from 6--16\% to $\approx 40$\% at 235~MHz.
As the \tsys corrections are scaling factors which are equally applied to all the observations and only depend on the frequency, they only affect the absolute values of the flux densities, but not their relative differences. This means that we can determine the dispersion of these data prior to applying the \tsys corrections. In this case, we obtain that the source is variable at 610 and 235~MHz with a confidence level of $35\sigma$ and $6\sigma$, respectively. At 610~MHz variability on day timescales is also detected (see MJD 53391--53392 and MJD 56492--56494 in Table~\ref{tab:ls5039-results}).

Therefore, we see the same behavior than at high frequencies: variability on timescales as short as one day, with a persistent emission along the years, and without clear signals of being orbitally modulated. Observing the 610~MHz data (Figure~\ref{fig:ls5039-gmrt-phase-flux}, top), a small orbital modulation seems to emerge, with a broad maximum located at $\phi \sim 0.8$--$0.2$ and a broad minimum at $\phi \sim 0.3$--$0.7$.
To guarantee that this variability is not due to calibration issues, we have monitored the flux densities of up to seven compact background sources in the field of LS~5039. Their flux densities vary in different ways, but none of them shows the same trend of variability as LS~5039.
Whether the orbital modulation suggested above for LS~5039 is produced by the poor sampling or by an intrinsic modulation of the source will remain unknown until further detailed observations are conducted.

The spectral indexes from Figure~\ref{fig:ls5039-gmrt-phase-flux} (bottom) provide a mean of $\alpha = 0.5 \pm 0.8$. Given the large uncertainties, we can not claim for variability at more than a 3-$\sigma$ level.
However, if we do not consider the uncertainties related to the \tsys corrections we do obtain a significant variability of the spectral index, which is exclusively produced by the two observed upper-limits.

\begin{figure}
	\centering
        \includegraphics[width=12cm]{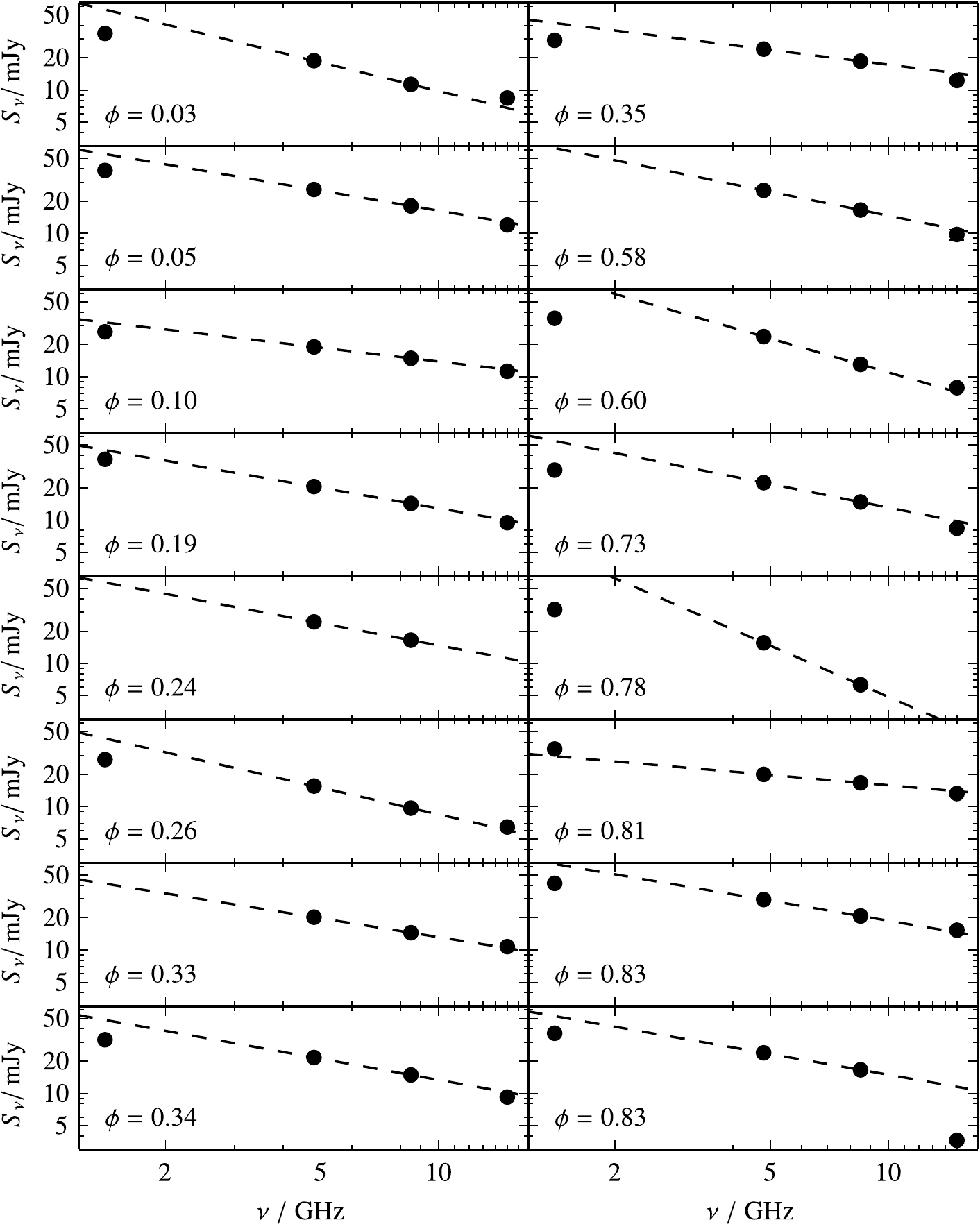}
        \caption[High-frequency spectra of LS~5039 from the VLA monitoring in 2002.]{Spectra of LS~5039 from the VLA monitoring in 2002 ordered with the orbital phase. The dashed line represents the power law derived from the 4.8 and 8.5~GHz data. Signals of a curved spectrum below $\sim$$2~{\rm GHz}$ are visible in most cases. At orbital phases $\sim$0.8 we observe the most extreme spectra (the steepest and the flattest ones).}
        \label{fig:ls5039-vla-freq-flux}
\end{figure}

\subsection{High frequency spectra ($>$1~GHz)}

Figure~\ref{fig:ls5039-vla-freq-flux} shows the spectrum of LS~5039 for each observation within the VLA monitoring in 2002. The spectra can be roughly fit with a power law, but they show a slight curvature below $\sim$$5~{\rm GHz}$ in most cases.
Given this curvature and the fact that at 15~GHz we can have a larger dispersion due to changing weather conditions, we have derived a power law for the remaining frequencies (4.8 and 8.5~GHz) to estimate the spectral index $\alpha$. The obtained values are in the range from $-0.35$ to $-1.1$, with an average value of $\alpha = -0.57\pm 0.12$, compatible with previously reported values. No signatures of orbital modulation are detected. However, around orbital phase 0.8 we observe the most extreme spectra, with the steepest and flattest ones.

\subsection{Non-simultaneous spectrum}

\begin{figure}
        \centering
        \includegraphics[width=10cm]{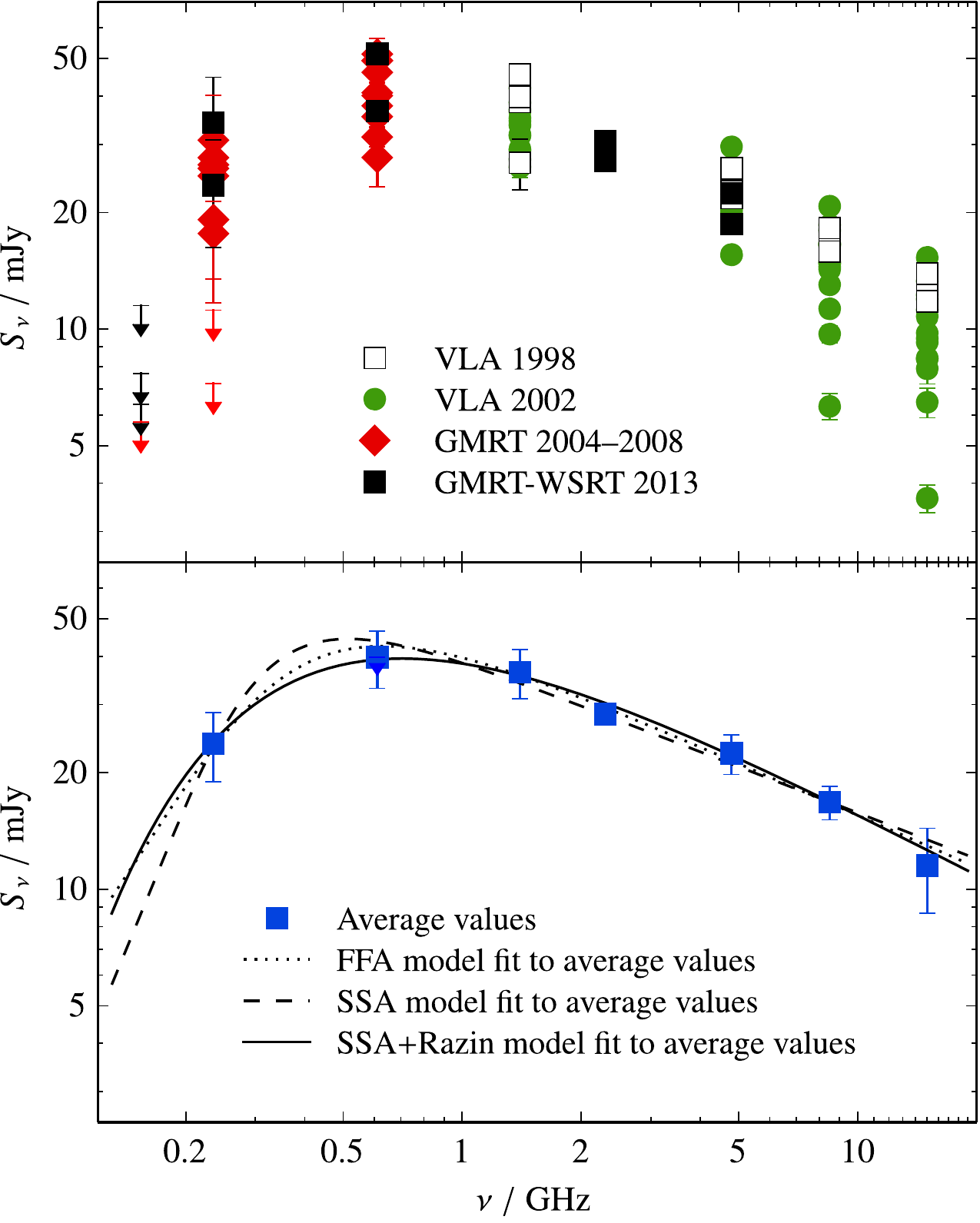}
        \caption[Non-simultaneous and average spectra of LS~5039 derived from all the data presented in this work.]{{\em Top}: non-simultaneous spectrum of LS~5039 obtained with all the data presented in this work (VLA 1998 \& 2002, GMRT 2004--2008, GMRT-WSRT 2013). Error bars represent 1-$\sigma$ uncertainties. Upper limits are plotted at the 3-$\sigma$ level (one at 235~MHz lies outside the figure limits).
        {\em Bottom}: average flux density value at each frequency. Error bars represent the standard deviations. The upper-limit at 235~MHz outside the figure limits has not been considered in the averaging. This average spectrum has been fit with three different absorption models: FFA, SSA, and SSA plus Razin effect (SSA+Razin). Mean square errors (within the symbol sizes, not shown in the figure) have been used as uncertainties in the fitting process (see text), while the data at 2.3~GHz and the upper-limits at 154~MHz have not been considered.}
        \label{fig:ls5039-all-freq-flux}
\end{figure}
Figure~\ref{fig:ls5039-all-freq-flux} (top) shows the obtained spectrum (0.15--15~GHz) from LS~5039 with all the data presented in this work (VLA monitorings in 1998 and 2002, the archival GMRT observations in 2004--2008, and the coordinated GMRT-WSRT campaign in 2013). All these data cover around 15 yr of observations. The source is always detected above 0.5~GHz. At 235~MHz the source is detected most of the times, but as already mentioned there are some upper limits. On 2004 August 3 ($\phi=0.18$) the rms of the image at this frequency is much higher than the flux density expected for the source and thus these data have not been taken into account in what follows. At 154~MHz the source is not detected in any observation, neither combining all the existing data. The spectrum is clearly curved, with a turnover around $\sim$$0.5~\mathrm{GHz}$.

We observe that the emission of LS~5039 along the years remains persistent with a moderate variability above the turnover frequency (below the turnover the uncertainties become larger). Given this behavior, we averaged the flux density of LS~5039 at each frequency, considering the 235~MHz upper-limits as detections with flux densities corresponding to three times the uncertainties. We show the average values in Figure~\ref{fig:ls5039-all-freq-flux} (bottom) and quote them in Table~\ref{tab:ls5039-mean-values}.
We observe a standard deviation below 25\% at all frequencies. The average value and the variability reported at 2.3~GHz (from WSRT data) are not as accurate as in the other cases because we only have two observations at such frequency.

\begin{table}
	\small
	\centering
    \caption[Average flux density values of LS~5039 at each frequency.]{Average value of the flux density ($S_{\nu}$) of LS~5039 and its relative variability ($\delta S_{\nu}\cdot S_{\nu}^{-1}$) at each frequency, considering all the data presented in this work. $N_{\rm obs}$ values have been considered at each frequency. In all cases we report a variability below 25\%. At 2.3~GHz we only have two observations to compute the mean value, so the amplitude of the variability is not relevant and cannot be compared with the rest of the data.}
        \begin{tabular}{c@{\hspace{+25pt}}c@{\hspace{+25pt}}r@{}c@{}l@{}@{\hspace{+25pt}}c}
            \hline\\[-10pt]
            $\nu\,/\ \mathrm{GHz}$ & $N_{\rm obs}$ & \multicolumn{3}{l}{$S_{\nu} /\ \mathrm{mJy}$} &
            $\delta S_{\nu}\cdot S_{\nu}^{-1} /\ \%$\\[+2pt]
            \hline
            0.235& 11 &23& $\,\pm\,$ &5 & 20\\
0.610& 12 &39& $\pm$ &6 & 16\\
1.4  & 17 & 36& $\pm$ &5 & 15\\
2.3  & 2  & 28.3& $\pm$ &1.6 & 6\\
4.8  & 22 & 22& $\pm$ &3 & 12\\
8.5  & 20 & 16.8& $\pm$ &1.7 & 10\\
15   & 18 & 11& $\pm$ &3 & 25\\

            \hline
        \end{tabular}
        \label{tab:ls5039-mean-values}
\end{table}

It is worth noting also that we obtained remarkable differences between the GMRT results presented here and those published in \citet{pandey2007}, \citet{godambe2008} and \citet{bhattacharyya2012}. All these differences are related to the \tsys corrections of GMRT data (explained in detail in \S\;\ref{app:tsys}). We first realized that \citet{godambe2008} and \citet{bhattacharyya2012} did not take into account these corrections, and thus all the flux densities from their data were under-estimated. In fact, the values we obtain before applying the \tsys corrections are compatible with the published ones in these two papers. In the case of \citet{pandey2007}, the origin of the discrepancy is different at each frequency. At 610~MHz, these authors did not apply the mentioned corrections (assuming that the contribution of the Galactic diffuse emission at such frequency is small, while we have found that it is relevant). At 235~MHz these authors obtained the \tsys corrections using the Haslam approximation, which overestimates the flux density in this case (see  \S\,\ref{app:tsys} and Table~\ref{tab:tsys-corrections}). These are the reasons why \citet{pandey2007} obtained a negative spectral index while we obtain a positive one at these low frequencies.

\subsection{Quasi-simultaneous spectra}

To avoid the problem of the non-simultaneity of the data in the previous spectrum, we conducted coordinated GMRT and WSRT observations in 2013 July 19 and 21. The results of these observations are shown in Figure~\ref{fig:ls5039-gmrt-freq-flux}. Although the two obtained spectra are qualitatively similar, again with a turnover around $\sim$0.5~GHz, we observe differences between them. The spectrum from 2013 July 19 exhibits a stronger emission and a pure power law behavior from 0.6 to 5~GHz, while on July 21 the spectrum is slightly curved. The data at 154~MHz were not simultaneous (taken every other day on 2013 July 18, 20 and 22) and provided upper limits on the flux density values of LS~5039. The combination of these three epochs does not improve remarkably the final upper limit, which is similar to the lowest one in Figure~\ref{fig:ls5039-gmrt-freq-flux}.
\begin{figure}[t]
        \centering
        \includegraphics[width=10cm]{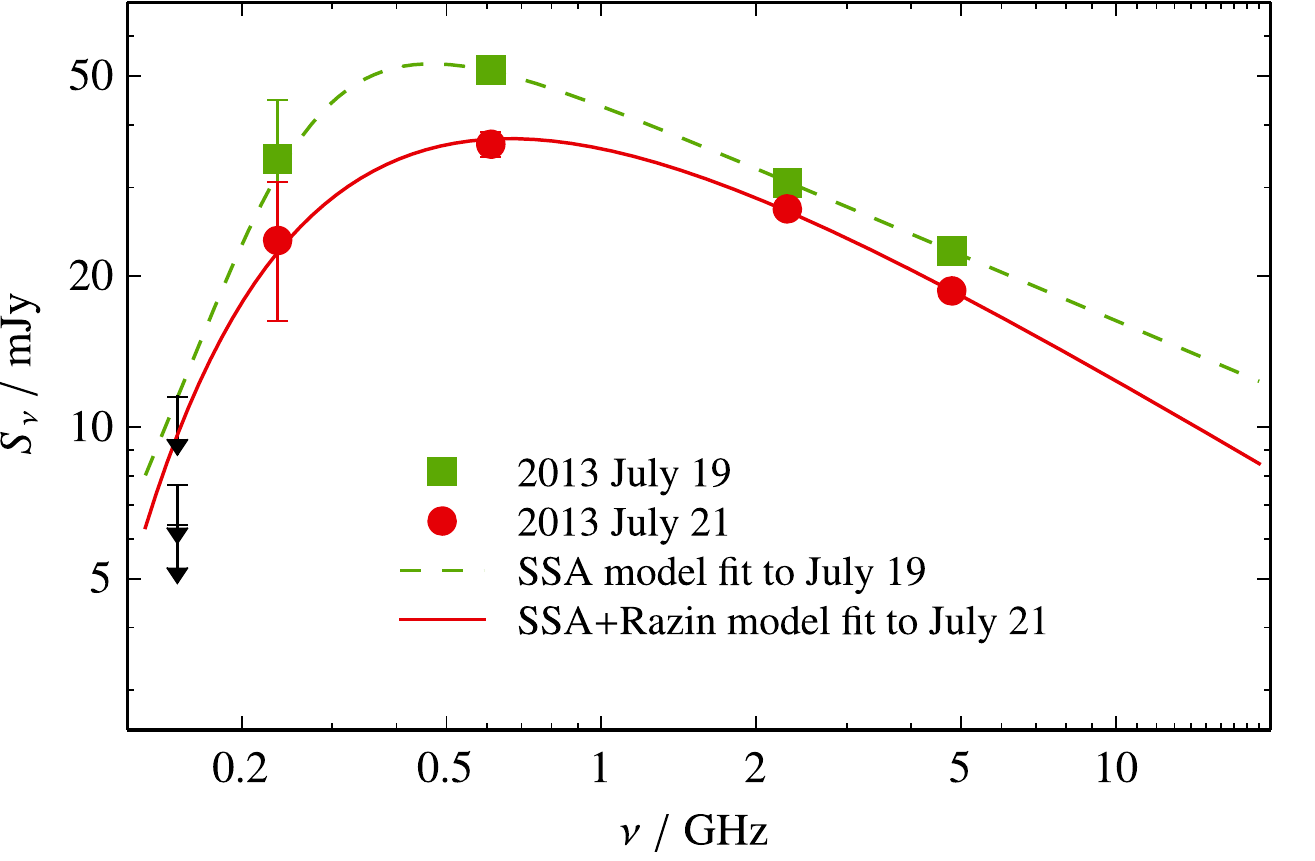}
        \caption[Quasi-simultaneous spectra of LS~5039 obtained with the coordinated GMRT-WSRT campaign in 2013.]{Quasi-simultaneous spectra of LS~5039 obtained with the coordinated GMRT-WSRT campaign in 2013. The data $< 1~\mathrm{GHz}$ were taken with GMRT and the data $> 1~\mathrm{GHz}$ with WSRT. The green squares represent the data of 2013 July 19 ($\phi\approx 0.9$) and the red circles the data of 2013 July 21 ($\phi\approx 0.4$). The black arrows show the 3-$\sigma$ upper limits from the 154~MHz GMRT data of 2013 July 18, 20 and 22. The green dashed line shows a SSA model fit to the July 19 data and the red solid line shows a SSA+Razin model fit to the July 21 data.}
        \label{fig:ls5039-gmrt-freq-flux}
\end{figure}

\section{Modeling the LS~5039 spectrum} \label{sec:ls5039-models}

We will now explain the different models that can fit the observed spectra of LS~5039 in the 0.15--15~GHz range and we will constrain some of the corresponding physical parameters. As discussed previously, the high frequency spectra above 1~GHz can be roughly fit with a power law with a negative spectral index of $\alpha \approx -0.5$. This, together with the compact radio emission seen at VLBI scales, has led to establishing the synchrotron nature of the radio source \citep{paredes2000}.
In addition, we have also unambiguously revealed the presence of a low-frequency turnover, which takes place at $\sim$$0.5~\mathrm{GHz}$.

To explain the observed radio spectrum of LS~5039 we have considered a simple model. The fact that the extended emission is only responsible for a small fraction of the total flux density \citep{moldon2012ls5039} allows us to consider as a first approximation a one-zone model. The absence of a clear orbital modulation in the radio flux density implies that the characteristics of the emitting region must be similar at any orbital phase from the point of view of the observer. The simplest model which verifies this condition is a spherically symmetric emitting region. For simplicity, we have also assumed that the emitting region is isotropic and homogeneous.
In this region we consider the presence of a synchrotron emitting plasma. The turnover below $\sim$$0.5~\mathrm{GHz}$ could be produced either by SSA, FFA or Razin effect. 

Considering a synchrotron emission from a power-law particle density distribution (as seen in \S\,\ref{sec:synchrotron-emission}), we obtain a flux density emission given by equation (\ref{eq:flux-density}), with
\begin{equation}\label{eq:emissivity}
    J_\nu\ \propto\ a(p)\,K\,B^{(p+1)/2}\,\nu^{(1 - p)/2},
\end{equation}
and $\kappa_{\nu} = \kappa_{\nu}^{\rm SSA} + \kappa_{\nu}^{\rm FFA}$ is the absorption coefficient, where
\begin{equation}
    \kappa_{\nu}^{\rm SSA}\ \propto\ K\,B^{(p+2)/2}\,b(p)\,\nu^{-(p+4)/2},
\end{equation}
and
\begin{equation}
    \kappa_{\nu}^{\rm FFA}\ \propto\ n_\mathrm{e}^2\,T_{\mathrm w}^{-3/2}\,\nu^{-2}.
\end{equation}

As the relativistic synchrotron emitting plasma is surrounded by a non-re\-la\-ti\-vis\-tic plasma from the stellar wind of the massive companion, the mentioned Razin effect would be also expected. In this case, we approximate the absorption in the emission by an exponential cutoff following equation (\ref{eq:razin-effect}).

Different combinations of the mentioned absorption mechanisms have been considered to explain the observed spectra of LS~5039, namely: SSA, FFA, SSA +FFA, SSA+Razin, FFA+Razin, and SSA+FFA+Razin. We have fit the data and obtained $\chi^2$ statistics from the residuals, from which we have computed the reduced value using the number of degrees of freedom (d.o.f.): $\chi_\mathrm{r}^2 \equiv \chi^2/\mathrm{d.o.f}$.

In the case of the non-simultaneous average spectrum we have taken into account all the data except the values at 2.3~GHz (given that we only have two observations at such frequency, the inferred mean value is not representative enough) and except the upper-limits at 154~MHz. In this case we observe that any of the considered models fit the data with $\chi_\mathrm{r}^2 \ll 1$, implying that their uncertainties are overestimated.
The reason is that these uncertainties reflect the variability of the source, and thus the dispersion of the data, but not the uncertainty of the mean value at each frequency.
Therefore, we need to consider the mean square errors, dividing the standard deviation at a given frequency by the square root of the number of measurements at such frequency.
When using the mean square errors, we observe that there are three models which can explain the spectrum: FFA, SSA and SSA+Razin (see Figure~\ref{fig:ls5039-all-freq-flux}, bottom, where mean square errors, within the size of the symbols, are not plotted). The $\chi_\mathrm{r}^2$ from these fits are 2.5, 4.6 and 1.9, respectively. The SSA+Razin model is thus the most accurate model to explain the LS~5039 average emission, and predicts a spectral index of $\alpha_\mathrm{SSA+Razin} = -0.58 \pm 0.02$ for the optically thin part of the spectrum.
Considering the different upper-limits at 154~MHz as possible flux density values at that frequency (i.e., a flux density of $3\sigma\pm\sigma$), we observe that the SSA+Razin is still the best model to reproduce the data (although if the assumed flux density at 154~MHz is as low as $\lesssim$7~mJy, a FFA+Razin seems to be needed).

In the case of the quasi-simultaneous spectra (Figure~\ref{fig:ls5039-gmrt-freq-flux}), we have followed the most conservative choice of considering the largest upper-limit at 154~MHz as the possible flux density of LS~5039 at such frequency ($12 \pm 4~\mathrm{mJy}$).
We have fit the two spectra using all the possible models mentioned above, and assessed their validity with a $\chi^2$ test.
A SSA+Razin model explains the spectra at both epochs, with $\chi_\mathrm{r}^2$ of $0.14$ and $0.48$ for 2013 July 19 and 21, respectively.
However, we note that on 2013 July 19 the SSA and the SSA+Razin models produce the same spectrum, because there is no significant contribution of the Razin effect at this epoch, and thus a SSA model fit produces a more significant explanation ($\chi_\mathrm{r}^2 = 0.07$, given the additional degree of freedom).
On 2013 July 21, a SSA model fit can not reproduce the observed spectrum ($\chi_\mathrm{r}^2 = 5.7$), which can be explained by a FFA model ($\chi_\mathrm{r}^2 = 1.4$). In addition, both the July 19 and 21 spectra are also reproduced by a FFA+SSA model (with $\chi_\mathrm{r}^2 = 0.014$ and $0.75$, respectively).
However, if we consider a more restrictive flux density of $7 \pm 2~\mathrm{mJy}$ at 154~MHz, we observe that the SSA+Razin model still reproduces satisfactorily both spectra with $\chi_\mathrm{r}^2 = 1.5$ and $0.40$, respectively (with a pure SSA model on July 19 as mentioned before, and thus $\chi_\mathrm{r}^2 = 0.83$ without the Razin contribution). On the other hand, with this lower flux at 154~MHz, the FFA+SSA model still explains the spectrum of July 19 ($\chi_\mathrm{r}^2 = 0.72$) but with a small contribution of FFA, while the July 21 is harder to be explained by this model ($\chi_\mathrm{r}^2 = 1.8$).

Therefore, the SSA+Razin model can explain both spectra. However, on 2013 July 19 the contribution of the Razin effect is negligible, and thus we observe a pure SSA.
The FFA model, in contrast, only explains the spectrum on 2013 July 21. The SSA+FFA can explain the two quasi-simultaneous spectra, although with a small contribution of FFA on July 19, but can not explain the average spectrum discussed in the previous paragraph.

In summary, a SSA+Razin model emerges as the most reasonable to explain the spectra of LS~5039, with a different contribution of the Razin effect, which might not be present at all, depending on the epoch.
We note that we only have 1--3 degrees of freedom in the fitting process, implying that in some cases it is subtle to distinguish between models producing similar spectra.

For the discussed models, the underlying physical parameters are coupled. Therefore, the number of free parameters is lower than the number of the physical ones. For the SSA model, we only have three parameters to fit: $p$, $P_1 \equiv \Omega\, B^{-1/2}$, and $P_2 \equiv K\,\ell\,B^{\frac{1}{2}\left(p+2\right)}$. For the SSA+Razin effect model, we have the three previous parameters and an additional one: $\nu_{\rm R} \equiv 20\,n_\mathrm{e}\,B^{-1}$. We can not decouple these physical parameters without making assumptions on some of them. Therefore, we prefer to work directly with the coupled parameters for the different spectra, which are quoted in Table~\ref{tab:ls5039-fits}. 

\begin{table}
\small
\centering
\caption[Parameters fit to the spectra of LS~5039.]{Parameters fit to the averaged data (A.S.) with a SSA+Razin model. The data on 2013 July 19 (J19) were fit with a SSA model and the data on 2013 July 21 (J21) with a SSA+Razin model. The parameters are: $p \equiv 1-2\,\alpha$, $P_1 \equiv \Omega\, B^{-1/2}$, $P_2 \equiv K\,\ell\,B^{(p+2)/2}$, and $\nu_{\rm R} \equiv 20\,n_\mathrm{e}\,B^{-1}$.}
\label{tab:ls5039-fits}
\begin{tabular}{l@{~~~~~}c@{~~~~~~}c@{~~~~~~}c@{~~~~~~}c}
	\hline\\[-10pt]
	Fit & $p$ & $P_1$ & $P_2$ & $\nu_{\rm R}$\\[-2pt]
	&& $\left(\mathrm{10^{-16}\,G^{-1/2}}\right)$ & $\left(10^{\,3}\,\mathrm{cm\,G^{(p+2)/2}} \right)$ & $\left(10^{\,8}\,\mathrm{Hz}\right)$\\[+4pt]
	\hline\\[-10pt]
	A.S.      & $2.16\pm0.04$ 	& $500\pm800$    & $3\pm5$      & $4.1\pm0.2$\\
	J19    & $1.867\pm0.014$	& $3.9\pm0.3$    & $(2.1\pm0.9)\times10^{6}$ & --\\
	J21    & $2.24\pm0.08$ 	& $200\pm600$	 & $0.4\pm1.7$  & $4.1\pm0.7$\\
	\hline
\end{tabular}
\end{table}

\section{Discussion}\label{sec:ls5039-discussion}

First we will discuss the physical implications of the modeling of the obtained radio spectra for LS~5039. Secondly, we will discuss the radio variability of the source, making a comparison with its multi-wavelength emission and the one observed in other gamma-ray binaries.

\subsection{Physical properties from the spectral modeling}

In this work we have reported multi-frequency data of LS~5039 that have allowed us to obtain a non-simultaneous average spectrum in the 0.15--15~GHz range, and two quasi-simultaneous spectra in the 0.15--5~GHz range, showing in all the cases a turnover at $\sim$$0.5~\mathrm{GHz}$.
The high-frequency emission is roughly explained by a power-law with a spectral index of $\alpha\approx-0.5$ (implying $p \sim 2$), which is typical of non-thermal synchrotron radiation, although some indications of curvature below 5~GHz are observed.
All the observed spectra of LS~5039 can be explained with the models which are summarized in Table~\ref{tab:ls5039-fits}. A SSA+Razin model explains the average spectrum and the 2013 July 21 spectrum, showing both similar properties (on the parameters $p \sim 2.2$ and $\nu_\mathrm{R} \approx 0.41~\mathrm{GHz}$), whereas on 2013 July 19, the spectrum is explained by a pure SSA model, with a negligible contribution of the Razin effect ($\nu_\mathrm{R} \approx 0$). In this case we observe a significant lower index $p = 1.867 \pm 0.014$, a 7-$\sigma$ deviation with respect the average spectrum. The $P_1$ and $P_2$ parameters can only be constrained for the 2013 July 19 spectrum  (see Table~\ref{tab:ls5039-fits}). Therefore, only in the case of $p$ and $\nu_{\rm R}$ we can claim for changes between the spectra obtained at different epochs. It seems that changes in the injection index but also in the electron density or the magnetic field are present in LS~5039 on timescales of about days.

The Razin effect is widely observed in CWBs, where there is a non-relativistic shock between the winds of two massive stars, with typical values of $\nu_{\rm R} \sim 2~\mathrm{GHz}$ \citep{vanloo2005thesis}. Interestingly, the $\nu_{\rm R}$ value inferred for LS~5039 on the average spectrum and on 2013 July 21 is only a factor of 5 smaller. The presence of the Razin effect in this gamma-ray binary, and the fact that shocks are present in CWBs, provides further support to the scenario of the young non-accreting pulsar, where shocks also take place, for LS~5039.
We remark that the presence of the Razin effect is also supported by the curvature reported on most of the LS~5039 spectra below 5~GHz, which is easily explained with the consideration of this effect but hardly explained by a SSA or FFA model.

The magnetic field $B$ can be estimated from the $P_1$ parameter obtained on the 2013 July 19 spectrum, by assuming a solid angle $\Omega$ for the emitting region, with a radius $\ell$ in the spherically symmetric case considered here.
From \citet{moldon2012ls5039} it is known that most of the radio emission arises from the compact core, which is not resolved with VLBI observations at 5~GHz, and thus has an angular size with a radius $\lesssim$$1~\mathrm{mas}$. At the $\sim$$3~\mathrm{kpc}$ distance of the source this represents $\sim$$3~\mathrm{AU}$, or about 10 times the semimajor axis of the binary system. We consider a solid upper-limit for the angular radius ($\ell\,d^{-1}$) of $\sim$$1.5~\mathrm{mas}$, and a lower-limit of $\sim$$0.5~\mathrm{mas}$ taking into account the orbital parameters and the absence of orbital modulation (which would be produced by the changing shock geometry and by absorption effects).
\begin{figure}[t]
        \centering
        \includegraphics[width=10cm]{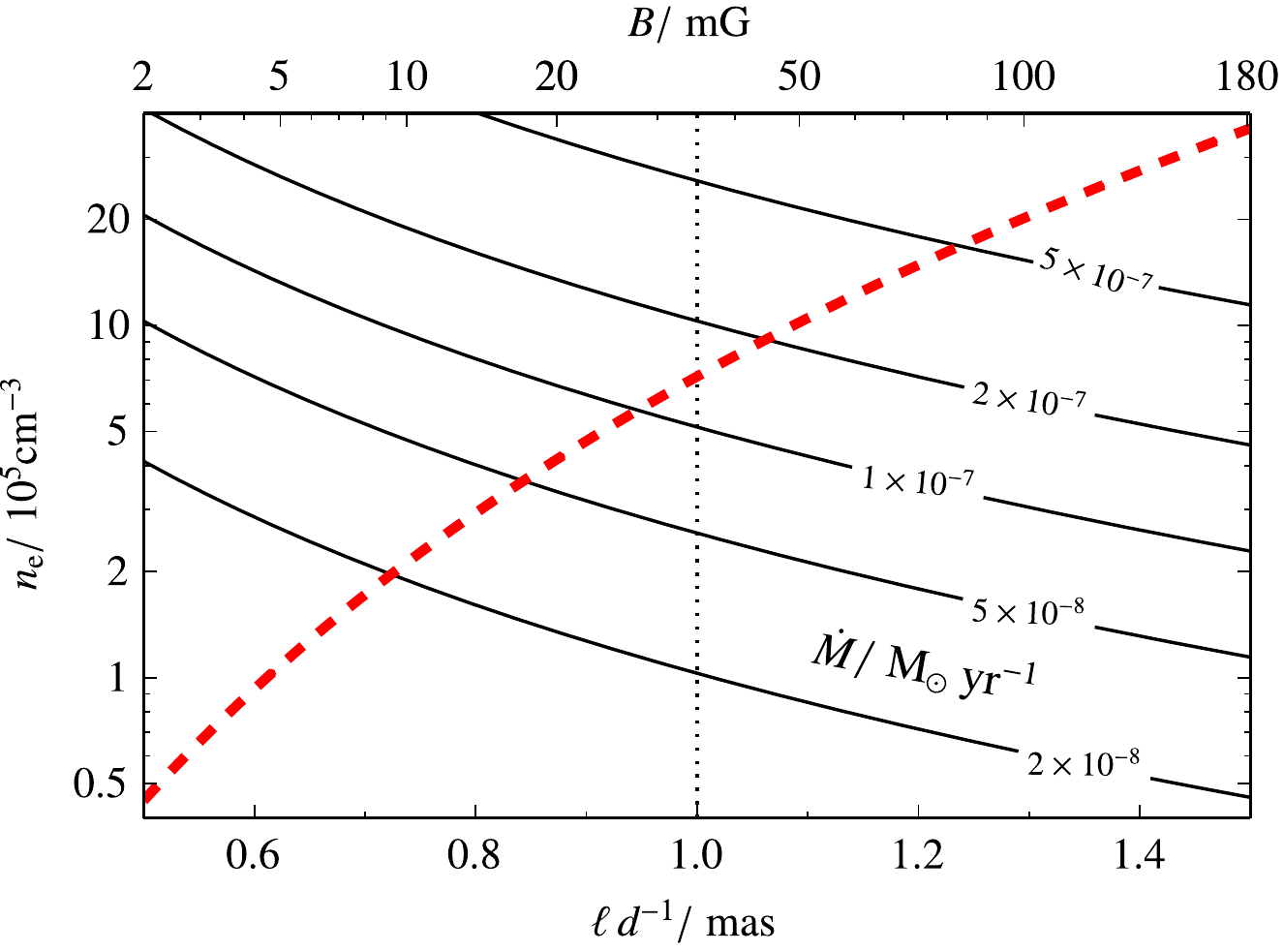}
        \caption[Electron number density for the non-relativistic plasma ($n_\mathrm{e}$) as a function of the angular radius of the emitting region ($\ell\, d^{-1}$) and of the magnetic field value ($B$).]{Electron number density for the non-relativistic plasma ($n_\mathrm{e}$) as a function of the angular radius of the emitting region ($\ell\, d^{-1}$, bottom axis) and of the magnetic field value ($B$, top axis). $B$ is determined from the angular radius and the $P_1$ parameter of the pure SSA fit to the 2013 July 19 data (see Table~\ref{tab:ls5039-fits}). The red dashed line denotes the $n_\mathrm{e}$ values inferred from the $\nu_\mathrm{R}$ parameter of the SSA+Razin fits (which is directly related to $B$). Solid black lines represent the $n_\mathrm{e}$ values obtained with different mass-loss rates $\dot M$. Values below the red dashed line imply an unrealistic mixing above 100\% of the non-relativistic wind inside the synchrotron radio emitting relativistic plasma. The vertical dotted line shows the central value of 1~mas.}
        \label{fig:ls5039-mdot}
\end{figure}
From the central value of $1~\mathrm{mas}$ we estimate $B \approx 35~\mathrm{mG}$.
For angular radius in the range of 0.5--1.5~mas we obtain a wide range of magnetic field values of 2--180~mG (see $x$-axes in Figure~\ref{fig:ls5039-mdot}).
These values range from much smaller to close to the $B\sim 200~\mathrm{mG}$ value inferred from the VLBI images at 5~GHz assuming equipartition between the relativistic electrons and the magnetic field \citep{paredes2000}. They are also encompassed with the 3--30~mG range quoted in \citet{bosch-ramon2009ls5039} using luminosity arguments for the lower limit and imposing FFA to derive the upper bound.
The obtained magnetic field value of $B \approx 35~\mathrm{mG}$ for 1~mas allows us to constrain the Lorentz factor of the electrons emitted at $0.5~\mathrm{GHz}$ to $\gamma\sim 60$, confirming that we are in the relativistic regime \citep{pacholczyk1970}.
Using the $\nu_\mathrm{R}$ value obtained from the SSA+Razin fits we derive the electron density for the non-relativistic plasma, $n_\mathrm{e}$, as a function of $B$ (red dashed line in Figure~\ref{fig:ls5039-mdot}), with values in the range of $4\times10^4$--$4\times10^6~\mathrm{cm^{-3}}$ for the 0.5--1.5~mas range, and of $7\times10^5~\mathrm{cm^{-3}}$ for the central value of 1~mas.
We can compare these values with the ones obtained from the stellar wind velocity and mass-loss rate. The velocity of the stellar wind is $v_\mathrm{w}\approx 2440~{\rm km\ s^{-1}}$ \citep{mcswain2004}.
Considering a mass-loss rate of $\dot{M} \approx 5 \times 10^{-7}~\mathrm{M_{\sun}\ yr^{-1}}$ \citep{casares2005ls5039} we estimate $n_\mathrm{e} \approx 2.6 \times10^6~\mathrm{cm^{-3}}$ for 1~mas, and thus a mixing of $\sim$$25\%$ of the non-relativistic wind inside the synchrotron radio emitting relativistic plasma.

\begin{figure}[t]
        \centering
        \includegraphics[width=10cm]{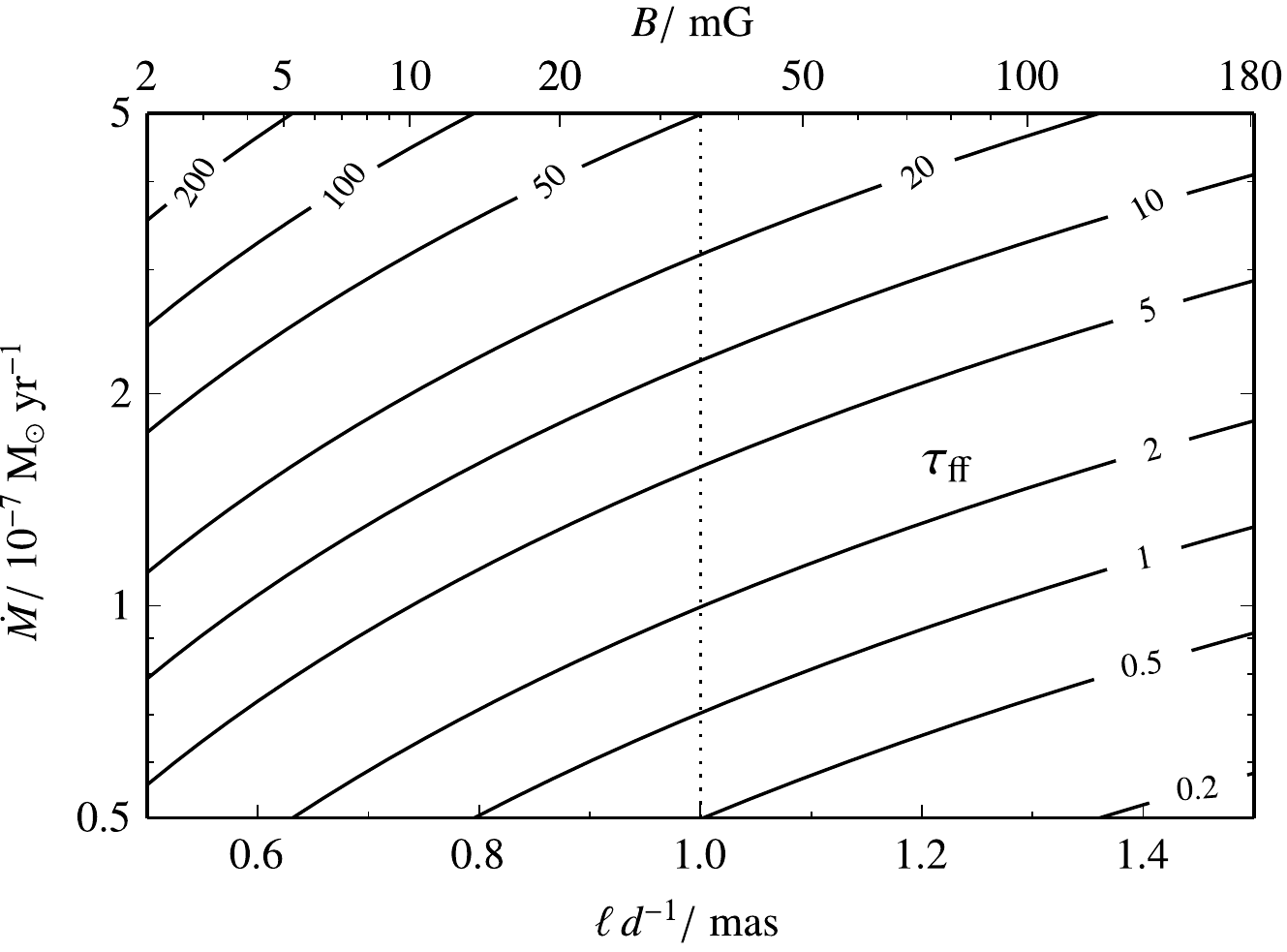}
        \caption[Mass-loss rate ($\dot M$) for different free-free opacities ($\tau_\mathrm{ff}$) as a function of the angular radius of the emitting region ($\ell\, d^{-1}$) and of the magnetic field value ($B$).]{Mass-loss rate ($\dot M$) for different free-free opacities ($\tau_\mathrm{ff}$) as a function of the angular radius of the emitting region ($\ell\, d^{-1}$, bottom axis) and of the magnetic field value ($B$, top axis). The relation between $\ell\, d^{-1}$ and $B$ is the same as in Figure~\ref{fig:ls5039-mdot}. Only the region below $\tau_\mathrm{ff}\sim1$ should be considered according to the observations, implying a low value for $\dot M$. The vertical dotted line shows the central value of 1~mas.}
        \label{fig:ls5039-tau}
\end{figure}
All these values allow us to estimate the free-free opacity with equation (\ref{eq:tauff}). Assuming a temperature for the stellar wind of $T_{\rm w}\approx1.3 \times 10^4~\mathrm{K}$ \citep{krticka2001} we obtain $\tau_{\rm ff} \approx 50$ for an angular radius of 1~mas (see Figure~\ref{fig:ls5039-tau}), which is not compatible with the presence of radio emission.
However, from optical spectroscopy there are indications of clumping, which might reduce the mass-loss rate by one order of magnitude down to $\dot{M} \approx 5 \times 10^{-8}~\mathrm{M_{\sun}\ yr^{-1}}$ (Casares et al.\, in preparation). This would yield $\tau_{\rm ff} \approx0.5$ (Figure~\ref{fig:ls5039-tau}), and thus the region would be optically thin to free-free opacity. This lower value of $\dot M$ would imply $n_\mathrm{e} \approx 2.6 \times 10^5~\mathrm{cm^{-3}}$, which is smaller than the value derived from the SSA+Razin model fits (see Figure~\ref{fig:ls5039-mdot}).
For this mass-loss rate, only angular radii of $\approx$$0.85~\mathrm{mas}$ or smaller would be supported (or the mixing would be above 100\%). At the same time, from Figure~\ref{fig:ls5039-tau} we observe that for sizes $\lesssim$$0.8~\mathrm{mas}$ we obtain $\tau_\mathrm{ff} \gtrsim 1$. Therefore, the inferred value for the radius of the emitting region is $\sim$$0.85~\mathrm{mas}$, implying a significant mixing of the non-relativistic wind inside the synchrotron radio emitting relativistic plasma, even close to $\sim$$100\%$. Another explanation in agreement with this value is the production of the radio emission by secondary electrons produced after photon-photon absorption within the stellar wind \citep{bosch-ramon2008}.

In summary, from the spectral modeling assuming a homogeneous one-zone emitting region we derive an angular radius of $\sim$$0.85~\mathrm{mas}$ ($\ell \sim 2.5~\mathrm{AU}$ at 3~kpc), yielding $B \sim 20~\mathrm{mG}$, and $n_\mathrm{e} \sim 4 \times 10^5~\mathrm{cm^{-3}}$. The mass-loss rate must be around $\dot M\sim 5 \times 10^{-8}~\mathrm{M_{\sun}\ yr^{-1}}$, i.e. about one order of magnitude smaller than the value reported in \citet{casares2005ls5039}, thus supporting the presence of a clumpy stellar wind. A high mixing with the relativistic plasma is supported.

\subsection{Variability}\label{sec:ls5039-variability}

In this work we have reported day to day variability, trends on week timescales and the absence of orbital variability.
The absence of orbital modulation in the radio emission of LS~5039 puts some constraints on the characteristics of the emitting region or radio core, in the sense that they can not change dramatically along the orbit or that the changes in different physical parameters should compensate each other, always as seen from the point of view of the observer.
In addition, the reported stability along the years of the average radio emission at all frequencies implies that this emitting region must be also stable on these timescales. On the other hand, hydrodynamical instabilities in the shocked material or in the outflow properties could explain more easily the observed variability on timescales of days and the trends observed on timescales of weeks \citep{bosch-ramon2009ls5039}.

In contrast to the radio emission, LS~5039 shows orbitally modulated flux at other wavelengths. The X-ray light-curve is periodic, exhibiting a maximum around orbital phases 0.7--0.8 (during the inferior conjunction of the compact object). It also exhibits small spikes at $\phi \approx 0.40$ and $0.48$ and a strong spike at $\phi \approx 0.70$, probably originated by geometrical effects. The X-ray emission, including these spikes, presents a long-term stability \citep{kishishita2009}.
The TeV light-curve is similar to the X-ray one, exhibiting a maximum around phase 0.7, a minimum flux at superior conjunction, and a strong spike but in this case at phase $\sim$0.8 \citep{aharonian2005ls5039}. The two light-curves are correlated, with the TeV variability ascribed to photon-photon pair production and anisotropic inverse Compton scattering and the X-ray variability to adiabatic expansion \citep{takahashi2009}.
The GeV light-curve, also periodic, exhibits the maximum emission around the periastron and superior conjunction ($\phi \sim 0.0$), with the minimum emission during the apastron and inferior conjunction ($\phi \sim 0.6$). Therefore, the GeV emission is almost in anti-phase with the X-ray and TeV emission, and is thought to reflect the change in the angle-dependent cross-section of the inverse Compton scattering process and in the photon density \citep{abdo2009}.

The absence of orbital modulation in radio is thus a particular case in the multiwavelength emission of LS~5039.
However, it is interesting to note the deviations from the general behavior observed around orbital phase 0.8 (as it happens at X-ray and TeV energies). Although in most cases they are not significant enough, we observe an increasing flux density at high frequencies around phase 0.8 (Figure~\ref{fig:ls5039-vla-phase-flux}), a higher flux density emission at 610~MHz starting just before phase 0.8 and extending until phase 0.2 (Figure~\ref{fig:ls5039-gmrt-phase-flux}), and a large dispersion in the high frequency spectral index close to phase 0.8 (Figure~\ref{fig:ls5039-vla-freq-flux}). Since the radio emission of low energy electrons might arise at scales significantly larger than the orbital system size, the fact that we see hints of variability during the inferior conjunction of the compact object at orbital phase 0.8 might suggest the presence of an additional component due to Doppler boosting or to a decrease in the absorption at radio frequencies when the cometary tail is pointing closer to the observer (as invoked by \citealt*{dhawan2006} to explain the radio outbursts of LS~I~+61~303).

To compare the behavior of LS~5039 with the one detected in the other known gamma-ray binaries, we summarize some of their properties in Table~\ref{tab:ls5039-other-binaries}. Given the reduced number of known sources and the large differences in their physical properties, a statistical comparison between them is impossible. However, it is remarkable that LS~5039 is the only case in which the radio emission is not orbitally modulated in a periodic or almost periodic way. At high energies (from X-rays to TeV) all of them, including LS~5039, show a periodic emission. That makes even more special the behavior of the radio emission of this source with respect to the rest of gamma-ray binaries. This behavior could be related to the fact that LS~5039 presents the shortest orbital period (3.9~d) and the smallest known eccentricity (0.35), which naturally imply a higher absorption of the inner part of the radio outflow.
1FGL~J1018.6$-$5856 can probably be considered as the physically most similar system to LS~5039, because both have similar massive stars (spectral type O6\,V versus O6.5\,V) and the shortest orbital periods (16.6 versus 3.9~d).
Unfortunately, the eccentricity of 1FGL~J1018.6$-$5856 is still unknown, and in the case of being high it could easily explain the observed orbitally modulated radio emission (e.g. due to changes in the absorption processes), contrary to the case of LS~5039.

\section{Summary and conclusions}\label{sec:ls5039-conclusions}

We have presented a coherent picture of the 0.15--15~GHz spectrum of LS~5039, solving the discrepancies of the low frequency data reported in previous publications. We have unambiguously revealed the presence of a curvature in most of the spectra below 5~GHz and a persistent turnover that takes place at $\sim$$0.5~\mathrm{GHz}$. As described in this work, the average spectrum of LS~5039 can be approximated by a simple model with one-zone emitting region, which can be considered homogeneous and spherically symmetric, radiating according to a SSA model and exhibiting evidence of Razin effect. The Razin effect, reported for the first time in a gamma-ray binary, explains the mentioned curvature below 5~GHz. As this effect is commonly observed in CWBs, it is expected to be present in case that the emission from gamma-ray binaries arises also from a shock originated by the winds of the companion star and the compact object, as it happens in the young non-accreting pulsar scenario.
We observe a certain stability for the parameters of the fits that are well constrained between the average spectrum and particular epochs. However, at other epochs the source shows a slightly different behavior that implies changes in the injection and in the contribution of the Razin effect (and thus changes in the electron density or the magnetic field).
The presence of FFA cannot be discarded in some of the spectra, although the SSA+Razin model is the only one which can explain all the observed spectra.
For angular radii in the range of 0.5--1.5~mas we derive magnetic field values of $B\sim 2$--$180~\mathrm{mG}$. A coherent picture within the one-zone modeling, considering reasonable values of free-free opacity, is obtained for an angular radius of 0.85~mas, $B \sim 20$~mG, $n_\mathrm{e} \sim 4 \times 10^5~\mathrm{cm^{-3}}$, and $\dot M \sim 5 \times 10^{-8}~\mathrm{M_{\sun}\ yr^{-1}}$. These values imply a significant mixing of the stellar wind within the relativistic plasma of the radio outflow. This is the first time that a coherent picture of the physical properties of the assumed one-zone emitting region is presented for LS~5039, including the magnetic field value of the radio emitting plasma.

We have also shown that the radio emission of LS~5039 is persistent with a small variability on day, week and year timescales. This variability is present in all the explored frequency range (0.15--15~GHz), and the relative variation in the flux densities is similar at all frequencies, exhibiting a standard deviation of $\sim$$10$--$25\%$. The absence of orbitally modulated variability constrains the characteristics of the radio emitting region. The observed variability would be produced by stochastic instabilities in the particle injection or in the shocked material, but not by geometrical effects due to the orbital motion. Although the persistence of the flux density was known at 1-yr timescales, we have extend this knowledge up to scales of $\sim$$15~\mathrm{yr}$.
At orbital phases $\sim$$0.8$ we have detected signatures of an increasing flux density trend at high frequencies, of the starting of an enhanced flux density emission at 610~MHz, and of a large dispersion in the high frequency spectra. Although these signatures are not significant, it is notable that they all happen at the same orbital phase when there is enhanced X-ray and TeV emission. Additional monitoring campaigns, specially at low frequencies but also at high frequencies could clarify if this behavior is the result of a reduced number of observations or if it is an intrinsic effect from the source due to Doppler boosting or changes in the absorption mechanisms possibly connected with the X-ray or TeV emission.

LS~5039 remains undetected at 154~MHz with the cumulative of $\sim$$17~\mathrm{hr}$ of GMRT observations, although according to the modeling its flux density should be close to our upper-limits. A detection with more sensitive interferometric observations would improve our knowledge about the absorbed part of the spectrum. We expect that the higher sensitivity of LOFAR would allow us to clearly detect the source for first time at such frequency. In the future, other facilities such us LWA, MWA, SKA-low, together with LOFAR, GMRT and the new low-frequency receivers at the VLA, will significantly improve the sensitivity and resolution of low frequency radio observations in the 10~MHz--1~GHz range, thus allowing for detailed studies of absorption in gamma-ray binaries.


%
%
%
%
\def\path{figures/lsi61303_figures}

\chapter[Variability at low frequencies of the gamma-ray binary LS~I~+61~303]{Variability at low frequencies of the gamma-ray binary LS~I~+61~303\vspace{-25pt}} \label{chap:lsi}





LS~I~+61~303 is a gamma-ray binary that exhibits an outburst at GHz frequencies each orbital cycle of $\approx$$26.5~\mathrm{d}$ and a superorbital modulation with a period of $\approx$4.6~yr. Its low-frequency emission has not been studied in detail up to now. At these low frequencies we would expect to see a modification of the outburst profile due to absorption processes, such as synchrotron self-absorption, free-free absorption, or Razin effect.
To unveil the low-frequency radio emission of LS~I~+61~303 we have performed a detailed study by analyzing all the archival GMRT data at 150, 235 and 610~MHz, and conducting regular LOFAR observations within the RSM at 150~MHz.
We have detected the source for the first time at 150~MHz, which is also the first detection of a gamma-ray binary at such a low frequency. We have obtained the light-curves of the source at 150, 235 and 610~MHz, all of them showing orbital modulation. The light-curves at 235 and 610~MHz also show the existence of superorbital variability. A comparison with contemporaneous 15-GHz data shows remarkable differences with these light-curves.
At 15~GHz we see clear outbursts, whereas at low frequencies we see variability with wide maxima. The light-curve at 235~MHz seems to be anticorrelated with the one at 610~MHz, implying a shift of $\sim$0.5 orbital phases in the maxima. We model the shifts between the maxima at different frequencies as due to the expansion of a one-zone emitting region assuming either free-free absorption or synchrotron self-absorption with two different magnetic field dependences. We always obtain a subrelativistic expansion velocity, in some cases being close to the stellar wind one. This work has been published in
\citet{marcote2015lsi61303}.

\section{Introduction}\label{sec:lsi-intro}

LS~I~+61~303 is a binary system comprising a young main-sequence B0~Ve star of 10--15~$\mathrm{M_{\sun}}$ and a compact object orbiting it with a period $P_{\rm orb} = 26.496 \pm 0.003~\mathrm{d}$ \citep{gregory2002} and an eccentricity $e = 0.72 \pm 0.15$ \citep{casares2005lsi61303}. The mass of the compact object remains unconstrained, although a neutron star would be favored in the case of a high inclination of the orbit ($\gtrsim$$25^{\degree}$), and a BH in the case of a low inclination \citep{casares2005lsi61303}. The system is located at $2.0 \pm 0.2~\mathrm{kpc}$ according to H~I measurements \citep{frail1991}, with coordinates \citep{moldon2012thesis}
$$\alpha = 2^{\rm h} 40^{\rm m} 31.7^{\rm s},\qquad \delta = +61^{\degree} 13' 45.6''.$$
Assuming the epoch of the first radio observation as the origin of the orbital phase, ${\rm JD_0} = 2\,443\,366.775$, the periastron passage takes place in the orbital phase range $0.23$--$0.28$ \citep{casares2005lsi61303, aragona2009}.
LS~I~+61~303 was initially identified as the counterpart of a gamma-ray source detected by {\em COS~B} \citep{hermsen1977}, and subsequently an X-ray counterpart was also detected \citep{bignami1981}. X-ray observations with {\em RXTE}/ASM revealed an orbital X-ray modulation \citep{paredes1997}. The system was also coincident with an EGRET source above 100~MeV \citep{kniffen1997} and finally it was detected as a TeV emitter by MAGIC \citep{albert2006}, which was later confirmed by VERITAS \citep{acciari2008}. An orbital modulation of the TeV emission was also found by MAGIC \citep{albert2009}. A correlation between the TeV and X-ray emission was revealed using MAGIC and X-ray observations, which together with the observed X-ray/TeV flux ratio, supports leptonic models with synchrotron X-ray emission and IC TeV emission (\citealt{anderhub2009}; but see \citealt{acciari2009} for other interpretations). Finally, {\em Fermi}/LAT reported GeV emission, orbitally modulated and anti-correlated with respect to the X-ray/TeV emission \citep{abdo2009}.
\begin{figure}
        \centering
        \includegraphics[width=9cm]{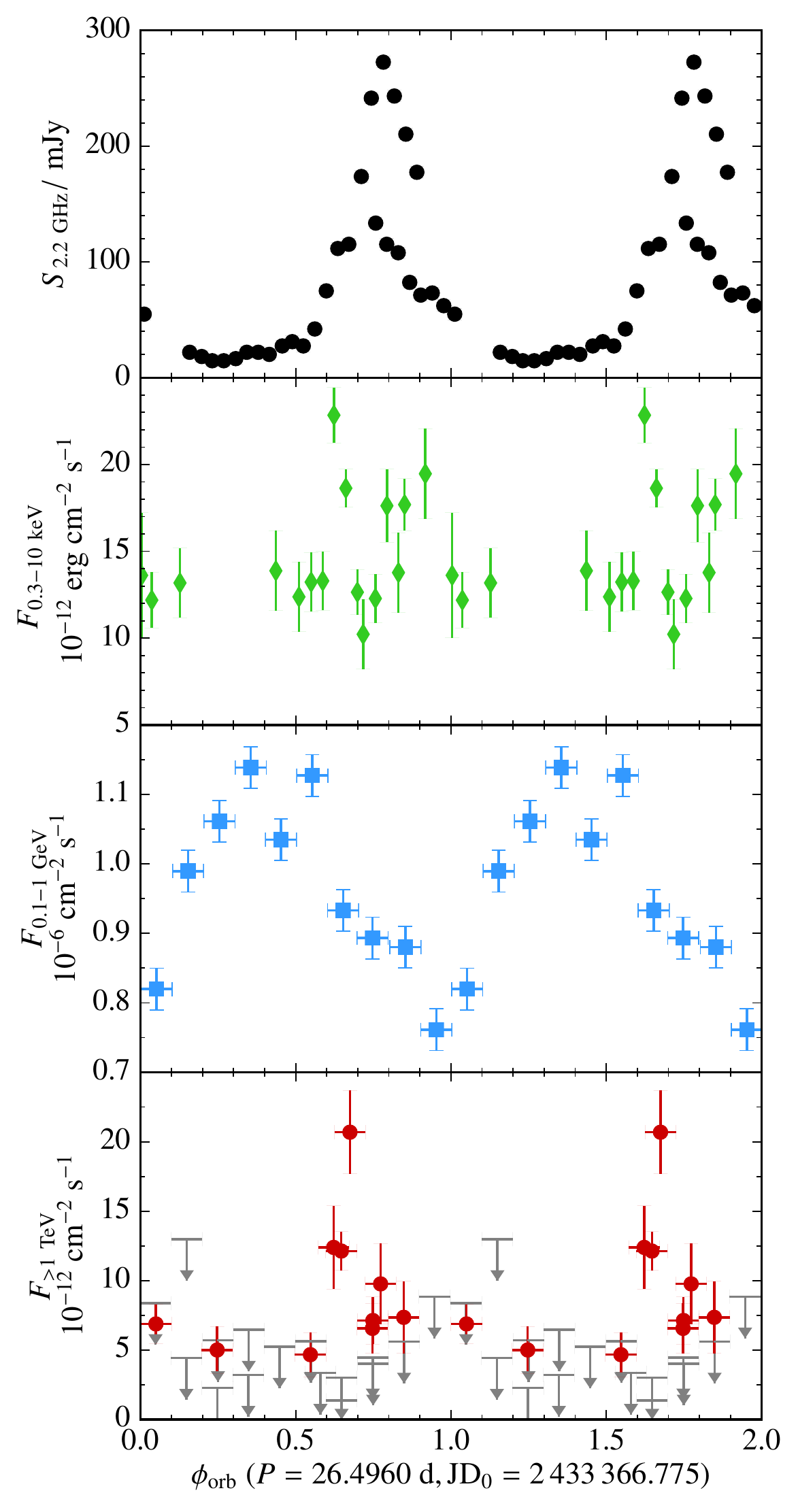}
        \caption[Multi-wavelength light-curves of LS~I~+61~303 folded with the orbital period at radio, X-rays, GeV, and TeV.]{Light-curves of LS~I~+61~303 folded with the orbital period at radio, X-rays, GeV, and TeV (from top to bottom, respectively). The data have been obtained from \citet{harrison2000} (radio and X-rays, in black circles and green diamonds, respectively), \citet{hadasch2012} (GeV, in blue squares), and \citet{acciari2011} (TeV: detections in red circles and upper-limits in gray) at similar superorbital phases ($\phi_{\rm so} \sim 0.1$), with the exception of the TeV ones, which cover almost all the superorbital phase range.}
        \label{fig:lsi61303-lihgtcurves-multiwavelength}
\end{figure}

The 1--10~GHz radio light-curve of LS~I~+61~303 is also orbitally modulated, showing a clear outburst each orbital cycle that increases the $\lesssim$\,50-mJy steady flux density emission up to $\sim$100--200~mJy. These outbursts are periodic, although changes in their shape and intensity have been reported from cycle to cycle (\citealt{paredes1990}; \citealt{ray1997}). \citet{strickman1998} studied the evolution of a single outburst with multifrequency observations in the range 0.33--23~GHz. The outburst is detected at all frequencies, but the flux density peaks first at the highest frequencies, and later at the lower ones. A low-frequency turnover was also suggested to be present in the range 0.3--1.4~GHz \citep{strickman1998}. Figure~\ref{fig:lsi61303-lihgtcurves-multiwavelength} shows the light-curves of LS~I~+61~303 at different wavelengths.

A long-term modulation is also observed at all wavelengths in LS~I~+61~303, the so-called superorbital modulation, with a period $P_{\rm so} = 1\,667 \pm 8~\mathrm{d}$ or $\sim$4.6~yr \citep{gregory2002}. This modulation was firstly found at GHz radio frequencies \citep{paredes1987, gregory1989, paredes1990}, affecting the amplitude of the non-thermal outbursts and the orbital phases at which the onset and peak of these outbursts take place, drifting from orbital phases of $\sim$0.45 to $\sim$0.95 \citep{gregory2002}. The source exhibits the minimum activity at GHz frequencies during the superorbital phase range of $\phi_{\rm so} \sim 0.2$--$0.5$, whereas the maximum activity takes place at $\phi_{\rm so} \sim 0.78$--$0.05$, assuming the same ${\rm JD_0}$ as for the orbital phase. A similar behavior is also observed in optical photomegftric and H$\alpha$ equivalent width observations that trace the thermal emission of the source \citep{paredes-fortuny2015}.
The origin of the superorbital modulation could be related to periodic changes in the circumstellar disc and the mass-loss rate of the Be star \citep{zamanov2013}, although other interpretations within the framework of a precessing jet are still discussed \citep[see][and references therein]{massi2014}.

At milliarcsecond scales LS~I~+61~303 has been observed and resolved several times \citep{massi2001, massi2004, dhawan2006, albert2008, moldon2012thesis}. \citet{dhawan2006} showed a changing morphology as a function of the orbital phase, resembling a cometary tail.
\citet{albert2008} observed similar morphological structures at similar orbital phases to those considered in \citet{dhawan2006}. Later on, \citet{moldon2012thesis} also reported this behavior: the morphology of LS~I~+61~303 changes periodically within the orbital phase. These morphological changes have been interpreted as evidence of the pulsar scenario \citep{dubus2006,dubus2013}, although other interpretations have also been suggested \citep[see][and references therein]{massi2012}. It must be noted that radio pulsation searches have also been conducted with unsuccessful results \citep{mcswain2011,canellas2012}.

At the lowest frequencies ($\lesssim 1~\mathrm{GHz}$), only a few observations have been published up to now. \citet{pandey2007} conducted two observations of LS~I~+61~303 at 235 and 610~MHz simultaneously with the GMRT. They observed a positive spectral index of $\alpha \approx 1.3$ in both epochs and reported variability at both frequencies at a 2.5 and 20-$\sigma$ level, respectively. Although \citet{strickman1998} also observed the source at 330~MHz three times during one outburst, given the large uncertainties of the obtained flux densities they could not infer variability at more than the $1$-$\sigma$ level.
At these low frequencies we should observe, as in the case of LS~5039 (see \S\,\ref{chap:ls}), the presence of different absorption mechanisms, such as SSA, FFA or the Razin effect. In addition, at these low frequencies we should also expect extended emission at about arcsec scales originating from the synchrotron emission from low-energy particles \citep{bosch-ramon2009ls5039,durant2011}.

In this work we perform the first deep and detailed study of the radio emission from LS~I~+61~303 at low frequencies through GMRT and LOFAR data to unveil its behavior along the orbit. We compare the results with contemporaneous high-frequency observations conducted with the Ryle Telescope (RT) and the Owens Valley Radio Observatory (OVRO) 40-m telescope. In Sect.~\ref{sec:lsi-observations} we present all the radio data analyzed in this work together with the data reduction and analysis processes. In Sect.~\ref{sec:lsi-results} we present the obtained results and in Sect.~\ref{sec:lsi-discussion} we discuss the observed behavior of LS~I~+61~303 in the context of the known orbital and superorbital variability. Our conclusions are presented in Sect.~\ref{sec:lsi-conclusions}.

\section{Observations and data reduction}\label{sec:lsi-observations}

To reveal the orbital behavior of LS~I~+61~303 at low radio frequencies we have analyzed data from different instruments and different epochs. These data include archival GMRT observations from a monitoring program in 2005--2006, as well as three additional archival observations from 2005 and 2008. They also include a LOFAR commissioning observation in 2011 and five LOFAR observations performed in 2013. RT and OVRO observations at 15~GHz, contemporaneous to these low-frequency observations, have also been analyzed to obtain a complete picture of the behavior of the source.
In Figure~\ref{fig:summary-obs} we summarize all these observations in a frequency versus time diagram. In the first seven columns of Table~\ref{tab:data} we show the log of the low-frequency observations.
\begin{figure}[!t]
    \centering
    \includegraphics[width=0.8\textwidth]{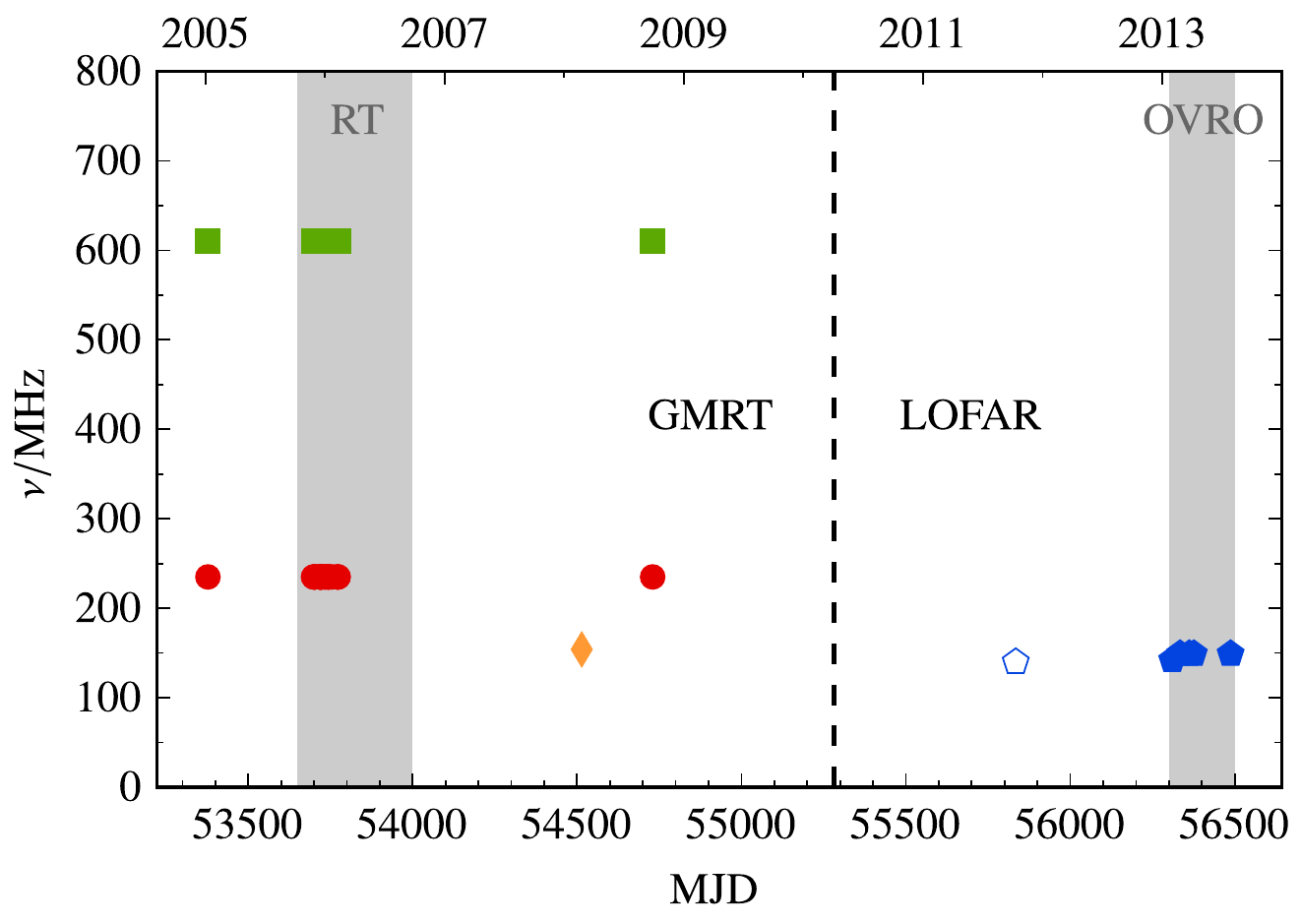}
    \caption[Summary of all the data of LS~I~+61~303 presented in this work in a frequency versus time diagram.]{Summary of all the data presented in this work in a frequency versus time diagram (Modified Julian Date, MJD, on the bottom axis and calendar year on the top axis). The vertical dashed line separates the GMRT data (left) from the LOFAR data (right). Green squares represent the 610-MHz GMRT observations, red circles correspond to the simultaneous GMRT observations at 235~MHz, the orange diamond shows the 154-MHz GMRT data and the blue pentagons represent the LOFAR observations at $\approx$150~MHz. The open pentagon corresponds to the commissioning LOFAR observation, which is not considered in the results of this work. The gray bands indicate the time intervals during which 15-GHz RT and OVRO observations used in this work were conducted.}
    \label{fig:summary-obs}
\end{figure}

\begin{sidewaystable}
\centering
\scriptsize
\def\s{~~~~~~}
\caption[Summary of the low-frequency data of LS~I~+61~303 presented in this work.]{Summary of the low-frequency data of LS~I~+61~303 presented in this work. Each row corresponds to one observation for which we show the facility used, the project code of the observation, the date (in calendar date and Modified Julian Date, MJD) of the center of the observation, the corresponding orbital phase $\phi_{\rm orb}$ (using $P_{\rm orb} = 26.4960~\mathrm{d}$ and ${\rm JD_0} = 2\,443\,366.775$), the corresponding superorbital phase (assuming $P_{\rm so} = 1\,667~\mathrm{d}$ and the same ${\rm JD_0}$), the total time on source ($t$), and the flux densities at each frequency with the 1-$\sigma$ uncertainty (3-$\sigma$ upper-limits in case of non-detections).}
\label{tab:data}
\begin{tabular}{l@{\s}c@{\s}c@{\s}c@{\s}c@{\s}c@{\s}c@{\s}r@{\;}c@{\;}l@{\s}r@{\;}c@{\;}l@{\s}r@{\;}c@{\;}l}
\hline\\[-8pt]
Facility & Project Code & Date & MJD & $\phi_{\rm orb}$ & $\phi_{\rm so}$ & $t$ &
\multicolumn{3}{c}{$S_{\approx150\,\mathrm{MHz}}$} &
\multicolumn{3}{c}{$S_{235\,\mathrm{MHz}}$} &
\multicolumn{3}{c}{$S_{610\,\mathrm{MHz}}$}\\
&&(dd/mm/yyyy)&&&& $(\mathrm{min})$&
\multicolumn{3}{c}{$(\mathrm{mJy})$} &
\multicolumn{3}{c}{$(\mathrm{mJy})$} &
\multicolumn{3}{c}{$(\mathrm{mJy})$}\\[+2pt]
\hline\\[-8pt]
 GMRT & \hphantom{4}\ 07PDA01$^{(1)}$ & 07/01/2005 & 53377.47 & 0.84& 0.01&30& &-& & 34 &$\pm$& 10 & 169 &$\pm$& 3\\
GMRT & 09PDA01 & 24/11/2005 & 53698.63 & 0.96 & 0.20& 82& &-& & \multicolumn{3}{c}{$<31$} & 89.9 &$\pm$& 1.3\\
GMRT & 09PDA01 & 25/11/2005 & 53699.71 & 0.00 & 0.20& 56& &-& & \multicolumn{3}{c}{$<30$} & 90 &$\pm$& 2\\
GMRT & 09PDA01 & 26/11/2005 & 53700.58 & 0.03 & 0.20& 82& &-& & 23 &$\pm$& 4 & 84.5 &$\pm$& 1.9\\
GMRT & 09PDA01 & 27/11/2005 & 53701.59 & 0.07 & 0.20& 164& &-& & 35 &$\pm$& 4 & 79.3 &$\pm$& 1.2\\
GMRT & 09PDA01 & 29/11/2005 & 53703.62 & 0.15 & 0.20& 80& &-& & 41 &$\pm$& 4 & 72.8 &$\pm$& 1.5\\
GMRT & 09PDA01 & 12/12/2005 & 53716.70 & 0.64 & 0.21& 60& &-& & 56 &$\pm$& 6 & 69.3 &$\pm$& 1.6\\
GMRT & 09PDA01 & 13/12/2005 & 53717.65 & 0.68 & 0.21& 70& &-& & 41 &$\pm$& 2 & 58.5 &$\pm$& 1.3\\
GMRT & 09PDA01 & 15/12/2005 & 53719.61 & 0.75 & 0.21& 80& &-& & 38 &$\pm$& 9 & 67.5 &$\pm$& 1.5\\
GMRT & 09PDA01 & 16/12/2005 & 53720.53 & 0.79 & 0.21& 524& &-& & 39.0 &$\pm$& 1.9 & 61.2 &$\pm$& 0.8\\
GMRT & 09PDA01 & 17/12/2005 & 53721.56 & 0.82 & 0.21& 130& &-& & 33 &$\pm$& 3 & 97.1 &$\pm$& 1.1\\
GMRT & 09PDA01 & 18/12/2005 & 53722.53 & 0.86 & 0.21& 216& &-& & 40 &$\pm$& 4 & 92.1 &$\pm$& 1.1\\
GMRT & 09PDA01 & 19/12/2005 & 53723.53 & 0.90 & 0.21& 80& &-& & 43 &$\pm$& 5 & 87.8 &$\pm$& 1.7\\
GMRT & 09PDA01 & 29/12/2005 & 53733.61 & 0.28 & 0.22& 80& &-& & 48 &$\pm$& 11 & 64.8 &$\pm$& 1.7\\
GMRT & 09PDA01 & 30/12/2005 & 53734.49 & 0.31 & 0.22& 80& &-& & &-& & 58.8 &$\pm$& 1.9\\
GMRT & 09PDA01 & 31/12/2005 & 53735.41 & 0.35 & 0.22& 56& &-& & 53 &$\pm$& 5 & 54 &$\pm$& 3\\
GMRT & 09PDA01 & 08/01/2006 & 53743.71 & 0.66 & 0.23& 41& &-& & \multicolumn{3}{c}{$<69$} & 26 &$\pm$& 3\\
GMRT & 09PDA01 & 09/01/2006 & 53744.36 & 0.68 & 0.23& 144& &-& & \multicolumn{3}{c}{$<46$} & 26.7 &$\pm$& 0.7\\
GMRT & 09PDA01 & 10/01/2006 & 53745.55 & 0.73 & 0.23& 82& &-& & 20 &$\pm$& 4 & 33.9 &$\pm$& 1.0\\
GMRT & 09PDA01 & 19/01/2006 & 53754.67 & 0.07 & 0.23& 82& &-& & &-& & 93.6 &$\pm$& 1.6\\
GMRT & 09PDA01 & 20/01/2006 & 53755.49 & 0.10 & 0.23& 70& &-& & 36 &$\pm$& 2 & 99.9 &$\pm$& 1.6\\
GMRT & 09PDA01 & 21/01/2006 & 53756.31 & 0.14 & 0.23& 48& &-& & &-& & 88.4 &$\pm$& 1.1\\
GMRT & 09PDA01 & 05/02/2006 & 53771.31 & 0.70 & 0.24& 60& &-& & 29 &$\pm$& 5 & 64 &$\pm$& 2\\
GMRT & 09PDA01 & 06/02/2006 & 53772.27 & 0.74 & 0.24& 82& &-& & \multicolumn{3}{c}{$<33$} & 75.9 &$\pm$& 1.4\\
GMRT & 09PDA01 & 07/02/2006 & 53773.61 & 0.79 & 0.24& 82& &-& & 22 &$\pm$& 5 & 90.3 &$\pm$& 1.7\\
GMRT & 13MPA01 & 18/02/2008 & 54514.28 & 0.74 & 0.69& 662& 52 &$\pm$& 11$^*$ & &-& & &-&\\
GMRT & 08DT051 & 20/09/2008 & 54729.90 & 0.88 & 0.82& 492& &-& & 140 &$\pm$& 2 & 337 &$\pm$& 3\\
LOFAR & L84298--L84317& 17/01/2013 & 56309.83 & 0.51 & 0.76&200 & 32 &$\pm$& 6$^\dagger$ & &-& & &-&\\
LOFAR & L89566--L89589& 10/02/2013 & 56333.71 & 0.41 & 0.78& 116& \multicolumn{3}{l}{$<24^{\dagger\dagger}$} & &-& & &-&\\
LOFAR & L100374--L100397& 10/03/2013 & 56361.67 & 0.47 & 0.80&116 & \multicolumn{3}{l}{$<28^{\dagger\dagger}$} & &-& & &-&\\
LOFAR & L107793--L107816& 24/03/2013 & 56375.58 & 0.99 & 0.80&116 & 77 &$\pm$& 9$^{\dagger\dagger}$ & &-& & &-&\\
LOFAR & L160534--L160557& 14/07/2013 & 56487.42 & 0.21 & 0.87&116 & 31 &$\pm$& 6$^{\dagger\dagger}$ & &-& & &-&\\
\hline
\end{tabular}
\flushleft {(1)~\citet{pandey2007}. $^*$ This observation was centred at 154~MHz with a bandwidth of 16~MHz. $^\dagger$ This observation was centred at 142~MHz with a bandwidth of 2.3~MHz. $^{\dagger\dagger}$ These observations were centred at 149~MHz with a bandwidth of 0.8~MHz.}
\end{sidewaystable}

\subsection{Archival GMRT observations}

The GMRT monitoring covers the observations performed between 2005 November 24 and 2006 February 7, at 235 and 610~MHz, simultaneously. These data sets include 25 observations spread over this time interval, all of them obtained with a single IF, with single circular polarization at 235~MHz (LL) and 610~MHz (RR). The 235 and 610~MHz data have bandwidths of 8 and 16~MHz, divided into 64 and 128 channels, respectively. These observations display a wide range of observing times, from 30~min to 11~h (see column 7 in Table~\ref{tab:data}). 3C~48, 3C~147 and 3C~286 were used as amplitude calibrators, and 3C~119 (0432$+$416) or 3C~48 as phase calibrators.

In addition to the previous monitoring, there are also three isolated archival GMRT observations. Two of these observations were carried out on 2005 January 7 and on 2008 September 20, at 235 and 610~MHz, using the same setup as that described above. The remaining observation was performed at 154~MHz on 2008 February 18, with dual circular polarization (LL and RR) and a bandwidth of 16~MHz divided into 128 channels. 3C~48, 3C~147 and 3C~286 were again used as amplitude calibrators, and 3C~119 was the phase calibrator for the three observations.

The GMRT data were calibrated and analyzed using standard procedures within AIPS, although Obit, ParselTongue and SPAM have also been used to develop scripts that call AIPS tasks to easily reduce the whole GMRT data set.
We loaded and flagged the raw visibilities in AIPS, removing telescope off-source times, instrumental problems or RFI. A first calibration using a unique channel, free of RFI, was performed on the data, and after that we bandpass calibrated them. A more accurate flagging process was performed later on and we finally calibrated the full data set in amplitude and phase, taking into account the previous bandpass calibration.
The target source was imaged and self-calibrated several times to correct for phase and amplitude errors that were not properly corrected during the previous calibration, using a Briggs robust weighting of zero in the cleaning process. As the GMRT field of view at these low frequencies covers a few degrees (between $\sim$5 and $1.5^{\degree}$ at 154 and 610~MHz, respectively), we needed to consider the full $uvw$-space during the imaging process and correct the final images for the primary beam attenuation. The GMRT observation performed on 2005 November 28 could not be properly calibrated and hence could not provide results, leading to a total of 24 flux density measurements. In the observations performed on 2005 December 30, 2006 January 19 and 21, the 235~MHz data could not be properly calibrated: the runs were relatively short and a large amount of data had to be flagged, thus providing clean images that were not good enough to self-calibrate the data, preventing the collection of reliable flux densities.

After this data reduction process, we also performed a correction of the system temperature, \tsys, for each antenna to subtract the contribution of the Galactic diffuse emission, relevant at low frequencies (see \S\,\ref{app:tsys}). The obtained \tsys corrections were directly applied to the flux densities of the final target images. We note that these corrections imply an additional source of uncertainty that does not affect the relative flux density variations from one epoch to another at a given frequency. The uncertainties introduced in the flux densities are close to the typical rms values obtained at 235 and 610~MHz, but at 154~MHz the uncertainties increased from $\approx$5\% to $\approx$20\%.

The measurement of the flux densities in each image was done using the {\tt tvstat} and {\tt jmfit} tasks of AIPS, both of them providing consistent values in all cases. We determined the rms noise of the images using {\tt tvstat} in a region around the source without including it nor any background source.

To guarantee the reliability of the measured flux densities we monitored the flux density values of four background sources detected in the field of view of LS~I~+61~303. The lack of a similar trend in all sources allows us to be confident that the variability described below is intrinsically related to LS~I~+61~303 and not due to calibration issues.

\subsection{LOFAR observations}

We have also analyzed several LOFAR observations conducted with the HBAs. First, we conducted a deep 6-h LOFAR observation at 140~MHz during its commissioning stage on 2011 September 30 to test the system and estimate the expected behavior of LS~I~+61~303 at these low frequencies. This observation was performed using 23 core stations plus 9 remote stations, with a total bandwidth of 48~MHz divided in 244 subbands (or IFs).
These data show a large rms noise level, with possible large uncertainties in the absolute flux density scale related to the calibration process. Hence we will not use these in this work (see preliminary results of this observation in \citealt{marcote2012}).

In 2013, we conducted five 3-hr LOFAR observations at around 150~MHz within the Radio Sky Monitor (RSM). The target source was always centered on transit, using 23 core stations plus 13 remote stations. In four of these observations we observed the source at 149~MHz with a total bandwidth of $\approx$$0.8~\mathrm{MHz}$ divided into 4 subbands. 3C~48 or 3C~147 were used as amplitude calibrators, observed in runs of 2 min interleaved in 11-min on-source runs. The observation on 2013 January 17 was conducted with a different setup: 20-min on-source runs centred at 142~MHz and divided into 12 subbands, with a total bandwidth of 2.4~MHz.

The LOFAR data were calibrated using standard procedures within the LOFAR Imaging Tools (LofIm, version 2.5.2, see \citealt{heald2010} and \citealt{vanhaarlem2013} for a detailed explanation).
The data were initially flagged using standard settings of the AOFlagger \citep{offringa2012}, and averaged in time and frequency with the LOFAR New Default Pre-Processing Pipeline (NDPPP) with an integration time of 10~s and 4 channels per subband.
In general, very bright sources did not need to be demixed from the target data sets, as they did not contribute significantly to the visibilities\footnote{The demixing process consists of removing from the target field visibilities the interference produced by the strongest off-axis sources in the sky, the so-called {\em A-team}: Cyg~A, Cas~A, Tau~A, Vir~A, Her~A and Hyd~A.}.

The calibration was performed on each subband individually with the BlackBoard Selfcal (BBS) package using standard settings, and the solutions were transferred to the target field. We used an initial sky model for the field of LS~I~+61~303 based on the VLA Low-Frequency Sky Survey (VLSS, \citealt{cohen2007}).
A manual flagging was performed using the {\tt casaviewer} and {\tt casaplotms} tasks from CASA and the imaging process was conducted with a development version of AWImager \citep{tasse2013}. The resulting model was used in a later phase self-calibration, and we imaged the data again. The last two tasks were performed recursively between 2 and 5 times until the solutions converged.
Due to the bright extended emission that is detected in the field of LS~I~+61~303 (located in the region of the Heart Nebula, see Figure~\ref{fig:image}, left) we only kept baselines larger than $0.2~\mathrm{k\lambda}$ (or 400~m) in subsequent analyses. Although the maximum baseline in the recorded data was $\sim$$85~\mathrm{k\lambda}$ (or 170~km), the baselines $\gtrsim$$8~\mathrm{k\lambda}$ (17~km) were removed during the data reduction (either during the manual flagging or during the self-calibration process). We used a Briggs robust weighting of zero again, because it produced the lowest noise level in the final images, and a pixel size of 10~arcsec\footnote{The beam is then undersampled, but allowed us to image a large field of view. However, the obtained synthesized beams and flux densities are stable with different pixel sizes.}.

We used the {\tt imstat} and {\tt imfit} tasks from CASA to measure the LS~I~+61~303 flux densities and the rms of the images.
To guarantee the reliability of the measured flux densities in the LOFAR images, we monitored the same four background sources chosen in the GMRT images. The lack of a similar trend in all sources, exhibiting consistent values with the ones inferred from the GMRT images, allows us again to be confident about the absence of flux calibration issues above the noise level in the LOFAR data.

We note that given the large field of view of these low-frequency images (either from the GMRT data or from the LOFAR ones), we detect differences in the ionospheric refractive effects for different regions across the field of view. These effects, in combination with the self-calibration cycles performed during the reduction process, produce slight displacements of the sources from their original positions. In general, it is common to end up with differences of up to the order of the synthesized beam size, although with differences in moduli and direction for different regions of the field.

\clearpage
		\thispagestyle{empty}
		\includepdf[fitpaper=true,width=\paperwidth,height=\paperheight]{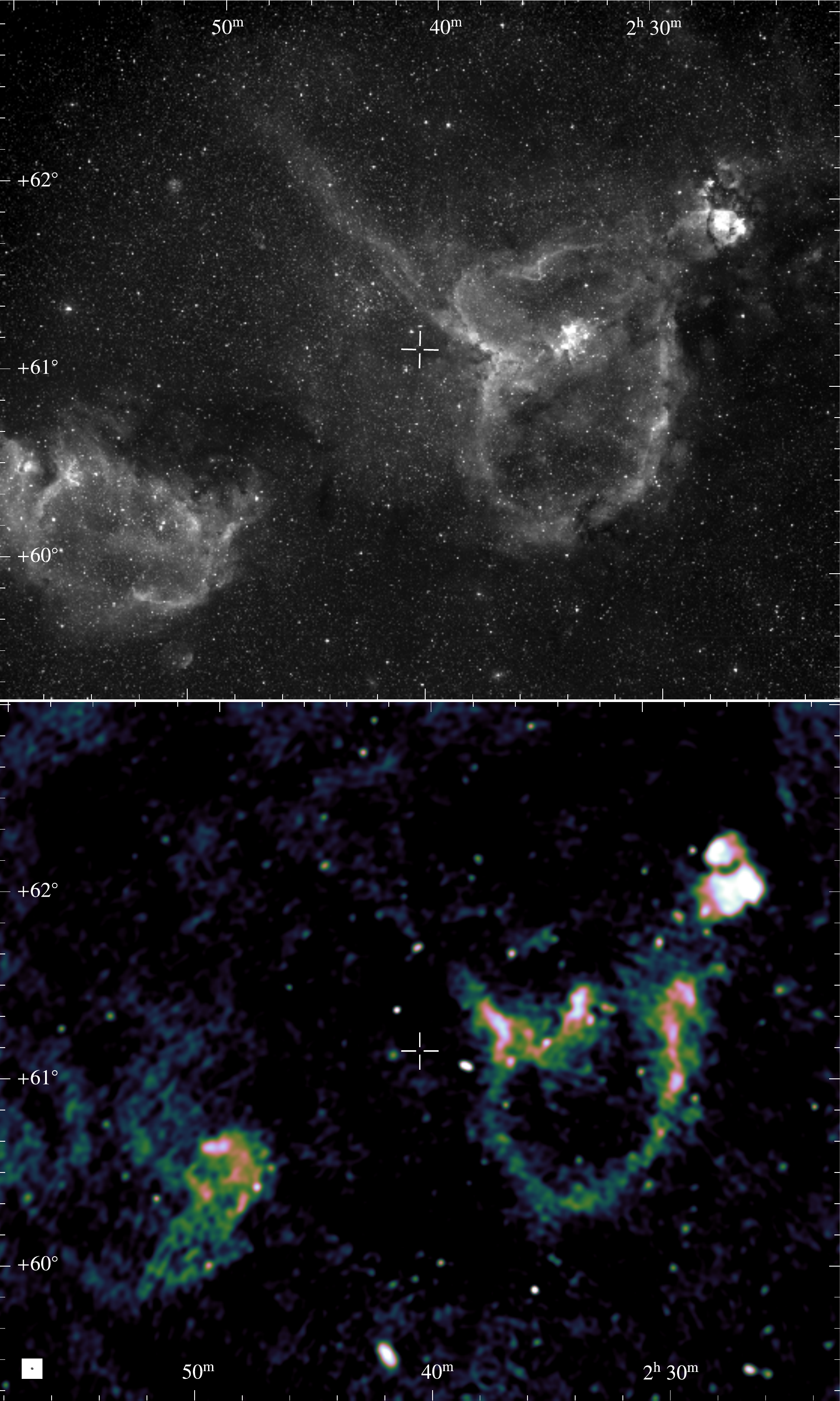}
\begin{figure}[!h]
	\includegraphics[width=\textwidth]{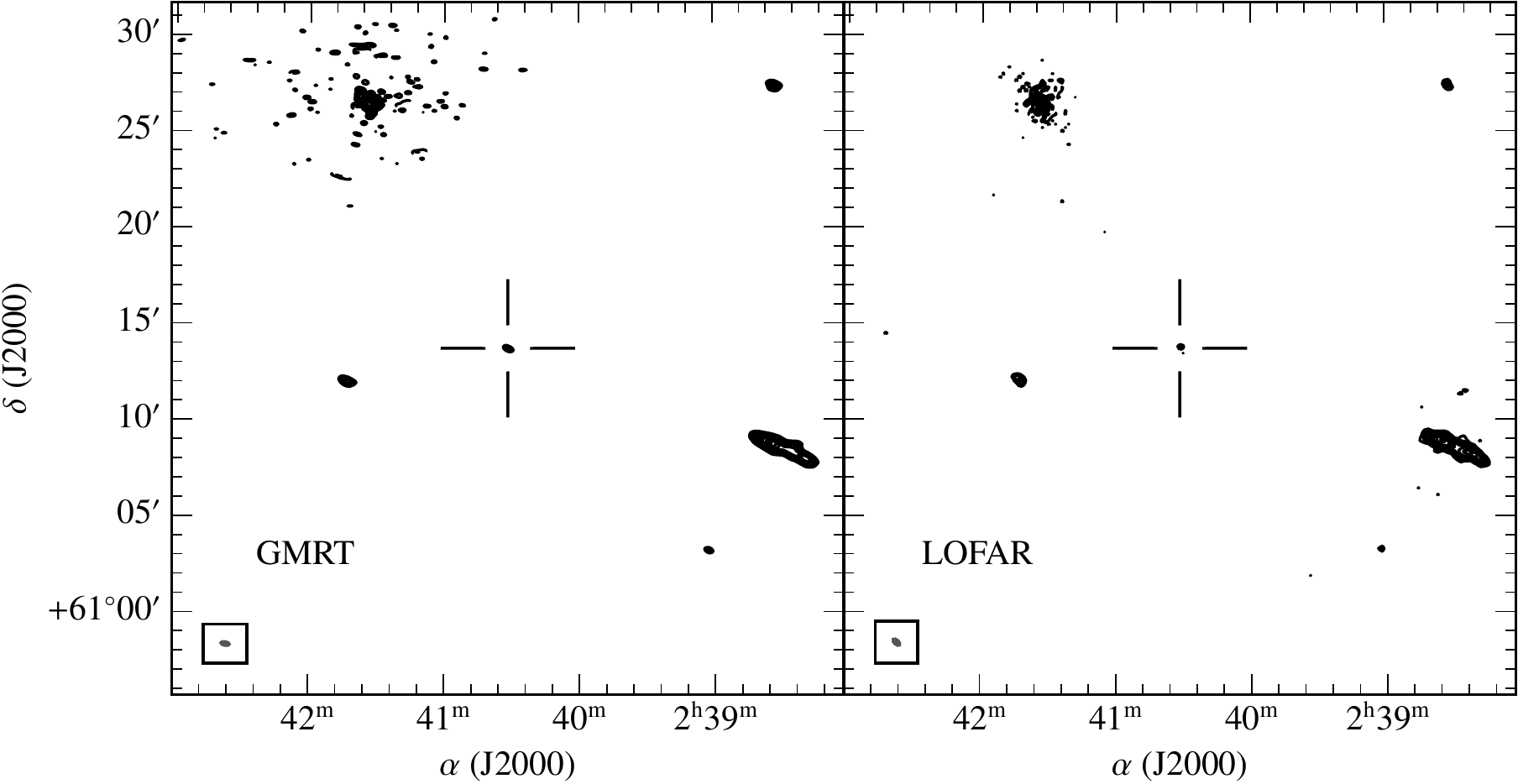}
    \caption[Images of LS~I~+61~303 as seen by LOFAR and GMRT at 150~MHz, and an optical image of the same region.]{{\em Left page}: field of LS~I~+61~303 seen by POSS~II \citep{reid1991} at optical red $F$ band (top), and by LOFAR at 149~MHz on 2013 July 14 without performing any cut in the $uv$-distance (bottom). The synthesized beam of the radio image, shown in the lower left corner, is $23 \times 15~\mathrm{arcsec^2}$ with a PA of $64^{\degree}$, and the rms is $8~\mathrm{mJy\ beam^{-1}}$ around the target source position. The positions of the sources present displacements of about half a synthesized beam, on average, with respect to the previously-recorded ones (see text). The location of LS~I~+61~303 is marked with the white cross on both images. The source is masked by extended emission from the Heart Nebula in this image. {\em Top:} zoom around the position of LS~I~+61~303 in the 154-MHz GMRT observation (top left) and in the 2013 July 14 149-MHz LOFAR observation applying the final cut at baselines below $0.2~\mathrm{k\lambda}$ (top right). The source is clearly detected in both cases. The synthesized beams, shown on the bottom left corner, are $27 \times 12~\mathrm{arcsec^2}$ with a PA of $80^{\degree}$, and $27 \times 15~\mathrm{arcsec^2}$ with a PA $= 49^{\degree}$, and the rms values are $11$ and $6~\mathrm{mJy\ beam^{-1}}$, respectively. Contours start at 5-$\sigma$ noise level and increase by factors of $2^{1/2}$. In the GMRT image (top left), we have applied a shift of 25~arcsec to recover the proper astrometric positions. No shift is required in the LOFAR image (top right).}
    \label{fig:image}
\end{figure}

\subsection{Complementary 15-GHz observations}

Two complementary observing campaigns that were conducted at the same epochs as the GMRT monitoring and the LOFAR observations have also been included in this work.

The Ryle Telescope \citep{pooley1997} has observed LS~I~+61~303 at 15~GHz over many years, with contemporaneous observations during the epoch at which the 2005--2006 GMRT monitoring was performed. The observations are centered at 15.2~GHz, recording Stokes $I+Q$ and a bandwidth of 350~MHz. The four mobile and one fixed antennas were arranged in a compact configuration, with a maximum baseline of 100~m.

The observing technique is similar to that described in \citet{pooley1997}. Given that we used baselines up to 100~m, we obtained a resolution in mapping mode of about 30~arcsec. The observations included regular visits to a phase calibrator (we used J0228+6721) to allow corrections for slow drifts of instrumental phase; the flux-density scale was established by nearby observations of 3C~48, 3C~147 or 3C~286.

Observations of LS~I~+61~303 were also conducted with the OVRO 40-m dish covering the epoch of the 2013 LOFAR observations. These data are part of a long-term monitoring that has been presented in \citet{massi2015}. The data were reduced by these authors following the procedures described in \citet{richards2011}.

\section{Results}\label{sec:lsi-results}

LS~I~+61~303 appears as a point-like source in all the images.
The resulting synthesized beams for the GMRT data range from $30 \times 14~\mathrm{arcsec^2}$ to $15 \times 5.4~\mathrm{arcsec^2}$, from 154 to 610~MHz, respectively. The synthesized beam for the LOFAR data is about $20 \times 15~\mathrm{arcsec^2}$. 
The left panel of Figure~\ref{fig:image} shows the field of LS~I~+61~303 obtained from the LOFAR observation conducted on 2013 July 14 without applying any $uv$-cut during the imaging process. The zoomed images (Figure~\ref{fig:image}, top) show the source as seen from the 154-MHz GMRT observation and from one of the LOFAR runs that we have analyzed in this work.
The flux density values of all the GMRT and LOFAR observations are shown in Table~\ref{tab:data}. In this section we discuss the light-curves obtained from the 235/610-MHz GMRT monitoring and from the $\approx$150-MHz LOFAR observations, including the 154-MHz GMRT data.

\subsection{Light-curves at 235 and 610~MHz}

\begin{figure}[!t]
    \centering
    \includegraphics[width=0.8\textwidth]{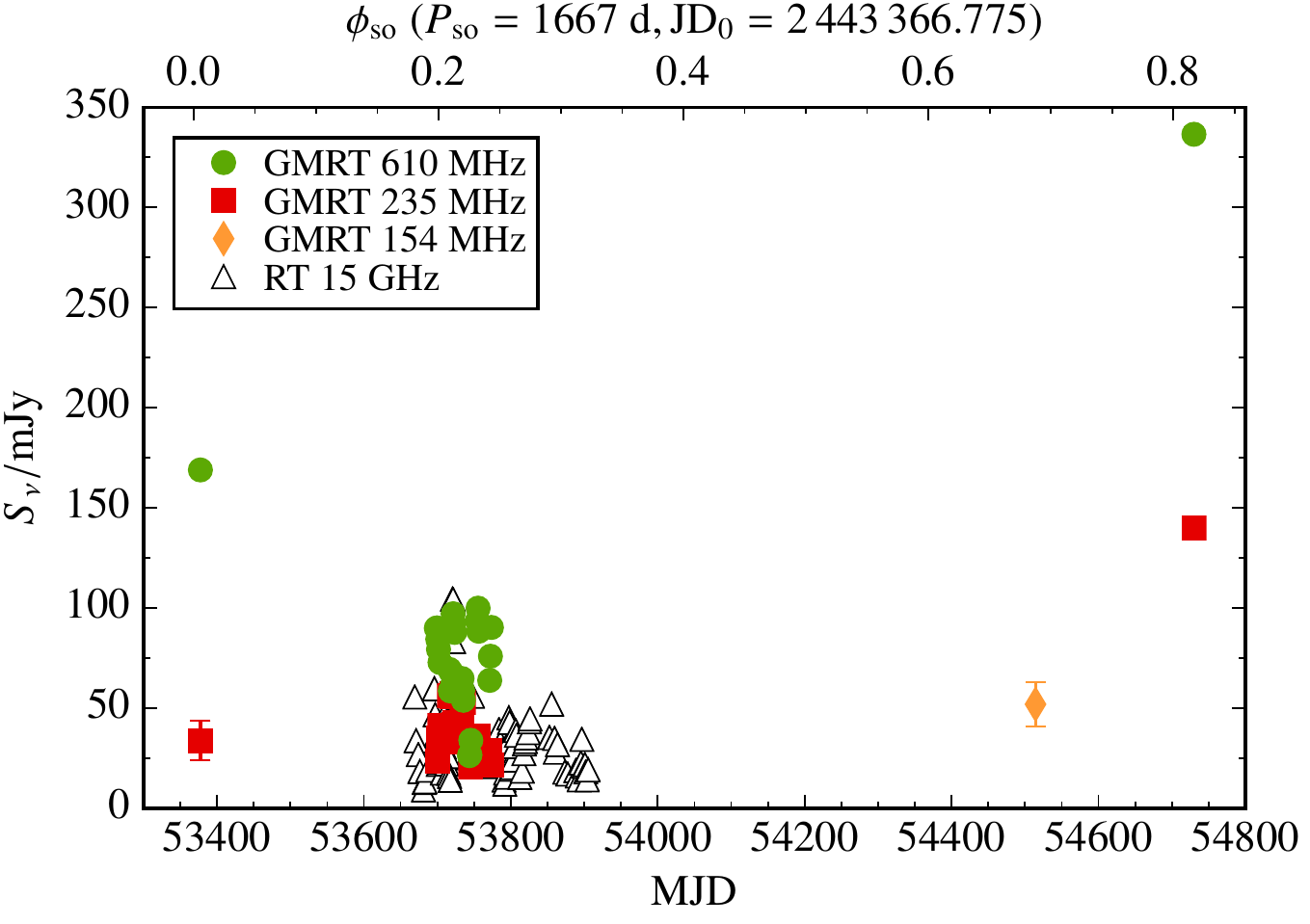}
    \caption[Flux density values of LS~I~+61~303 as a function of the MJD obtained from all the analyzed GMRT data.]{Flux density values of LS~I~+61~303 as a function of the MJD obtained from all the analyzed GMRT data (2005--2008). We show the superorbital phase, $\phi_{\rm so}$, on the top $x$-axis. The green circles represent the 610-MHz data, the red squares the 235-MHz data, and the orange diamond the 154-MHz ones. The gray triangles represent the daily averages of the RT data at 15~GHz. Error bars represent 1-$\sigma$ uncertainties. We note that the source exhibits larger flux densities during the isolated GMRT observations (at superorbital phases of $\phi_{\rm so}\approx 0.80$--$0.95$) than during the GMRT monitoring ($\phi_{\rm so} \approx 0.2$).}
    \label{fig:gmrt-mjd-all}
\end{figure}
The dual 235/610-MHz mode in the GMRT observations allows us to observe the state of LS~I+61~303 and its evolution simultaneously at these two frequencies. Figure~\ref{fig:gmrt-mjd-all} shows the flux densities of all the analyzed GMRT observations (the GMRT monitoring at 235 and 610~MHz and the three isolated GMRT observations) as a function of the MJD (bottom axis) and the superorbital phase ($\phi_{\rm so}$, top axis).
We observe that the flux densities obtained in the GMRT monitoring at 610~MHz (conducted at $\phi_{\rm so} \approx 0.2$, around the minimum activity of LS~I~+61~303 at GHz frequencies) are significantly lower than the ones obtained in the other two observations (conducted at $\phi_{\rm so} \approx 0.8$--$0.0$, during the maximum activity). We also observe this effect at 235~MHz with respect to the observation at $\phi_{\rm so} \approx 0.8$ (but not at $\phi_{\rm so} \approx 0.0$). This indicates the existence of the superorbital variability at frequencies below 1~GHz.

\begin{figure}[!t]
    \centering
    \includegraphics[width=0.8\textwidth]{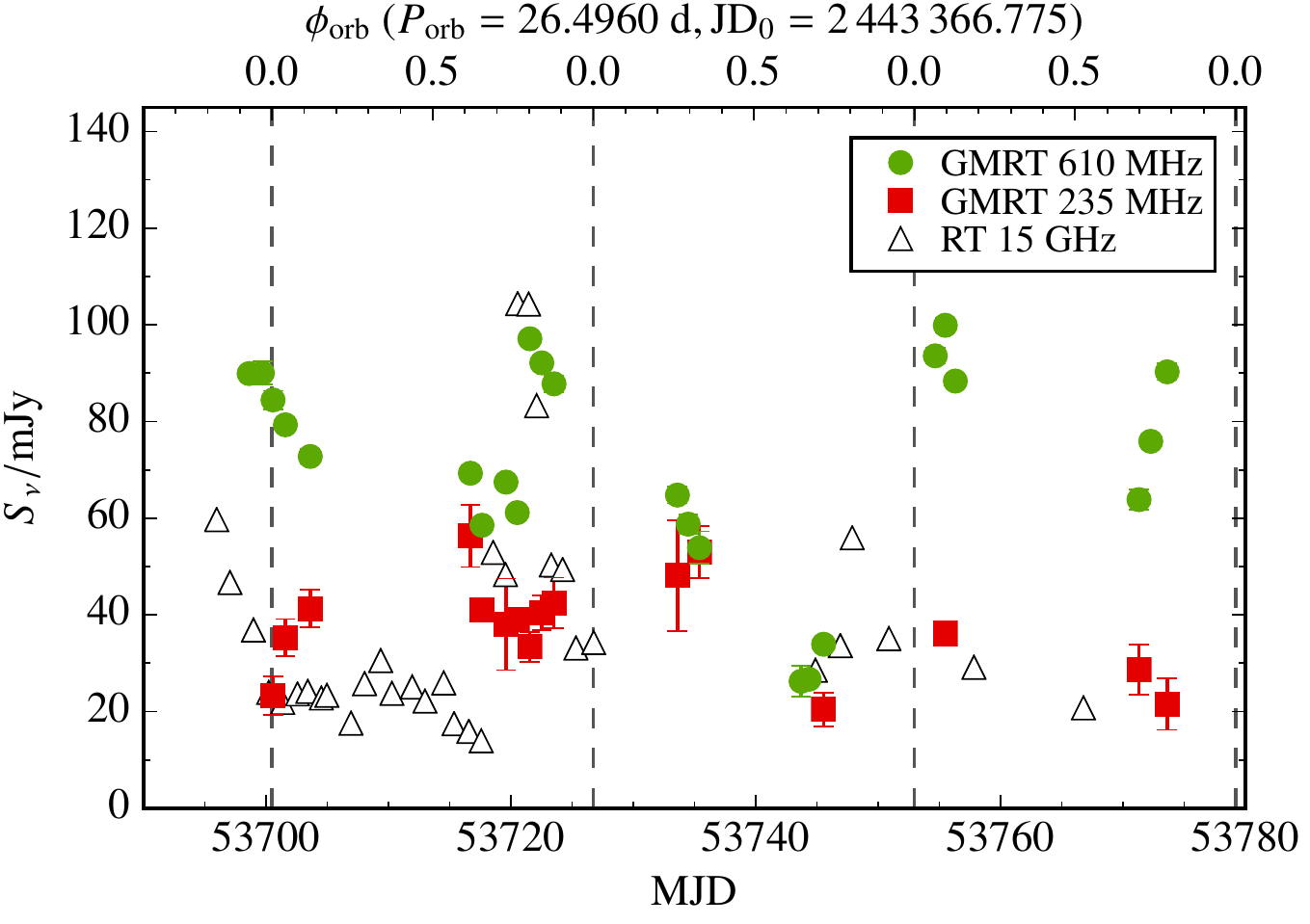}
    \caption[Same as Figure~\ref{fig:gmrt-mjd-all} but zooming in on the GMRT monitoring observations.]{Same as Figure~\ref{fig:gmrt-mjd-all} but zooming in on the GMRT monitoring observations. In this case we show the orbital phase, $\phi_{\rm orb}$, on the top $x$-axis. The vertical dashed lines show the epochs at which $\phi_{\rm orb} = 0$. We observe the presence of enhanced emission at 610~MHz coincident with the outbursts at 15~GHz but with a slower decay (see MJD~52695--52705, 53720--53725). In contrast, at 235~MHz we observe a smaller degree of variability, with no clear orbital trends. The arrows represent the 3-$\sigma$ upper-limits.}
    \label{fig:gmrt-mjd}
\end{figure}
Figure~\ref{fig:gmrt-mjd} focuses on the GMRT monitoring only, showing the 610 and 235-MHz light-curve of LS~I~+61~303 along three consecutive orbital cycles. The 610-MHz data exhibit variability with maxima roughly coincident with the outbursts observed in the 15-GHz RT data (e.g.\ see $\mathrm{MJD} \sim 53720$). However, at 610~MHz we observe that the decay of the emission is slower (e.g.\ compare the decays in MJD~53695--53705, 53720--53725). The average flux density during the whole monitoring is $70~\mathrm{mJy}$, with a standard deviation of $20~\mathrm{mJy}$ and a significant variability. For these variability analyses (and in the rest of this work) we have taken the most conservative choice to consider the 3-$\sigma$ upper-limits as the possible flux density value of the source, assuming $S_{\nu} \approx 3\sigma \pm \sigma$.
At 235~MHz we observe a smaller degree of variability, with no clear correlation with the previous one. In this case we infer an average flux density of $37 \pm 8~\mathrm{mJy}$ and a variability at a 6-$\sigma$ confidence level.

Folding these data with the orbital phase (Figure~\ref{fig:gmrt-phase}, top), we clearly see that the radio emission is orbitally modulated.
At 610~MHz we observe that the enhanced emission takes place in the range of $\phi_{\rm orb} \sim 0.8$--$1.1$, whereas at 15~GHz the maximum occurs at $\phi_{\rm orb} \approx 0.8$ and shows a faster decay.
At 235-MHz we observe that the flux density increases between $\phi_{\rm orb} \sim 0.0$ and $0.4$. This phase range is followed by an interval without data between phases 0.4 and 0.6, after which we observe the largest flux density value followed by a fast decrease in the flux density. The increase and decrease of the flux density are traced by 4--5 data points in each case. Therefore, we observe that the maximum emission probably occurs in the range between $\phi_{\rm orb} \approx 0.3$ and $\approx$$0.7$, which is the location of the minimum at 610~MHz. In fact, the 235~MHz and the 610~MHz light-curves are almost anticorrelated.
Figure~\ref{fig:gmrt-phase} (bottom) shows the spectral index $\alpha$ (determined from the 235 and 610~MHz data) as a function of the orbital phase. The spectral index is also orbitally modulated, following essentially the 610-MHz flux density emission.
From the isolated GMRT observations we obtain a spectral index coincident with the modulation observed from the GMRT monitoring, despite the much larger flux density values (see open hexagons in Figure~\ref{fig:gmrt-phase}, bottom).
We note that Figure~\ref{fig:gmrt-phase} shows data from different orbital cycles that have been poorly sampled. In addition, we remark that the outbursts seen at GHz frequencies exhibit changes from cycle to cycle. Therefore, we note that the profile seen in the folded light-curve is different than the one we would obtain in a single orbital cycle.

\subsection[Light-curve at $\approx$150~MHz]{Light-curve at \boldmath{$\approx$}150~MHz}

\begin{figure}[!t]
    \centering
    \includegraphics[width=0.8\textwidth]{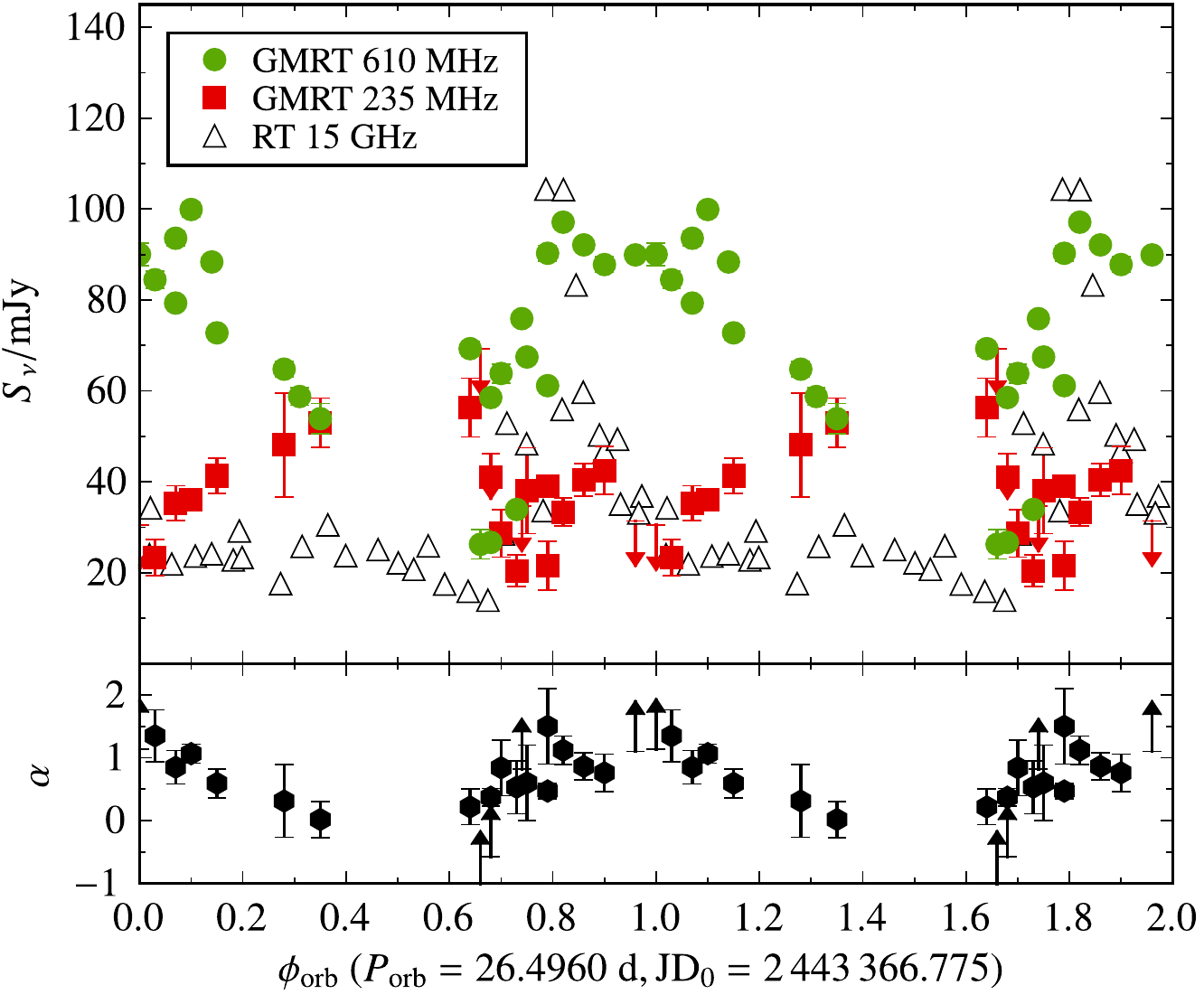}
    \caption[Folded light-curve of LS~I~+61~303 at 235 and 610~MHz with the orbital period, and the obtained spectral indexes. 15-GHz RT data are also shown for comparison.]{{\em Top:} folded light-curve with the orbital period of the data shown in Figure~\ref{fig:gmrt-mjd}. The 610~MHz data show a quasi-sinusoidal modulation with enhanced emission at $\phi_{\rm orb} \approx 0.8$--$1.1$, whereas the 15-GHz outbursts take place at $\phi_{\rm orb} \approx 0.8$ with a fast decay. The 235~MHz light-curve is almost anticorrelated with the one observed at 610~MHz. 
    {\em Bottom:} spectral index $\alpha$ derived from the GMRT data presented above. Open hexagons represent the spectral index from the isolated GMRT data (not shown on top). Error bars represent 1-$\sigma$ uncertainties and the arrows represent the 3-$\sigma$ upper/lower-limits.}
    \label{fig:gmrt-phase}
\end{figure}
Several data sets at a frequency around 150~MHz have been analyzed: one GMRT observation at 154-MHz, one LOFAR observation at 142~MHz and four LOFAR observations at 149~MHz. In the following, we will refer to all these observations generically as 150-MHz observations. We note that the differences between these frequencies would not imply any significant change in the flux density values of LS~I~+61~303 for reasonable spectral indices.

The 150-MHz GMRT data (obtained at a superorbital phase of $\phi_{\rm so} \approx 0.69$) reveal for the first time a detection of LS~I~+61~303 at this frequency\footnote{We note that LS~I~+61~303 was inadvertently detected at $\approx$150~MHz in a study of the Galactic emission made by \citet{bernardi2009} with WSRT. The source is marginally detected in the Figures.~2 and 3 of \citet{bernardi2009}, close to the edge of the primary beam. At these positions the flux density measurements are much less reliable, and thus we will not discuss these data here. We also note that the positions of the sources in the mentioned figures present a general displacement of about 5~arcmin to the East direction with respect to the real positions.}, being a point-like source with a flux density of $52 \pm 11~\mathrm{mJy}$.

\begin{figure}[!t]
    \centering
    \includegraphics[width=0.8\textwidth]{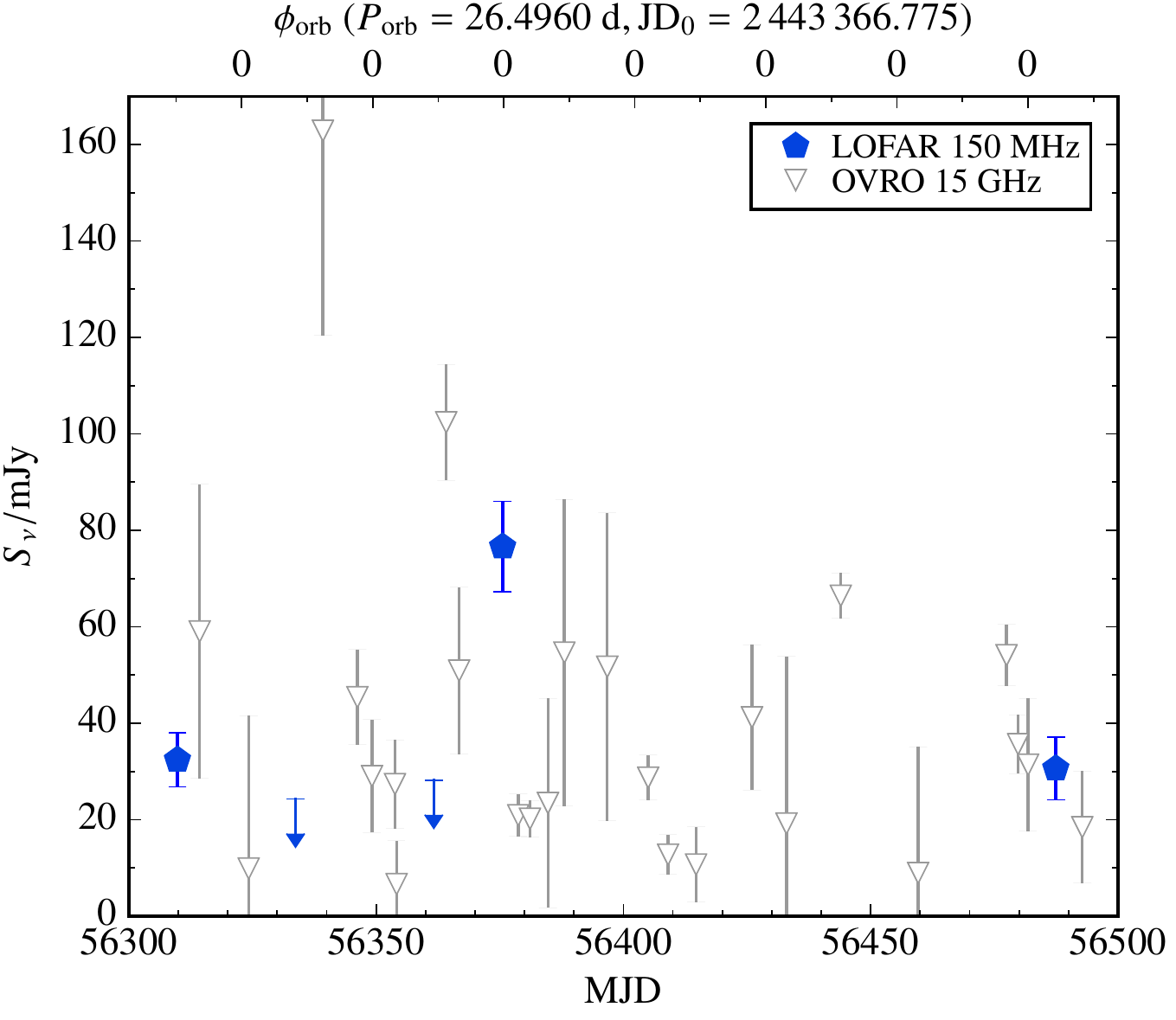}
    \caption[Flux density values of LS~I~+61~303 as a function of the MJD obtained from the 150-MHz LOFAR observations.]{Flux density values of LS~I~+61~303 as a function of the MJD obtained from the 150-MHz LOFAR observations (conducted at $\phi_{\rm so} \approx 0.8$). We show the orbital phase, with labels every $\phi_{\rm orb} = 0$, on the top $x$-axis. We detect variability and a strong increase in the flux density of LS~I~+61~303 of at least a factor 3 in only 14~d (between MJD~56361 and 56375). The open triangles represent the contemporaneous OVRO data at 15~GHz.}
    \label{fig:lofar-mjd}
\end{figure}
The 150-MHz LOFAR observations taken within the RSM in 2013 ($\phi_{\rm so} \approx 0.76$--$0.87$) allow us to obtain a light-curve of the source on week timescales (spread along 7 orbital cycles). These results are shown in Figure~\ref{fig:lofar-mjd}. With these data we clearly detect variability and a strong increase in the flux density of at least a factor 3 in only 14~d between MJD~56361 and 56375 (approximately half of the orbital period).
Folding these data with the orbital period, and adding the GMRT data at approximately the same frequency, we obtain the light-curve shown in Figure~\ref{fig:lofar-phase}. We note that all these observations were conducted at similar superorbital phases.
We observe a kind of baseline state with flux densities around $30~\mathrm{mJy}$ on top of which there is a stronger emission between $\phi_{\rm orb} \approx 0.7$--$1.0$ (given the reduced coverage of the orbit, we certainly cannot constrain this range and it could actually be wider). Despite the poor sampling, the onset of the outburst at 150~MHz could take place at the same orbital phase as in the contemporaneous 15-GHz OVRO data, although the maximum of the outburst appears to be delayed at 150~MHz.
The average flux density for these 150-MHz observations is $35 \pm 16~\mathrm{mJy}$ (considering the upper-limits as the possible flux density value of the source: $S_{\nu} \approx 3\sigma \pm \sigma$).

\begin{figure}[!t]
    \centering
    \includegraphics[width=0.8\textwidth]{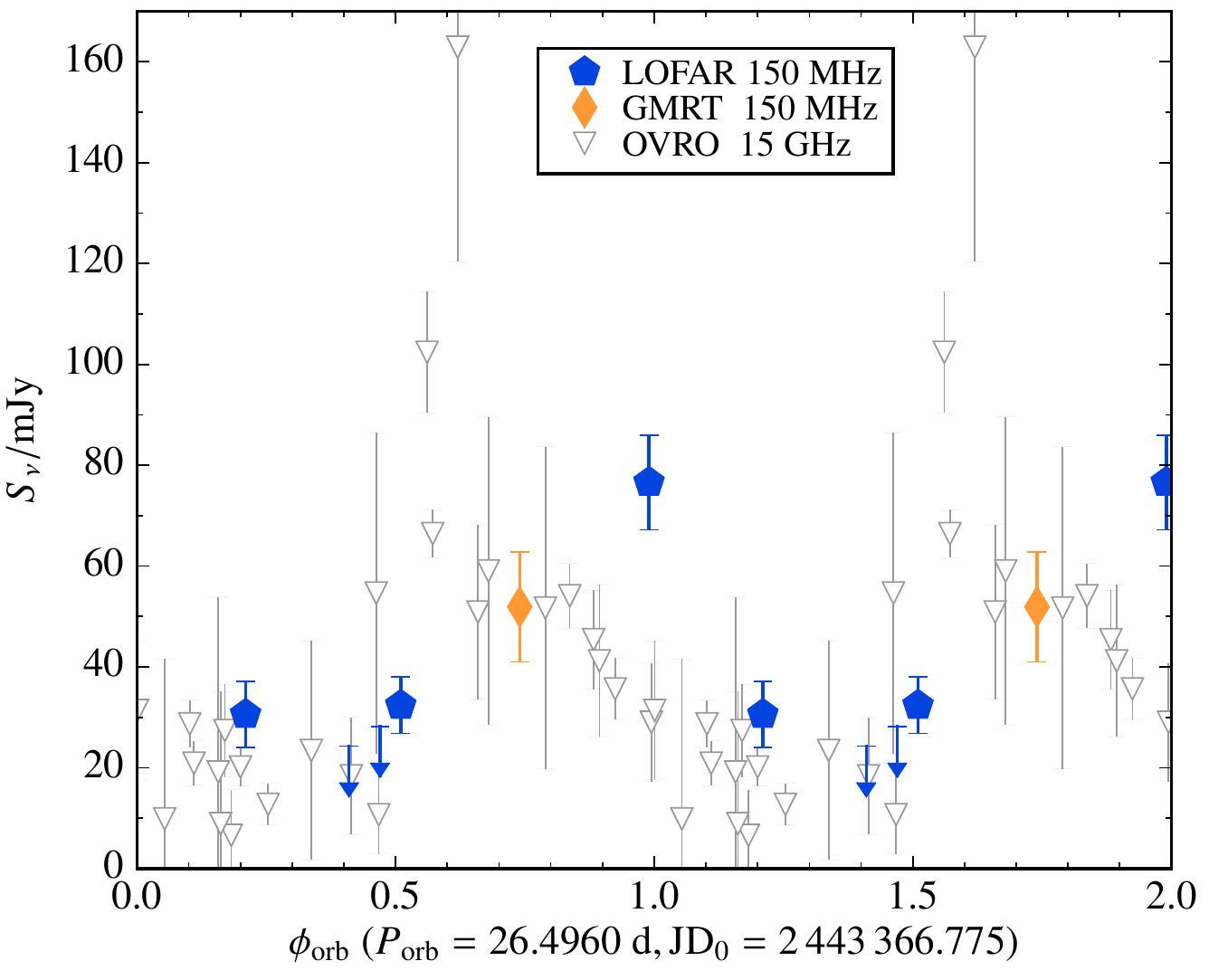}
    \caption[Folded light-curve with the orbital period from the data shown in Figure~\ref{fig:lofar-mjd} plus the 150-MHz GMRT observation.]{Folded light-curve with the orbital period from the data shown in Figure~\ref{fig:lofar-mjd} plus the 150-MHz GMRT observation (orange diamonds). The flux density is orbitally modulated, exhibiting an enhanced emission at $\phi_{\rm orb} \approx 0.7$--$1.0$. Due to the poor sampling, we can not determine the delay of the peak emission between 150~MHz and 15~GHz, although a delay seems to be observed between these frequencies.}
    \label{fig:lofar-phase}
\end{figure}

\section{Discussion}\label{sec:lsi-discussion}

We have presented here data of the gamma-ray binary LS~I~+61~303 at low radio frequencies, from 150 to 610~MHz.
This is the first time that a gamma-ray binary is detected at 150~MHz, although previous searches have been performed in other sources \citep[see][or \S\,\ref{chap:ls}, for the case of LS~5039]{marcote2015ls5039}. A multifrequency monitoring conducted with the GMRT in 2005--2006 (at a superorbital phase of $\phi_{\rm so} \approx 0.2$) shows significant variability at 610~MHz with the maximum emission at orbital phases of $\phi_{\rm orb}\approx 0.8$--$1.1$. This variability is roughly coincident with the outbursts observed at 15~GHz, but with a significantly wider shape, a delay of about 0.2 orbital phases, and a slower decay. However, at 235~MHz we show that the maximum emission of the source occurs in the range between $\phi_{\rm orb} \approx 0.3$ and $\approx0.7$. The light-curve is thus almost anti-correlated with respect to the 610-MHz one. The LOFAR observations were conducted, in contrast, at a superorbital phase of $\phi_{\rm so} \approx 0.8$. In this case we observe a behavior similar to the one observed at 610~MHz: a large variability with the maximum emission taking place at orbital phases between $0.7$ and $1.0$.

In this section we discuss the observed behavior at 150, 235 and 610~MHz and its relationship with the superorbital modulation. It must be noted that the compact VLBI radio emission detected at GHz frequencies represents $\gtrsim$90\% of the total flux density of LS~I~+61~303, leaving small room for extended radio emission \citep{paredes1998}. For this reason, and also because of the limited number of frequencies with available light-curves (and the use of data from different orbital cycles), we assume a one-zone model with an homogeneous emitting region. We consider here the two most probable absorption mechanisms: free-free and synchrotron self-absorption (there are not enough data to consider the Razin effect such as in LS~5039, \S\,\ref{chap:ls}). We infer the physical changes required in the system to explain the observed delay between the orbital phases at which the maximum emission takes place at different frequencies. 

From Figure~\ref{fig:gmrt-mjd-all} we observe that the superorbital phase still has an important role at these low frequencies, as we detect a much larger flux density at high $\phi_{\rm so}$ than at low ones. Given that we only have two isolated observations at high superorbital phases, we cannot determine what was the state of the source during these observations (maximum of the orbital variability, minimum or in between). Therefore, we can not accurately estimate the increase of the emission as a function of the superorbital phase.

A correlation between changes in the thermal emission of the circumstellar disk, observed from optical photometry and H$\alpha$ observations, and the superorbital modulation of the non-thermal emission has been reported \citep[][and references therein]{paredes-fortuny2015}. A larger amount of material in the disk (higher values of H$\alpha$ equivalent width) is observed at superorbital phases coincident with the maximum emission at GHz frequencies ($\phi_{\rm so}\sim 0.8$--$0.0$). This larger amount of material in the disk might imply more target material for a shock with the putative pulsar wind. This could potentially lead to a more efficient particle acceleration and stronger emission, but also to a larger absorption at low radio frequencies. The presence of stronger radio emission and a clear variability at 150~MHz at these superorbital phases indicates that the emission increase, probably produced as a result of the more efficient particle acceleration, seems to dominate over the decrease due to enhanced absorption at these low radio frequencies.

The positive spectral indices obtained from the GMRT monitoring (Figure~\ref{fig:gmrt-phase}, bottom) confirm the suggestion of a turnover between 0.3 and 1.4~GHz made by \citet{strickman1998}, and we constrain it to be in the 0.6--1.4~GHz range.
Assuming that the spectrum at low frequencies is dominated by FFA we can estimate the radius of the emitting region with the condition that the free-free opacity, given by equation~(\ref{eq:tauff}), is
\begin{equation}
    \tau_{\nu}^{\rm FFA} \approx 30 \dot M_{-7}^2 \nu_{\rm GHz, max}^{-2} \ell_{\rm AU}^{-3} v_{\rm W, 8.3}^{-2} T_{\rm W, 4}^{-3/2} = 1.
\label{eq:tauff1}
\end{equation}
Assuming a luminosity of $L \sim 4.65~\mathrm{L_{\sun}}$, typical from B0~V stars, we derive a mass-loss rate of $\dot M_{-7} \sim 0.5$ \citep{howarth1989}, for which we will assume reasonable values in the range of $\dot M_{-7} \sim 0.2$--$1$. Considering the effective temperature of $T_{\rm eff} \sim 28\,000~\mathrm{K}$, also expected for this type of star \citep{cox2000}, we deduce a wind velocity of $1\,500 \pm 500~\mathrm{km\ s^{-1}}$ (or $v_{\rm W, 8.3} \sim 0.75$, \citealt{kudritzki2000}). We also expect a wind temperature of $\sim$10\,000~K at a distance of the order of the apastron \citep{krticka2001}, and a turnover frequency $\nu_{\rm GHz, max} \sim 1$ (as mentioned before). With these values we estimate a radius for the spherically emitting region of $2.4_{-1.1}^{+1.7}~\mathrm{AU}$.

We can compare this result with the displacements of the peak positions obtained with VLBA observations along one orbital cycle by \citet{dhawan2006}.
From their Figure~4 we observe displacements as large as $\sim$2.5 and $\sim$2~mas ($\sim$5 and $\sim$4~AU) at 2.2 and 8.4~GHz, respectively, which in our spherically symmetric model imply radii of $\sim$2.5 and $\sim$2~AU.
Although we are using a spherical model, we note that the radius that we have derived is clearly compatible with these values, as expected considering that the emission has to be produced far away enough from the massive star to avoid being in the optically thick part of the spectrum due to free-free absorption.

On the other hand, we can consider that the emitting region is expanding, and thus the turnover frequency evolves along the time. In this case we consider that the transition from an optically-thick to an optically-thin region would produce the delay between the different maxima observed at 235 and 610~MHz, as previously reported for other binary systems \citep{ishwarachandra2002}. From equation (\ref{eq:tauff1}) we can determine the radius of the emitting region when the turnover is located at a frequency $\nu$:
\begin{equation}
    \ell_{\nu} = 30^{1/3}\, \dot M_{-7}^{2/3}\, v_{\rm W, 8.3}^{-2/3}\, T_{\rm W,4}^{-1/2}\, \nu^{-2/3}.
    \label{eq:ellff}
\end{equation}
With this relation we can infer the velocity of the expanding emitting region, or expansion velocity, assuming it to be constant for simplicity:
\begin{equation}
    v = \frac{\Delta \ell}{\Delta t} = \frac{\ell_{\nu_2}-\ell_{\nu_1}}{\Delta \phi P_{\rm orb}},
    \label{eq:vff}
\end{equation}
where $\Delta \phi$ is the shift in orbital phase for the maximum at a frequency $\nu_2$ and $\nu_1$, and $P_{\rm orb}$ is the orbital period.
We can consider for simplicity the case that $\dot M_{-7}$, $v_{\rm W, 8.3}$ and $T_{\rm W,4}$ remain constant during this expansion.
Figures~\ref{fig:gmrt-mjd} and \ref{fig:gmrt-phase} show that the peak of the emission at 610~MHz is located somewhere between $\phi_{\rm orb} \sim 0.8$ and $1.1$. In parallel, we observe that the flux density at 235~MHz remains increasing at $\phi_{\rm orb} \sim 0.4$, and in the range of 0.6--0.7 the flux density is already decreasing. Therefore, the maximum emission probably takes place in the range of 0.3--0.7. These values imply that the maximum emission exhibits a shift of $\approx$$0.2$--$0.9$ in orbital phase between the two frequencies. Although the uncertainty in this shift is large, we can assume a shift of $\sim$0.5 to provide a rough estimation of the expansion velocity for the radio emitting region. Assuming this value and with all the previously mentioned data, we infer that the emitting region should expand by a factor of $2.0 \pm 0.5$ during this delay of $\approx$0.5 in orbital phase ($\approx$$13~\mathrm{d}$) from 610 to 235~MHz, implying a constant expansion velocity of $v_{\rm FFA} = 350 \pm 220~\mathrm{km\ s^{-1}}$.
If we also consider the shifts in orbital phase with respect to the maximum at 15~GHz,
we can deduce, from equations (\ref{eq:ellff}) and (\ref{eq:vff}), the expected delay in orbital phase as a function of the velocity, $v_{\rm FFA}$, and the frequency, $\nu$:
\begin{equation}
    \Delta\phi_{\rm FFA} =  \frac{30^{1/3}\, \dot M_{-7}^{2/3}\, v_{\rm W, 8.3}^{-2/3}\, T_{\rm W,4}^{-1/2}}{v_{\rm FFA} P_{\rm orb}} \left(\nu_{\rm GHz}^{-2/3} - 15^{-2/3} \right).
    \label{eq:phiff}
\end{equation}
Figure~\ref{fig:shift} shows the velocity curves derived from this equation, where a fit to the considered shifts at 235 and 610~MHz with respect to 15~GHz produces a constant velocity of $v_{\rm FFA} = 660 \pm 280~\mathrm{km\ s^{-1}}$. This velocity is a factor of $\sim$2 smaller than the one of the stellar wind.
However, the derived velocity depends strongly on the assumed mass-loss rate, which is unconstrained from the observational point of view. Figure~\ref{fig:mdot-velocity} shows the derived expansion velocity of the emitting region for different values of the mass-loss rate (in Figure~\ref{fig:shift} we assumed $\dot M_{-7} = 0.5$). We note that for mass-loss rates of $\dot M_{-7} \approx 1$ (which is still possible) the derived velocities overlap with the range of possible stellar wind velocities. In any case, these velocities are much lower than the relativistic velocity of the putative pulsar wind (see discussion in \citealt{bogovalov2012}).
\begin{figure}[!t]
    \centering
    \includegraphics[width=0.8\textwidth]{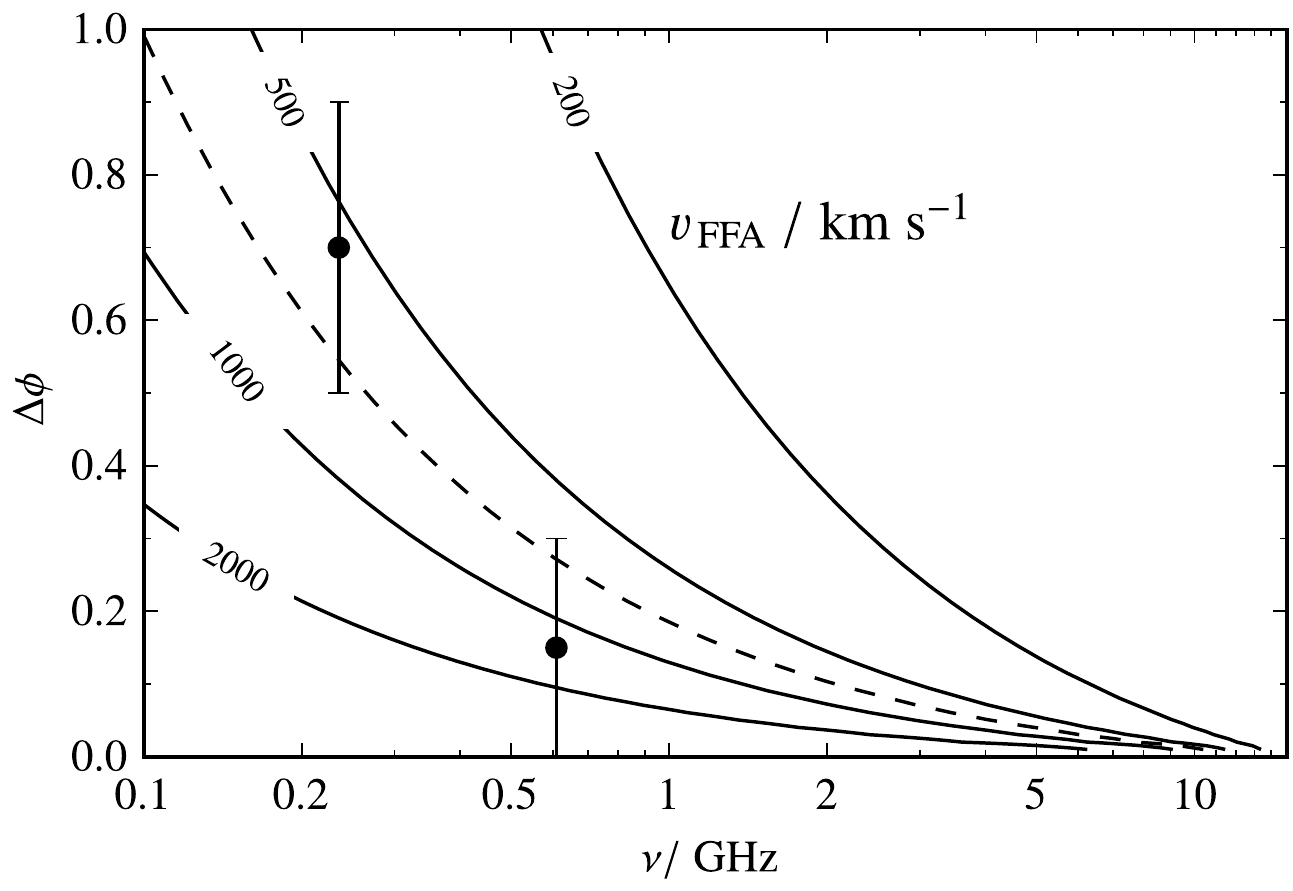}
    \caption[Shift in orbital phase ($\Delta \phi$) expected for the maxima in the flux density emission between a frequency $\nu$ and 15~GHz for different velocities of expansion of the radio emitting region ($v_{\rm FFA}$) assuming that FFA is the dominant absorption process.]{Shift in orbital phase ($\Delta \phi$) expected for the maxima in the flux density emission between a frequency $\nu$ and 15~GHz for different velocities of expansion of the radio emitting region ($v_{\rm FFA}$) assuming that FFA is the dominant absorption process and following equation (\ref{eq:phiff}). The black circles represent the shifts observed at 235 and 610~MHz with respect to 15~GHz in the folded light-curve. We fit the data with an expansion velocity of $\sim$$700~\mathrm{km\ s^{-1}}$ (dashed line).}
    \label{fig:shift}
\end{figure}

\begin{figure}[!t]
    \centering
    \includegraphics[width=0.8\textwidth]{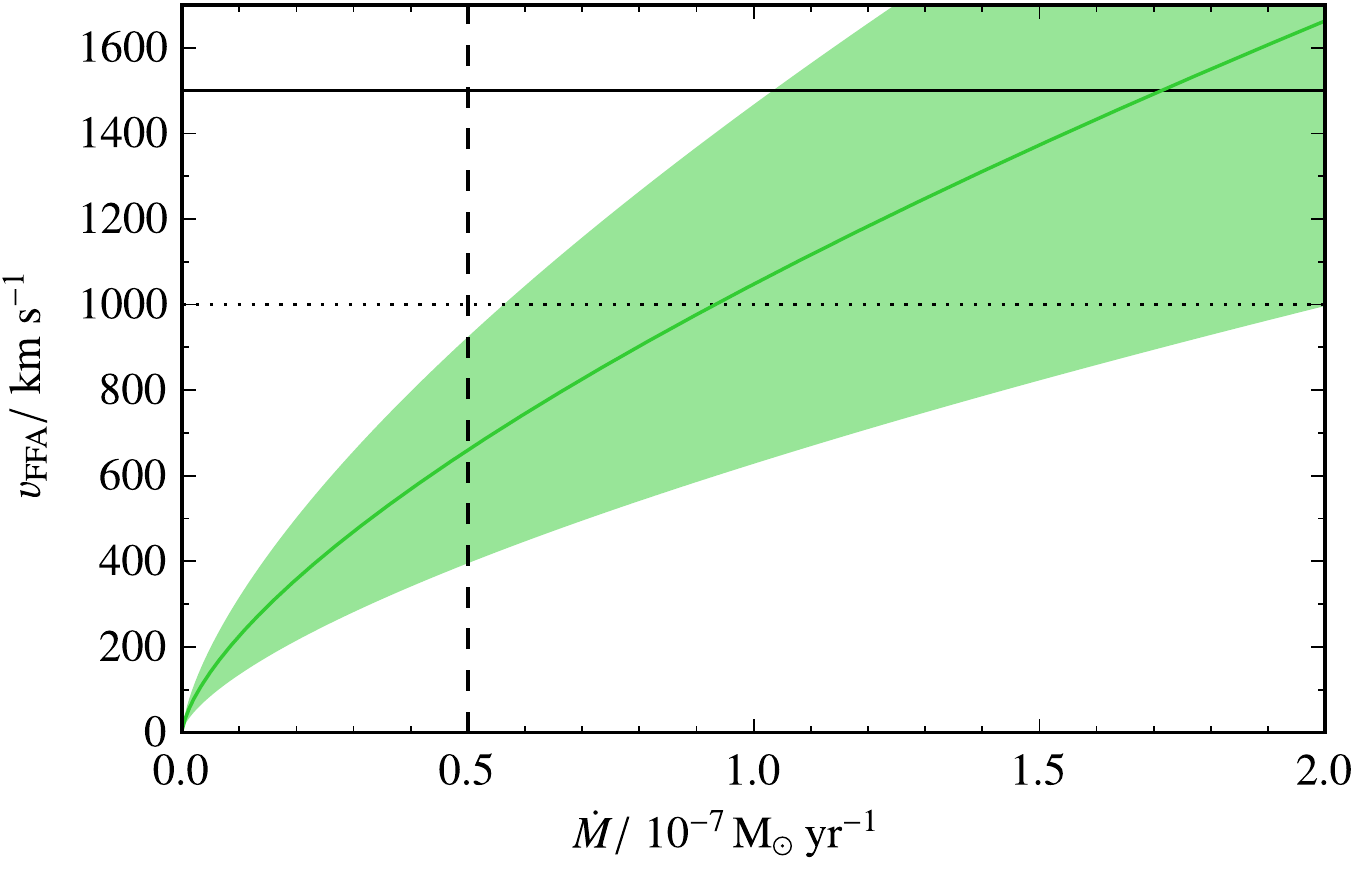}
    \caption[Velocities, $v_{\rm FFA}$, of the expanding emitting region as a function of the mass-loss rate, $\dot M$, derived from the observed delay between the maxima at 235 and 610~MHz with respect to 15~GHz, assuming a free-free absorbed region.]{Velocities, $v_{\rm FFA}$, of the expanding emitting region as a function of the mass-loss rate, $\dot M$, derived from the considered delay between the maxima at 235 and 610~MHz with respect to 15~GHz, assuming a free-free absorbed region. The green line represents the mean derived velocity and the light green shaded region represents the possible velocity values by considering the uncertainties in the orbital phase shifts between different frequencies, and in the stellar wind velocity. The vertical dashed line denotes the derived mass-loss rate. The horizontal lines represent the mean and lower value of the stellar wind velocity (solid and dotted line, respectively).}
    \label{fig:mdot-velocity}
\end{figure}
\citet{dhawan2006} estimated the outflow velocity along the orbit at 2.2 and 8.4~GHz, obtaining maximum values of $\sim$$7\,500~\mathrm{km\ s^{-1}}$ near periastron and $\sim$$1\,000~\mathrm{km\ s^{-1}}$ near apastron. Therefore, the expansion velocities derived here are also close to the one derived by these authors near apastron.

If instead of FFA we consider SSA as the dominant absorption process at low frequencies, we can estimate the properties of the emitting region in the optically thick to optically thin transition. In this case, from equation~(\ref{eq:taussa}) we obtain the condition
\begin{equation}
    \tau_{\nu}^{\rm SSA} = 3.354 \times 10^{-11} (3.54 \times 10^{18})^{p} K B^{(p+2)/2} b(p) \nu_{\rm GHz}^{-(p+4)/2} \ell_{\rm AU} = 1.
    \label{eq:taussa1}
\end{equation}
Assuming $p \sim 2$ (i.e. $\alpha \sim -0.5$ in the optically thin region, similar to the average spectral index observed in LS~I~+61~303 above $\sim$2~GHz) we would expect that the quantity $K\,B^2\,\ell$ decreases a factor of $\sim$18 during the delay of 0.5 in orbital phase between the maximum emission at 235 and 610~MHz. 
These three quantities ($K, B, \ell$) are coupled in the equation above and thus we can not estimate them separately. All of them would a-priori change along the orbit, but since we assume an expanding region from the same population of accelerated particles, $K$ should remain constant (provided that losses are not significant). 
Additionally, one can consider different dependences of $B$ as a function of $\ell$: $B \sim \ell^{-2}$ (as in a spherical expansion) or $B \sim \ell^{-1}$ (as in a conical or toroidal expansion, as it happens either in a relativistic jet from the microquasar scenario or in a cometary tail from the young non-accreting pulsar scenario). With these considerations we expect an expansion factor of $\sim$2.6 (if $B \sim \ell^{-2}$) or $\sim$18 (if $B \sim \ell^{-1}$).
To derive the expansion velocity of the emitting region, we can assume the radius of $\ell = 2.4_{-1.1}^{+1.7}~\mathrm{AU}$ that we have obtained at 1~GHz in the FFA case, which is also compatible with the results from \citet{dhawan2006}.
With these assumptions, and the equations (\ref{eq:vff}) and (\ref{eq:taussa1}), we obtain the shifts in orbital phase between the maximum at a frequency $\nu$ and at 15~GHz:
\begin{equation}
    \Delta\phi_{\rm SSA} =  \frac{\nu_{\rm GHz}^3 - 15^3}{v_{\rm SSA} P_{\rm orb} 3.354 \times 10^{-11} (3.54 \times 10^{18})^{2} K B^{2} b(2)}.
    \label{eq:phissa}
\end{equation}
Figure~\ref{fig:shiftSSA} shows the derived velocity curves as a function of the frequency and the orbital phase shift for the SSA case. In the case of $B \sim \ell^{-2}$ we can fit the data with a constant velocity of $v_{\rm SSA} = 1\,000 \pm 200~\mathrm{km\ s^{-1}}$, where the uncertainties result from the propagation of the uncertainties in $\ell$ at 1~GHz. The fitted velocity is also compatible with the one of the stellar wind. In the case of $B \sim \ell^{-1}$ the data are best-fitted with a constant velocity of \new{$17\,000 \pm 2\,000~\mathrm{km\ s^{-1}}$}, which is significantly faster than the stellar wind one. In any case, both velocities in the SSA scenario are clearly slower than the relativistic velocity of the putative pulsar wind or the relativistic jet.
We note that if energy losses were considered, the derived expansion velocities would be even lower than the ones discussed.
A dependence of $B \sim \ell^{-1}$ would be more plausible given the {\em conical} geometry of the current scenarios, based either on a cometary tail or a jet. However, at least in the young non-accreting pulsar scenario, the emitting region at low frequencies could exhibit a more toroidal or spherical geometry after mixing with stellar wind, in which case a dependence closer to $B \sim \ell^{-2}$ would be expected.

We show that either FFA or SSA (with different magnetic field dependences) can explain the delays of the low-frequency emission of LS~I~+61~303 with expansion velocities of $\sim$$1\,000$ or $\sim$$17\,000\ \mathrm{km\ s^{-1}}$, depending on the considered model.
On the one hand, a large expansion velocity of $\sim$$17\,000\ \mathrm{km\ s^{-1}}$ could be accommodated either in the young non-accreting pulsar scenario or in the microquasar one. In the first scenario the obtained velocity would indicate that the stellar wind does not dominate the expansion of the emitting region. In the second one, this velocity could be an estimate of the lateral expansion velocity of the jet.
On the other hand, low expansion velocities of $\sim$$1\,000\ \mathrm{km\ s^{-1}}$ appear too low to represent the lateral expansion velocity of the jet, while they could be obtained in the young non-accreting pulsar scenario if the stellar wind dominates the expansion of the emitting region.

We can also compare the delays predicted by our models with the simultaneous data taken between 2.6 and 15~GHz in 2012, $\phi_{\rm so} \sim 0.6$ \citep{zimmermann2015}. These data show that the peak of the emission at these frequencies takes place at roughly the same orbital phases, setting a delay $\lesssim$0.05 within this frequency range. Although this delay is compatible with the three fitted models within uncertainties, we note that the FFA and SSA with $B \sim \ell^{-2}$ models predict delays closer to 0.1 orbital phases, while delays close to zero could only be reproduced by the SSA with $B \sim \ell^{-1}$ model. However, we stress that the superorbital phase of the 2012 observations is 0.6, different to the one of the GMRT monitoring (0.2), and thus the orbital variability as a function of the frequency could be different.
\begin{figure}[!t]
    \centering
    \includegraphics[width=0.8\textwidth]{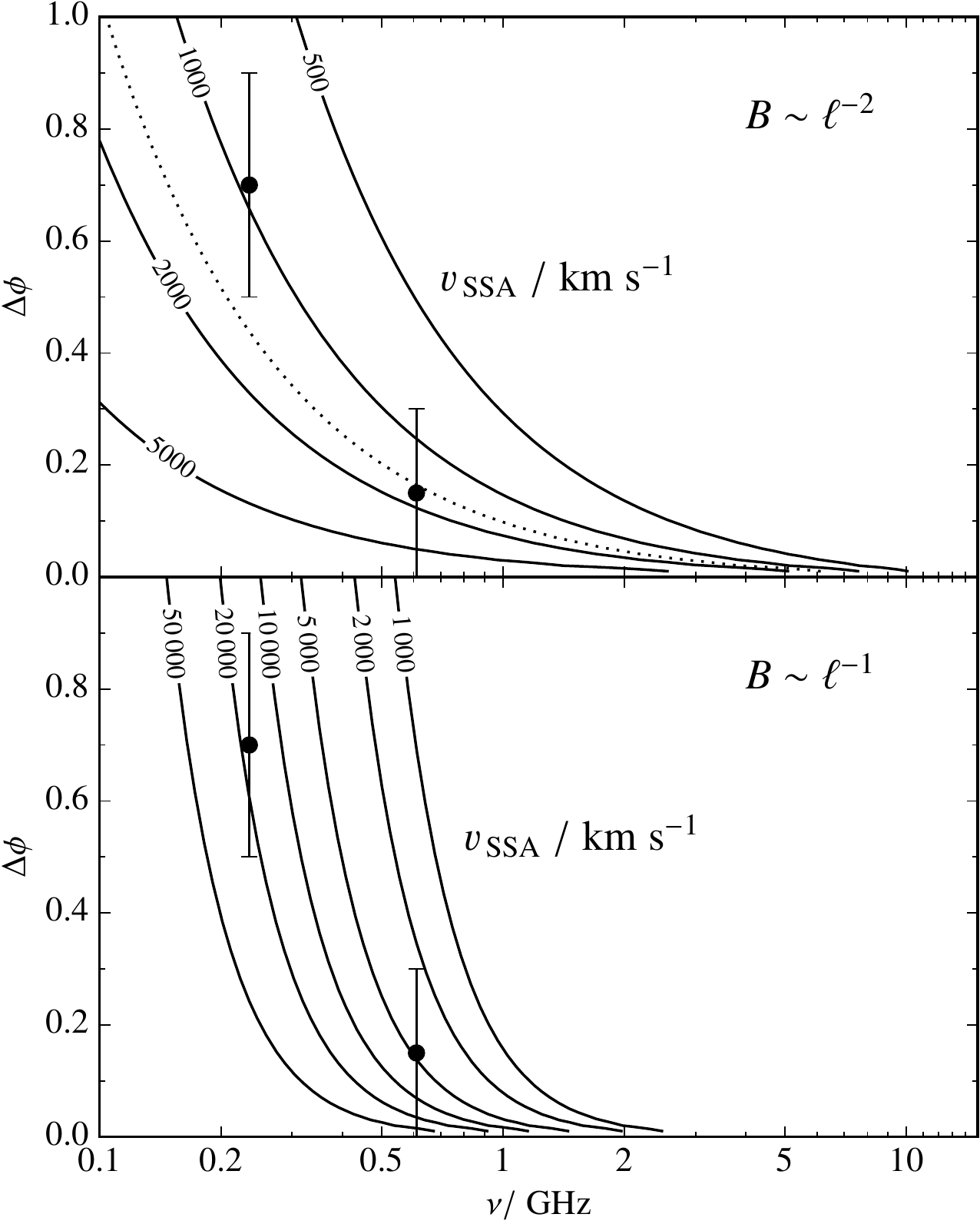}
    \caption[Same as Figure~\ref{fig:shift} but assuming that SSA is the dominant absorption process.]{Same as Figure~\ref{fig:shift} but assuming that SSA is the dominant absorption process and following equation (\ref{eq:phissa}). Two different dependences of $B$ have been considered: $B \sim \ell^{-2}$ (top) or $B \sim \ell^{-1}$ (bottom). The dotted line on the top panel represents the mean value of the derived wind velocity. We note that with the dependence $B\sim \ell^{-2}$ (top) we can easily explain the data with an expansion velocity of $\sim$$1\,000~\mathrm{km\ s^{-1}}$, while with the dependence $B \sim \ell^{-1}$ (bottom) a velocity of $\sim$$17\,000~\mathrm{km\ s^{-1}}$ is required.}
    \label{fig:shiftSSA}
\end{figure}


Extrapolating these results to 150~MHz, we infer an orbital phase delay of about 0.9 (assuming FFA) or $\sim$1.0 (assuming SSA with $B \sim \ell^{-2}$). In the case of SSA with $B \sim \ell^{-1}$ we would expect a mixing between different outbursts.
However, these results cannot be directly compared with the obtained light-curves at 150~MHz. These data were taken at different superorbital phases ($\phi_{\rm so} \sim 0.9$ instead of $\sim$0.2), and we observe a larger flux density emission and a larger variability at these 150~MHz data than the ones expected from an extrapolation of the 235 and 610-MHz light-curves. This implies, as mentioned before, that the superorbital phase still plays a significant role at low frequencies.
In any case, the observed shift at 150~MHz with respect to the 15-GHz data is not well constrained and it could be between $\sim$0.0 and $\sim$0.5 orbital phases, with the possibility of observing a full cycle shift (and thus between 1.0 and 1.5). We note that these values are roughly compatible with the ones derived from the 235 and 610-MHz data assuming a one-cycle delay. However, we cannot discard the possibility of being observing a delay of only $\sim$0.0--0.5 orbital phases, as we would expect less absorption at these high superorbital phases, which would also explain why we see a large orbital variability.


\section{Conclusions}\label{sec:lsi-conclusions}

We have detected for first time a gamma-ray binary, LS~I~+61~303, at a frequency as low as $150~\mathrm{MHz}$. This detection establishes the starting point to explore the behavior of gamma-ray binaries in the low frequency radio band, which will allow us to unveil the absorption processes that can occur in their radio spectra or light-curves.
Additionally, we have obtained light-curves of LS~I~+61~303 at 150, 235 and 610~MHz, observing orbital and superorbital variability in all cases. In the folded light-curve with the orbital period we observe quasi-sinusoidal modulations with the maxima at different orbital phases as a function of the frequency. The observed delays between frequencies seem to be also modulated by the superorbital phase. The flux density values are also modulated by the superorbital phase, with the source displaying a stronger emission at $\phi_{\rm so} \sim 1$.

We have modeled the shifts between the maxima at different frequencies as due to the expansion of a one-zone spherically symmetric emitting region assuming either free-free absorption or synchrotron self-absorption with two different magnetic field dependences.
The derived expansion velocities are clearly subrelativistic and in some cases close to the stellar wind one. Both the young non-accreting pulsar scenario and the microquasar scenario could accommodate the obtained highest velocities ($\sim$$17\,000\ \mathrm{km\ s^{-1}}$), although the first scenario seems to be favored in the case of low velocities ($\sim$$1\,000\ \mathrm{km\ s^{-1}}$), which are close to the stellar wind one.

The limited amount of data acquired up to now precludes detailed modeling to establish the origin of the variability at different frequencies and epochs.
Further multi-epoch observations of LS~I~+61~303 with LOFAR are needed. These data would allow us to determine the light-curve of LS~I~+61~303 folded in orbital phase and study its changes as a function of the superorbital phase. A good coverage of a single orbital cycle is mandatory to obtain a reliable profile of the variability of LS~I~+61~303 due to the significant differences observed between outbursts at different orbital phases.
Future simultaneous multifrequency observations with the GMRT, LOFAR and the VLA at different superorbital phases would allow us to study the light-curve of LS~I~+61~303 and its dependence with the frequency. These data could unveil the changes in the physical parameters that characterize the emitting region and the absorption processes required to explain the superorbital modulation.
Finally, the use of the International stations in LOFAR observations (longer baselines) would allow us to search for the extended emission at arcsec scales that is expected to arise at low frequencies.

%
%
%
%

\chapter[Radio emission decrease in the gamma-ray binary HESS~J0632+057]{Radio emission decrease in the gamma-ray binary HESS~J0632+057\vspace{-25pt}} \label{chap:hess}

The gamma-ray binary HESS~J0632+057 exhibits an orbitally modulated X-ray light-curve with a main and a secondary X-ray outburst. Existing radio observations were taken mostly around the main X-ray outburst, with only a couple observations at other orbital phases. Previous EVN observations performed in 2011 during and just after the main X-ray outburst reveal an extended radio emission and a decay in the total radio flux density. A similar behavior during the secondary X-ray outburst was expected: a new radio outflow could be produced, probably with a different orientation due to the different orbital phases. To unveil the radio emission of HESS~J0632+057 during this secondary X-ray outburst, we observed the source at orbital phase $\phi = 0.76$ simultaneously with the EVN and WSRT. The use of both arrays was expected to allow us to determine the changes of the radio emission at different angular scales. However, we have only found two upper-limits to the emission of HESS~J0632+057, both from the WSRT and from the EVN data. The EVN data point out a strong decrease in the radio emission of at least one order of magnitude with respect to the main X-ray outburst, which indicates a different variability of the compact radio emission compared to the X-ray one. However, the lack of simultaneous X-ray data prevents us from properly understand what is the origin of this decrease. These results have been published in the proceedings of a poster shown in the {\em 12th European VLBI Network Symposium and Users Meeting} \citep{marcote2014hess}.

\section{Introduction}

HESS~J0632+057 is a binary system comprising a B0~Vpe star of $13$--$19~\mathrm{M_{\sun}}$ and a compact object of $1.3$--$7.1~\mathrm{M_{\sun}}$ \citep{aragona2010}, with coordinates
$$ \alpha = 06^{\rm h}\ 32^{\rm m}\ 59.257^{\rm s},\qquad \delta = +05^{\circ}\ 48'\ 01.16''$$
at the epoch of MJD~55\,607 \citep{moldon2011hess}, being located $\sim$1.5~kpc away from the Sun. The orbit of the system shows a high eccentricity of $e = 0.83 \pm 0.08$ \citep{casares2012} and an orbital period of $315_{-4}^{+6}~\mathrm{d}$ \citep{aliu2014}.

\begin{figure}[!t]
	\centering
	\includegraphics[width=0.8\textwidth]{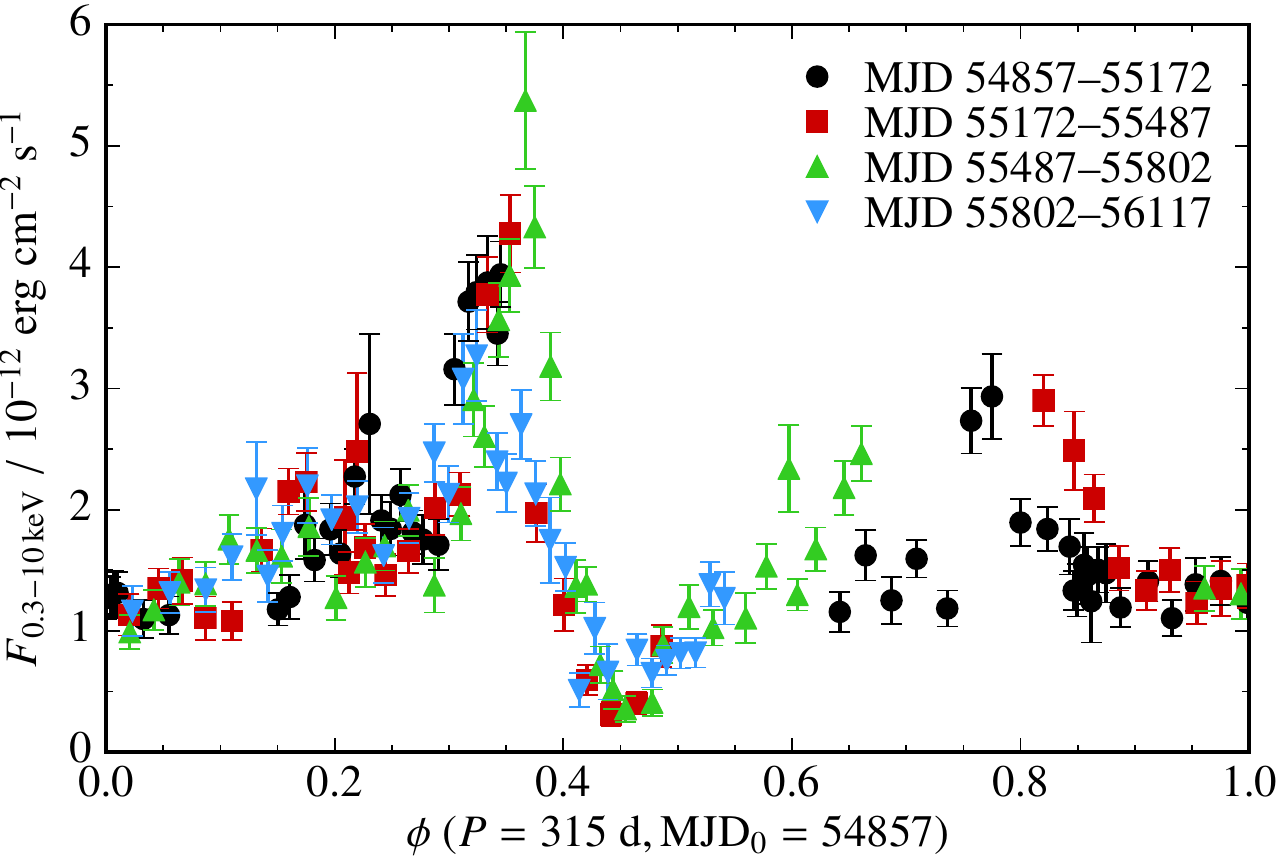}
	\caption[X-ray light-curve of HESS~J0632+057 folded with the orbital period.]{X-ray light-curve of HESS~J0632+057 folded with the orbital period from {\em Swift}-XRT data. Different orbital cycles are indicated by different colors and markers. Adapted from \citet{aliu2014}.}
	\label{fig:hess-X-ray-lightcurve}
\end{figure}
HESS~J0632+057 was discovered by the H.E.S.S.\ Collaboration as a point-like TeV source \citep{aharonian2007}. Subsequently, variability in the TeV emission was reported by \citet{acciari2009}. A variable X-ray counterpart \citep{acciari2009,hinton2009} and also a variable radio counterpart \citep{skilton2009} were found later. The massive B0~Vpe star MWC~148 was confirmed as the optical counterpart of the TeV source \citep{hinton2009}. As it is shown in Figure~\ref{fig:hess-X-ray-lightcurve}, the X-ray emission exhibits a main outburst in the orbital phase range of 0.3--0.4 followed by an X-ray dip at orbital phases 0.43--0.46, and a secondary outburst between orbital phases 0.6 and 0.9 (phase $\phi = 0$ is arbitrarily defined at MJD~54\,857 according to \citealt{bongiorno2011}, while the periastron takes place at orbital phase 0.97 following \citealt{casares2012}). The maximum of the main X-ray outburst changes slightly in amplitude and phase from cycle to cycle, while the X-ray dip remains stable. The X-ray profile of the main outburst and the following dip resembles to the light-curve observed in $\upeta$-Carinae, which is known to be originated by wind collisions. However, in HESS~J0632+057, most of the X-ray emission exhibits a non-thermal origin \citep{falcone2010}. In contrast to the main X-ray outburst and the dip, the orbital phases at which the secondary X-ray outburst takes place vary significantly from cycle to cycle (see Figure~\ref{fig:hess-X-ray-lightcurve}). The earliest outburst has been reported at $\phi \approx 0.6$, whereas the latest one has been observed at $\approx$0.8 \citep{aliu2014}.

We note that HESS~J0632+057 has not been detected yet at GeV energies, making the system unique among all the known gamma-ray binaries \citep{caliandro2013}, as it is shown in Table~\ref{tab:ls5039-other-binaries}.
The orbital variability of the TeV emission is clearly correlated with that of the X-ray emission, suggesting that both are originated by the same population of accelerated particles \citep{aliu2014}.
These authors show that simple one-zone leptonic models explain this correlation assuming that the X-ray emission is produced by synchrotron emission of the accelerated particles, whereas the TeV emission is produced by IC scattering of the UV radiation originated in the companion star by the same accelerated particles.

The SED of HESS~J0632+057 is very similar to the one of LS~I~+61~303, but one order of magnitude fainter \citep{hinton2009}. The eccentricities of both systems are similar, and both host a B0 star with a circumstellar disk, although the orbital period is about one order of magnitude larger in HESS~J0632+057.
We note that the distance to HESS~J0632+057 is smaller than to LS~I~+61~303 ($\sim$1.4 versus 2.0~kpc, respectively), implying that the source must be intrinsically fainter. This is probably related to the larger separations between the companion star and the compact object.

\subsection{The radio emission of HESS~J0632+057}

HESS~J0632+057 has been previously explored with radio interferometers at GHz frequencies.
\citet{skilton2009} conducted six GMRT observations at 1.3~GHz, and three contemporaneous VLA observations at 5.0~GHz, all of them covering the main X-ray outburst and the subsequent dip in the same orbital cycle (from orbital phases $\approx$0.3 to 0.55). Whereas the 1.3-GHz GMRT data show a stable emission of about $0.67~\mjy$, the three 5-GHz VLA data show significant variability, with an average flux density of $\approx$$0.28~\mjy$. However, we note that this variability is not correlated with the X-ray light-curve.
These authors also conducted one observation at 5 and 8~GHz with the VLA during the secondary X-ray outburst. Two upper-limits were inferred from these data, although their 3-$\sigma$ values are close to the flux density values detected in the previous observations at 5~GHz.

Two EVN observations were conducted in 2011 during, and just after, the main outburst, revealing an extended emission with an one-sided structure that extends $\sim$50~mas \citep{moldon2011hess}. The position of the peak of the emission suffers a displacement of about 14~mas (21~AU) between the two observations, aligned with the direction of the extended emission, and implying a projected expansion velocity of $\sim$$1\,200~\mathrm{km\ s^{-1}}$. HESS~J0632+057 shows a flux density emission of $0.41 \pm 0.09~\mjy$ during the peak of the main X-ray outburst, and $0.18 \pm 0.03~\mjy$ during the subsequent dip. A significant decrease in the radio emission is thus observed, following the X-ray behavior.
This decay, which is not seen in the GMRT data analyzed by \citet{skilton2009} at roughly the same frequency, suggests the presence of an extended radio outflow that is filtered out at VLBI scales.

A similar behavior could be expected during the secondary X-ray outburst. However, almost all the existing radio data were taken around the main one. We decided to conduct a new EVN observation to track the evolution of the extended emission reported by \citet{moldon2011hess} in another part of the orbit, during the secondary X-ray outburst. In Sect.~\ref{sec:hess-obs} we present this observation and the obtained results, and we discuss them in Sect.~\ref{sec:hess-discussion}. Finally, we state the conclusions of this work in Sect.~\ref{sec:hess-conclusions}.

\section{EVN observation and results} \label{sec:hess-obs}

We conducted a 10-hr EVN observation on 2014 February 20 (at $\phi = 0.76$, during the secondary X-ray outburst) using the following 14 antennas: Ef, Jb, Kn, Wb, Mc, On, Tr, Sh, Ur, Hh, Sv, Zd, Bd and Ro (project code EM111). The observation was conducted at 1.6~GHz, using eight IFs, with dual circular polarization (RR and LL), each of them with a bandwidth of 16~MHz divided in 32 channels.
We used J0619+0736 as phase calibrator, which together with 0528+134 and DA~193 were used as fringe finders. 
The data were correlated at JIVE. WSRT (Wb) data were also recorded separately to obtain an image sensitive to larger angular scales. In this case, 64 channels were recorded with full circular polarization (RR, LL, RL, and LR).

We reduced the data within AIPS and Difmap \citep{shepherd1994}\footnote{\url{ftp://ftp.astro.caltech.edu/pub/difmap/difmap.html}}. After an initial flagging and automatic ionospheric corrections, we fringe fit and bandpass calibrated the data. We produced a model for the phase calibrator with Difmap, after several cycles of imaging and self-calibration. This model allowed us to produce a more accurate calibration of the data. We transferred its solutions to the target source data and we produced the final image in AIPS again.
We note that the data from the Sv antenna could not be recorded, and thus data of only 13 antennas are present in the final dataset.

Figure~\ref{fig:hess-image} shows the images obtained with WSRT and the EVN around the field of HESS~J0632+057. The source is not detected above the noise level, establishing 3-$\sigma$ upper-limits of $0.7~\mathrm{mJy\ beam^{-1}}$ (for the WSRT data) and $30~\mathrm{\upmu Jy\ beam^{-1}}$ (for the EVN data).
\begin{figure}[!t]
	\includegraphics[width=\textwidth]{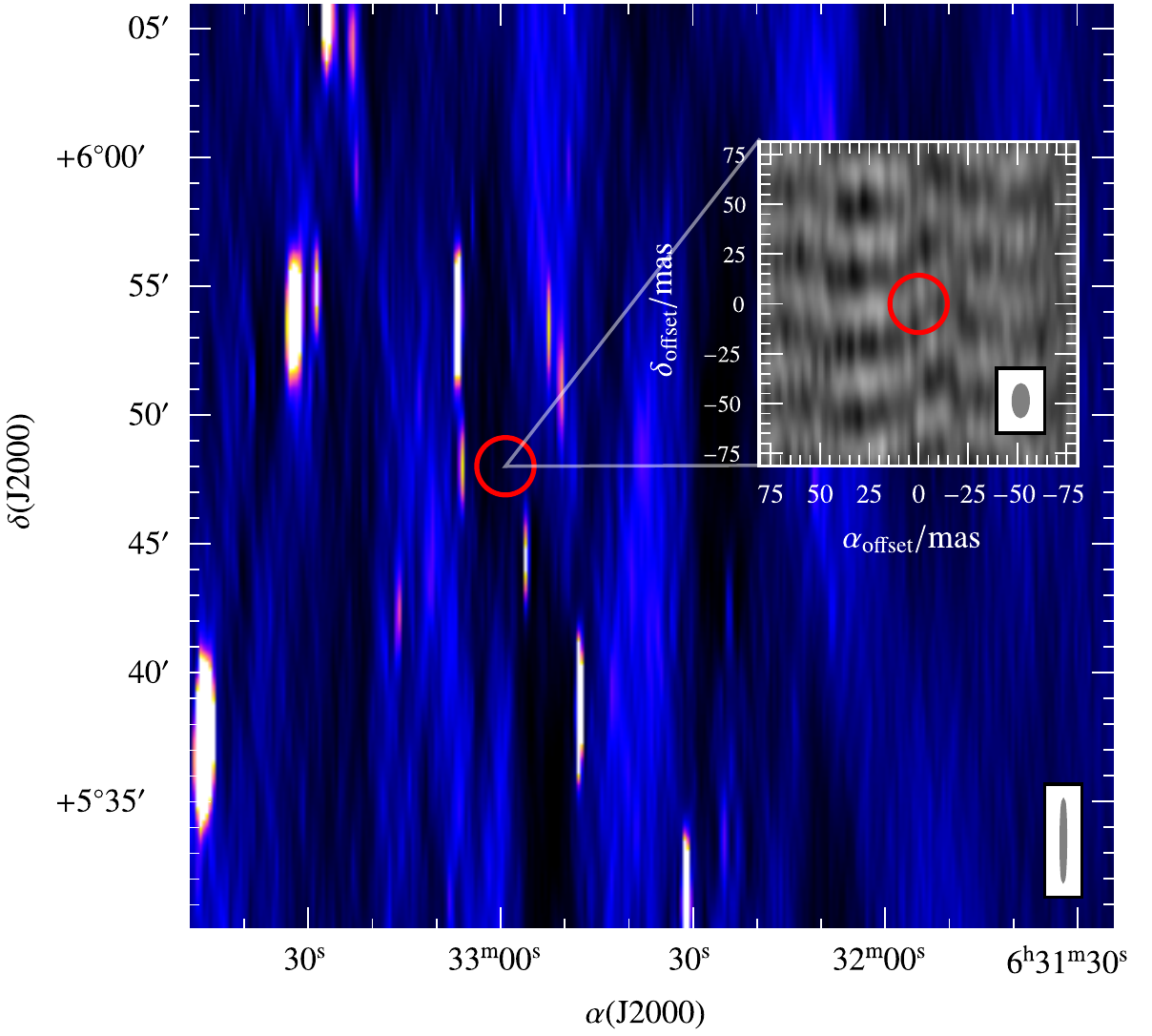}
	\caption[Images of the field of HESS~J0632+057 as seen from the WSRT data and from the EVN data.]{Images of the field of HESS~J0632+057 as seen from the WSRT data and from the EVN data (zoom). Radio emission from the source is not detected in any of them, with 3-$\sigma$ upper-limits of $\sim $$0.7~\mathrm{mJy\ beam^{-1}}$ and $30~\mathrm{\upmu Jy\ beam^{-1}}$, respectively. The position of HESS~J0632+057 is denoted by the red circle,  and the synthesized beams are shown on the bottom right corner of each image.}
	\label{fig:hess-image}
\end{figure}
Figure~\ref{fig:hess-lightcurve} (top) shows the radio light-curve of the source folded with the orbital period. In addition to our new radio data (highlighted by a vertical pale red line, at orbital phase $0.76$), the light-curve includes all the available radio data at a similar frequency (i.e.\ the previously mentioned GMRT data at 1.3~GHz from \citealt{skilton2009} and the EVN data from \citealt{moldon2011hess}). The X-ray light-curve is also shown for comparison.

We observe a decrease in the flux density values of at least one order of magnitude with respect to the previous EVN detections published by \citet{moldon2011hess}. With the WSRT data we have obtained a 3-$\sigma$ upper-limit at the level of the previous GMRT detections reported by \citet{skilton2009}, and thus we cannot conclude if the total emission of HESS~J0632+057 is weaker at this epoch.
\begin{figure}[!t]
	\centering
	\includegraphics[width=0.9\textwidth]{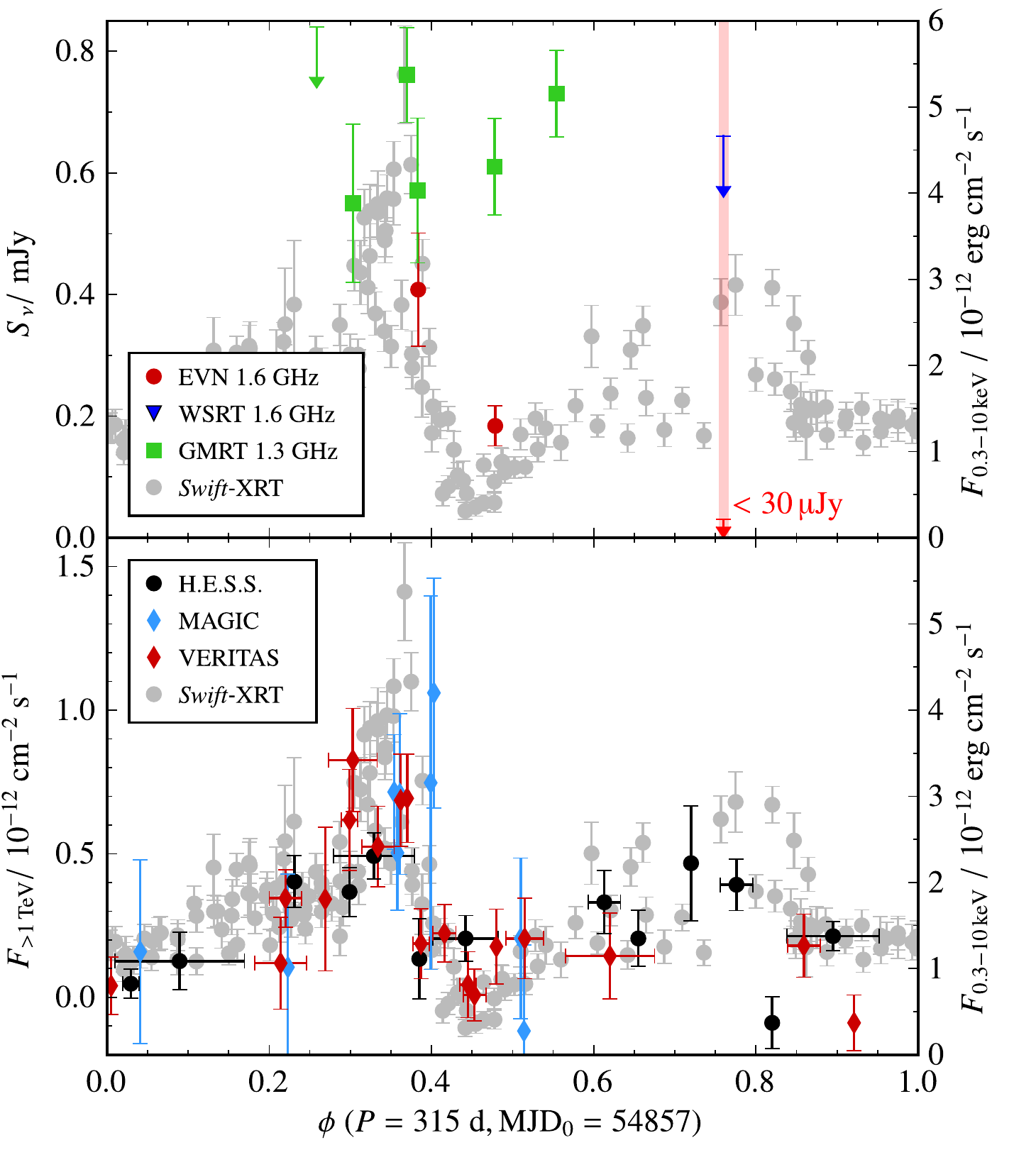}
	\caption[Light-curves of HESS~J0632+057 folded with the orbital period at radio frequencies (the data presented in this work together with GMRT and EVN data from literature), X-rays and TeV.]{{\em Top:} radio light-curve of HESS~J0632+057 folded with the orbital period. The upper-limits inferred in this work from the EVN and WSRT data (red and blue arrow, respectively) are highlighted by the vertical pale red line at orbital phase 0.76. We also show the 1.3-GHz GMRT data from \citet{skilton2009} represented by green squares, and the 1.6-GHz EVN data from \citet{moldon2011hess} by red circles. The gray circles represent the X-ray data shown in Figure~\ref{fig:hess-X-ray-lightcurve} (also shown on bottom). Error bars represent the 1-$\sigma$ uncertainties and arrows represent 3-$\sigma$ upper-limits. {\em Bottom:} TeV light-curve of HESS~J0632+057 folded with the orbital period from the data presented by \citet{aliu2014}.}
	\label{fig:hess-lightcurve}
\end{figure}

\section{Discussion} \label{sec:hess-discussion}

The obtained upper-limit from the EVN data reveals a strong decrease in the radio emission of HESS~J0632+057 of at least one order of magnitude with respect to the one observed during the main X-ray outburst. Considering the radio/X-ray behavior observed during the main X-ray outburst, this strong decrease was somehow unexpected, although we do not have contemporaneous X-ray data during the cycle of the EVN/WSRT data.
The ratio between the radio and the X-ray flux around the main X-ray outburst is $\sim$$10^{-6}$. We observe from Figure~\ref{fig:hess-lightcurve} that this ratio is roughly constant along the outburst decay if we consider exclusively the EVN observations by \citet{moldon2011hess}.
Considering the non-simultaneous X-ray data from {\em Swift}, and assuming the same X-ray emission also during our EVN observation, the ratio goes down to $\lesssim$$10^{-7}$ during the secondary X-ray outburst. We note that the TeV/X-ray flux ratio remains roughly constant along all the orbital phases (see Figure~\ref{fig:hess-lightcurve}, bottom).

The observed reduction in the radio/X-ray flux ratio from one epoch to the other could be explained by different behaviors.
On the one hand, the radio and the X-ray emission could arise from different populations of accelerated particles. Therefore, no correlation would be expected from the emission of the two populations. However, this scenario is unlikely, as we see in other systems that the same population appears to originate the radio and the X-ray emission.
On the other hand, we need to consider that the radio and the X-ray data were not simultaneous. This fact can lead into two different possibilities. First, the emission of HESS~J0632+057 could be much fainter than the one observed in previous orbital cycles. We have seen a large variability from cycle to cycle, with changes in amplitude and shape in the X-ray light-curve, particularly during the secondary outburst.
Secondly, we could have observed the source before the start of the secondary outburst. No radio outflow would thus have been originated yet. Comparing the three secondary X-ray outbursts observed by \citet{aliu2014}, see Figure~\ref{fig:hess-X-ray-lightcurve}, we note that one took place at orbital phase $\approx$0.6, another one in the phase range $0.75$--$0.80$, and in the last one they only observed the decrease of the emission, at phases $\approx$$0.85$ (and thus the outburst could have taken place at $\approx$0.8). Given that the radio outflow is expected to be produced just after or simultaneous to the X-ray outburst, this is a feasible possibility.
In case of a late outburst, the radio outflow would have not been produced yet. In such case, the radio emission produced by the previous outflow, originated during the main X-ray outburst, would be faint enough to be undetectable in our data.

\section{Conclusions} \label{sec:hess-conclusions}

The 2014 EVN upper-limit during the secondary X-ray outburst implies a strong decrease, of at least one order of magnitude, in the radio emission of HESS~J0632 +057 compared to the previous EVN observations reported by \citet{moldon2011hess} during the main X-ray outburst. This strong decrease indicates a different variability of the compact radio emission compared to the X-ray one. However, the large variability observed in the secondary X-ray outburst from cycle to cycle, together with the fact that the radio and X-ray data were not simultaneous, implies that we do not know if the X-ray outburst had actually occurred at the epoch of the EVN observations.
In case of a late outburst, it is possible that the radio outflow had not been originated yet, and thus no strong radio emission would be observed.

Simultaneous radio and X-ray observations are thus mandatory to understand the behavior of HESS~J0632+057, and to determine the possible connection between the radio and X-ray emission. Furthermore, a better coverage of the radio light-curve of the source is necessary to unveil its radio emission along the orbit. Repeated multiwavelength observations during several orbital cycles could reveal possible shifts in orbital phase observed in the secondary X-ray outburst and wether these same shifts are present in the radio light-curve.

%
%
%
%

\chapter{The discovery of the colliding wind binary HD~93129A} \label{chap:hd}

Radio observations are an effective tool for discovering particle acceleration regions in colliding-wind binaries (CWBs) through detection of synchrotron radiation.  
Only five wind-collision regions (WCRs) have been resolved to date at radio frequencies on mas angular scales. HD~93129A, a prototype of the very few known O2~I stars, is a promising target for study. Recently, a second massive, early-type star about 50~mas away was discovered, and a non-thermal radio source was detected in the region. Preliminary long-baseline array data suggest that a significant fraction of the radio emission from the system comes from a putative WCR.
We seek evidence that HD~93129A is a massive binary system with colliding stellar winds that produce non-thermal radiation through spatially resolved images of the radio emitting regions.
We conducted observations with the LBA to resolve the system at mas angular scales. ATCA data have also been analyzed to derive the total radio emission. We compiled and analyzed optical astrometric data from the historical Hubble Space Telescope ({\em HST}) archive to obtain absolute and relative astrometry of the stars with mas accuracy.
The optical astrometric analysis leads us to conclude that the two stars in HD~93129A form a gravitationally bound system. The LBA radio data reveal an extended bow-shaped non-thermal source between the two stars, which is indicative of a WCR. 
The wind momentum-rate ratio of the two stellar winds is estimated from these data, providing a rough estimation of the mass-loss rate ratio.
The ATCA data show a point-like source with a change in the flux density level between 2003--2004 and 2008--2009, which are modeled with a non-thermal power-law spectrum with spectral indices of $\alpha = -1.03 \pm 0.09$ and $-1.21 \pm 0.03$, respectively.
These results have been published in {\colorexpandedcite Benaglia, Marcote et al.}\ (\citeyear{benaglia2015}) and \citet{marcote2014hd93129a}. The contribution of the author of this thesis has been focused on the reduction and analysis of the LBA radio data.

\section{Introduction}

HD~93129A is the brightest source in the most crowded part of the young open cluster Trumpler~14, in the Carina nebula, and it has been studied for several decades. It was firstly classified as an O3 star \citep{walborn1982}, though was later reclassified as an O2~If* \citep{walborn2002}. \citet{mason1998} suggested that HD~93129A was a speckle binary, and \citet{nelan2004} confirmed the presence of two components, Aa and Ab, from {\em HST} observations with the Fine Guidance Sensors (FGS) instrument. The Ab component is an O3.5~V star, currently located at a distance of about 55~mas with respect to Aa (140~AU at the distance of 2.5~kpc at which the system is located). 
These authors also provided the PA between both components \citep[but see][]{nelan2010}, and \citet{maiz2008} detected proper motion along the radial direction between both components by analyzing multiepoch {\em HST} data with a relative velocity of about $2~\mathrm{mas\ yr^{-1}}$. This velocity is much higher than the observed velocities between stars in the field, supporting the binary nature of the system. These authors suggested a highly elliptical and/or inclined orbit, but without determining the orbital parameters. Table~\ref{tab:hd-parameters} shows the main parameters of HD~93129A that are relevant along this work.

\begin{table}[b!]
	\small
    \begin{center}
        \caption[Adopted parameters of HD~93129A that are relevant in this work.]{Adopted parameters of HD~93129A that are relevant in this work.}
        \label{tab:hd-parameters}
        \begin{tabular}{lrll}
            \hline\\[-10pt]
            Parameter   & Value & Unit & Ref. \\[+2pt]
            \hline\\[-10pt]
            Aa spectral type & O2~If*& & \citet{walborn2002}\\
           Ab spectral type & O3.5~V& & \citet{benaglia2006} \\
            $\Delta m (m_{\rm Ab} - m_{\rm Aa})$   & $0.90 \pm 0.05$ & mag & \citet{nelan2010} \\
            Wind terminal velocity$^\dag$  & $3200 \pm 200$ & $\mathrm{km\ s^{-1}}$ & \citet{taresch1997} \\
            System mass & $200 \pm 45$ & $\mathrm{M_{\sun}}$ & \citet{maiz2008}\\
            Effective temperature$^\dag$ & 42500 & K & \citet{repolust2004} \\
            System luminosity & 6.2 & $\log(L /\ \mathrm{L_{\sun}})$ & \citet{repolust2004} \\
            X-ray luminosity$^{\dag\dag}$ & $1.3 \times 10^{-7}$ & $L_{\rm bol}$ & \citet{cohen2011} \\
            Distance & 2.5 & kpc & \citet{walborn1995} \\
            \hline\\[-10pt]
            \multicolumn{4}{l}{$^\dag$ Derived assuming a single star.}\\
            \multicolumn{4}{l}{$^{\dag\dag}$ Derived from the energy band 0.5--8 keV.}
    \end{tabular}
    \end{center}
\end{table}

The relative motion of the two stars can be traced from archival {\em HST}/FGS observations taken from 1996 to 2009, together with VLT/NACO and VLTI/PIONIER observations reported by \citet{sana2014}. Figure~\ref{fig:hd-positions} shows the relative astrometry from all these measurements. The oldest ones present large uncertainties because the position angle of the system was unknown, and the two components were only resolved along only one of the two FGS axes. For later observations, rough estimations of the position angle avoided this problem, allowing more accurate measurements.
\begin{figure}[t]
    \centering
    \includegraphics[width=0.6\textwidth]{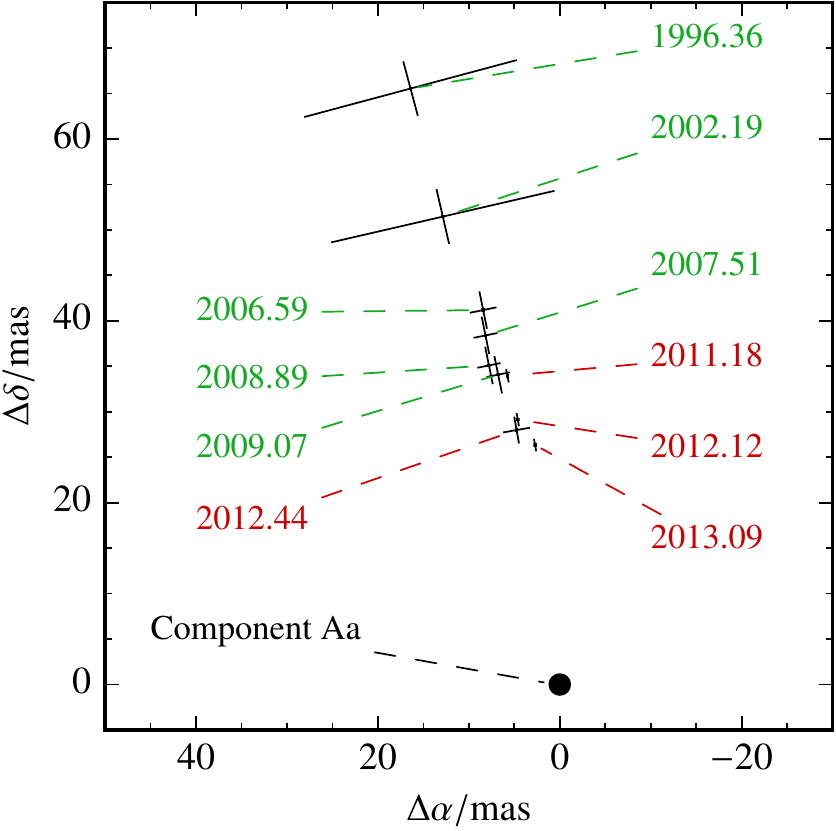}
    \caption[Relative positions of the two components of HD~93129A.]{Relative positions, including error bars, of HD~93129Ab with respect to Aa, derived from archival {\em HST}/FGS data (green labels) and from VLT/NACO and VLTI/PIONIER data published by \citet[red labels]{sana2014}. The primary component, Aa, is placed at the origin.}
    \label{fig:hd-positions}
\end{figure}

Radio emission coincident with the position of HD~93129A is detected in the range of 1.4--24.5~GHz with flux densities of 1--10~mJy \citep{benaglia2006}. The spectrum shows a power-law emission with a negative spectral index, typical from synchrotron emission.

At X-rays, \citet{cohen2011} determined that the X-ray emission observed for HD~93129A is dominated by a thermal component originated by the individual stellar winds in the system. Moreover, \citet{gagne2011} reported a ratio between the X-ray and bolometric luminosities of $L_{\rm X} / L_{\rm bol} \approx 1.3 \times 10^{-7}$. This value suggests that only a small contribution of the X-rays could come from a putative CWR. However, for very early-type stars with strong and dense winds, such as HD~93129A, the individual X-ray emission should be somewhat fainter than suggested by the canonical ratio of $L_{\rm X} / L_{\rm bol} \sim 10^{-7}$ \citep{debecker2013Ostars,owocki2013}.

HD~93129A is fairly close to the {\em Fermi} source 1FGL~J1045.2$-$5942. This source is associated with $\upeta$-Carinae, the only CWB detected at \g-rays up to now \citep{reitberger2015}. It seems very unlikely that HD~93129A contributes to the emission detected by {\em Fermi}, as clarified by \citet{abdo2010}. At VHE, no source has been found around the position of HD~93129A in the TevCat catalog\footnote{http://tevcat.uchicago.edu}. Therefore, we conclude that there is no known \g-ray emission associated with HD~93129A.

With the firm possibility of having a binary nature, HD~93129A became as a candidate to be one of the earliest, hottest and most massive binary systems discovered up to now in the Galaxy. The radio emission could be originated by a WCR, which could only be revealed with high-resolution VLBI observations and accurate astrometric measurements. We have reduced and analyzed optical {\em HST}/FGS observations and ATCA and LBA radio observations to characterize the source, unveiling the presence of a WCR. In Sect.~\ref{sec:hd-observations} we present the analyzed data. In sect.~\ref{sec:hd-results} we detail the results obtained from the optical and radio observations, which are discussed in Sect.~\ref{sec:hd-discussion}. Finally, we state the conclusions of this work in Sect.~\ref{sec:hd-conclusions}.

\section{Multiwavelength campaign}\label{sec:hd-observations}

Multiwavelength observations are required to unveil the nature of HD~93129A. High-resolution radio data can resolve the radio emission. High resolution optical data are also necessary to determine the positions of the two stars of the system, comparing them to the radio emission. In the case of a WCR, we would expect a bow-shaped radio structure located between both stars. To test this possibility, we analyzed {\em HST}/FGS data and conducted two LBA radio observations. ATCA observations have also been analyzed to infer the spectrum of HD~93129A and its possible variability along the time.

\subsection{Optical observations with the Hubble Space Telescope}

To improve the absolute astrometric position of HD~93129A (only the positions from the Tycho catalog were available previously, with uncertainties of about $0.1''$), we have reanalyzed the archival {\em HST}/FGS data. The FGS instrument on the {\em HST} is composed of two-channel white-light shearing interferometers that provide positional measurements of guide stars to enable accurate pointing and stabilization of the {\em HST}. The FGS is capable of resolving sources down to $\sim$15~mas when operated in its high angular transfer mode (TRANS), or performing sub-mas astrometry when operated in position mode (POS). See \citet{nelan2014} for a description of FGS and its operation.

HD~93129A was observed with {\em HST}/FGS in eight epochs between 2006 and 2009 in TRANS and POS mode. The first mode provides the component separation, position angle, and magnitude difference at each epoch, whereas the second mode supplies the relative positions of the field stars and composite HD~93129A system. The combination of both modes provides the position of the two components, Aa and Ab, with respect to the field of stars at different epochs, and allows us to infer the proper motions of the system.

To compare the FGS coordinates of HD~03129Aab to the radio source, it is necessary to convert the relative FGS positions to the absolute International Celestial Reference System coordinates (ICRS). We used the PPMXL catalog\footnote{PPMXL is a catalog of positions, proper motions, 2MASS and optical photometry of 900 million stars and galaxies, aiming to be complete down to about magnitude $V = 20$  for the full-sky. It is the result of a re-reduction of USNO-B1 together with 2MASS to the ICRS.} \citep{roeser2010} for the coordinates and proper motions of the seven reference stars as input to a model that finds the right ascension, $\alpha$, and declination, $\delta$, of the target source.
The FGS observations were obtained between 2006 and 2009, while the PPMXL coordinates are at the epoch J2000. The proper motions of the reference stars must be taken into account for the most accurate correlation of the FGS data with the radio observations. However, the proper motions provided by PPMXL have large uncertainties. As it is standard in FGS astrometry \citep{nelan2013, benedict2007}, the PPMXL proper motions are input considering uncertainties into the model, which combines the FGS astrometry from the eight epochs. The model outputs the best fitting proper motion for each reference star, based upon the FGS measurements but constrained by the PPMXL input values.

\subsection{LBA radio observations}

Two LBA observations on HD~93129A have been performed up to now. First, HD~93129A was observed on 2007 June 22 at 2.3~GHz as part of the eVLBI experiment vt11D3, using three antennas: Parkes, Mopra, and ATCA with a total observing time of 10~h and 3-hr of time-on-source. PKS~0637$-$752 (J0635$-$7516) was used as amplitude calibrator and J1047$-$6217 as phase calibrator.
\begin{figure}[t]
    \centering
    \includegraphics[width=.9\textwidth]{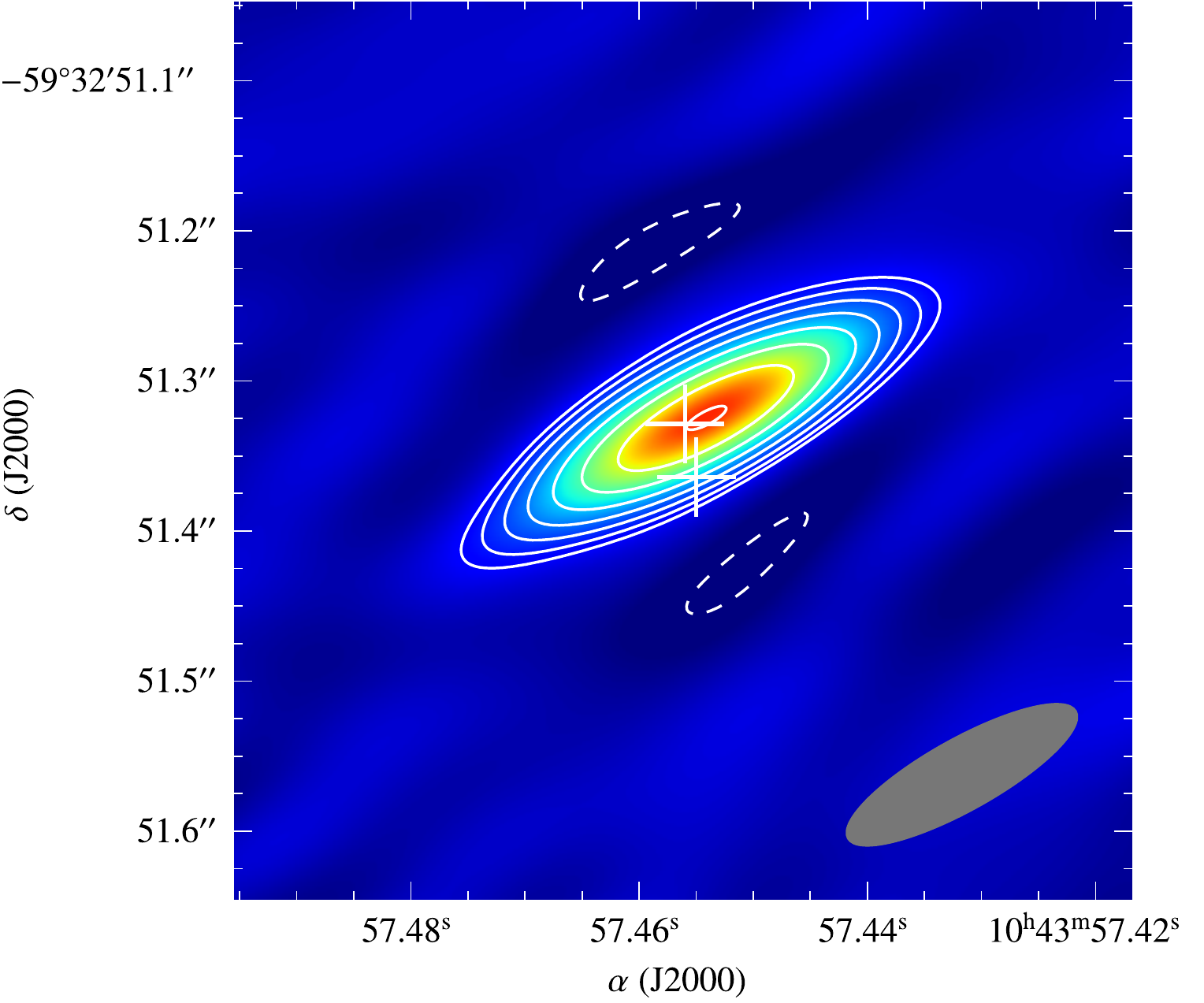}
    \caption[Image of HD~93129A from the 2007 LBA data.]{Image of HD~93129A from the 2007 LBA data. The optical positions of the two stars are marked with crosses (Aa is the southern one, and Ab the northern one). The rms of the image is $0.1~\mjybeam$. Contours start at 3-$\sigma$ noise level and increase by factors of $2^{1/2}$. The synthesized beam, shown on the bottom right corner, is $0.2 \times 0.05~{\arcsec}^2$, with a $\mathrm{PA} = 121^\degree$.}
    \label{fig:hd-lba2007}
\end{figure}
Standard VLBI calibration procedures were used, obtaining a synthesized beam of $0.2 \times 0.05~{\arcsec}^2$, $\mathrm{PA} = 31^\degree$.
The results of this observation were published in \citet{benaglia2010}, who reported a point-like source with a flux density of $\sim$$3~\mjy$ (see Figure~\ref{fig:hd-lba2007}).

A more detailed LBA observation was conducted on 2008 August 6 at 2.3~GHz. Five antennas were used this time: Parkes, Mopra, ATCA, Ceduna, and Hobart (project code V191B). ATCA was used in tied-array mode (i.e. all ATCA antennas were recording as a single station). The baselines range from 100 to 1\,700~km, providing an angular resolution of $\sim$15~mas and a better $uv$-coverage than in the previous observation. Figure~\ref{fig:hd-uvcoverage} shows the $uv$-coverage for this observation. The total observing time was 11~h, with 3.2~h of time-on-source. J1047$-$6217 was used as phase calibrator, and PKS~0637$-$752, 1549$-$790, J1051$-$6518, and J1023$-$6646 were used as fringe finders.
\begin{figure}[t]
    \centering
    \includegraphics[width=0.8\textwidth]{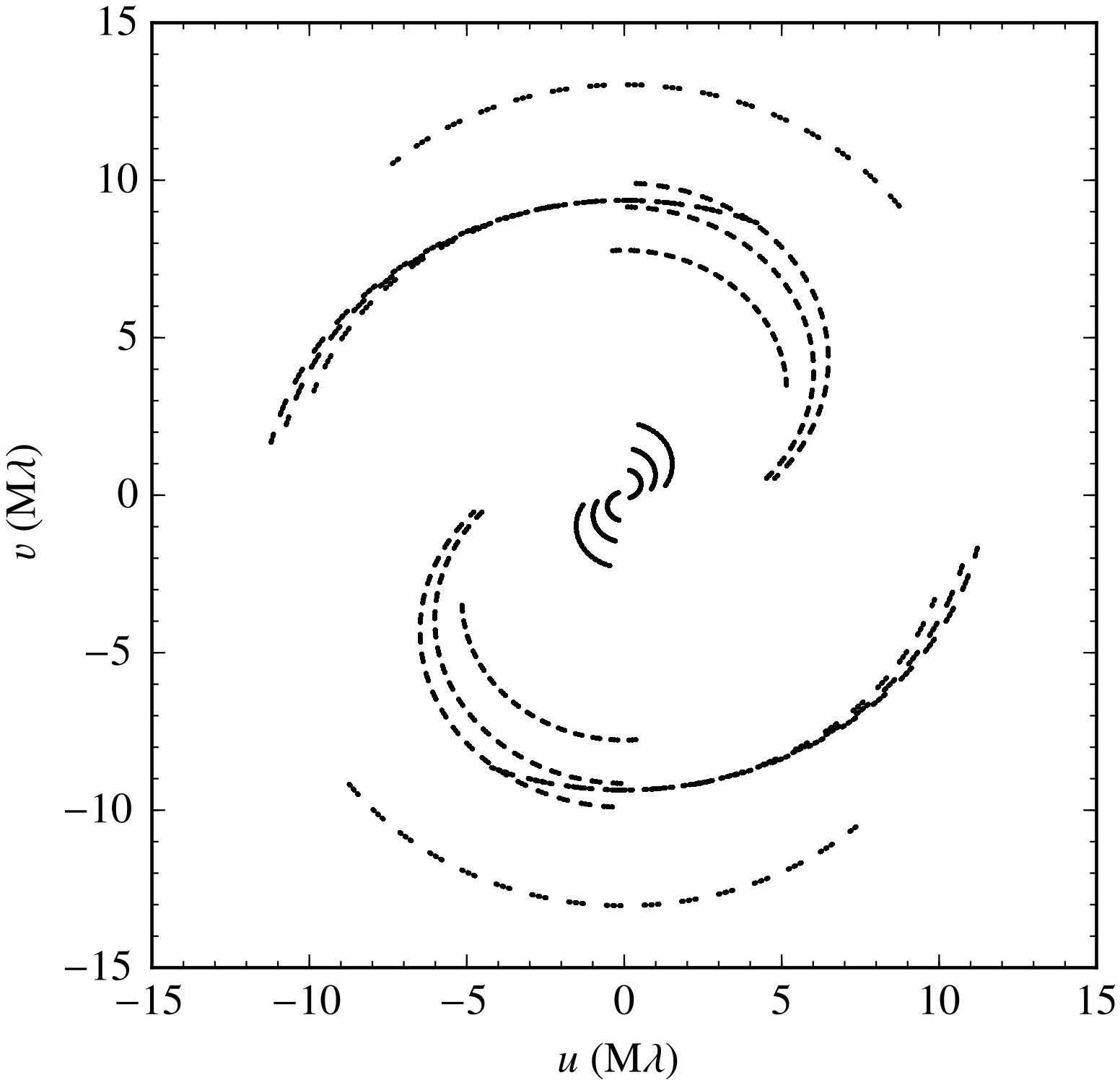}
    \caption[$uv$-coverage of the LBA observation of HD~93129A on 2008 August 6]{$uv$-coverage of the LBA observation of HD~93129A on 2008 August 6. Five antennas were used (Parkes, Mopra, ATCA, Ceduna, and Hobart) in a 11-hr observation. HD~93129A was observed during 3.2~h. We note the relatively poor $uv$-coverage of the LBA data.}
    \label{fig:hd-uvcoverage}
\end{figure}
The data were recorded at 256~Mbps provided by four IFs of 16-MHz bandwidth at each of the two recorded circular polarizations. We correlated the data with the DiFX correlator, using 128 channels per IF, 125-kHz wide each, and an integration time of 2~s. The data from the ATCA antennas were also correlated as an independent interferometer to obtain the total flux density of the source (the LBA correlated data are not sensitive to the short baselines).

The data reduction was performed using AIPS. Bad data were flagged using telescope a-priori flag files and information provided in the observing logs. A manual flagging was also performed to remove the remaining wrong data. The amplitude calibration was performed using the antenna system temperatures, and phase solutions on the calibrators were obtained using the {\tt fring} task from AIPS. 
The fringe finders and the phase calibrator were used for the bandpass calibration. We firstly produced an accurate model of the phase calibrator through several self-calibration and imaging cycles within Difmap. The phase solutions and the amplitude scale were then transferred to the target source, which was imaged. No self-calibration was attempted due to the faintness of the target source.

\subsection{ATCA radio observations}

All the archival ATCA observations of HD~93129A were analyzed to obtain a spectrum of the source. In addition to the data presented by \citet{benaglia2006} and the observations associated with the LBA projects mentioned here, we found an unpublished project (C1726) that observed the source at 4.8 and 8.6~GHz during 3.3~h. We summarize in the first three columns of Table~\ref{tab:hd-atcafluxes} all these observations (project code, date, and frequency). In all these data HD~93129A appears as a point-like source.

\begin{table}[t]
    \small
    \begin{center}
        \caption[Radio flux densities of HD~93129A at different frequencies obtained with ATCA and LBA data.]{Radio flux densities of HD~93129A at different frequencies obtained with ATCA and LBA data. We include the project code, the data, and the used flux density calibrator name with its flux density value.}
        \begin{tabular}{lc r@{}c@{}l r@{\,}c@{\,}l c r@{}c@{}l}
        \hline\\[-10pt]
        Array/project & Date & \multicolumn{3}{c}{$\nu$} & \multicolumn{3}{c}{S$_\nu$} & Flux calibrator& \multicolumn{3}{c}{$S_{\rm \nu}^{\rm cal}$} \\
        &(dd/mm/yyyy) & \multicolumn{3}{c}{(GHz)} & \multicolumn{3}{c}{(mJy)} & & \multicolumn{3}{c}{(Jy)}\\[+2pt]
            \hline\\[-10pt]
            ATCA/C678$^\dag$\\
                  & 28/01/2003  & 4&.&8 & 4.1&$\pm$&0.4 & 1934$-638$ &2&.&84\\
                  & 28/01/2003  & 8&.&6 & 2.0&$\pm$&0.2 & 1934$-638$ & 5&.&83 \\
                  & 20/12/2003  & 1&.&4 & 9.4&$\pm$&0.9 &1934$-638$ &14&.&98 \\
                  & 20/12/2003  & 2&.&4 & 7.8&$\pm$&0.4 &1934$-638$ &11&.&59 \\
                  & 05/05/2004  & 17&.&8 & 1.8&$\pm$&0.15 & Mars & \multicolumn{3}{c}{--}\\
                  & 05/05/2004  & 24&.&5 & 1.5&$\pm$&0.35 & Mars & \multicolumn{3}{c}{--}\\
            ATCA/V191B\\
                  & 06/08/2008 & 2&.&3 & 7.5&$\pm$&0.11 &0637$-$752 &  5&.&32 \\
            LBA/V191B\\
                  & 06/08/2008  & 2&.&3 & 2.9&$\pm$&0.51 & -- & \multicolumn{3}{c}{--}\\
            ATCA/C1726\\
                  & 18/01/2009 & 4&.&8  & 5.6&$\pm$&0.3 & 1934$-638$ & 5&.&83\\
                  & 18/01/2009  & 8&.&6 & 2.9&$\pm$&0.3 & 1934$-638$ &2&.&86\\
        \hline\\[-10pt]
        \multicolumn{4}{l}{$^\dag$ \citet{benaglia2006}}
        \end{tabular}
        \label{tab:hd-atcafluxes}
    \end{center}
\end{table}

\section{Results}\label{sec:hd-results}

\subsection{Optical astrometry of Aa and Ab}

The data from the {\em HST}/FGS observations allow us to determine the positions and proper motions of the HD~93129Aa and Ab components. The absolute positions inferred for the epoch of 2008 August 6 (when the last LBA observation was conducted) are:
\begin{align*}
    \alpha_{\rm Aa},\ \delta_{\rm Aa} &= 10^{\rm h}43^{\rm m}57.455^{\rm s},\ -59^{\degree}32' 51.36''\\
    \alpha_{\rm Ab},\ \delta_{\rm Ab} &= 10^{\rm h}43^{\rm m}57.456^{\rm s},\ -59^{\degree}32' 51.33'',
\end{align*}
where the uncertainty in each coordinate is $\pm 27~\mathrm{mas}$, dominated by the PPMXL catalog uncertainties in the position and proper motion of the seven FGS reference stars surrounding HD~93129A. However, the relative position between Aa and Ab is well determined, leading to a separation of $36 \pm 1~\mathrm{mas}$ and $\mathrm{PA} = 12 \pm 1^{\degree}$. The proper motions derived from the FGS data are:
\begin{align*}
    \mu_\alpha^{\rm Aa} \approx -8.4~\mathrm{mas\ yr^{-1}}, &\qquad \mu_\delta^{\rm Aa} \approx 2.6~\mathrm{mas\ yr^{-1}}\\
    \mu_\alpha^{\rm Ab} \approx -9.0~\mathrm{mas\ yr^{-1}}, &\qquad \mu_\delta^{\rm Ab} \approx 0.0~\mathrm{mas\ yr^{-1}}.
\end{align*}

\subsection{Resolving the radio source}

\begin{figure}
    \centering
    \includegraphics[width=0.75\textwidth]{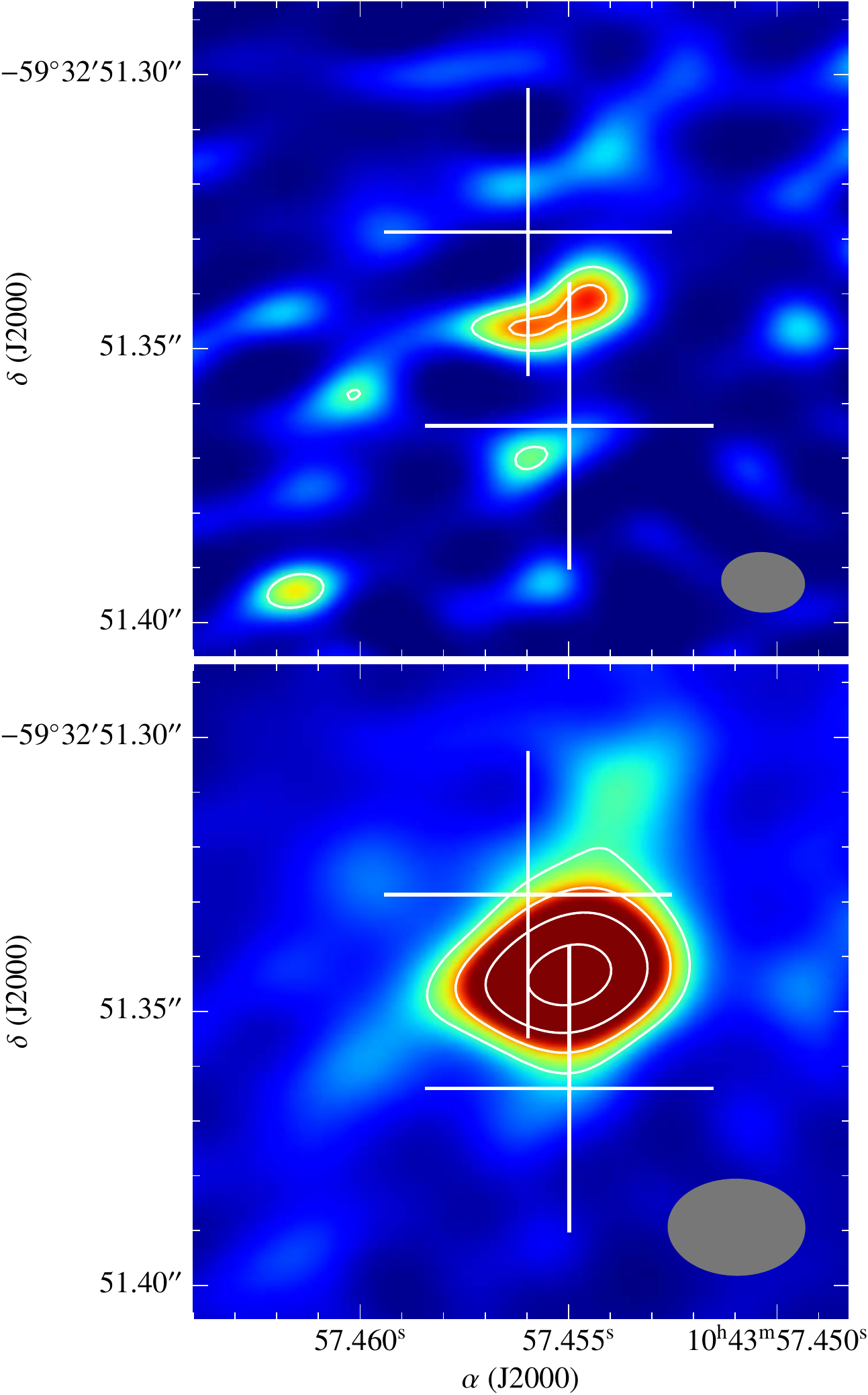}
    \caption[Images of HD~93129A obtained from the 2008 August 6 LBA observation at 2.3 GHz.]{Images of HD~93129A obtaiend from the 2008 August 6 LBA observation at 2.3 GHz with high angular resolution ({\em top}) and with a tappering to increase the weight of the shorter baselines ({\em bottom}). The rms are $0.21$ and $0.13~\mjybeam$, and the synthesized beams, shown on the bottom right corner of each image, are $15 \times 11~\mathrm{mas^2}$ with a PA of $85^{\degree}$, and $31 \times 23~\mathrm{mas^2}$ with a PA of $89^{\degree}$, respectively. Contours start at 3-$\sigma$ noise level and increase by factors of $2^{1/2}$. The crosses denote the absolute position of the Aa and Ab components of the system and the corresponding 1-$\sigma$ uncertainties at the epoch of the radio observation. We note that although the absolute positions of both components are not well constrained, the relative position between them exhibits an uncertainty of only 1~mas in distance and 1$^\degree$ in PA.}
    \label{fig:hd-lba-image}
\end{figure}
Figure~\ref{fig:hd-lba-image} shows the radio images obtained from the 2008 August 6 LBA observation at 2.3~GHz. In the high-resolution image we observe a bow-shaped radio emission with a flux density of $1.5 \pm 0.5~\mjy$. In the low-resolution one, the structure is poorly resolved, and we obtain a larger flux density value of $2.9 \pm 0.5~\mjy$, with a peak flux density of $1.8 \pm 0.2~\mjybeam$, assuming a two-dimensional Gaussian fit. The larger flux density value in this last image is expected as we are recovering part of the extended emission which is detected only with the shortest baselines.
The position of the radio emission lies between the astrometric positions of the stars derived in the previous section. In addition, the bow-shaped emission is extended in a direction perpendicular to the line joining the two stars (which is well constrained to be $\mathrm{PA} = 12 \pm 1^{\degree}$). A centroid fit from the observed radio emission in the high resolution image (see Figure~\ref{fig:hd-lba-image}, top) shows a position of $\alpha_{\rm C} = 10^{\rm h}43^{\rm m}57.5462^{\rm s}$, $\delta_{\rm C} = -59^\degree 32' 51.339''$. A systematic uncertainty of 1.1~mas in $\alpha$ and 2.8~mas in $\delta$, due to the position error of the phase reference calibrator in the second International Celestial Reference Frame (ICRF2), sets an absolute uncertainty in the position of the radio source of about 3~mas.

The phase calibrator was slightly resolved on the longest baselines, showing a total flux density of $0.99 \pm 0.02~\mathrm{Jy}$. This value can be compared with the simultaneous ATCA data, which shows a flux density for the same source of $1.300 \pm 0.002~\mathrm{Jy}$. This discrepancy can be explained either by emission that is filtered out in the LBA data due to the lack of short baselines, or by an incorrect amplitude scale of the LBA data (typically, the use of \tsys to calibrate the amplitudes can introduce an uncertainty of about 10\%).

\subsection{Radio spectrum from ATCA data}

\begin{figure}[t]
    \centering
    \includegraphics[width=10cm]{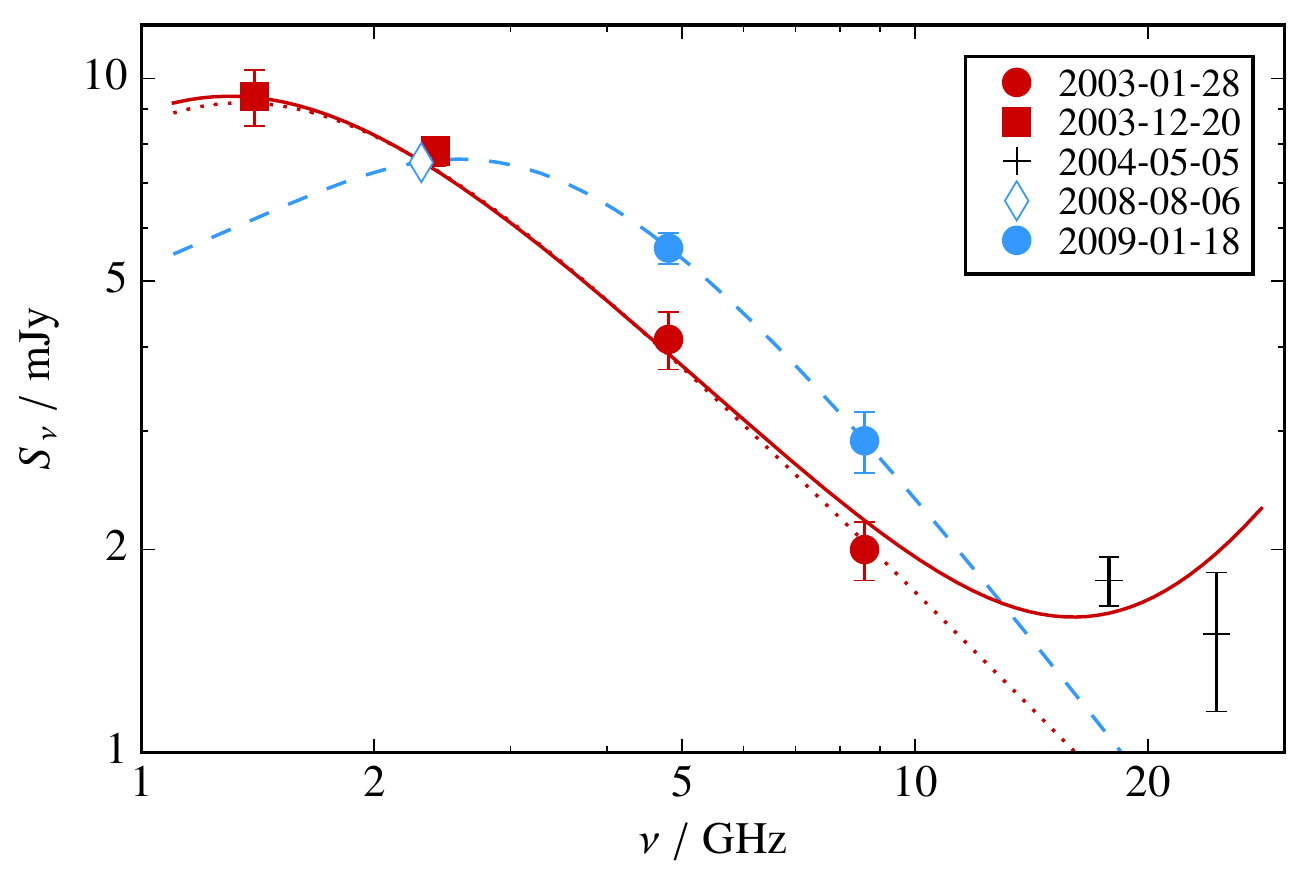}
    \caption[Spectra of HD~93129A obtained from the ATCA data presented in Table~\ref{tab:hd-atcafluxes} at different epochs.]{Spectra of HD~93129A obtained from the ATCA data presented in Table~\ref{tab:hd-atcafluxes} at different epochs. The data from 2003 have been fit with two different models: a FFA model (red dotted line), and a FFA plus a thermal component, considering also the 2004 data (red solid line). The 2008--2009 data have been fit with a FFA model (blue dashed line).}
    \label{fig:hd-atcaspectrum}
\end{figure}
The flux densities obtained from the ATCA data (Table~\ref{tab:hd-atcafluxes}) cover frequencies from 1.4 to 24~GHz and epochs between 2003 and 2009.
Simultaneous observations at 4.8 and 8.6~GHz were performed in January 2003 and in January 2009, which allow us to estimate the spectral index at the two epochs. Simultaneous observations at 1.4 and 2.4~GHz were also conducted in December 2003. Figure~\ref{fig:hd-atcaspectrum} shows the spectra derived from all the ATCA data. The non-thermal origin of the radio emission at frequencies below $\sim$10~GHz is clear due to the obtained negative spectral index. However, a thermal component is expected to be dominant at $\gtrsim 20~\mathrm{GHz}$.
Comparing the two epochs at which there are 4.8 and 8.6-GHz data, we observe an increase from 2003 to 2009 in the total flux density emission of HD~93129A.

\section{Discussion}\label{sec:hd-discussion}

\subsection{On the binary nature of HD~93129A}

One question that still needs to be answered is if HD~93129Aa and Ab form a gravitationally bounded binary system. The system is being monitored in the OWN Survey project \citep{barba2010}. Preliminary radial velocity curves show slight variations although no period has been determined yet.
In the simplest scenario, assuming that the two stars are in the plane of the sky, we estimate a separation of only 66~AU, which implies, together with the relative proper motions, a bounded system.
Based on the {\em HST}/FGS data, the separation between the two components between 1996 and 2009 has been decreasing at approximately $2.4~\mathrm{mas\ yr^{-1}}$, but VLTI data report a faster motion of about $4.2~\mathrm{mas\ yr^{-1}}$ between 2011 and 2013 \citep{sana2014}. This is consistent with a scenario where the two stars are approaching periastron.
Furthermore, the existence of a WCR coincident with the region between the two stars makes highly improbable that HD~93129A is not a bounded binary system.

\subsection{Orbit estimation}

The relative motion observed between the components Aa and Ab supports that HD~93129A is a gravitationally bound system. However, the current relative positions (see Figure~\ref{fig:hd-positions}) cover a small part of the orbit and thus are inadequate for performing a standard determination of orbital elements. No definitive radial velocity variation has been reported, which is not surprising given the wide separation and apparent near linear trajectory of the stars. Nevertheless, a first approximation of an orbital fit was attempted. An algorithm that minimizes the weighted square distance between the measured data points and the fit orbit (Casco \& Villa, private communication) was used to determine a preliminary set of orbital parameters. Our results suggest an orbital period of the order of $200~\mathrm{yr}$, an eccentricity larger than $0.9$, and a semimajor axes of about $37$ and $93~\mathrm{mas}$, for components Aa and Ab, respectively. The fit solution also points to a periapsis argument of about $220^{\degree}$ and an inclination of about $103^{\degree}$. Figure~\ref{fig:hd-orbit} shows the fit to the orbit. We note that this fit only provides a rough idea of the orbit. The large uncertainties in the positions and the small coverage of the orbit can produce extremely different orbital solutions. Actually, the available data is still roughly compatible with a straight trajectory.
In any case, this fit provides a preliminary basis to organize future dedicated observational campaigns, and we estimate that the next periastron passage would take place in 2024.
\begin{figure}[t]
    \centering
    \includegraphics[width=0.7\textwidth]{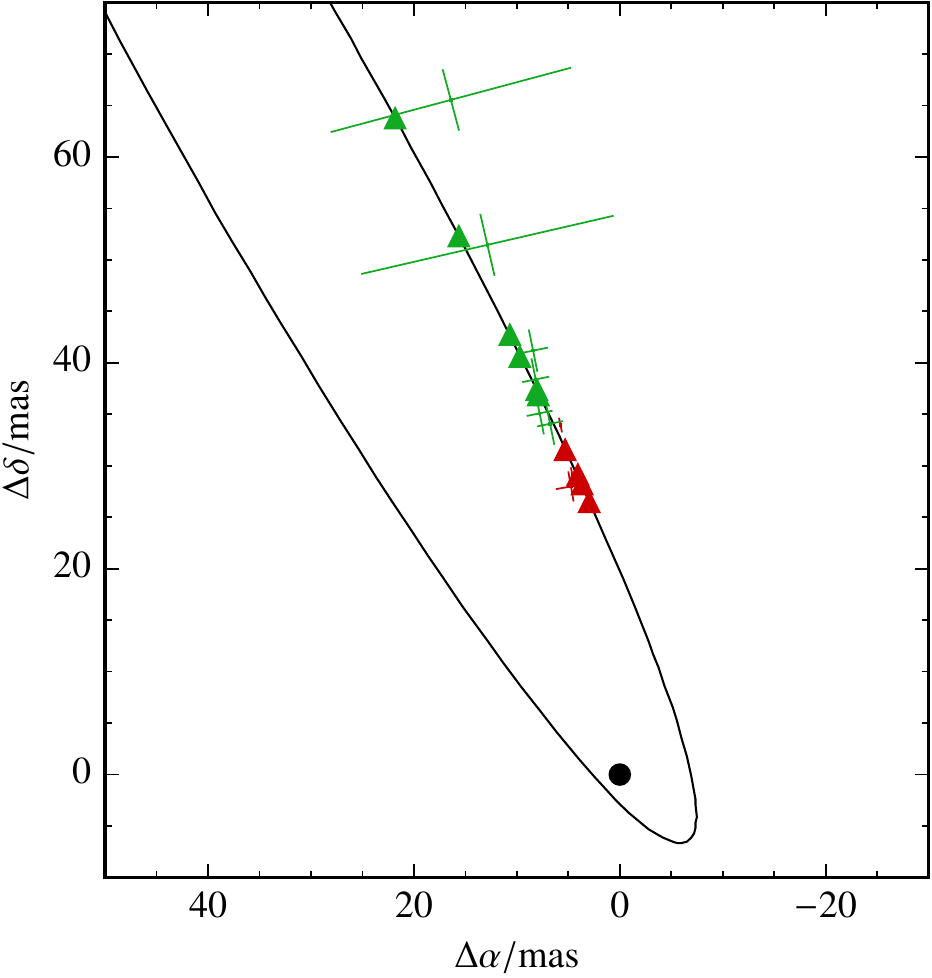}
    \caption[Preliminary fit for the orbit of HD~93129A using the data shown in Figure~\ref{fig:hd-positions}.]{Preliminary fit for the orbit of HD~93129A using the data shown in Figure~\ref{fig:hd-positions}. The triangles represent the derived position of Ab from the fit at the epoch of each observation. The black circle at the origin of coordinates represents the Aa star.}
    \label{fig:hd-orbit}
\end{figure}

\subsection{Estimating the wind-momentum rates}

The image obtained from the 2008 LBA data (Figure~\ref{fig:hd-lba-image}, top) shows a bow-shaped radio emission, slightly curved around the Ab component. This shape resembles the one observed in other CWBs, such as WR~140 \citep{dougherty2005}, WR~146 \citep{oconnor2005}, or Cyg~OB2~\#5 \citep{ortizleon2011}. Assuming that the radio source is centered on the stagnation point of the WCR, the wind momentum rates ratio, $\eta$, can be expressed as
\begin{equation}
    \eta = \left( \frac{R_{\rm b}}{R_{\rm a}} \right)^2 = \frac{\dot M_{\rm b} v_{\rm b}}{\dot M_{\rm a} v_{\rm a}}, \label{eq:eta-winds}
\end{equation}
where $R_{i}$ is the distance between the WCR and the $i$-star, $\dot M_i$ is the mass-loss rate, and $v_i$ is the wind velocity \citep{usov1992}. We note that $\eta$ is independent of the inclination of the orbit.

\begin{figure}[t]
    \centering
    \includegraphics[width=9cm]{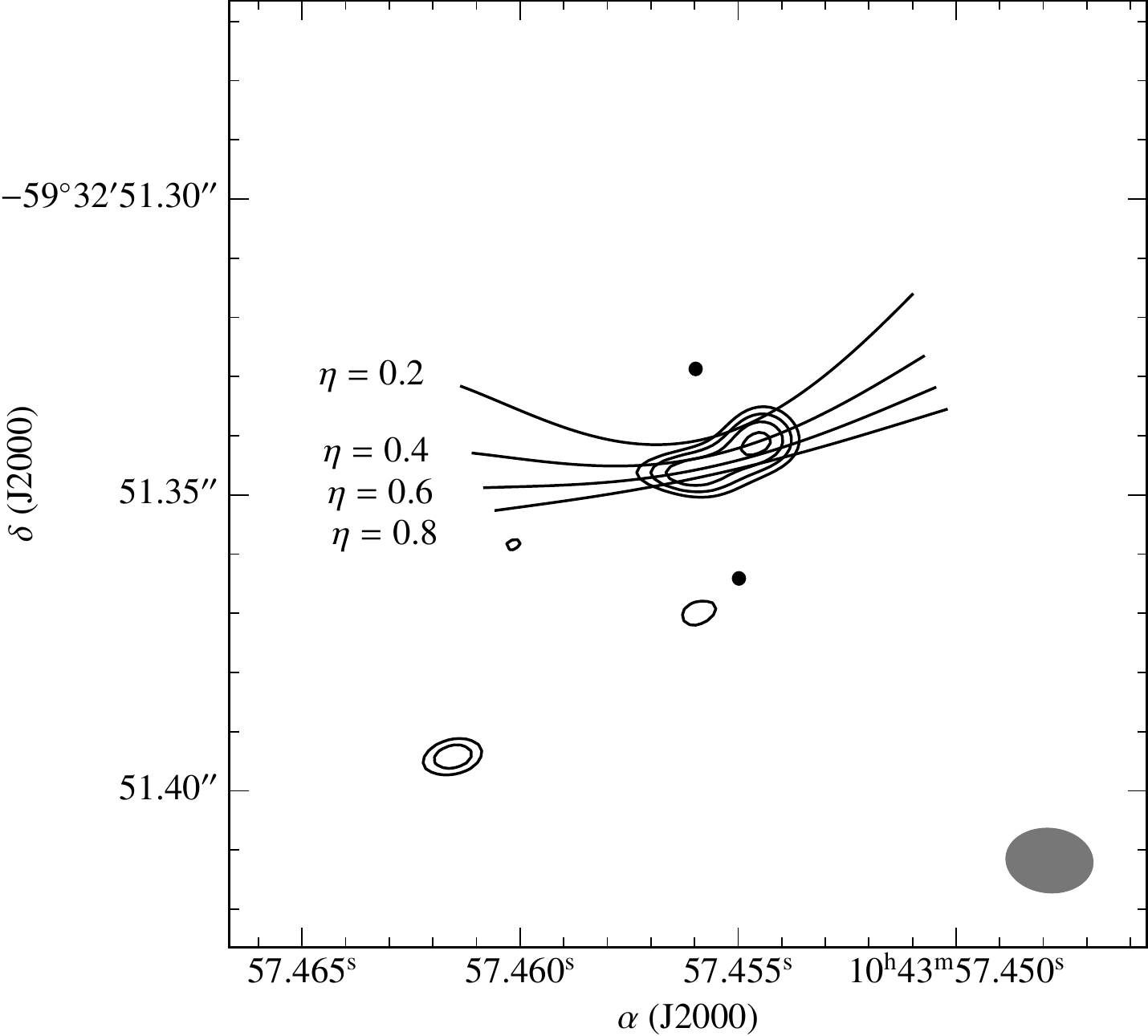}
    \caption[Same image as Figure~\ref{fig:hd-lba-image} (top), with the theoretical contact discontinuities for different values of $\eta$.]{Same image as Figure~\ref{fig:hd-lba-image} (top), with contours starting at 3$\sigma$ level and increase of $1~\mjybeam$. We show the theoretical contact discontinuities for different values of $\eta$. The black circles represent the positions of the Aa and Ab stars. We note that the relative position between the two stars is well constrained, although a global offset of up to $30$~mas could take place between these positions and the radio image. Values of $\eta \sim 0.5$ roughly match with the morphology of the observed emission.}
    \label{fig:hd-eta}
\end{figure}
We can use the observed shape of the WCR to estimate $\eta$. As \citet{pittard2006} showed for WR~140, the radio emission of the WCR is close to the stagnation point, and the opening angle has not reached the asymptotic value. They also noted the challenge of identifying the stagnation point relative to the stellar components due to opacity effects. In the best case, we can assume that the contact discontinuity (CD) between the two stellar winds follows the radio emission \citep{canto1996}. Figure~\ref{fig:hd-eta} represents the shape of the CD for different values of $\eta$ relative to the detected radio emission. The relative positions between the two stars are well constrained, so only a global offset with respect to the radio image could take place. We observe that, again, values of $\eta$ between $\sim$0.4 and $\sim$0.6 are the best match to the observed radio emission. We estimate thus $\eta \sim 0.5$. The stellar winds of the two stars are assumed to have reached their terminal velocities before colliding, considering the wide orbit, and we also assume that the terminal velocities of both components are similar. Therefore, we can observe from equation~(\ref{eq:eta-winds}) that $\eta$ reduces to a mass-loss rate ratio, leading $\dot M_{\rm b} \sim 0.5 \dot M_{\rm a}$.

\subsection{The mass-loss rates of HD~93129A}

The determination of the mass-loss rate, $\dot M$, can be conducted through a wide number of methods, each one with same advantages and/or limitations \citep[see e.g.][for a review]{puls2008}. The mass-loss rate of HD~93129A has already been estimated in different works. \citet{taresch1997} estimated it by fitting the H$\alpha$ profile but also using a more complete spectral synthesis approach in the ultraviolet, assuming that HD~93129A was an isolated star. Both approaches provide consistent values of about $2 \times 10^{-5}~\mathrm{M_{\sun}\ yr^{-1}}$. In case of the presence of two very similar components, these authors considered that these values should be reduced by a factor $2^{3/4} \approx 1.7$. Also using H$\alpha$ line diagnostics, \citet{repolust2004} derived a value of about $2.4 \times 10^{-5}~\mathrm{M_{\sun}\ yr^{-1}}$, noting that this result could suffer from contamination by a companion star.

\citet{benaglia2004} inferred the mass-loss rate through 4.8 and 8.6-GHz ATCA radio data. They derived a spectral index $\alpha = -1.2 \pm 0.3$, considering the presence of thermal and non-thermal emission. As shown by \citet{wright1975}, the mass-loss rate of a star with a thermal wind can be expressed in terms of the fraction of thermal emission that contributes to the radio flux density. Considering this fraction at a given frequency, $f_{\rm T}$, it is possible to express the mass-loss rate as
\begin{equation}
    \dot M = f_{\rm T}^{3/4} 7.2 \times 10^{-5}~\mathrm{M_{\sun} yr^{-1}},
\end{equation}
at 8.6~GHz and for a stellar distance of 2.5~kpc.

An estimate of the thermal emission from the two stellar winds can be obtained from the radio data presented by \citet{benaglia2006} from ATCA observations at 17.8 and 24.5~GHz. At these frequencies, non-thermal emission is negligible and can be disregarded to zeroth order. The total flux density is then the sum of the thermal emission from the two stellar winds and is characterized by a spectral index of $\alpha \approx +0.6$ \citep{wright1975}. Consequently, the thermal contribution to the flux at lower frequencies can be derived by extrapolation and assuming no variations between 2003 and 2004. This leads to a thermal contribution at 8.6~GHz of 1.2~mJy, compared with a total flux of $2.00 \pm 0.15~\mjy$ (see Table~\ref{tab:hd-atcafluxes}). Therefore, $f_{\rm T} \approx 0.6$ and thus $\dot M \approx 4.9 \times 10^{-5}~\mathrm{M_{\sun}\ yr^{-1}}$. Table~\ref{tab:hd-masslossrates} summarizes all the estimations of the mass-loss rate of HD~93129A performed to date.

\begin{table}
    \small
    \begin{center}
        \caption[Mass-loss rate values derived for HD~93129A from different methods.]{Mass-loss rate values derived for HD~93129A from different methods.}
        \label{tab:hd-masslossrates}
        \begin{tabular}{l@{~~~}l@{~~~}cr@{}c@{}l}
            \hline\\[-10pt]
            Reference  & Method & \multicolumn{4}{c}{$\dot M$}\\
            & & \multicolumn{4}{l}{($10^{-5}\ \mathrm{M_{\sun}\ yr^{-1}}$)}\\[+2pt]
              \hline\\[-10pt]
            \citet{taresch1997} & H$\alpha$ profile  & \ \hspace{.5cm}\ &1&.&8\\
            \citet{taresch1997} & Ultraviolet lines  & &2&.&08\\
            \citet{repolust2004}& H$\alpha$ profile  & &2&.&36\\
            \citet{benaglia2004}& 8.6~GHz radio flux & &7&.&2$f_{\rm T}^{3/4}$\\
            This Work & Separating T and NT fluxes & &4&.&9\\
            \hline
       \end{tabular}
 \end{center}
\end{table}
The main conclusion here is that the mass-loss rate determinations are challenging when dealing with a binary system. Considering the large number of binaries among O-type stars \citep[see e.g.][and references therein]{sana2014}, many observational determinations of $\dot M$ should certainly be viewed with caution, including the ones of HD~93129A.

\subsection{Flux density variability of HD~93129A}

Figure~\ref{fig:hd-atcaspectrum} shows an increase of $\sim$37\% in the flux density of HD~93129A between 2003 and 2009 at 4.8-GHz, and of $\sim$45\% at 8.6~GHz. These data indicate a significant increase in the flux density between the two epochs. Such a trend is indeed anticipated if the two stars are approaching the periastron passage (as it has been reported in other CWBs such as WR~140, \citealt{dougherty2005}). A synchrotron component that dominates over the thermal emission from the stellar winds of the two stars can explain the measured flux densities. We modeled the data with a power-law spectrum with a low-frequency cutoff produced by FFA. We firstly fit the data from 2003 and 2004, taking into account the thermal contribution at high frequencies. We assume that the flux density should not vary significantly within this interval of time, given the separation between Aa and Ab. From these data we infer the presence of a turnover at $\approx$1.4~GHz, and a spectral index of $\alpha = -1.03 \pm 0.09$.
The data from 2008 and 2009 have been fit with the same model of synchrotron emission plus FFA. No thermal component has been considered given that there are no contemporaneous data at high frequencies. In this case, we observe a higher emission at frequencies $\sim$5--10~GHz, the presence of a turnover at higher frequencies: $\sim$3~GHz, and a bit steeper spectrum, with $\alpha = -1.21 \pm 0.03$.

The observed drift of the turnover frequency to higher frequencies could also be indicative of a stronger and denser WCR. During the approach to the periastron passage, we expect an increase of material in the WCR, producing a stronger radio 0emission. However, this would also lead to more absorption at low frequencies.

\section{Summary and conclusions}\label{sec:hd-conclusions}

The LBA observation of the massive binary HD~93129A shows an extended and curved radio emission with a flux density of $\sim$3~mJy at 2.3~GHz. Following a detailed analysis of high-resolution {\em HST} and VLTI data, we provide compelling evidence that the radio emission is coincident with the expected position of a wind interaction region between components Aa and Ab, suggesting a wind-momentum ratio of $\sim$$0.5$.

Archival ATCA observations in the range of 1--25~GHz, reduced again in a uniform way, show that the flux density increased between 2003 and 2009, being both epochs well fit by a power-law spectrum with a negative spectral index. Similar increases in flux have been reported in other CWBs as the two stars approach to periastron. Therefore, the results presented in this work lend significant and additional support to the idea that WCRs are the sites where relativistic particles are accelerated in CWBs.

VLBI observations of CWBs specifically help to quantify the properties of the non-thermal radio emission. Such measurements are crucial for models aimed to reproduce the particle acceleration and non-thermal physics at work in these objects. Future VLBI observations of HD~93129A across a wide frequency range would allow us to determine more accurate properties of the derived spectrum, hence the properties of the relativistic electron populations involved in the synchrotron emission process, and potentially reveal associations between flux variations and changes in the properties of the WCR. Additional VLBI observations, specially during the putative periastron passage, would reveal the evolution of the WCR during the approach of the two stars. Significant changes in the morphology and emission are expected, as observed in other CWBs.
In parallel, new ATCA observations, preferably simultaneous across all the observing frequencies, are necessary to obtain accurate total flux densities. Repeated observations during the approach to the putative periastron passage would reveal whether the flux increases as observed in other CWBs.

Despite HD~93129A has not been reported to be a VHE \g-ray source, observations with the forthcoming Cherenkov Telescope Array (CTA), in particular during the periastron passage, will surely provide useful constraints to help understanding the high-energy phenomena in these objects.
However, a more accurate knowledge of the orbit is necessary to estimate the time of, and separation at, periastron to model its possible multiwavelength emission. A similar behavior to the one observed in $\upeta$-Carinae, with an enhanced emission around periastron, could be expected if the two stars are close enough.

\chapter{Searching for new gamma-ray binaries} \label{chap:new}
%
%
%
%

In this Chapter we present the results obtained for two sources that were postulated to be gamma-ray binaries, TYC~4051-1277-1 (Sect.~\ref{sec:tyc}) and MWC~656 (Sect.~\ref{sec:mwc}). We have conducted radio observations to unveil their natures and their possible connection to high-energy emission.

\section{The candidate TYC~4051-1277-1} \label{sec:tyc}

The reduced population of known gamma-ray binaries does not allow us to discriminate what kind of physical processes are really common to the whole population or which are related with particularities of each source.
The low number of sources that have been discovered up to now precludes to perform statistical studies.
For this reason, new searches for new gamma-ray binaries are quite important, although this is a difficult task. Given that this kind of sources displays emission along the full electromagnetic spectrum, a usual method to find new ones is the cross-identification of sources across public catalogs at different wavelengths. This kind of searches is not only helpful for the gamma-ray binaries, but for all kind of sources displaying multiwavelength emission. However, the comparison between different catalogs is not straightforward (e.g.\ one has to deal with extremely different resolutions and intrinsic problems at each wavelength). 

Whereas in the optical range there is a huge amount of catalogs with accurate stellar positions, spectral types, or spectra, at the other wavelengths the amount of data is not so large.
At radio wavelengths, there are few surveys with enough resolution and sensitivity, such as the NRAO VLA Sky Survey at 1.4~GHz (NVSS: \citealt{condon1998}), the Westerbork Northern Sky Survey at 330~MHz (WENSS: \citealt{rengelink1997}), and the TIFR GMRT Sky Survey at 150~MHz (TGSS\footnote{\url{http://tgss.ncra.tifr.res.in}}).
The LOFAR Multifrequency Snapshot Sky Survey at 60 and 150~MHz (MSSS: \citealt{heald2012}), which is currently being conducted, will be a new addition to these radio surveys in the coming years. At X-rays one can find surveys like the Second {\em ROSAT} All Sky Survey, \citep{boller2014} at soft X-ray energies, or the {\em XMM-Newton} Serendipitous Source Catalogue \citep[2XMMi-DR3;][]{watson2009}, observing at energies between 0.2 and 12~keV. At GeV there is the Third {\em Fermi}/LAT Source Catalog \citep[3FGL;][]{fermi2015}, whereas at TeV there is the H.E.S.S.\ Galactic Plane Survey \citep[HGPS;][]{carrigan2013}.

In this section we present the results of a cross-identification between the catalog of Luminous Stars in the Northern Milky Way (LS, \citealt{hardorp1959})\footnote{\url{http://cdsarc.u-strasbg.fr/viz-bin/Cat?III/76A}} and the WENSS and NVSS radio surveys. In this cross-identification we obtained two positive coincidences, and we conducted dedicated optical and radio observations to validate them. We will focus only in one of these two coincidences and in the radio data analyzed by the author of this thesis. A full discussion of these results has been published in {\colorexpandedcite Mart\'i, Luque-Escamilla, Casares, Marcote et al.}\ (\citeyear{marti2015}).

\subsection{Cross-identifying possible target sources} \label{sec:cross-identifying}

Two fundamental properties can be highlighted from the known gamma-ray binaries: all of them host an early-type massive star and they also exhibit non-thermal radio emission. Considering these two properties, we conducted a cross-i\-den\-ti\-fi\-ca\-tion between known luminous early-type stars (mainly from the LS catalog) and radio sources detected in WENSS and NVSS.
We compared the positions of sources detected in the three catalogs, considering as possible coincidences any separation below $2''$. This value was chosen due to the existing astrometric uncertainties in the three catalogs.
Additional stellar catalogs with spectral class information, such as the one compiled by \citet{voroshilov1985}, were also used. 
Early-type stars (the ones seen in gamma-ray binaries), and sources with non-thermal radio emission were only considered as possible candidates. The estimation of the spectral index, to determine its possible non-thermal origin, was performed by comparing the two radio catalogs. Only sources detected in both, WENSS and NVSS, were thus considered. Most of the radio sources detected in WENSS were also detected in the NVSS, but not viceversa. Therefore, our identification is mainly limited by the WENSS survey.

Given the limited resolution of the surveys, during this type of cross-i\-den\-ti\-fi\-ca\-tion we clearly get a probability to obtain chance coincidences. To estimate this probability we carried out simulations using Monte Carlo techniques with simulated WENSS populations. For each simulation we counted the chance coincidences between the LS stars and the simulated WENSS population within a few arc-seconds. We finally conducted $10^5$ simulations, obtaining about 20 coincidences, and thus a probability of chance coincidence for a given LS star of $\sim$$2 \times 10^{-4}$. This value implies that we would expect $\sim$1 chance coincidence, on average, in our comparison.

Two coincidences were found in this cross-identification: the star TYC~4051-1277-1, which was compatible with the radio sources WNB~0231.5+6222 and NVSS~J023529\-+623520 (see Figure~\ref{fig:tyc-possii-nvss}); and the star TYC~3594-2269-1, compatible with the radio sources WNB~2130.2+4736 and NVSS~J213203+474948. In both cases the positions were compatible within 1-$\sigma$ of the astrometric error and the (non-simultaneous) radio spectra exhibit spectral indexes between $\simeq$$-0.6$ and $-0.9$, and thus show a clear non-thermal origin. In addition, TYC~4051-1277-1 is located close to an unassociated Fermi gamma-ray source 2FGL~J0233.9+6238c.
\begin{figure}[t]
\begin{center}
	\includegraphics[width=0.85\textwidth]{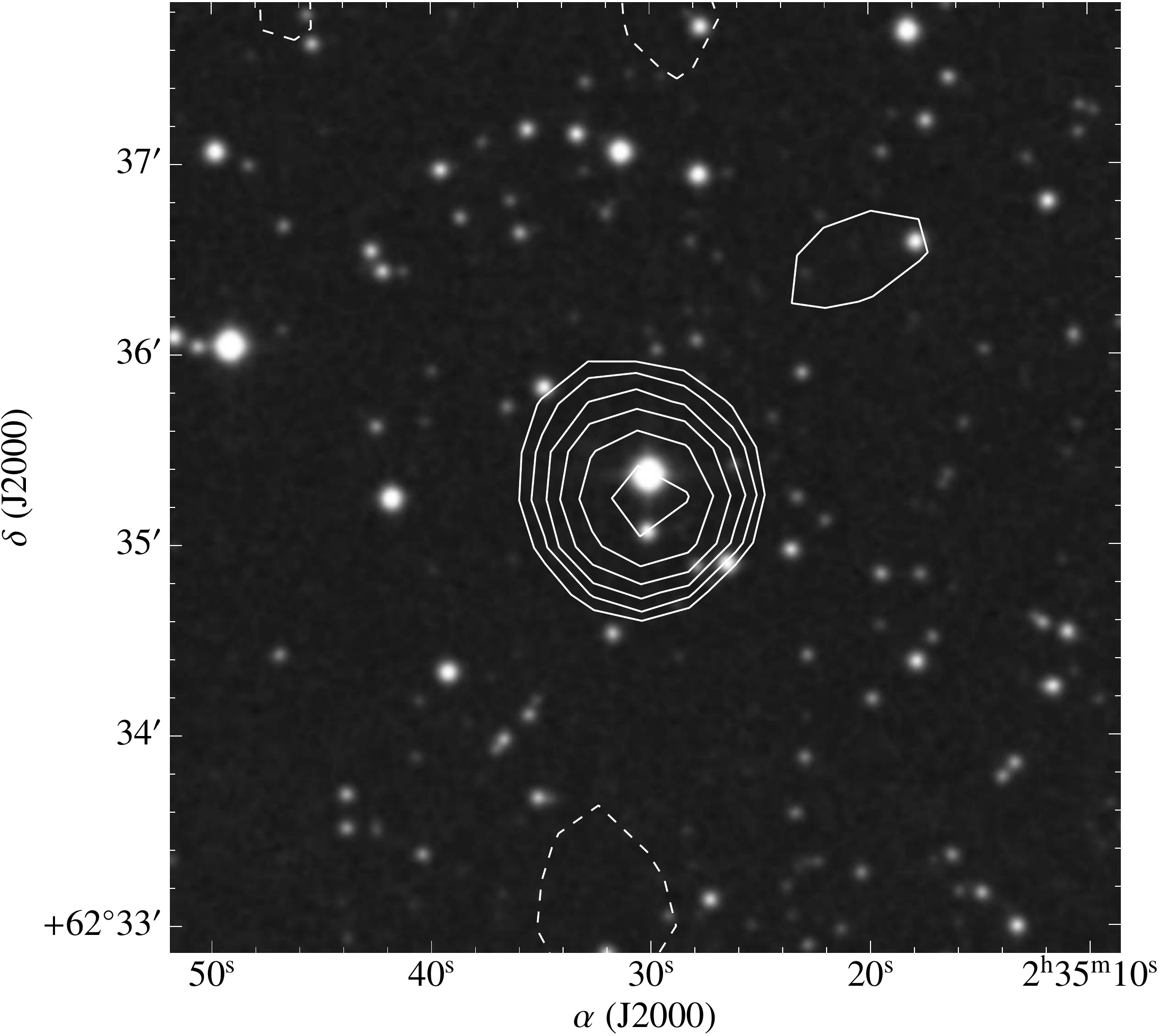}
	\caption[Field of TYC~4051-1277-1 as seen by the Second Palomar Observatory Sky Survey (POSS~II) in the $F$ band (red).]{Field of TYC~4051-1277-1 as seen by the Second Palomar Observatory Sky Survey (POSS~II: \citealt{reid1991}) in the F band (red). The radio map from the NVSS is superimposed in contours. TYC~4051-1277-1 is detected as the brightest star in the center of the image, with a magnitude $\sim$11. The NVSS~J023529+623520 radio source is almost centered at the same position, with a flux density of $\sim$$6.8~\mathrm{mJy}$.}
	\label{fig:tyc-possii-nvss}
\end{center}
\end{figure}

Taking into account the information from the two previous paragraphs, the matches found are not statistically significant, and they could just represent chance coincidences. Despite of that, we started a detailed exploration of these systems to clarify if they were real coincidences or not. In the following, we will only focus on the TYC~4051-1277-1 association, the only one that the author of this thesis has explored at radio frequencies.

\subsection{Optical observations}

A spectroscopic optical observation of TYC~4051-1277-1 was conducted with the 4.2-m William Herschel Telescope (WHT) at the Roque de los Muchachos Observatory, in La Palma (Spain), on 2012 October 25 during 1\,800~s. The obtained spectrum shows dominant absorption lines corresponding to the Balmer series. A classification of the spectrum leaded in a spectral type of B9~III for the star (which had been previously classified as B9~V).

Optical photometric observations were also conducted to obtain the light-curve of the source. These observations were performed by the robotic 0.5-m telescope Fabra-ROA Montsec (TFRM, \citealt{fors2013}) at the Observatori Astron\`omic del Motsec, in Lleida (Spain). The observations were taken along 81 nights between 2012 July 31 and 2014 February 5. The source was observed around 20 times per night with short exposure times of 25--45~s.
From these data we detect variability at 2-$\sigma$ confidence level, with a possible period of about 6.08~d which was not statistically significant. Therefore, the hints of periodicity would indicate a binary nature for this source.

\subsection{Radio observations and results}

Although the radio source is detected in both, WENSS and VLSS, dedicated observations with radio interferometers would provide higher resolution and a more accurate astrometric position.


Service radio observations with WSRT were conducted on 2012 November 5 to explore the possible association of TYC~4051-1277-1 with NVSS~J023529 +623520 (project code S12B003). The source was observed at 2.3 and 4.9~GHz during $10.5~\mathrm{h}$ with 8 IFs of 128 channels each, with a total bandwidth of 128~MHz. Runs of 15~min at each frequency were interleaved to guarantee the optimal $uv$-coverage for both frequencies. We used 3C~48 and 3C~147 as amplitude calibrators at the beginning and at the end of the observation, respectively. Given that a phase calibrator was not used during the observation, we could not perform precise astrometric measurements in the obtained images.

\begin{figure}
\begin{center}
	\vspace{-8pt}
	\includegraphics[width=0.77\textwidth]{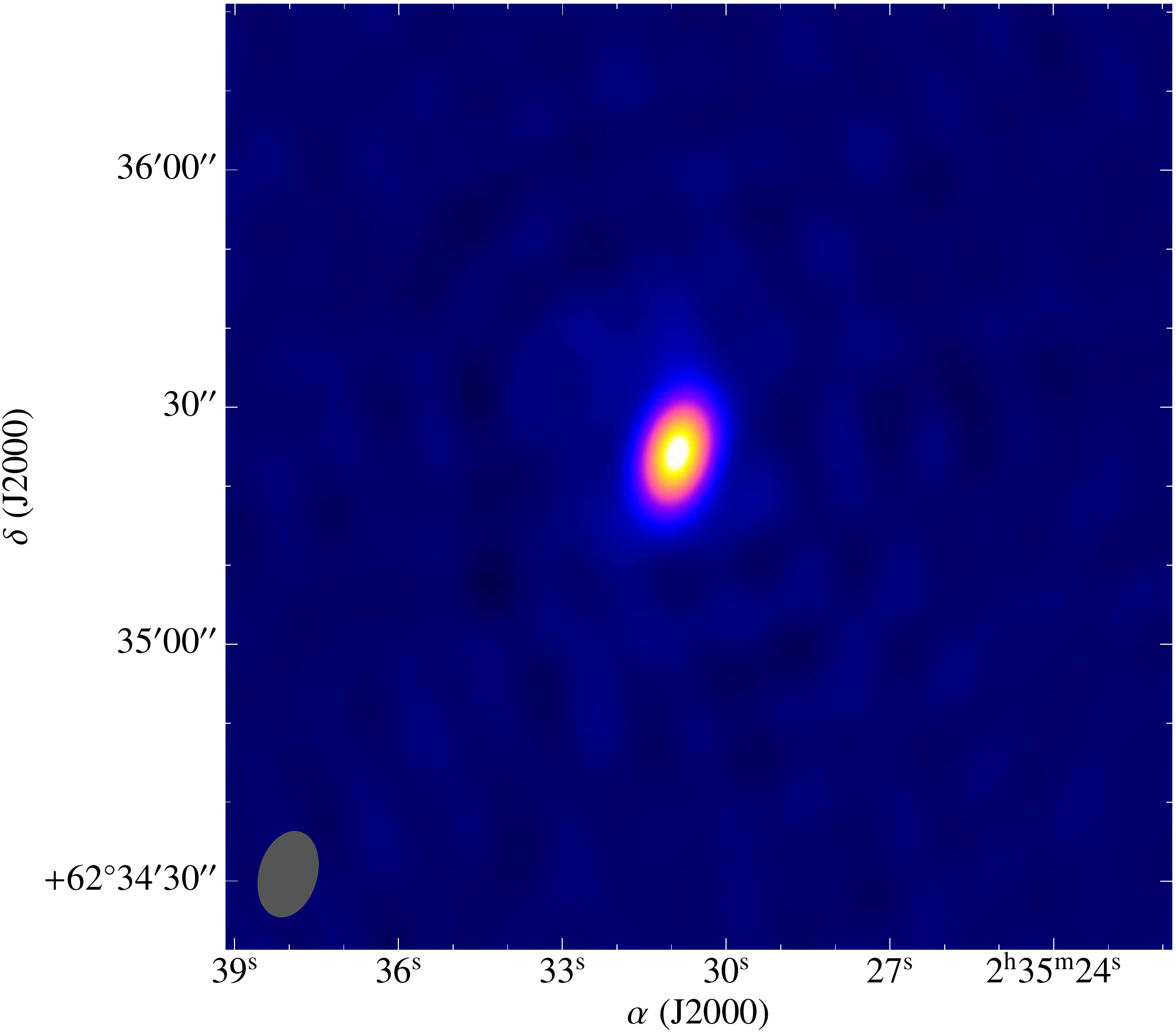}\\[+5pt]
	\includegraphics[width=0.77\textwidth]{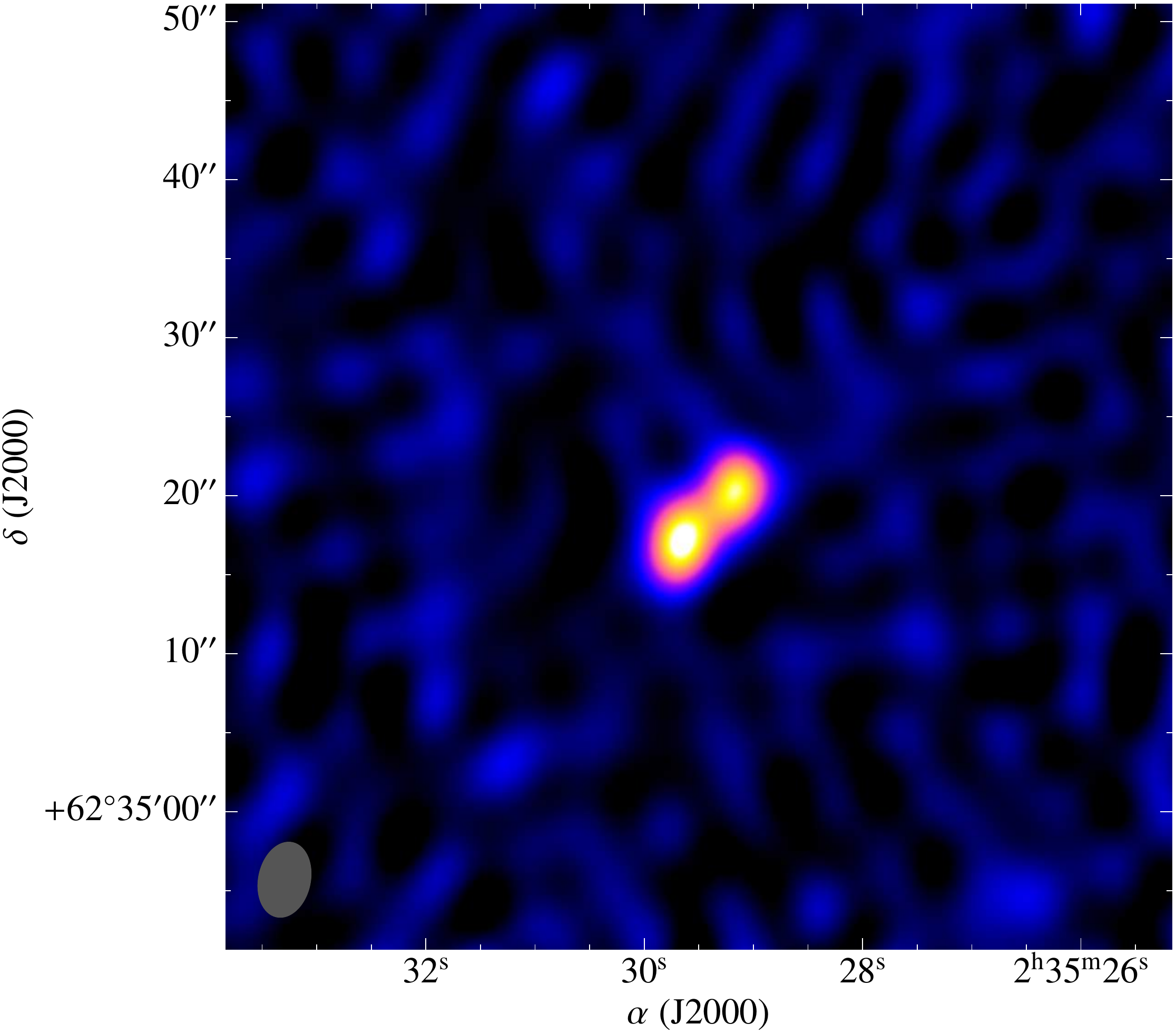}\\[-6pt]
	\caption[Field of TYC~4051-1277-1 observed with WSRT at 2.3~GHz (top) and 4.9~GHz (bottom).]{Field of TYC~4051-1277-1 observed with WSRT at 2.3~GHz (top) and 4.9~GHz (bottom). At 2.3~GHz we detect a point-like source with a flux density of $6.23 \pm 0.13~\mathrm{mJy}$. However, at 4.9~GHz the source is resolved in a double-lobed shape with a total flux density of $3.19 \pm 0.09~\mathrm{mJy}$. The synthesized beams are shown on the bottom left of each image. Astrometry is not accurate due to absence of phase calibrator during the observation, showing a large deviation between both images. For this reason we do not show the position of TYC~4051-1277-1. We note that the two images show different scales.}
	\label{fig:tyc-wsrt}
\end{center}
\end{figure}
Figure~\ref{fig:tyc-wsrt} shows the obtained images at 2.3~GHz and 4.9~GHz. The 2.3-GHz image presents a synthesized beam of $10.7 \times 7.2~\mathrm{arcsec^{2}}$ with a PA of $-16^{\degree}$, and a rms of $100~\ujybeam$. We detect a point-like source with a flux density of $6.23 \pm 0.13~\mathrm{mJy}$.
The 4.9-GHz image presents a synthesized beam of $4.8 \times 3.2~\mathrm{arcsec^{2}}$ with a PA of $-12^{\degree}$, and a rms of $80~\ujybeam$. In this case, we resolve the source into a double-lobed shape with a total flux density of $3.19 \pm 0.09~\mathrm{mJy}$. Assuming two Gaussian components, the individual flux densities for each lobe are $1.61 \pm 0.08~\mathrm{mJy}$ and $1.63 \pm 0.08~\mathrm{mJy}$, for the northwest and the southeast component, respectively. From the total flux density detected at both frequencies, we estimate a spectral index of $\alpha \approx -0.88 \pm 0.17$, typical of non-thermal synchrotron emission.

One short 30-min VLA observation was also conducted on 2013 October 20 to provide an accurate astrometry for the radio source (project code 13B-032). The observation was performed with the B configuration of the array at 5.5~GHz. The new WIDAR correlator was used, observing with 16 IFs of 64 channels each, with a total bandwidth of 2~GHz. 3C~48 was used as amplitude and bandpass calibrator, and J0228+671 as phase calibrator. The source was observed only during 5~min, but given the large bandwidth of the new VLA setup, we could reach enough sensitivity for our purposes. The obtained synthesized beam for this observation was $1.10 \times 0.82~\mathrm{arcsec^2}$ with a position angle of $-28.7^{\,\degree}$.

\begin{table}[t]
	\small
	\begin{center}
	\caption[Flux density values measured for the radio source NVSS J023529 +623520 from the WSRT and VLA data.]{Flux density values measured for the radio source NVSS~J023529+623520 from the WSRT and VLA data. At 2.2~GHz we did not resolved the source, whereas at 4.9/5.5~GHz we separate it into two components (the northwest one, NW, and the southeast one, SE). Two-dimensional gaussians were fitted to each component to measure the flux density values. In the case of the 2.2-GHz data and the measurement of total flux density at 4.9~GHz we have used the {\tt imstat} task of CASA (see \S\,\ref{sec:reduction-measuringfluxes}).}
	\begin{tabular}{lccc}
		\hline\\[-10pt]
		Facility & $\nu /\ \mathrm{GHz}$ & Component &   $S_{\nu} /\ \mjy$\\[+2pt]
		\hline\\[-10pt]
		WSRT & 2.2 & --   & $6.23 \pm 0.13$\\
		WSRT & 4.9 & total & $ 3.19\pm 0.09$\\
		WSRT & 4.9 & NW & $ 1.61\pm 0.08$\\
		VLA 	   & 5.5 & NW & $1.06 \pm 0.05$\\
		WSRT & 4.9 & SE  & $ 1.63\pm 0.08$\\
		VLA 	   & 5.5 & SE  & $1.27 \pm 0.04$\\
		\hline
	\end{tabular}
	\label{tab:tyc-data}
	\end{center}
\end{table}
The VLA data\footnote{We note that the VLA data was neither reduced nor analyzed by the author of this thesis. However, the discussion presented in this work is supported by both, the WSRT and the VLA data.} resolved again the source into a double-lobed shape, with flux densities of $1.06 \pm 0.05~\mjy$ and $1.27 \pm 0.04~\mjy$ (for the northwest and the southeast component, respectively). Table~\ref{tab:tyc-data} summarizes the flux density values obtained from all the mentioned radio data (from the WSRT and VLA observations). We could also infer an accurate astrometry of the two lobes, obtaining the following positions:
\begin{align*}
	{\rm NW:} &\quad 02^{\rm h} 35^{\rm m} 29.061 \pm 0.003^{\rm s},\quad +62^\degree 35' 20.19 \pm 0.02''\\
	{\rm SE:} &\quad 02^{\rm h} 35^{\rm m} 29.634 \pm 0.001^{\rm s},\quad +62^\degree 35' 16.53 \pm 0.02''
\end{align*}
for the northwest (NW) and the southeast (SE) components. These positions are $\sim$5~arcsec away with respect to the known (optical) position of TYC~4051-1277-1, and thus, the radio source is not compatible with the position of the star (see Figure~\ref{fig:tyc-vla}).
\begin{figure}[!t]
\begin{center}
	\includegraphics[width=.8\textwidth]{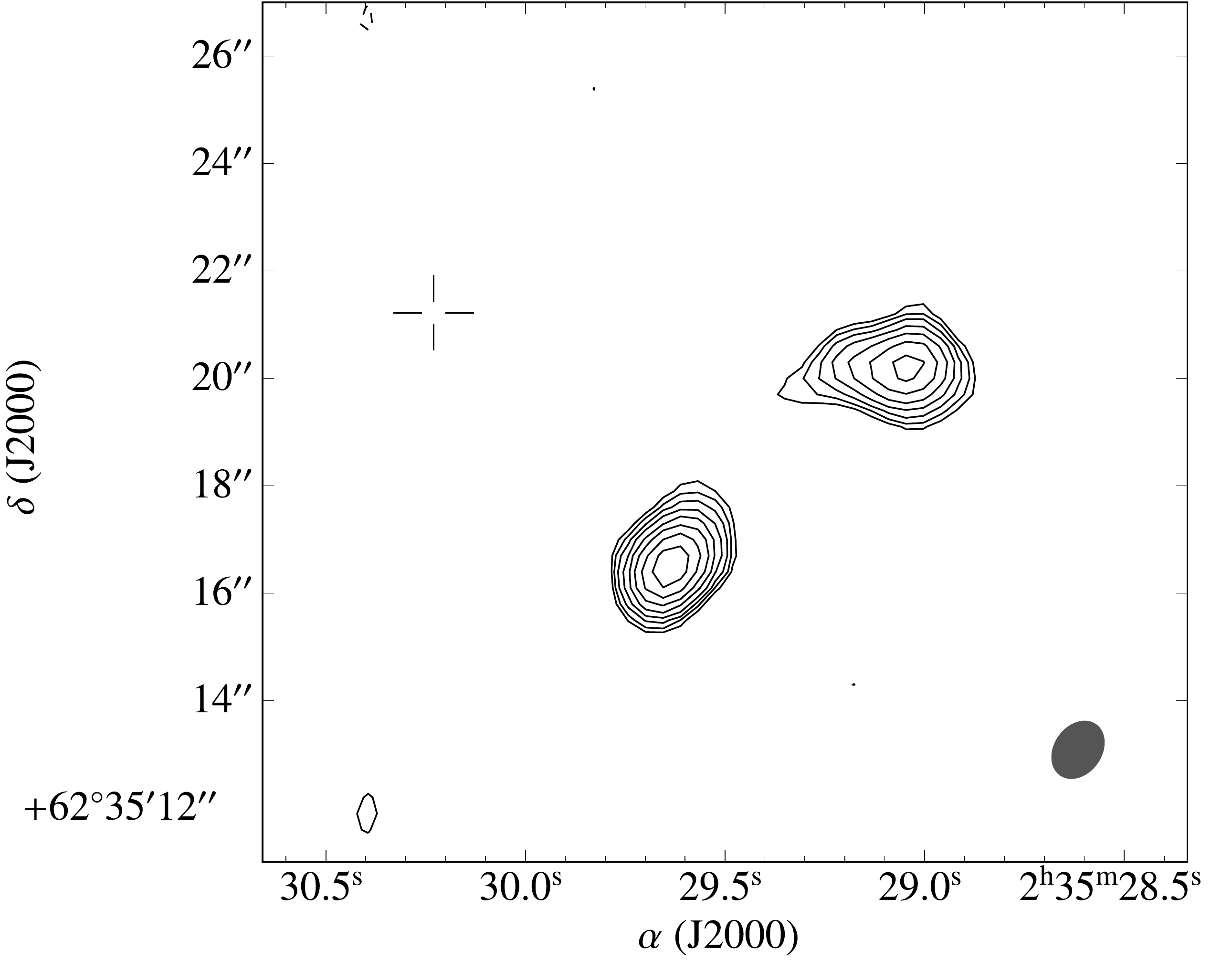}
	\caption[Field of TYC~4051-1277-1 observed with the VLA at 5.5~GHz.]{Field of TYC~4051-1277-1 observed with the VLA at 5.5~GHz. The rms of the image is $20~\ujybeam$, and the contours start at 3-$\sigma$ noise level and increase by factors of $2^{1/2}$. The synthesized beam, shown on the bottom right of the image, is $1.11 \times 0.82~\mathrm{arcsec^2}$ with a PA of $-29^{\degree}$. The cross represents the optical position of the TYC~4051-1277-1 star. Adapted from \citet{marti2015}.}
	\label{fig:tyc-vla}
\end{center}
\end{figure}

\subsection{Discussion}

The images obtained by the WSRT and the VLA data unambiguously show that the radio source NVSS~J023529+623520 (or WNB~0231.5+6222) is not associated with the TYC~4051-1277-1 star. We have obtained an upper-limit of $0.06~\mjy$ for the radio emission of the star, provided by the VLA observation. The fact that the source exhibits a two-lobed shape, together with the non-thermal spectral index of $\alpha \approx -0.88 \pm 0.17$, is a strong indicatiion that the source is likely a radio galaxy with bent jets (see Figure~\ref{fig:tyc-vla}). A comparison with the field observed by the Two Micron All Sky Survey \citep[2MASS;][]{skrutskie2006} in search of a point-like infrared core in between the two lobes provided null results.

We observe a significant discrepancy between the flux densities reported from the WSRT and the VLA data for the two components of the radio source, with lower values in the VLA data. Although there is a slight difference between the observed central frequencies for each array (4.9 and 5.5~GHz, respectively), we would estimate a flux density of $1.45 \pm 0.08~\mjy$ for both components at 5.5~GHz, based on the obtained WSRT flux density values and the derived spectral index. We obtain a significant deviation between the predicted value and the one obtained from the VLA image for the NW component (at more than 3-$\sigma$).
This implies that the source could present extended emission which is filtered out with the longer VLA baselines, but still detected with the WSRT ones.

The possible variability and photometric period of $\sim$6~d observed in TYC~4051-1277-1 from the optical data is not statistically significant and more data are required to confirm this behavior. In any case, this variability could also be associated with a slow low-amplitude pulsation that is common in B-type stars.

\subsection{Conclusions}

We have presented an approach to identify new early-type luminous stars displaying non-thermal emission, which is typical of the known gamma-ray binaries, by cross-identifying catalogs of sources detected in optical and radio. The negative results obtained here for the possible association of the TYC~4051-1277-1 star with the radio source NVSS~J023529+623520 are expected from a statistical point of view, as discussed in \S\,\ref{sec:cross-identifying}. We note that we also obtained negative results for the other possible association discussed at the beginning of the section \citep[see][]{marti2015}. However, this detailed study was required to unambiguously reject or accept the obtained possible associations.

Despite the null results, we have obtained accurate positions for the non-thermal radio emission and the possible extragalactic origin of the radio source NVSS\ J023529\-+623520. We have not detected radio emission associated with TYC~4051-1277-1, but we have provided an improved spectral classification of the star. A possible, but not significant, optical periodic variability has been reported for TYC~4051-1277-1, and a more extended optical campaign is required to clarify this point and the origin of this variability, in case of being confirmed.

The cross-identification of sources between different catalogs is thus a powerful tool to discover new sources that display emission at different wavelengths. However, all the obtained associations must be considered with caution. Catalogs with higher resolution will allow us to reduce the probabilities of change coincidences.

%
%
%
%
%

\section{MWC~656, the first Be/BH system} \label{sec:mwc}

The Be star MWC~656 became an interesting source to the high-energy community when it was proposed to be the possible counterpart of a \g-ray flare detected by {\em AGILE} from an unknown source. After discovering its binary nature, MWC~656 became a new gamma-ray binary candidate. Subsequent multiwavelength observations revealed the presence of a BH as a compact object, and MWC 656 was confirmed as the first known system to host a Be star and a BH.
In this section we present the work done to discover the possible radio counterpart of MWC~656 and the results obtained.

\subsection{Chronicle of a discovery}

The \g-ray satellite {\em AGILE} detected on 2010 July 25--26 a point-like source flaring with a flux of $15 \times 10^{-7}~\mathrm{cm^{-2}\ s^{-1}}$ above 100~MeV, leading to a significance above 5$\sigma$ \citep{lucarelli2010}. This source was designed as AGL~J2241+4454, with coordinates $\alpha = 22^{\rm h} 41^{\rm m}$, $\delta = 44^{\degree} 50'$ (or Galactic coordinates of $l = 100.0^{\degree}$, $b = -12.2^{\degree}$). This position exhibits a large uncertainty of about $\pm 0.6^{\degree}$.
Despite the {\em AGILE} significance, {\em Fermi}/LAT could not confirm the detection. From simultaneous data they inferred a conservative upper-limit on the integral flux above 100~MeV of $3.0 \times 10^{-7}~\mathrm{cm^{-2}\ s^{-1}}$ at 95\% confidence level\footnote{\url{http://fermisky.blogspot.com.es/2010/07/}}.

\citet{williams2010} suggested two possible counterparts compatible with the position of the gamma-ray detection. On one hand, the probable quasar RX\ J2243.1+4441, discovered by its X-ray emission but with an unidentified optical counterpart \citep{brinkmann1997}. On the other hand, the Be star HD~215227 ($\alpha = 22^{\rm h} 42^{\rm m} 57^{\rm s}$, $\delta = 44^{\degree} 43' 18''$, also known as MWC~656, as we will refer to the system in the following). \citet{williams2010} reported a period of $60.37 \pm 0.04~\mathrm{d}$ through optical photometric observations, suggesting the presence of a companion orbiting this star. They determined that the circumstellar disk contributes with $\approx$$33\%$ of the total optical flux emission (without specifying the wavelength range), and the observed changes in the optical emission could possibly be related with changes in the disk produced by the passage of the companion.
From spectral observations these authors also inferred changes in the H$\upgamma$ emission line on day timescales. These changes are unexpectedly fast for typical Be stars, suggesting again perturbations of the circumstellar disk probably produced by the presence of the companion.

This system presented similarities with the gamma-ray binary LS~I~+61~303: both display a relatively similar period ($\sim$$60~\mathrm{d}$ versus $\sim$$26~\mathrm{d}$), both host a Be star, and the changes observed in the optical photometry and spectral lines in MWC~656 resemble somehow the ones observed in LS~I~+61~303. Therefore, the high-energy community became quickly interested in this source to unveil if it was actually a binary, or if it could be associated with the {\em AGILE} detection, and thus be a new addition to the reduced population of gamma-ray binaries.

\subsection{Studying its multiwavelength emission}

Several multiwavelength campaigns were focused on the MWC~656 star and its possible association with the {\em AGILE} detection. Optical observations were conducted to obtain the orbital parameters of the binary system and the evolution of the circumstellar disk. X-ray observations with {\em XMM-Newton} and TeV observations with MAGIC were conducted to determine the possible high energy emission of this system. Radio observations were also conducted to discover the radio counterpart of the system and study its radio emission in case of detection. Although we have been working only in this last part (the radio data), we will discuss first the observations conducted at different wavelengths and their results, given that a multiwavelength knowledge is mandatory to fully understand the nature of these kinds of sources.

\subsubsection{Optical observations}

\citet{casares2012} confirmed the binary nature of MWC~656 from H$\alpha$ emission line observations. They inferred an eccentricity of 0.4 for the orbit by assuming the previously reported orbital period of $60.37 \pm 0.04~\mathrm{d}$ \citep{williams2010}, and provided a rough estimation of the mass of the companion of $\approx$$2.7$--$5.5~\mathrm{M_{\sun}}$, which is compatible with a neutron star or a black hole.

An ongoing optical campaign with the TFRM is being carried out since 2012 May 12 to obtain an accurate light-curve of MWC~656. Preliminary results for four months (two orbital cycles) were published by \citet{paredesfortuny2012}, showing a sinusoidal light-curve with the maximum at orbital phase 0 and a semiamplitude of $0.024 \pm 0.001~\mathrm{mag}$ when folded with the 61.37-d period, compatible with the results reported by \citet{williams2010}. We note that the origin of the orbital phase is defined arbitrarily at $\mathrm{MJD_0} = 53242.8$, coinciding with the maximum of the optical emission \citep{williams2010}.

Later on, \citet{casares2014} conducted additional detailed optical observations, improving the radial velocity light-curve of the Be star and refining its classification, resulting in a B1.5--B2~III star. These results constrained the mass of the compact object to $3.8$--$6.9~\mathrm{M_{\sun}}$, and thus unambiguously indicating the presence of a black hole. A distance of $2.6 \pm 0.6~\mathrm{kpc}$ was also derived for the system. The Fe\,{\sc ii} and He\,{\sc ii} lines in the optical spectrum of the source indicate the presence of a decretion disk around the main star (the circumstellar disk) but also an accretion disk around the BH.

\subsubsection{X-ray observations}

X-ray observations were conducted with {\em XMM-Newton}, showing a faint X-ray source compatible with the optical position of MWC~656 at 2.4$\sigma$ in the $0.3$--$5.5~\mathrm{keV}$ band \citep{munar2014}.
The obtained spectrum reveals the presence of two components: an absorbed thermal black body component, compatible with the one expected from isolated Be stars, and a non-thermal power-law component that dominates above 0.8~keV.
The inferred total X-ray luminosity is $L_{\rm X} = (3.7 \pm 1.7) \times 10^{31}\ \mathrm{erg\ s^{-1}}$, being the thermal component $L_{\rm bb} = (2.1_{-1.5}^{+2.8}) \times 10^{31}\ \mathrm{erg\ s^{-1}}$ and the non-thermal component $L_{\rm pow} = (1.6_{-0.9}^{+1.0}) \times 10^{31}\ \mathrm{erg\ s^{-1}}$. This non-thermal component implies a luminosity of $(3 \pm 2) \times 10^{-8}~{L_{\rm edd}}$, where $L_{\rm edd}$ is the {\em Eddington luminosity}, being compatible with a BH in quiescence state.

\subsubsection{High Energy and Very High Energy \g-ray observations}

{\em Fermi}/LAT could not confirm the mentioned {\em AGILE} detection, and even with an analysis of 3.5-yr of cumulative observations, {\em Fermi}/LAT reports an upper-limit of $9.4 \times 10^{-10}\ \mathrm{cm^{-2}\, s^{-1}}$ at 90\% confidence level \citep{mori2013}.
However, a detailed analysis reveals that most of the {\em AGILE} photons were received in a time slot not covered by {\em Fermi}. The first {\em Fermi} scan available after this flare, about 1-h later, also shows a peak in the emission, with a value just below the 3-$\sigma$ significance limit \citep{alexander2015}. 
Recently, nine more flaring events, registered by {\em AGILE} between 2007 and 2014, have been reported by \citet{munar2015}.
Unfortunately, almost all these flares took place at zenith distances above $50^{\degree}$ as seen by {\em Fermi}. At these high zenith distances, the effective area of the {\em Fermi} satellite decreases significantly, which might explain the absence of detections from this satellite.

The MAGIC telescopes have also observed the source to constrain its VHE emission \citep{aleksic2015mwc656}. Two observations were performed on May--June 2012 and June 2013 (just after the periastron passage).
The June 2013 observation was part of a multiwavelength campaign together with the {\em XMM-Newton} observation discussed in the previous subsection and published by \citet{munar2014}.
Neither the cumulative data nor the daily analysis report a significant emission of MWC~656, setting an upper-limit for the flux above 300~GeV of $2 \times 10^{-12}~\mathrm{cm^{-2}\ s^{-1}}$ at a 95\% confidence level. This flux value corresponds to a luminosity of $L_{\rm > 300\;MeV} < 10^{33}~\mathrm{erg\ s^{-1}}$ assuming a typical power-law model with a photon index of $\Gamma = 2.5$ \citep{aleksic2015mwc656}.
An extrapolation of the emission observed initially by {\em AGILE} to VHE would predict a luminosity of $L_{\rm > 300\;MeV} \sim 2 \times 10^{34}~\mathrm{erg\ s^{-1}}$, which is excluded by the mentioned upper-limit. However, as the {\em AGILE} detection would be related with a flare, it can not be compared with the steady state during the MAGIC observations (no flares have been reported during these observations).

\subsection{Searching for the radio counterpart}

MWC~656 was initially considered as a new gamma-ray binary candidate.
Under this initial consideration, we expected a radio flux density for the source between 0.1 and 200~mJy at GHz frequencies (the range displayed by the known gamma-ray binaries, which are located at a similar distance than MWC~656).
Therefore, the source should be detectable by the current radio observatories. We note that this was the motivation at the time that the radio observations were conducted, although nowadays we know that MWC~656 is the first known Be/BH HMXB, as we have explained above.

\citet{moldon2012thesis} conducted three radio observations with the EVN at orbital phases of 0.82, 0.17 and 0.38 during 2011, but the source remained undetected with a lowest upper-limit of $30~\ujybeam$ at 3-$\sigma$ noise level. However, we had two main considerations to these results to continue observing the source. First, most of the known gamma-ray or X-ray binaries exhibit variable radio emission. Therefore, at the time when the EVN observations were conducted, the source could present an inactive or quiescent state, or fainter radio emission. At other epochs, or different orbital phases, MWC~656 could display a stronger emission (as happens in LS~I~+61~303 or PSR~B1259--63, with a radio emission that changes dramatically as a function of the orbital phase). Secondly, the EVN observations were not sensitive to angular scales larger than $\sim$$100~\mathrm{mas}$ (or $\sim$$300~\mathrm{AU}$ considering the distance of $\sim$$2.6~\mathrm{kpc}$). In case of an extended emission larger than such angular scales, the EVN observations could not detect it. However, we note that in any of the explored gamma-ray binaries we do not see extended emission at scales larger than these ones. For all these reasons, we continued the search of the radio counterpart of MWC\ 656.

\subsubsection{WSRT observations}

We requested service time to observe MWC~656 with WSRT (project code S11B/ 013). We conducted five runs  at 1.4~GHz, performed between 2011 December 26 and 2012 February 17, with observing times between 4 and 7~h. 8 IFs (with XX and YY linear polarizations) of 64 channels each were used in all the observations, with a total bandwidth of 20~MHz. We used 3C~286, 3C~147, 3C~48 and/or CTD93 as amplitude calibrators.

These WSRT data reveal the presence of a strong quasar, RX~J2243.1+4441 (the other proposed candidate to be the counterpart of the {\em AGILE} flare), located at a distance of few arcmin to the East of the MWC~656 position (see Figure~\ref{fig:mwc656-wsrtimage}). MWC~656, however, remains undetected in all the images with a lowest 3-$\sigma$ upper-limit of $180~\ujybeam$ (see Table~\ref{tab:mwc656-radio} for the values obtained in all the radio images).
We note that the detected quasar strongly affects all the field of view due to the poor $uv$-coverage and the low resolution of the WSRT data.
Therefore, a better $uv$-coverage and higher resolution is required to correctly clean the structure of the quasar and obtain rms values closer to the theoretical ones (we estimated theoretical rms values of about $15~\ujybeam$ for these data, but we obtained $60~\ujybeam$).

\begin{figure}[!t]
	\begin{center}
		\includegraphics[width=10cm]{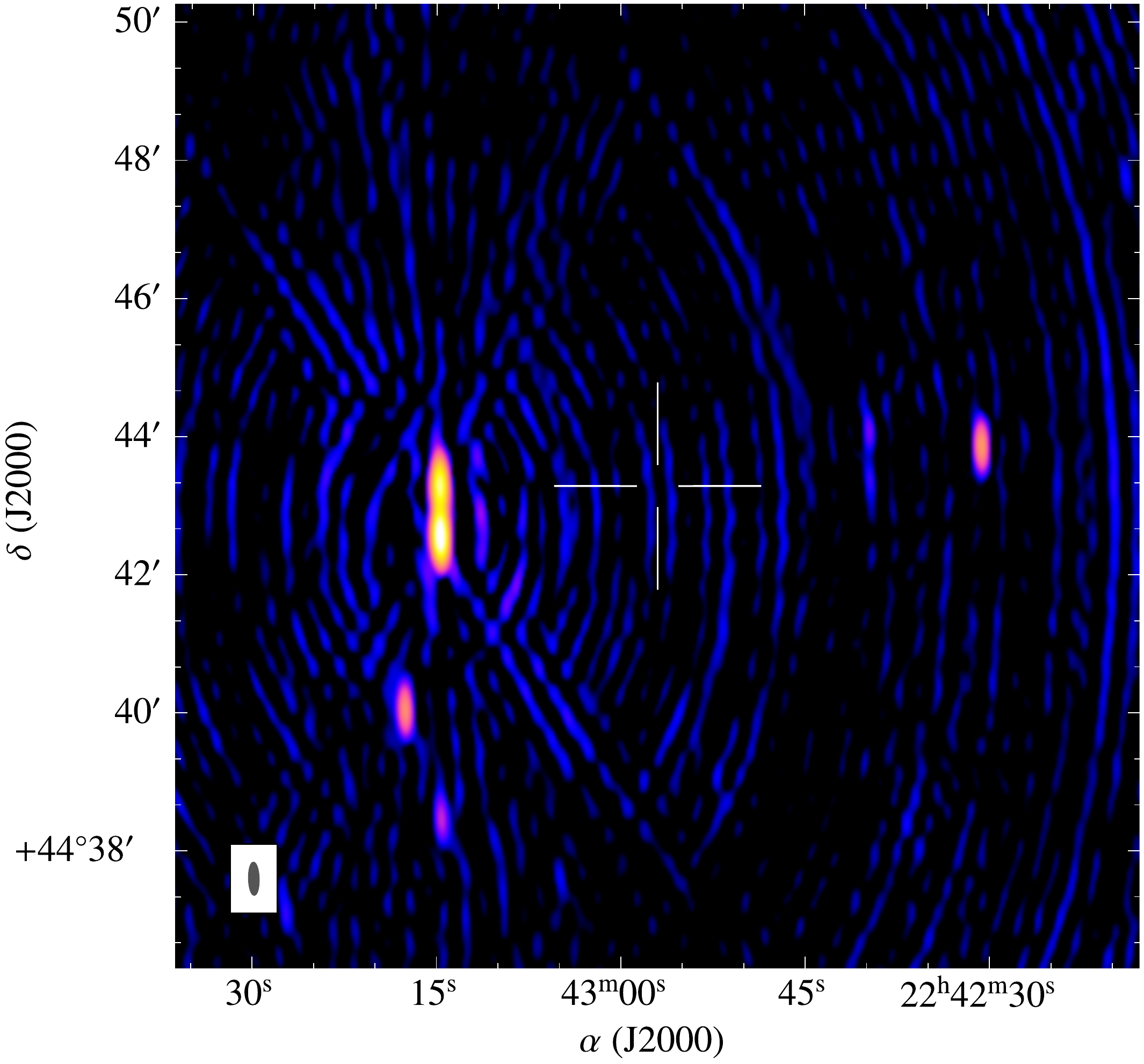}
		\caption[Field of MWC~656 observed with WSRT at 1.4~GHz on 2012 January 14.]{Field of MWC~656 observed with WSRT at 1.4~GHz on 2012 January 14. The white cross represents the position of MWC~656, although we do not detect any radio emission above the rms of $60~\ujybeam$ (setting a 3-$\sigma$ upper-limit of $180~\ujybeam$). We observe strong interferences produced by the bright quasar RX~J2243.1+4441 located to the East of MWC~656 as a consequence of the poor $uv$-coverage and the low resolution of the image. The synthesized beam, shown on the bottom left corner of the image, is $28 \times 8.5~\mathrm{arcsec^2}$ with a PA of $-180^\degree$.}
		\label{fig:mwc656-wsrtimage}
	\end{center}
\end{figure}

\def\qq{~~~~}
\begin{table}[t]
	\small
	\begin{center}
	\caption[Summary of the radio observations of MWC~656 conducted with WSRT and the VLA.]{Summary of the radio observations of MWC~656 conducted with WSRT and the VLA. We show the facility, the date and MJD of the central observation time, the central frequency $\nu$ and the total observing time $t$, the corresponding orbital phase $\phi_{\rm orb}$ (using $P = 60.37~\mathrm{d}$ and $\mathrm{MJD_0} = 53\,242.7$), and the inferred 3-$\sigma$ upper-limit for the undetected radio emission of MWC~656.}
	\begin{tabular}{c@{\qq}c@{\qq}c@{\qq}c@{\qq}c@{\qq~}c@{~~~~~}c}
	\hline\\[-10pt]
	Facility & Date & MJD & $\nu$ & $t$ & $\phi_{\rm orb}$ & 3-$\sigma$ upper-limit\\
	 & (dd/mm/yyyy) & & {\footnotesize(GHz)} & {\footnotesize(h)} & & {\footnotesize($\ujybeam$)}\\[+1pt]
	\hline\\[-10pt]
	WSRT & 26/12/2011 & 55921.78 & 1.4 & 6.5 & 0.38 & 210\\
	WSRT & 03/01/2012 & 55929.64 & 1.4 & 6.7 & 0.51 & 200\\
	WSRT & 14/01/2012 & 55940.60 & 1.4 & 6.4 & 0.69 & 180\\
	WSRT & 11/02/2012 & 55968.52 & 1.4 & 6.8 & 0.15 & 200\\
	WSRT & 17/02/2012 & 55975.52 & 1.4 & 7.0 & 0.27 & 180\\
	VLA-A& 05/10/2012 & 56205.19 & 3.0 & 1.0 & 0.07 & 65\\
	VLA-A& 15/10/2012 & 56215.33 & 3.0 & 1.0 & 0.24 & 55\\
	VLA-A& 06/12/2012 & 56267.21 & 3.0 & 1.0 & 0.10 & 40\\
	\hline
	\end{tabular}
	\label{tab:mwc656-radio}
	\end{center}
\end{table}

\subsubsection{VLA observations}

To avoid this problem, we conducted VLA observations in its A configuration (project code 12B-061), which guaranteed us a better $uv$-coverage and a high resolution. We observed MWC~656 at 3.0~GHz, the optimal frequency band to obtain the best signal-to-noise ratio (assuming a typical power-law emission for the source with $\alpha \approx -0.5$), and a field of view large enough to include both, MWC~656 and the quasar RX~J2243.1+4441. 1-h runs were conducted at three different epochs: on 2012 October 5, 15, and on 2012 December 6. We observed with 16 IFs of 64 channels each one, with a total bandwidth of 2~GHz. We used 3C~48 as amplitude calibrator and J2202+4216 as phase calibrator.

In these images, we properly resolve the quasar RX~J2243.1+4441 (shown in Figure~\ref{fig:mwc656-vlaimage}), observing a point-like core, a jet towards South, and two lobes originated by the shocked material at the end of the two putative jets (the detected one and the one expected in the opposite direction). With the high resolution of the VLA data we could properly clean the map and thus reach lower rms values, avoiding the presence of strong interferences such as the ones observed in the WSRT images. However, MWC~656 remains undetected in all the images, with a lowest upper-limit of $40~\ujybeam$ at 3-$\sigma$ noise level.
In Table~\ref{tab:mwc656-radio} we summarize the results obtained for all the observations, indicating the obtained 3-$\sigma$ upper-limit in each run.
We note that the theoretical rms values for the VLA data were $\sim$6\,--$7~\ujybeam$. Due to 4 IFs that were lost at the edge of the band in the data, we would expect a bit larger values, of $\sim$$9~\ujybeam$. However, we have obtained rms values of about 13--$21~\ujybeam$. Hence, lower values could be still reached with a better $uv$-coverage, implying longer runs.

\begin{figure}
	\begin{center}
		\includegraphics[width=\textwidth]{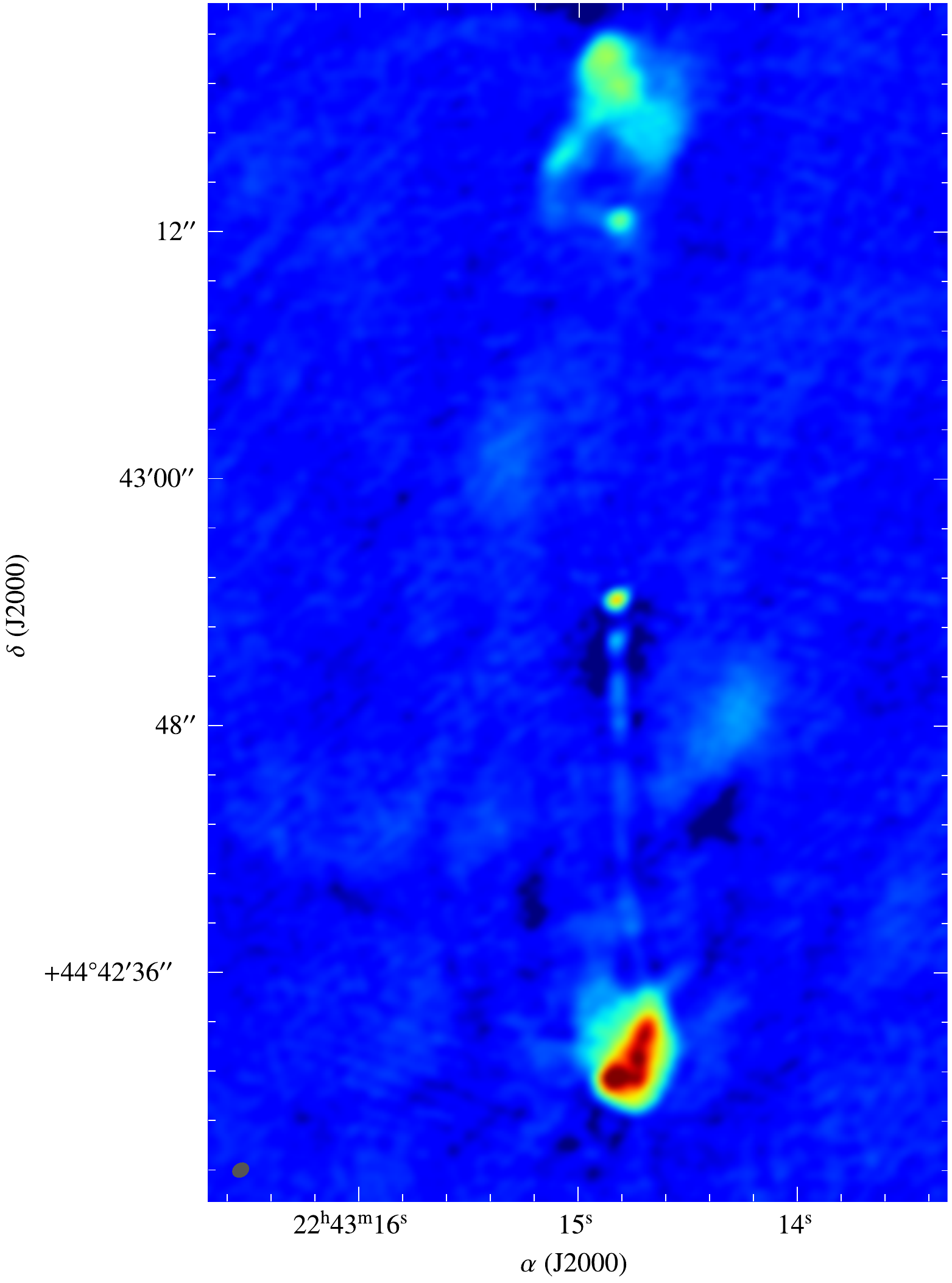}
		\caption[Zoom around the bright quasar RX~J2243.1+4441 that is detected in the field of MWC~656.]{Zoom around the bright quasar RX~J2243.1+4441 that is detected in the field of MWC~656. The image corresponds to the VLA observation conducted on 2012 October 15 at 3~GHz. The synthesized beam, shown on the bottom left corner, is $0.71 \times 0.54\ \mathrm{arcsec^2}$ with a PA of $-74^{\degree}$.}
		\label{fig:mwc656-vlaimage}
	\end{center}
\end{figure}

\subsection{Current understanding of MWC~656}

MWC~656 is the first known Be star hosting a BH companion, establishing a new type of binary system.
Until the publication of \citet{casares2014}, 81 Be X-ray binaries had been discovered. Before the discovery of MWC~656, 48 Be X-ray binaries hosted a neutron star and no one hosted a confirmed black hole. This was known as the missing Be/BH X-ray binary problem. From the current evolutionary models we expect a dominant number of Be/NS systems over the Be/BH ones, with ratios of $\sim$$10$--$50$ \citep{belczynski2009}.
This numbers point out that for the known population of Be/NS systems we would expect $\sim$$0$--$2$ Be/BH systems. Therefore, the discovery of MWC~656 is consistent with these numbers and we would not expect to find out many of these systems.
However, we note that most of the Be/NS systems have been discovered by its X-ray emission. In contrast, MWC~656 has been discovered by the claimed gamma-ray flare, exhibiting a low X-ray emission. The search of these Be/BH binary systems could be thus biased, and it could be even harder to detect them.

According to the latest results, MWC~656 consists of a Be star that presents a Keplerian circumstellar (decretion) disk, and a black hole that exhibits an accretion disk around it. The material from the circumstellar disk falls to the accretion disk of the BH, forming a hot spot on it (Figure~\hyperlink{fig:representation-mwc656}{8.6} shows a representation of this scenario). The expected mass transfer is low (with maximum values of $\sim$$10^{-11}~\mathrm{M_{\sun}\ yr^{-1}}$), which would originate extremely long periods between outbursts, and transient activity almost suppressed in between \citep{casares2014}. Figure~\ref{fig:mwc656-sed} shows the SED of MWC~656 obtained to date.

The source remains undetected at radio frequencies, and we cannot confirm that MWC~656 is actually a radio emitter. Although the radio observations analyzed in this work have not improved the upper-limits reported by \citet{moldon2012thesis}, we have improved the coverage of the orbit, excluding emission above $\sim$$50$ $\upmu\mathrm{Jy}$\ $\mathrm{beam^{-1}}$ in most part of it. Therefore, radio emission stronger than such value becomes now improbable, unless the source experiences a flaring activity. We note that the observed \g-ray flaring activity by {\em AGILE} could also produce a radio flaring activity.

The dim X-ray flux emission that has been observed indicates that MWC~656 exhibits a quiescent state, with an accretion highly inefficient. We know that the BH LMXBs in the low hard state, including the quiescent state, exhibit a correlation between the radio and X-ray luminosity (as suggested and confirmed by \citealt{gallo2003} and \citealt{falcke2004}). Assuming that this correlation could still be valid for BH HMXBs, and hence for MWC~656, \citet{munar2014} estimated the ratio between the measured X-ray luminosity of MWC~656 and the current upper-limits at radio frequencies by using the two most accurate determinations of the correlation, reported by \citet{gallo2012} and \citet{corbel2013}.\fullpagecover{current page.north west}{xshift=5cm,yshift=-3cm, text width=0.4\textwidth}{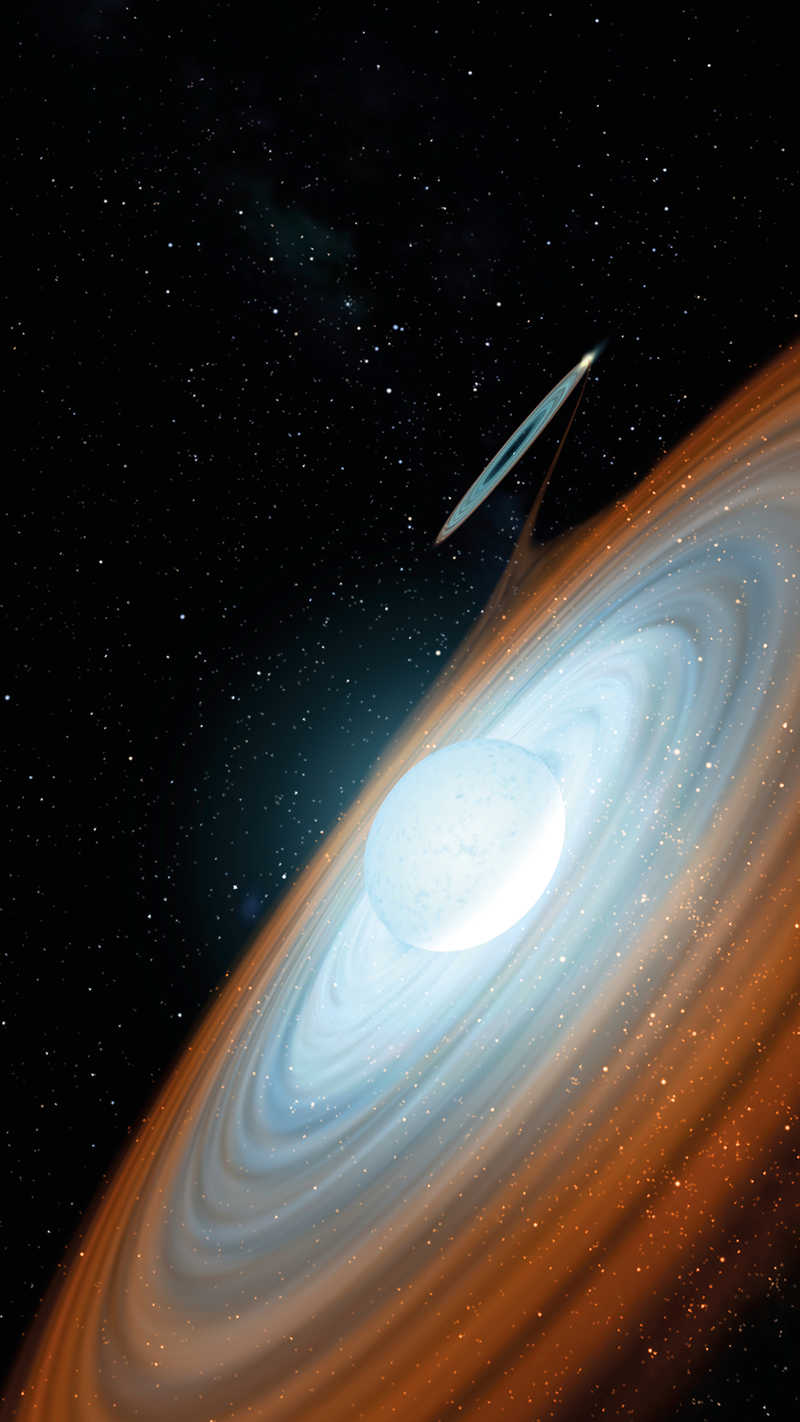}{\hypertarget{fig:representation-mwc656}{}\label{fig:representation-mwc656}\addtocounter{figure}{1}\addcontentsline{lof}{figure}{\number\value{chapter}.\number\value{figure}\quad Representation of the MWC~656 system.}{\bf Figure~\number\value{chapter}.\number\value{figure}.} Representation of the MWC~656 system. We observe the Be star on the bottom together with its decretion disk. The material falls to the accretion disk that is orbiting the black hole (on top), forming a hot spot.\\
\copyright\ Gabriel P\'erez - SMM (IAC).}{0.8}{width=\paperwidth}
\noindent In such case, we would expect a radio flux density at 8.6~GHz of about $8_{-4}^{+6}~\ujy$ (derived using the correlation obtained by \citealt{gallo2012}) or $13_{-4}^{+5}~\ujy$ (from the correlation derived by \citealt{corbel2013}). In any case, we would expect a radio emission below, but close, to the current upper-limits.

A multiwavelength observation with simultaneous data at X-rays (using the {\em Chandra} satellite) and at radio frequencies (using the VLA) will be conducted in July 2015. The source will be observed in the 8--12~GHz band during 6~h, with the A configuration of the VLA. The theoretical rms is $2~\ujybeam$. At these frequencies, the quasar will be fainter than at 3~GHz and the resolution of the image will be higher, and the $uv$-coverage of the data will be much better than in the previous observations. Therefore, we expect to obtain rms values close to the theoretical ones in this observation. These values should be low enough to detect the putative radio emission of MWC~656, provided it follows the mentioned correlation.
\begin{figure}[t]
	\begin{center}
		\includegraphics[width=\textwidth]{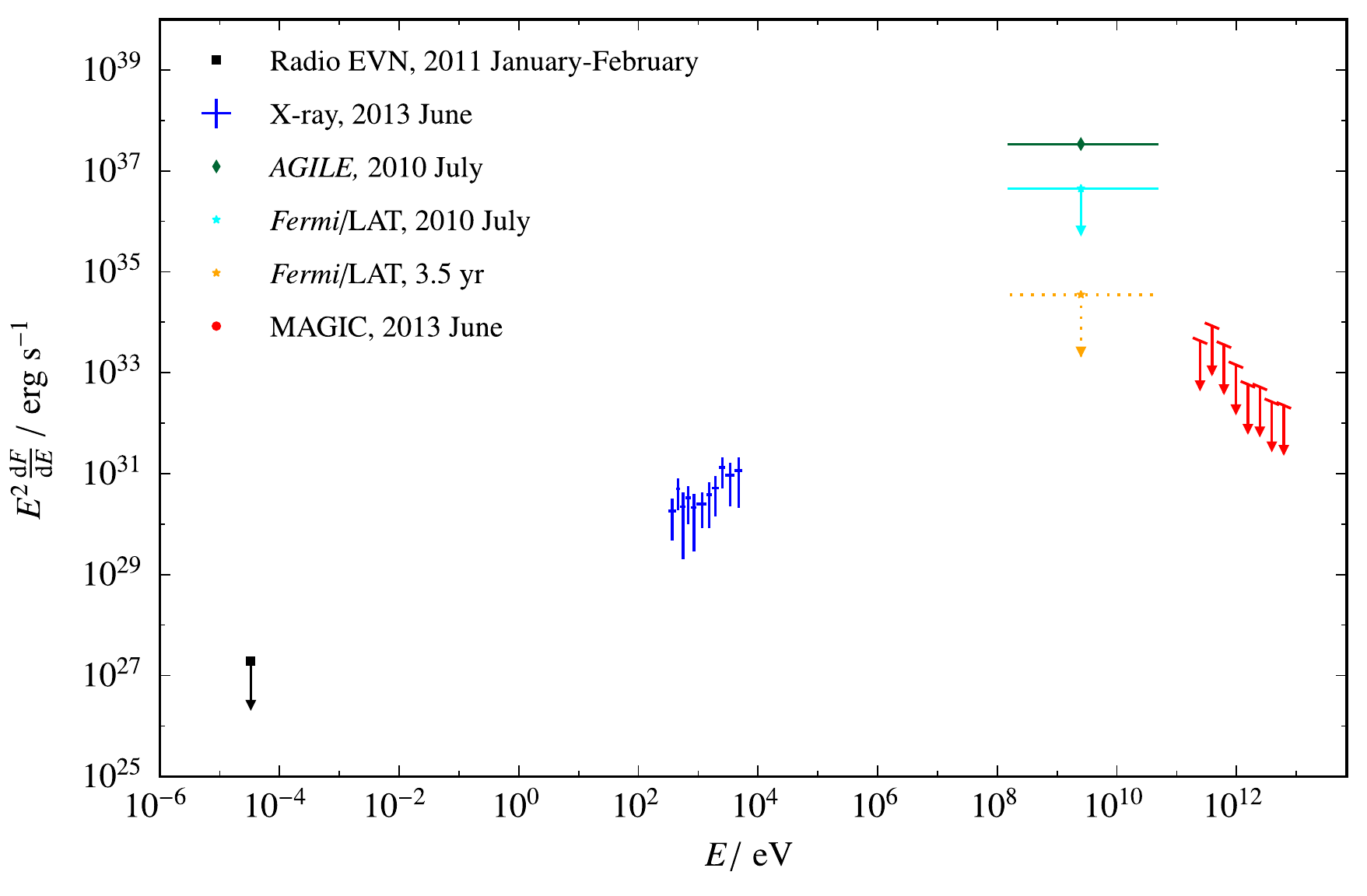}
		\caption[SED of MWC~656 including the available data up to 2014.]{SED of MWC~656 including the available data up to 2014. We include the stringent radio upper-limit from the EVN observations \citep{moldon2012thesis}, the {\em XMM-Newton} X-ray observations from \citet{munar2014}, the {\em Fermi}/LAT upper-limits \citep{mori2013}, the {\em AGILE} detection at HE \citep{lucarelli2010}, and the VHE upper-limits from MAGIC \citep{aleksic2015mwc656}. This figure has been adapted from \citep{aleksic2015mwc656}.}
		\label{fig:mwc656-sed}
	\end{center}
\end{figure}

\subsection{On the quasar RX~J2243.1+4441 in the field of view of MWC~656}

The detected quasar RX~J2243.1+4441 in the field of MWC~656, and clearly resolved in the VLA data (e.g.\ Figure~\ref{fig:mwc656-vlaimage}), was initially considered as the other possible candidate to be the origin of the {\em AGILE} flare \citep{williams2010}. After the discovery of the Be/BH system in MWC~656, all attention turned to this binary system. However, we note that the origin of the \g-ray flare is still unknown.
For that reason, it is worth to study the emission of RX~J2243.1+4441 along the three obtained VLA images. These data could allow us to determine if the radio emission of the quasar is variable on timescales of 10~d and 2~months.
Due to its extragalactic origin, the spatial scales of the source imply that only the compact core component can exhibit variability on these timescales. From the structure seen in Figure~\ref{fig:mwc656-vlaimage}, we have studied the variability of the compact core and the South lobe. The North lobe has not been studied due to its morphological complexity, which makes difficult to provide accurate flux density values. We note that a significant variability for the core emission could point out the possibility of flares, which could be associated with the {\em AGILE} detections.

We summarize the obtained measurements in Table~\ref{tab:mwc656-quasar-results}. The core component seems to be stable along the three observations, exhibiting a flux density of $\approx$$2.2~\mjybeam$.
\begin{table}[t]
	\small
	\begin{center}
	\caption[Flux density values obtained from the core and south lobe components of the quasar RX~J2243.1+4441 from the VLA observations performed at 3.0~GHz.]{Flux density values obtained from the core and south lobe components of the quasar RX~J2243.1+4441 shown in Figure~\ref{fig:mwc656-vlaimage} from the VLA observations at 3.0~GHz. We provide the flux density values of the core fitting a Gaussian component with {\tt imfit}, and the peak and integrated flux density from the South lobe using {\tt imstat}.}
	\begin{tabular}{ccccc}
	\hline\\[-10pt]
	Date & MJD & $S^{\rm core}$ & $S_{\rm peak}^{\rm south\ lobe}$ & $S_{\rm int}^{\rm south\ lobe}$\\[+2pt]
	(dd/mm/yyyy) & & ($\mjy$) & $(\mjybeam)$ & $(\mjy)$\\[+2pt]
	\hline\\[-10pt]
	05/10/2012 & 56205.19 & $2.22 \pm 0.04$ & $3.2$ & $37.5 \pm 1.0$\\
	15/10/2012 & 56215.33 & $2.21 \pm 0.02$ & $6.3$ & $76.5 \pm 1.0$\\
	06/12/2012 & 56267.21 & $2.06 \pm 0.02$ & $3.9$ & $44 \pm 5$\\ 
	\hline
	\end{tabular}
	\label{tab:mwc656-quasar-results}
	\end{center}
\end{table}
However, we obtain surprising results for the South lobe. Either the integrated or the peak flux densities show a significant increase of about a factor of two for the emission from 2012 October 5 to October 15. The lobe increased from the original $37.5 \pm 1.0~\mjy$ to $76.5 \pm 1.0~\mjy$. On December 6, the flux density emission decreased again, obtaining values close to the initial ones.
We note that this lobe can be split in three point-like components plus an extended emission (see Figure~\ref{fig:mwc656-vlaimage}). We have performed a rough estimation of the flux density emission for these components. However, the flux density increase cannot be directly associated with any of these individual components.

The reported variability on $\sim$10~d timescales cannot be explained by any mechanism, given that the lobe must present a size of tens of kpc or even Mpc. However, the stability of the core component guarantees that this variability cannot be related with calibration issues or systematic errors in the flux density values of each image. At this point, the presence of a Galactic source, confused within the quasar lobe, becomes as the most reasonable answer. The nature of this hypothesized source must be resolved in the future. The new VLA observation to be conducted at 10~GHz could provide new clues about this source due to its higher resolution.

We have also searched for optical data available for the field of view of RX\ J2243.1+4441 in the USNO, 2MASS, and Sloan Digital Sky Survey (SDSS) catalogs, but these data do not provide any possible counterpart (no optical sources are detected around the position of the quasar or any of the lobes). Furthermore, the available X-ray {\em XMM-Newton} data shows that all the X-ray emission comes from the core of the quasar, and significant X-ray emission is not observed from any of the lobes \citetext{P.\ Munar-Adrover, private communication}.

\subsection{Concluding remarks}

The {\em AGILE} \g-ray flare has unveiled an interesting region of the sky that needs to be explored in depth. It has lead to the discovery of the first Be/BH binary source. A quasar which is resolved at radio frequencies, emits at X-rays, but is still unassociated to an optical counterpart, could also be responsible of the mentioned \g-ray flare. The presence of a variable Galactic radio source, confused with one of the lobes of the quasar, is also hypothesized from the latest radio results.

It is clear the needed of additional multiwavelength observations to study these two (or three) sources in detail. The planned observation with {\em Chandra} and the VLA can provide useful data to unveil part of these unresolved questions.

%
%
%
%

\chapter{Concluding remarks} \label{chap:conclusions}

In this thesis we have studied several high-energy binary systems at radio frequencies, either with connected interferometers (such as GMRT, LOFAR, VLA and WSRT) or with very long baseline radio interferometers, VLBI (such as the EVN and LBA). We have studied the gamma-ray binaries LS~5039 and LS~I~+61~303 at low radio frequencies, determining their spectra and light-curves. The gamma-ray binary HESS~J0632+057 has been explored with VLBI observations, although we only provide upper-limits on its radio emission. The new colliding wind binary HD~93129A has been discovered through VLBI radio observations and optical data. Finally, we have also performed radio observations on two new sources that were hypothesized to be gamma-ray binaries, TYC~4051-1277-1 and MWC~656: whereas the first one is unassociated with the observed radio emission, the second one is the first Be/BH binary system known and remains undetected in radio.

The gamma-ray binaries had barely been studied at low radio frequencies before the results presented in this thesis. All the detailed studies previously published were focused at GHz frequencies, where the spectrum is unabsorbed. However, we have demonstrated that data at these low frequencies, where some absorption processes start to be relevant, can provide new clues in the determination of the physical properties of these binary systems. A study of the absorption mechanisms through the obtained radio spectra allows us to constrain the regions at which the radio emission is produced. The knowledge obtained at radio frequencies allows a better modeling of the multiwavelength emission of these sources.

In the case of colliding wind binaries, we have seen how VLBI observations can reveal the extended bow-shaped non-thermal radio emission that is originated between the two stars. VLBI observations are thus critical to unveil the nature of the radio emission. An accurate astrometry, not only in the radio data but also at other wavelengths, is necessary to confirm the existence of a wind collision region in these cases. The results presented in this work support the idea that wind-collision regions are the sites where relativistic particles are accelerated in colliding wind binaries.

Interestingly, radio data can contribute significantly to the knowledge of high-energy astrophysical sources. The study of high-energy binary systems through its radio emission can characterize the emitting region and the properties of the systems.
The higher resolution and sensitivity compared to other wavelengths allow us to resolve this region, leading to morphological studies of the systems. We can determine the evolution of the extended emission produced in gamma-ray binaries or the wind collision region originated in colliding wind binaries.
The low-frequency observations, which have been barely considered up to now, together with high-frequency and VLBI observations allow us to obtain a coherent picture of the origin and nature of the radio emission in all these systems. The addition of an array such as LOFAR, with long baselines and high sensitivity, together with the GMRT, the return of the VLA at low radio frequencies, and the upcoming SKA, will produce important results in the coming years.

\backmatter
\titleformat{\chapter}[block]{\setstretch{1.0}\usefont{T1}{kurier}{b}{n}\selectfont\huge\bfseries}{}{}{\noindent #1}
\cleardoublepage\phantomsection
\bibliographystyle{mn2e}
\addcontentsline{toc}{chapter}{\bibname}
{\small\bibliography{bibliography.bib}}

\cleardoublepage\phantomsection
\addcontentsline{toc}{chapter}{List of Figures}
\listoffigures
\cleardoublepage\phantomsection
\addcontentsline{toc}{chapter}{List of Tables}
\listoftables
%
%
%
%

\chapter{List of publications}

\begin{itemize}
	\renewcommand{\labelitemi}{$\circ$}
	\item {\bf Marcote, B.}, Rib\'o, M., Paredes, J.~M., Ishwara-Chandra, C.~H., Swinbank, J.~D., Broderick, J.~W., Markoff, S., Fender, R., Wijers, R.~A.~M.~J., Pooley, G.~G., Stewart, A.~J., Bell, M.~E., Breton, R.~P., Carbone, D., Corbel, S., Eisl\"offel, J., Falcke, H., Grie{\ss}meier, J.-M., Kuniyoshi, M., Pietka, M., Rowlinson, A., Serylak, M., van der Horst, A.~J., van Leeuwen, J., Wise, M.~W., Zarka, P., {\em Orbital and superorbital variability of LS~I~+61~303 at low radio frequencies with GMRT and LOFAR}, 2016, MNRAS, 456, 1791
	\item {\bf Marcote, B.}, Rib\'o, M., Paredes, J.~M., Ishwara-Chandra, C.~H., {\em Physical properties of the gamma-ray binary LS 5039 through low and high frequency radio observations}, 2015, MNRAS, 451, 4578
	\item Benaglia, P., {\bf Marcote, B.}, Mold\'on, J., Nelan, E., De Becker, M., Dougherty, S.~M., Koribalski, B., {\em A radio-map of the colliding winds in the very massive binary system HD 93129A}, 2015, A\&A, 579, A99
	\item Mart\'i, J., Luque-Escamilla, P~L., Casares, J., {\bf Marcote, B.}, Paredes-Fortuny, X., Rib\'o, M., Paredes, J.~M., N\'u\~nez, J., {\em In quest of non-thermal signatures in early-type stars}, 2015, Ap\&SS, 356, 277
	\item {\bf Marcote, B.}, Benaglia, P., Mold\'on, J., Nelan, E., De Becker, M., Dougherty, S.~M., Koribalski, B., {\em Discovering the colliding wind binary HD 93129A}, 2014, in proceedings of the 12th European VLBI Network Symposium and Users Meeting. Cagliari, Italy, PoS(EVN2014)57
	\item {\bf Marcote, B.}, Mold\'on, J., Rib\'o, M., Paredes, J.~M., Paragi, Z., {\em The changing morphology of the radio outflow of HESS J0632+057 along its orbit}, 2014, in proceedings of the 12th European VLBI Network Symposium and Users Meeting. Cagliari, Italy, PoS(EVN2014)95
	\item {\bf Marcote, B.}, Rib\'o, M., Paredes, J.~M., Ishwara-Chandra, C.~H., Swinbank, J., Broderick, J., Fender, R., Markoff, S., Wijers, R., {\em GMRT and LOFAR low frequency observations of the gamma-ray binaries LS~5039 and LS~I~+61~303}, 2014, in proceedings of The Metrewavelength Sky Conference. Pune, India, Bull. Astr. Soc. India
	\item {\bf Marcote, B.}, Rib\'o, M., Paredes, J.~M., Swinbank, J., Broderick, J., Fender, R., Markoff, S., Wijers, R., {\em First LOFAR observations of gamma-ray binaries}, 2012, in proceedings of HIGH ENERGY GAMMA-RAY ASTRONOMY: 5th International Meeting on High Energy Gamma-Ray Astronomy, AIP Conf. Proc., 1505, 374
\end{itemize}

\chapter{Agradecimientos}

Con esta tesis se cierra un ciclo más. Un ciclo de cuatro años que desde luego ha representado una etapa importante, tanto a nivel profesional como personal. 

En primer lugar, dar las gracias a mis directores de tesis, los doctores Josep Maria Paredes y Marc Ribó. Creo poder decir con total sinceridad que he tenido la suerte de llegar a un grupo increíble. Josep Maria, gracias por guiarme durante estos cuatro años en un mundo para mí desconocido. Por hacerme ver siempre la dirección hacia la que debíamos ir.
Marc, gracias por las ilimitadas horas perfeccionando el trabajo hecho, hasta que conseguíamos encajar cada pieza del puzzle. Trabajo duro, pero que me ha valido para mejorar increíblemente estos años.
Gracias a los dos por todo lo que me habéis enseñado y ayudado. Y en lo personal, gracias también. Porque un doctorado no sólo se mide en el trabajo realizado si no también en el ambiente que ha existido. Y siempre habéis estado ahí, no como directores, si no también como compañeros.

En segundo lugar, obviamente, los agradecimientos de esta tesis van para mis padres. Que siempre me han apoyado, incluso desde la distancia. Ya saben que los quiero mucho, aunque no lo suela decir muy a menudo. A ellos va dedicada esta tesis sin duda. Sin vosotros no habría llegado hasta aquí.

Estos años también me han permitido conocer a mucha gente que debería aparecer aquí. Empezando por el Departamento, el DAM. Creo que no podría empezar de otra forma si no es recordando al héroe particular, José Ramón\dots aunque siempre seas simplemente JR. Porque haces que ``todo simplemente funcione'', aunque nadie sepa cómo! y siempre tienes una sonrisa y una solución a cualquier problema. Y a las otras dos secretarias, Montse y Rosa, que hacen todos los trámites del DAM posibles, que no es poco.
Después están nuestros IT Crowd personales, Gabi y Jordi, que aunque no se suela ver vuestro trabajo, cuando faltáis todo empieza a fallar. Gracias por las conversaciones de café sobre las últimas novedades en móviles, ordenadores, etc.

También al resto de nuestro grupo de Altas Energías, los que están en el DAM y los que están ya diseminados por el mundo. Gracias a Valentí Bosch-Ramon por sus aportaciones a nivel teórico. Y también a Simone Migliari, Victor Zabalza, Pol Bordas, Kazushi Iwasawa, Javier Moldón, Pere Munar y Roberta Zanin.

A continuación está toda la gente con la que he tenido el placer de convivir estos años.
Algunas personas que se han ido o han regresado al DAM, otros que se han doctorado mientras yo estaba aquí, o el relevo generacional que ha entrado en los últimos años y que permitirá que el DAM siga su curso.

Empezando por antigüedad, pero sin decir cuánta para que no se enfaden!, a Rosa Rodríguez y a Neus Águeda, las chicas solares. Por brillar con luz propia dentro del DAM. Neus, por descubrirnos alimentos que no sabíamos ni que existían, y Rosa, siempre organizando cosas para no caer en la monotonía.

Después vienen aquellos que he visto culminar, para después dejarnos y seguir sus vidas. No puedo empezar por otro que no sea Javier Mold\'on, el que inició mis pasos en el complicado mundo de la radio astronomía, guiándome a través de los oscuros designios de Aips. Todo lo que he aprendido de la reducción de datos comenzó con tus enseñanzas, además de hacerme ver que con Python, y una taza de té, se puede hacer cualquier cosa. También a nivel personal, por todo lo que hemos compartido estos años, ya sea en Barcelona, Groningen, o en cualquier parte del mundo.
A Pere Munar, reciente doctor y que nos ha abandonado también! se te hecha de menos en el despacho, aunque ya sabes que desde el primer minuto ya tenías sustituto. Siempre admiraremos tu determinación en presentar resultados sin resultados, con gráficas sin un sólo punto. Pero también en lo personal, por estar siempre dispuesto a ayudar, echar una mano, o pasar un rato tomando algo. Y por hacernos ver que es imposible entender a un mallorquín!
Seguiré por Laura Darriba, por demostrarnos que siempre se puede seguir avanzando por muy negra que parezca la situación, y por hacer más amenas las tardes de trabajo hasta altas horas de la noche, pegándonos la afición por los tés junto a Javi. Por ser invencible a cualquier juego de mesa en el que  que participes, aunque siempre planeemos todos en tu contra. Cuatro años pidiendo una ventana y cuando por fin la tienes nos dejas! espero que te vaya todo muy bien y no te perdamos la pista.
A Jordi Viñas, que desde que te has cambiado de bando estás desaparecido de un lado para otro del mundo\dots al menos sabemos que te va bien todo! gracias por conservar siempre la alegría con independencia de la situación. Espero que disfrutes tu nueva etapa.
Javi, Lau, Jordi, gracias por todos los ratos que hemos pasado juntos jugando a las cartas, a juegos de mesa o simplemente cenando. Se echan de menos las batallas al {\em Catán} o echando ``la última'' en buena compañía.
A Josep Maria Masqué y Pau Frau, la otra sección radio del DAM. Espero que os vaya todo bien también!
Y toca el turno de los {\em Gaiers}, Maria Monguió y Santi Roca. Maria, por tu sentido del humor, y tus chistes malos malos malísimos no, peores aún! pero que siempre nos terminan sacando una sonrisa, y Santi, por mantener la incógnita de cómo comiendo tan poco se puede ser tan activo! Que disfrutes el cambio de aires.
De los del pasillo de enfrente, territorio ICC, a Héctor Gil y Andreu Ariño, también se recuerda vuestra ausencia a la hora de comer.

Pasemos a la siguiente generación de doctorandos, aquellos que han llegado después de mí al departamento. Xavier, primero compartiendo despacho en las galeras y después ya en el despacho más molón de la {\em part vella}. Ha sido ya bastante tiempo aunque haya pasado muy rápido. Gracias por tu compañía, la he disfrutado sinceramente. Y por pluriemplear tu ordenador permitiéndonos que corramos nuestros scripts. Sigue por el buen camino Pythonero! harás grandes cosas y te llevará lejos.
De compañero de despacho a compañero de despacho\dots Victor, gracias por demostrarnos que un madrileño también puede sobrevivir en Barcelona! te perdonamos que seas un GNUPlotero a cambio de tu compañía.
Laia! gracias por ser la alegría del DAM, y los buenos momentos que nos haces pasar tanto a la hora de comer como fuera del DAM.
E Ignasi, con tu dialecto {\em ignasí} indescifrable que nos obliga a todos a prestar la máxima concentración para poder entenderte.
Roger, todo formalidad aunque siempre sueltes la chispa después! Carmen, por estar siempre con una sonrisa, día tras día.
Y termino con los dos últimos fichajes del grupo. Dani, con quien he podido compartir varios shifts de MAGIC, Keep Calm y que vaya todo bien, que has empezado fuerte! Y finalmente N\'uria, espero que tengas un doctorado emocionante y entretenido. Te deseo lo mejor.

Y a todos los demás integrantes del departamento, aunque no vengáis reflejados directamente aquí, pero con los que he compartido clases, o simplemente conversaciones de pasillo, gracias por la convivencia que hemos tenido.

Aunque tampoco me olvido de la parte menos personal, {\em folre}, compañero de tesis desde el principio hasta el final, y que ha permitido todo el trabajo que se ha hecho aquí. Con el que he pasado tanto buenos como malos momentos. Nos costó un tiempo, pero finalmente conectamos casi a la perfección gracias a GNU/Debian, y el maravilloso entorno gráfico i3. Un entorno productivo sin igual al unirlo al uso de Vim y Python. Por el camino quedó los malos momentos, como cuando un disco duro falla perdiendo parte del trabajo de los últimos meses. ¡Buen recordatorio de que siempre hay que tener una copia de seguridad a mano! Y también al café, nuestro mejor amigo y la mejor maquina de convertir energía en trabajo.

Durante esta tesis he tenido el placer de poder hacer tres estancias largas, todas ellas en Holanda, aunque repartidas entre el API y JIVE, (Amsterdam y Dwingeloo, así que con un contraste máximo entre ambas). De todas las veces que he estado en el API guardo un recuerdo fantástico. Primero dar gracias a Sera Markoff y John Swinbank, por dejarme trabajar con vosotros, involucrándome en LOFAR, y después a todas las personas que facilitaron mi estancia allí. Desde Milena, la secretaria y que siempre viene para alegrar el día a todo el mundo, hasta Martin, por ser la referencia para casi cualquier cosa, y llevarnos a los mejores sitios para cenar, tomar algo en el molino, o simplemente enseñarnos nuevos juegos de cartas. Theo, Montse, Dario, Yvette, Antonia, Óscar, Evert, Gijs, Alexander, Diego, Rik\dots hay muchas personas con las que he pasado muy buenos momentos allí! De JIVE (y ASTRON) gracias a Zsolt Paragi por enseñarme todo lo relevante de VLBI, aprendí mucho durante mi estancia ahí. Y al resto de gente con quien pude estar allí, Javi, Carme, Iván, Giuseppe, Ciriaco, Maura y Elizabeth\dots también gracias.
También a Ishwara Chandra por acogerme en el NCRA durante varias semanas, permitiéndome aprender todos los detalles de la reducción de datos de GMRT. Y a Breezy, por hacer de guía local. A Paula Benaglia, por permitirme participar en el interesante campo de las binarias con colisión de vientos.

También recordar a toda la gente que únicamente veo de congreso en congreso, pero con los que he pasado buenos momentos. Rubén, siempre recordaremos tus lágrimas indicando que la comida de Socorro hay que comerla con precaución! Anahi, Florent, Eugenia, siempre es un placer encontrarnos por ahí. Y a Rebecca, no desesperes que ya te falta poco también! y lo vas haciendo genial. Espero ciertamente el siguiente congreso donde nos volveremos a ver, cada vez en un sitio distinto pero disfrutando igual.

Un capítulo a parte merecen los shifts de MAGIC. Tres shifts en La Palma dan para muchas historias, casi para escribir un libro completo, ya que nunca te aburres. Primero dar las gracias primero al equipo local: Eduardo ¡maestro!, Javier y Martin. Los shifters siempre agradecerán eternamente que estéis localizables las 24 horas, 365 días. Y nunca poniendo mala cara o contestando mal. Preparados para solventar cualquier problema que no sepamos o que sea grave. Gracias por hacer los shifts más fáciles, y haciendo que disfrutemos y conozcamos también la {\em isla bonita}.
Villi, gracias por enseñarnos todo lo que un operador debe saber, siendo un shift leader fantástico. Fue un placer conocerte y compartir tanto tiempo contigo. A Irene y Michelle, qué se puede decir. Pasamos un shift lo mejor posible, mezclando trabajo, risas, excursiones y películas. Todo por no volvernos locos allí arriba. Michelle, porque aunque fue caótico fue una estancia fantástica. Eso sí, nos quedó claro que en la cocina no podíamos estar los dos a la vez! y que hay que comprobar el nivel de gasolina antes de empezar la subida\dots eso cuando el coche no nos deja tirados en mitad de la nada, anocheciendo!
El segundo shift, ya con toda la responsabilidad, iba a ser más tranquilo. Pero, Tomislav, nos tocó todo lo peor aquella noche. Una alarma de {\tt cosy} nunca es buena señal, pero en ese caso, como vimos después, era todavía peor. La estructura de MAGIC-1 había dicho basta de repente! Creo que no me equivoco si digo que ése fue el peor momento de estos cuatro años. Una noche y día interminables, llamando al equipo local, intentando dejar MAGIC en una posición salvable, y ante la luz del Sol activando el protocolo de emergencia del ORM para no incendiar La Palma! Al final, y gracias al trabajo de todo MAGIC, se pudo recuperar M1 en pocos meses, y esperemos que siga así durante mucho tiempo.
Y mi último shift, Dani, Gareth, John, Simona, Damir y Mónica, gracias por ese tiempo juntos. Y a Simona y Roberta, por enseñarnos el {\em italian style}! Y al resto de la colaboración, en especial a Alicia, Dani, Jezabel y Rubén, que siempre es entretenido encontrarnos.

Pero antes de despedirme, recordar también los orígenes, todo lo que ha habido y que ha permitido que llegase hasta aquí. Las pequeñas cosas que van cambiando los hilos sutilmente. Entré en física, como muchos, por mi pasión por la astronomía, que nació por un afortunado mini libro de bolsillo. Y se afianzó gracias a Arturo Bravo, mi profesor de instituto, que me introdujo en la astronomía a nivel de aficionado. Y a la asociación de astronomía donde pasé grandes momentos. Y ya en la carrera, gracias a toda la gente con la que he compartido alegrías y penurias, Paula, siempre disfrutando contigo dentro y fuera de la universidad, Julia, aunque siempre vengas con algún problema por resolver!, Aníbal, que aunque ya no bajes de $\mathbb{R}^n$ siempre se agradecen tus discusiones científicas, Ricardo, Isa, Diego, Juanra, Sonia,\dots y a todos los que me dejo! También a M$^{\rm a}$ Isabel, que empezamos a la vez la aventura del doctorado, aunque tú en química, y pronto lo acabarás también, ánimo!
También a Xavier Barcons, mi director de trabajo de fin de carrera, por introducirme en la investigación. Aunque haya pasado de extragaláctico a Galáctico, y de rayos X a radio, me ha servido de mucho y guardo un buen recuerdo del trabajo que realizamos juntos.

Y por último, a ti, Erica, por estar todo este tiempo conmigo y aguantarme, especialmente durante los últimos meses de tesis!\\[+1cm]

{\flushright Benito Marcote Mart\'in\\ Barcelona 2015\\}


\chapter{Acknowledgements}

A period is closed with this thesis. A four-year period that has represented a really important stage in my life, both professionally and personally.

First, I owe my deep and sincere gratitude to my supervisors, Dr.\ Josep M.\ Paredes and Marc Rib\'o. I can firmly say that I have been lucky to arrive at a fantastic group. Josep Maria, thanks for guiding me during these years into a world unknown to me, always finding the right direction.
Marc, thanks for the endless time improving the work, until all the pieces fit the puzzle. Hard work, but I have improved a lot these years thanks to it. Thanks to both of you for your teachings. Personally I thank you too. Because these years I have enjoyed my stay at the department, the DAM.

Secondly, I thank to my parents. Because they have always supported me, even from the distance. They know how much I love them, although I do not often say it to them. This thesis is dedicated to them. Without their support I would not have gone so far.

Throughout this thesis I have met many people I will definitely remember. I shall start from the department, the DAM. I cannot start with other person but José Ramón, although he is always simply JR. Because with you the sentence ``it just works'' becomes right, although nobody knows how you do it! and you always welcome people with a smile and the solution for each problem.
I also thank to the other two secretaries, Montse and Rosa, because they make all the formalities possible, which is not a relaxed work. After that, I thank to our own IT Crowd, Gabi and Jordi. Although your work is many times invisible, when you are not there everything fails. Thanks also for all the coffee discussions about the last news on technology.

I remember the rest of our group of High-Energy Astrophysics too, to those who are in the DAM, and those who left. Thanks Valentí Bosch-Ramon for your theoretical explanations. And thanks also to Simone Migliari, Victor Zabalza, Pol Bordas, Kazushi Iwasawa, Javier Moldón, Pere Munar and Roberta Zanin.

Next, all the people I have lived with these years.
Some people have left, other ones have returned to the DAM\dots few of them have doctorated while I was here, and another ones will be the next generation of PhD, keeping alive the DAM.

I will start by seniority, although without mentioning the age! to Rosa Rodríguez and Neus Águeda, the solar girls. Because they bright with their own lights. Neus, because with you we discover new eatable things, and Rosa, always organizing activities to avoid monotony.

After them, I will thank to those who have already finished their PhD before me and have left. I must start with Javier Mold\'on, who helped me at the first steps into radio astronomy. You always guided me through the dark paths of Aips. Everything I have learned is thanks to you. You also showed me that we could do everything with Python, and a coup of tea. I thank you also in the personal way. Because of everything we have shared these years, either in Barcelona, Groningen or wherever. 
To Pere Munar, a recent Doctor who has just left! I miss you at the office, although you had substitute from the first minute. We will always admire your determination to present results without results, showing figures without a single point! On the personal level, because you are always ready to help, lend a hand, or go for a drink. And show us that it is sometimes impossible to understand people from Majorca!

I continue with Laura Darriba, thanks because you demonstrated us that people can always keep going, even facing dark situations. And because with you, the long working days that remained until night were more pleasant. Because of you and Javi, half of the DAM are now drinking tea! And also because you are invincible to table games, even if all of us are planning against you.
Four years asking for a window and when you get it you left! I hope everything goes fine and I would like to not miss you.
To Jordi Viñas, who has not stopped in the same place too long since left the DAM. At least we know that everything goes smooth. Thanks for your happiness even in the bad moments.
Javi, Lau, Jordi, thanks for all the moments we have spent playing cards, table games, or simply having dinner. I miss the battles with the {\em Catan} or playing ``the last one'' in good company.
To Josep Maria Masqué and Pau Frau, the other radio astronomers in the DAM. I hope you are enjoying your current positions!
It is time for {\em Gaiers}, Maria Monguió and Santi Roca. Maria, thanks because of your sense of humor, always telling jokes that are not really really bad, they are even worse than that! but they always make us laugh. And Santi, because you keep the unknown on how you can eat as little and be so active! enjoy your new place.
To those from the other side, from the ICC, Héctor Gil and Andreu Ariño, because there is an absence during the lunch time without you.

I will continue with the next generations, those who have arrived after me to the department. First, Xavier, you have been my office mate all this time, in the office without windows, and in the new one, the best office from the old DAM. It has been a lot of time, although it has gone pretty fast. Thanks for your company, I have really enjoyed it. And also because you always allow us to use some cores from your computer to run our scripts! Continue in the good road of Python, you will do big things.
From office mate to office mate, Victor! thanks for demonstrate us that someone from Madrid can survive in Barcelona. We forgive you your love for GNUPlot because of your company.
Laia! thanks for being the happiness of the DAM, and the great times we spend with you, inside or outside the DAM.
And Ignasi, with your {\em ignasian} dialect, uncrabbed to all of us!
Roger, you are the best example of formality, although later you always start the fight\dots and Carmen, always with a smile! every single day.
And I finish with the last signings from our group. Dani, I have spent few MAGIC shift with you, Keep Calm and enjoy your PhD. You have started too strong! And N\'uria, I hope your PhD will be exciting and entertaining, with a lot of science to do.

To everyone else, even if you are not mentioned here, but I have spent some teaching classes or just talking in the hallway, thanks for these years.

I do not forget the less personal part, {\em folre}, my partner from the beginning to the end. All this work has been done thanks to you. We have spent good and bad times. The beginning was hard, but finally, we have almost achieved perfection using GNU/Debian and the wonderful i3 desktop environment, together with Vim and Python. In the past it was few bad times, as when a hard disk breaks, loosing the work from the last months, you always need a backup! and thanks to the man's best friend, the coffee, the perfect machine to convert energy into work.

I have spend part of my thesis in other institutes during relatively long periods. All of them in The Netherlands, although spread at the API and JIVE (Amsterdam and Dwingeloo, quite strong contrast). I keep a fantastic remember from my time at API. First of all, thanks to Sera Markoff and John Swinbank to allow me to work with you, learning how to use and help LOFAR. And later on, to all those people I have met there. From Milena, the secretary, brightening the day to everyone, to Martin, because your are the reference to everything there. And you have showed us the best places to have dinner, the windmill, or just new card games. Theo, Montse, Dario, Yvette, Antonia, Óscar, Evert, Gijs, Alexander, Diego, Rik\dots there are many people with who I have spent great times there! From JIVE (and ASTRON) thanks to Zsolt Paragi for teaching me how to deal with VLBI data. I learned too much during my stay there. And also to the rest of the people I met: Javi, Carme, Iván, Giuseppe, Ciriaco, Maura and Elizabeth\dots
I thanks to Ishwara Chandra for welcoming at the NCRA during some weeks, teaching me how to deal with the GMRT data. And thanks Breezy, for play as local guide. Thanks to Paula Benaglia, who involved me participating in her interesting work on colliding-wind binaries.

I do not forget that people I only meet from one conference to another, but those who I have spent great times with. Rubén, I will always remember your tears because of the Socorro's food. You must eat with caution! Anahi, Florent and Eugenia, I always enjoy meet you again. And Rebecca, be patient that you have almost finished! and you are doing a great job. Certainly I wait for the next conference where we will see each other again, each time in a different place but always enjoying.

A separate chapter should correspond to the MAGIC shifts. Three shifts performed at La Palma are an ideal source of stories, because you never get bored up there. First, thanks to the local team: Eduardo, maestro!, Javier and Martin. Shifters will be always grateful to you for being operative 24/7. You never answer us with a bad word, being ready to solve any problem that we have found during the night. Thanks for making easy the shifts, and showing us the {\em isla bonita}.
Villi, thanks for showing us everything that an operator must know. That shift was fantastic! I really enjoyed your company. To Irene and Michelle, what can I say. We spent a grateful shift, mixing work with pleasure. Everything to avoid to become crazy up there. Michelle, because it was chaotic, but it was a fantastic stay. Although it was clear that only one of us should be at the kitchen at the same moment! and now we always check the oil level before going up. But it does not matter when the car stops in the middle of nowhere, when it is getting dark!

The second shift should have become easier. But, Tomislav, that night was the worst night ever. An alarm from {\tt cosy} is never a good signal, but we could not imagine that this time it was because the structure of MAGIC-1 had broken! A really long night, and the next day, without rest. Thanks to the local team, who came quite fast. We required to activate the emergency protocol\dots but everything went fine. And my last shift, Dani, Gareth, John, Simona, Damir and Mónica, thanks for all the time we spent together. To Simona and Roberta, thanks for showing us the {\em italian style}! And to the rest of the MAGIC collaboration, specially to Alicia, Dani, Jezabel and Rubén, thanks for the good times.

But before finishing, I want to remember the origins, everything that has happened before and influenced me to arrive here.
The little things that change you subtly. I decided to study physics because of my love for astronomy, as many of you. That become stronger thanks to my high school teacher, Arturo Bravo, who showed me the charming of astronomical observations. During my degree, thanks to all people with who I have shared joys and hardships. Paula, I have always enjoyed stay with you, inside or outside university. Julia, although you always came with a problem! Aníbal, although you are now at $\mathbb{R}^n$, the talks about physics are fantastic with you. Ricardo, Isa, Diego, Juanra, Sonia,\dots and all others! Thanks also to M$^{\rm a}$ Isabel. We started our PhD at the same time, and you are close to end it too!
I thanks to Xavier Barcons, my director of the bachelor thesis. You introduced me into research. Even when I have changed from extragalactic to Galactic, and from X-ray to radio, I really enjoyed the work we performed together.

And the last one, to you, Erica, because of all the time we have spent together.

\newpage
{\small
This research has been conducted at the Departament d'Astronomia i Meteorologia and Institut de Ci\`encies del Cosmos (ICC), Universitat de Barcelona (UB).
I acknowledge the financial support by the FPI fellowship from the Spanish Ministerio de Econom\'ia y Competitividad (MINECO) under grant BES-2011-049886, and by the FPI stays that have allowed me to stay in foreign institutions, under grants EEBB-I-12-05280, EEBB-I-13-06412, and EEBB-I-14-07963. We also acknowledge support by the MINECO under grants AYA2010-21782-C03-01, AYA2013-47447-C3-1-P, FPA2010-22056-C06-02, and FPA2013-48381-C6-6-P. We also acknowledge the Catalan DEC grant 2014 SGR 86.

We thank the staff of the VLA, GMRT, LOFAR, WSRT, EVN and LBA who made the observations presented along this work possible.

The Very Long Array is operated by the USA National Radio Astronomy Observatory, which is a facility of the USA National Science Foundation operated under co-operative agreement by Associated Universities, Inc.

The Giant Metrewave Radio Telescope is run by the National Centre for Radio Astrophysics of the Tata Institute of Fundamental Research.

The Low Frequency Array was designed and constructed by ASTRON, with facilities in several countries that are owned by various parties (each with their own funding sources), and that are collectively operated by the International LOFAR Telescope (ILT) foundation under a joint scientific policy.

The Westerbork Synthesis Radio Telescope is operated by the ASTRON (Netherlands Institute for Radio Astronomy) with support from the Netherlands Foundation for Scientific Research (NWO).

The European VLBI Network is a joint facility of European, Chinese, South African and other radio astronomy institutes funded by their national research councils.

The Long Baseline Array is part of the Australia Telescope National Facility which is funded by the Commonwealth of Australia for operation as a National Facility managed by CSIRO.

This research made use of the NASA Astrophysics Data System. Data reduction was performed using AIPS, CASA, Obit, ParselTongue and Difmap. The thesis was typeset in \LaTeX\ by the author and all the figures presented along this work have been done using Python and its open source plotting libraries {\tt matplotlib} and {\tt aplpy}.\\
}

{\flushright Benito Marcote Mart\'in\\ Barcelona 2015\\}

\pagestyle{empty}
\cleardoublepage
\newpage
\ \\
\cleardoublepage
\includepdf{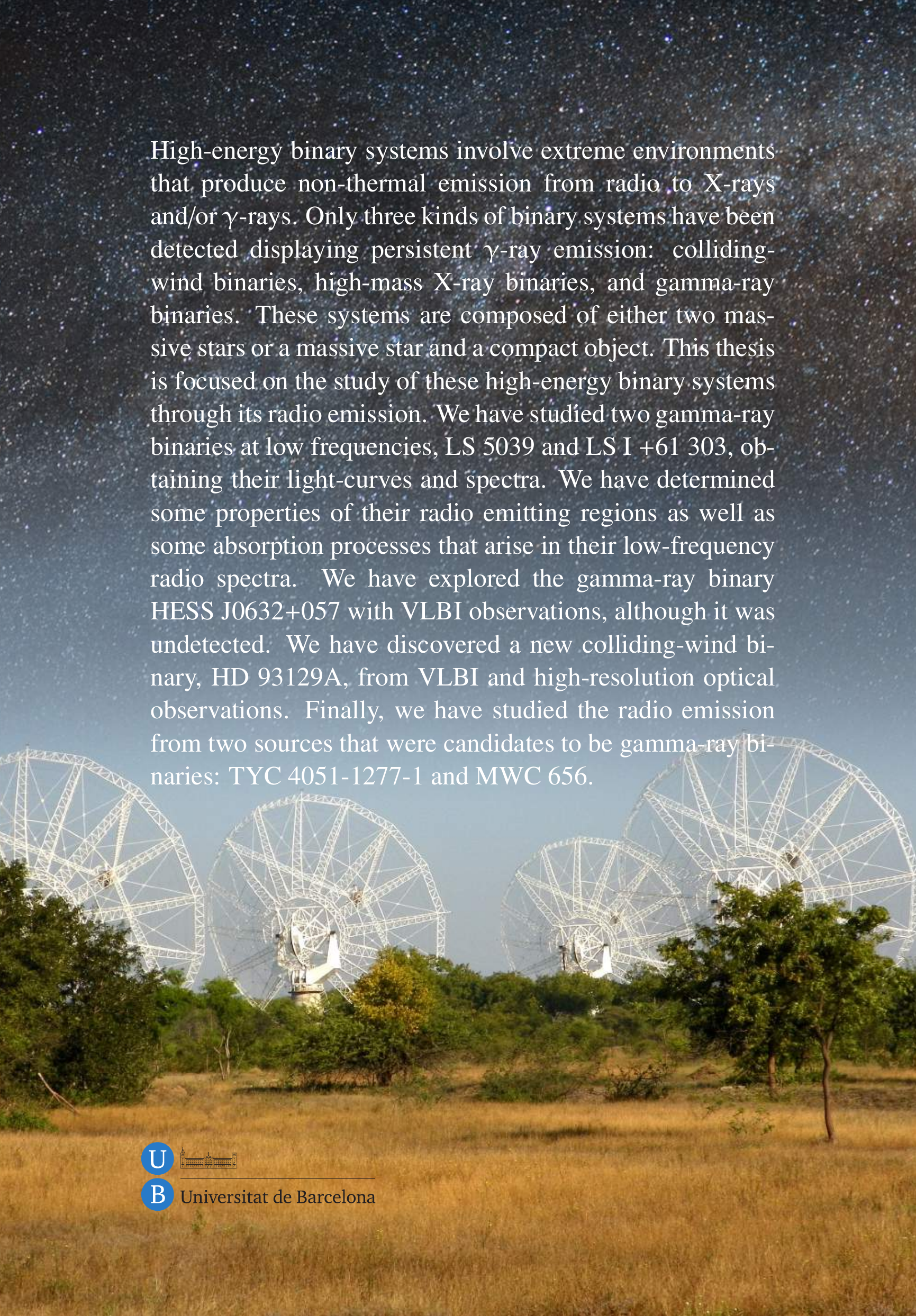}

\end{document}